\documentclass[preprint,10pt]{elsarticle}
\usepackage{lineno,hyperref}
\modulolinenumbers[5]

\journal{}

\biboptions{sort&compress}
\usepackage[margin=2.5cm]{geometry}
\usepackage{bm}
\usepackage[cal=boondoxo]{mathalfa}
\usepackage{amssymb}
\usepackage{amsmath}
\usepackage{amsthm}
\usepackage{mathdefs}
\usepackage{amsfonts}
\usepackage{amsmath}
\usepackage{mathtools}

  {
      \theoremstyle{plain}
      
  }
\DeclareMathAlphabet{\mathcal}{OMS}{cmsy}{m}{n}
\usepackage{algorithm}
\usepackage{algpseudocode}
\usepackage{color}

\usepackage{listings}
\usepackage{xcolor}

\definecolor{codegreen}{rgb}{0,0.6,0}
\definecolor{codegray}{rgb}{0.5,0.5,0.5}
\definecolor{codepurple}{rgb}{0.58,0,0.82}
\definecolor{backcolour}{rgb}{0.95,0.95,0.92}

\lstdefinestyle{mystyle}{
    backgroundcolor=\color{backcolour},   
    commentstyle=\color{codegreen},
    keywordstyle=\color{magenta},
    numberstyle=\tiny\color{codegray},
    stringstyle=\color{codepurple},
    basicstyle=\ttfamily\footnotesize,
    breakatwhitespace=false,         
    breaklines=true,                 
    captionpos=b,                    
    keepspaces=true,                 
    numbers=left,                    
    numbersep=5pt,                  
    showspaces=false,                
    showstringspaces=false,
    showtabs=false,                  
    tabsize=2
}

\begin{document}

\begin{frontmatter}

%\title{Multidimensional Quadrature for
%Nonlinear Inverse Problems: A Variational Bayesian Inference Approach with A Gaussian Mixture Model}
\title{Variational Inference for Nonlinear Inverse Problems via\\ Neural Net Kernels: Comparison to Bayesian Neural Networks,\\ Application to Topology Optimization}

\author[affil1]{Vahid Keshavarzzadeh\corref{mycorrespondingauthor}}
\cortext[mycorrespondingauthor]{Corresponding author}
\ead{vkeshava@sci.utah.edu}

\author[affil1,affil2]{Robert M. Kirby}
\ead{kirby@sci.utah.edu}

\author[affil1,affil3]{Akil Narayan}
\ead{akil@sci.utah.edu}

\address[affil1]{Scientific Computing and Imaging Institute, University of Utah}
\address[affil2]{School of Computing, University of Utah}
\address[affil3]{Department of Mathematics, University of Utah}

%A typical Bayesian inference provides a mean behavior and is incapable of revealing the true distribution of underlying parameters mainly due to the standard likelihood evaluation via summing/averaging over all available data points and incorporating arbitrarily specified noise.
\begin{abstract}
Inverse problems and, in particular, inferring unknown or latent parameters from data are ubiquitous in engineering simulations. A predominant viewpoint in identifying unknown parameters is Bayesian inference where both prior information about the parameters and the information from the observations via likelihood evaluations are incorporated into the inference process. In this paper, we adopt a similar viewpoint with a slightly different numerical procedure from standard inference approaches to provide insight about the localized behavior of unknown underlying parameters. We present a variational inference approach which mainly incorporates the observation data in a point-wise manner, i.e. we invert a limited number of observation data leveraging the gradient information of the forward map with respect to parameters, and find true individual samples of the latent parameters when the forward map is noise-free and one-to-one. For statistical calculations (as the ultimate goal in simulations), a large number of samples are generated from a trained neural network which serves as a transport map from the prior to posterior latent parameters. Our neural network machinery, developed as part of the inference framework and referred to as Neural Net Kernels (NNK), is based on hierarchical (deep) kernels which provide greater flexibility for training compared to standard neural networks. We showcase the effectiveness of our inference procedure in identifying bimodal and irregular distributions compared to a number of approaches including a maximum a posteriori probability (MAP)-based approach, Markov Chain Monte Carlo sampling with both Metropolis-Hastings and Hamiltonian Monte Carlo algorithms, and a Bayesian neural network approach, namely the widely-known Bayes by Backprop algorithm via a pedagogical example. We further apply our inference procedure to two and three dimensional topology optimization problems where we identify the latent parameters in the random field elastic modulus, modeled as a Karhunen-Lo\'{e}ve expansion, considering the high dimensional design iterate-displacement pair as training data. As future research, we discuss the application of this inference approach in identifying constitutive models in nonlinear elasticity and development of fast linear algebraic solvers for large scale similar, i.e. finite element-based inverse problems calculations. 
\end{abstract}

\begin{keyword}
Bayesian Inference, Variational Inference, Neural Network-based Kernels, Bayesian Neural Networks
%\texttt{elsarticle.cls}\sep \LaTeX\sep Elsevier \sep template
%\MSC[2010] 00-01\sep  99-00
\end{keyword}

\end{frontmatter}

\section{Introduction}

\subsection{Background} The problem of inferring parameters of a mathematical model from observations has significant practical interest. This particular inverse problem has been tackled from two major distinct perspectives: a) the deterministic view which tries to find the optimal parameters by minimizing the prediction misfit and b) the probabilistic view which uses prior knowledge about the parameters' probability distribution and updates this knowledge via an inference procedure known as Bayesian inference to find the posterior distribution. The latter view accommodates: 1) the inherent uncertainty in the observations also known as data noise as well as model parameters and 2) the potential ill-posedness of the inverse problems.   In practice within the Bayesian framework, one is typically interested in computing the statistical moments of the model's quantity of interest with respect to the posterior distribution and drawing samples from the posterior distribution e.g. in reliability estimation. Both of these tasks are onerous since the posterior distribution often has a non-obvious form. Markov Chain Monte Carlo (MCMC) sampling approaches~\cite{Gelman13, Liu04} have been devised to facilitate the sampling from the posterior distribution. MCMC sampling which entails a random walk however comes with the price of evaluating a large number of samples to find a rather small set of accepted samples. Hamiltonian Monte Carlo (HMC) \cite{Neal11}, in particular tackles this issue by traversing the important regions of posterior probability in a deterministic fashion via a Hamiltonian dynamics simulation informed by gradient of the posterior. The computational cost is still notable especially in physics simulations where each evaluation entails an expensive large-scale computation. As another class of approaches, variational Bayesian inference~\cite{Bishop13} formulates the Bayesian inference as a deterministic optimization problem on the Kullback–Leibler (KL) divergence between the target distribution and a simpler parameterized distribution. In variational inference the KL divergence is transformed to the evidence lower bound form which consists of the prior-dependent complexity cost and the data-dependent likelihood cost~\cite{Blundell15}. Variational inference plays a significant role in machine learning as these optimization approaches are scalable to large datasets in contrast to slower MCMC techniques. 

Aside from the computational complexity, a practical issue with these inference approaches is their inability to predict the true distribution of the underlying parameter. The MCMC approaches at best provide samples from the posterior probability which is formulated in a Bayesian setting. Often, and as will be demonstrated, the Bayesian posterior distributions are mainly suitable for describing the average behavior of the unknown underlying parameter and subsequently the model output. Variational inference can suffer from the same issue if the data-dependent (likelihood) part is computed and optimized considering all data points (observations). Indeed, the formulation of the inference procedure and the manner according which the data is assimilated in the process (through the likelihood function) has the most significant impact on the identification of the true underlying parameter. It should also be emphasized that an inverse problem is generally construed to be ill-posed as the inverse map may yield many possible solutions. An in-depth discussion of this issue from a measure-theoretic perspective is provided in~\cite{Breidt11}. When the forward map is many-to-one, the identification of true underlying uncertainty is impossible. We, however, specifically aim to tackle this problem and develop a practical computational framework which is shown to be more successful in identifying challenging distributions in comparison with standard existing approaches. As a result, our approach outperforms generic Bayesian neural nets as widely-known probabilistic surrogates in terms of predicting the statistical distribution of model outputs.   

\begin{figure}[!h]
\centering
\includegraphics[width=1.7in]{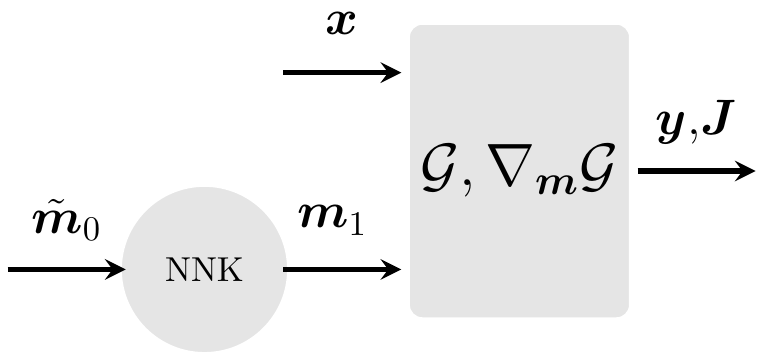} \hspace{0.5cm}
\includegraphics[width=2.85in]{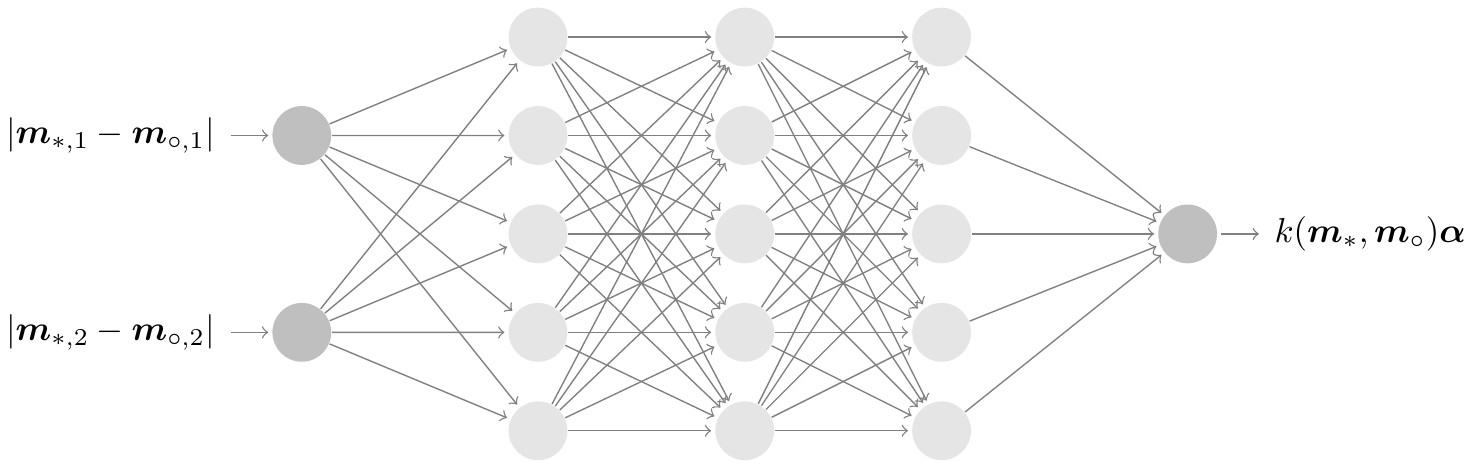}\\
\includegraphics[width=2.5in]{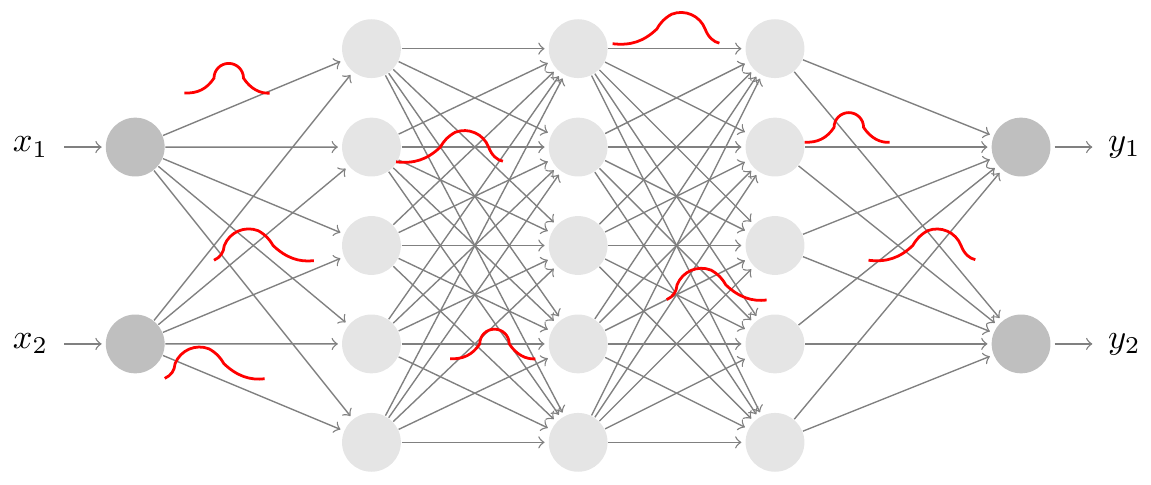}
\caption{\small{Top row: schematic representation of our inference procedure where $\bm J=[\partial \mathcal{G}/\partial \bm m_1] \in \mathbb{R}^{n \times k}$ is the Jacobian, $\tilde{\bm m}_0$ are the permuted prior samples and $\bm m_1$ are the posterior samples of the uncertain parameter (left), the neural net kernel (NNK) with optimizable $\bm \alpha \in \mathbb{R}^{n_{\ast} \times n}$ (right). In this plot $\bm m_{\ast,i}$ denotes the $i$th dimension of $\bm m_{\ast}$. Bottom row: a standard Bayesian neural network (BNN) with uncertain weights. In both cases (top and bottom rows), a probabilistic surrogate between $\bm x$ and $\bm y$ is trained; however, as opposed to the simulation-based procedure on top, a black-box BNN does not provide any insight about the underlying uncertain parameters.}}\label{schematic}
\end{figure}

Our inference procedure, depicted schematically in Figure~\ref{schematic}, is summarized in three main steps which are explained in details in Section~\ref{Sec3}: 
\begin{itemize}
\item Sample-wise optimization of the forward model: This step stems from our variational inference formulation which mainly puts emphasis on the data-dependent term and is explained in Section~\ref{Sec2}. The result of this step yields a small number of true samples of the underlying parameter used for training a neural net.
\item Permutation of the prior samples: The randomly distributed prior samples are permuted according to their distance to the optimized samples of the underlying parameter to facilitate the neural network training.  
\item Training of a specialized deterministic neural network for generating a large number of posterior samples: The surrogate, referred to as NNK, is developed based on hierarchical kernels which encode the relationship between samples as the neural net function of their distances. The prediction is in the form of  $\bm m_{\ast_1} = k(\bm m_{\ast_0}, \bm m_{\circ}) \bm \alpha$ where $ k(\bm m_{\ast_0}, \bm m_{\circ})$ is a neural network-based kernel, $\bm m_{\ast_0}$ are the prior samples (denoted by $0$ subscript), $\bm m_{\ast_1}$ (denoted by $1$ subscript) are the posterior samples and $\bm \alpha$ are optimizable coefficients which enhance the approximation power of the deterministic surrogate. $\bm m_{\circ}$ in this work are chosen as randomly distributed points following the same distribution as prior samples $\bm m_{\ast_1}$. 
\end{itemize}

A notable aspect of our approach is the inference on the space of samples rather than fictitious analytical distributions. Our approach can be practically used to infer about the underlying parameters when the parametric form of the forward model and its sensitivity are known. In other words, in cases when a computer model such as a finite element model is considered  a viable representation of reality, the real observations such as displacements acquired via accurate sensors could be practically used in conjunction with our approach to infer about the distribution of unknown underlying parameters. In what follows, we briefly mention a number of related works in the general context of Bayesian inference as well as a more detailed description of (and comparison to) three existing research papers in the literature which address similar issues in this paper.

\subsection{Related Work} Several researchers have developed computational approaches for Bayesian inference and posterior approximation. The literature is vast on these topics. As examples, the authors in~\cite{Bui-Thanh13, CHEN2017147, Schillings2020, Schillings_2013, Ganther2016, CHEN2016470, ELMOSELHY20127815, Spantini2018, pmlr-v37-rezende15, LIU16, NEURIPS2018_fdaa09fc, CHEN2019NIPS} have addressed the problem of posterior characterization and statistical computation with respect to posterior by deterministic quadrature approaches, Gaussian posterior approximation and transport maps. In a similar vein, authors in~\cite{stuart2004, Girolami2011, Martin2012, Bui-Thanh14, LAN201681, BESKOS2017327, Cui2015, Constantine2016, Oliver2017, Wang2018, Bardsley2020} have tackled the issue of efficient MCMC sampling by developing methods on Langevin and Hamiltonian MCMC, dimensionality reduction MCMC and randomized/optimized MCMC, etc.  The issue of large-scale forward computation, as an indispensable part of inverse problems is addressed in~\cite{MARZOUK2007560, MARZOUK2009, Schwab_2012, CHEN2017147, Farcas20, WANG200515, lieberman2009, Nguyen10, Lassila13, Cui2015, CHEN2016470, CHEN2020NIPS, Dodwell2019, Teckentrup2015, Scheichl2017, Peherstorfer2018, Farcas20} via polynomial approximation, model reduction (greedy reduced basis) and multifidelity/multilevel modeling. It is also worth mentioning that part of our approach which relies on optimization of forward model can benefit from mutifidelity approximation to accelerate the forward model and adjoint computation
~\cite{Akil_MF, VK_IJNME, VK_stress, VK_comp, Panos21}. Our work in this paper addresses similar issues which were considered in the following three papers: \emph{Variation inference with normalizing flows}~\cite{pmlr-v37-rezende15}: In this work the authors use neural networks to approximate irregular posterior distributions. A part of our approach adopts the same idea, i.e. we take advantage of the approximation power of neural networks to train a transport map however the aforementioned work considers analytical distributions and performs the identification in the space of analytical distributions rather than samples which are usually observed in the real setting. \emph{Stein variational gradient descent}~\cite{LiuStein2016}: This work discusses a variational inference approach with a distinct feature; the authors considers Stein identity as opposed to KL divergence to formulate the variational inference procedure. The derivation of the method is based on a new theory which connects the derivative of KL divergence with Stein's identity. It is shown (via a numerical example) that the approach is capable of approximating a bimodal distribution in one dimension however as mentioned above these approaches are mainly considered in the setting of analytical distributions. It appears that utilization of these approaches in conjunction with standard likelihood evaluation from samples will result in the same averaged behavior for underlying parameter. \emph{Bayesian inference with optimal maps}~\cite{ELMOSELHY20127815}: Our development in this paper is mainly motivated by this paper in terms of training a transport map. The foundation of their approach is based on finding transport map such that the inverted map applied to the observation data matches the prior distribution. The transport map is also considered in the form of hierarchical polynomials which is similar to the transport map in this paper. A number of practical engineering examples are provided which demonstrates the success of the approach on identification of relatively straight-forward uncertain parameters. This work also follows standard form of likelihood function which may hamper its successful application for non-trivial distributions. 

\subsection{Contributions of This Paper}
Our contribution in this work is twofold: 1) We present a numerical procedure which explicitly takes advantage of gradient information in simulations. By explicit, we mean the ability to directly optimize the forward model for the underlying uncertain parameters as opposed to approaches such as HMC where the gradient information is implicitly used in an MCMC fashion for evolving a dynamic process.  As a result, we demonstrate successful identification of underlying parameter when the forward map is noise-free and one-to-one. For many-to-one maps where the identification of latent parameters are less accurate, we demonstrate that our procedure yield very accurate system outputs as the optimization always minimizes the discrepancy to system output.  2) The approach in this paper can be a practical replacement to back-box probabilistic surrogates such as Bayesian neural networks (BNN) when the parametric forward model and its sensitivity are known. A  BNN approach for uncertainty quantification in data-driven materials modeling is presented in~\cite{OLIVIER2021114079}. BNNs can only predict the output and provides no insight on the underlying parameter. Our simulation-based approach is inherently more informative than a BNN as the knowledge of forward model  guides the identification of unknown parameters.  It is shown that our procedure outperforms a standard BNN on model prediction. Our approach can also be scaled to large problems as the optimization is done using established linear algebra procedures. On the other hand, for large input-output data, the utilization of BNN requires dimensionality reduction. Our approach can handle those large scale data with no further modifications. In fact the dimensionality reduction is effectively performed in the optimization stage and subsequently the computations are performed in the possibly low-dimensional latent parameter space.

\textbf{Inherent challenges in inverse problems and our approach:} \emph{Consideration of noise}: As mentioned at the beginning of this section, Bayesian inference approaches naturally  accommodate noise in their formulation (in the form of a weighted $\ell_2$ norm in the likelihood function); however, from a practical perspective the amount of noise is unknown. Also the widely-used form of additive noise is an assumption which may not be realistic. Our sample-wise inference approach is less amenable to noise consideration however we show how to practically model noise and perform statistical sample-wise optimization in the numerical examples. \emph{Non-unique solution to the inverse problem}: This inherent challenge (also briefly mentioned) limits the ability to find true underlying uncertain parameters. However, an advantage of our approach which matches the model output in a sample-wise manner is more accurate prediction of model output compared to standard inference procedures irrespective of the correct identification of the latent parameters. \emph{Highly nonlinear maps}: Our approach adopts a transport map which transforms randomly distributed prior samples (i.e. uninformed samples) to particular posterior samples. These generic transformations are often highly nonlinear. In one of our examples, we consider the shape of $U$ as the distribution of posterior samples in addition to bimodal (and trimodal) distributions. A viable tool for approximating such generic maps is neural network which we take advantage of in this work. To facilitate the training of these highly nonlinear maps, we permute the prior samples according to the distribution of the limited number of posterior samples (which we find in the first step of our procedure) and demonstrate appreciable improvement in our numerical examples.

This paper is organized as follows. In Section~\ref{Sec2} we briefly discuss the mathematical setting of Bayesian and variational Bayesian inference. In particular, we discuss the variational inference formulation which we consider for optimization of the forward model in this paper. In Section~\ref{Sec3} we discuss the algorithmic details of the three steps in our inference procedure, namely the optimization of forward model, permutation of optimized samples and training via neural net kernels, NNK. Section~\ref{Sec4} presents the numerical examples on an illustrative problem including comparisons to MAP-based approaches, MCMC  sampling and Bayesian neural network, in addition to inference on random field elastic modulus in two and three dimensional topology optimization problems. Finally, in Section~\ref{Sec5} we discuss the concluding remarks including plans/ideas for future research.

\section{Bayesian Inference}\label{Sec2}

\subsection{Notation}
%$\bm X$ defined on a physical domain $D$ which is an open and bounded subset of $\mathbb{R}^l$ ($l=1,2,3$) with Lipschitz boundary $\partial D$. 

We use bold characters to denote matrices, vectors, and multivariate quantities. For example, $\bm x \in \mathbb{R}^l$ indicates a vector of variables in the domain of a multivariate function. We will write $\bm{x}^{(1)}, \ldots \bm{x}^{(n)}$ to denote a collection of points in $\mathbb{R}^l$, and each point $\bm{x}$ has components $\bm{x} = \left( x_1, \ldots, x_l\right)$. Given a finite-dimensional Euclidean set $M \subset \mathbb{R}^d$ for some ${d} \in \mathbb{N}$, the uncertain parameter which we seek to find its posterior measure is denoted by $\bm m \in  M$. The model output $\bm y$ is also defined on a finite-dimensional set $Y \subset \mathbb{R}^n$ for some $n \in \mathbb{N}$. We denote the parameter to observable map by $\mathcal{G}: X \times M \rightarrow Y$ which maps the parameter $\bm m \in M$ to the observation $\bm y \in Y$ i.e. $\bm y = \mathcal{G}(\bm x, \bm m)$. The map involves another set of input $\bm x \in X \subset \mathbb{R}^l$ which is specifically considered in our computational setting to mimic the generic setting of probabilistic surrogates between $\bm x$ and $\bm y$ such as BNN.  The most expensive part of the computation is the evaluation of $\mathcal{G}$ and its gradient with respect to $\bm m$ which involves several linear solves associated with the discretized form of a partial differential equation (PDE).

\subsection{Bayesian Inference}\label{Sec2_1}
 %_{\bm m \in X}
% \begin{align*}
% \bm m \equiv \mathop{\textrm{argmin}}_{\bm m \in \bm X} \frac{1}{2} \| \bm y - \mathcal{G}(\bm m)\|^2_{\bm Y}
% \end{align*}
The inverse problem has significant practical interest as this problem is inherent in the engineering design. It can be cast as finding a suitable parameter $\bm m$ to a mathematical model denoted by $\mathcal{G}$ given output (or observation) $\bm y$ i.e. solving for $\bm m \in M$, given $\bm y \in Y$ from the following equation
 \begin{align}\label{eq_inverse1}
 \bm y= \mathcal{G}(\bm m).
 \end{align}
In the above equation, it is noted that we remove the input parameter $\bm x$ as we only discuss the standard Bayesian setting at this juncture. We introduce the input parameter $\bm x$ later in Subsection~\ref{Sec2_2}. This inverse problem is usually challenging to solve as it maybe ill-posed or the solution might be sensitive to the observation data $\bm y$. A deterministic approach to solve this problem is to replace it with a least-square problem using an appropriate norm on $Y$ (typically $\ell^2$ norm) i.e. $\bm m \equiv \mathop{\textrm{argmin}} \frac{1}{2} \| \bm y - \mathcal{G}( \bm m)\|^2_{Y}$. Solving the above optimization problem might be still challenging as it may yield non-unique solutions. This issue can be circumvented by solving a regularized problem 
\begin{align*}
\bm m \equiv \mathop{\textrm{argmin}}_{\bm m \in  M} \left(\frac{1}{2} \| \bm y - \mathcal{G}( \bm m)\|^2_{Y} + \frac{1}{2} \|\bm m - \bm m_0 \|^2_{M} \right).
\end{align*}
In this treatment to the problem, the choice of norms for minimization as well as the point $\bm m_0$ are somewhat arbitrary. On the other hand the richer or more comprehensive approach is to solve this problem in a statistical fashion i.e. via a Bayesian approach. Bayesian approach finds a probability measure on $M$ which describes the relative probability of different states $\bm m$ given observation $\bm y$. The probabilistic framework is comprehensive as it considers that the observation is subject to noise. In this probabilistic context, the model equation~\eqref{eq_inverse1} often takes the form
\begin{align}\label{eq_inverse1_prob}
\bm y= \mathcal{G}(\bm m) + \eta
\end{align}  
where $\eta$, a mean zero random variable whose statistical properties might be known but its actual value is unknown, is referred to as \emph{obsevational noise}. 

In the Bayesian context, the prior belief about $\bm m$ is described in terms of a probability measure $p_0(\bm m)$ and the goal is to find the posterior probability measure $p_1(\bm m)$. The probability of $\bm y$ given $\bm m$ can be expressed in terms of the probability density function of noise:
\begin{align*}
p(\bm y|\bm m)= p(\bm y-\mathcal{G}(\bm m)).
\end{align*} 
Using the Bayes formula, the probability density function of posterior $p_1(\bm m)$ is obtained via 
\begin{align}\label{BayesPost}
p_1(\bm m) = \displaystyle \frac{p(\bm y - \mathcal{G}( \bm m))p_0(\bm m)}{\displaystyle \int_{M} p(\bm y-\mathcal{G}( \bm m))p_0(\bm m)d \bm m}
\end{align} 
which can be written as a proportionality with its constant only dependent on $\bm y$, i.e. $p_1(\bm m) \propto p(\bm y-\mathcal{G}( \bm m))p_0(\bm m)$. This relation between the prior and posterior measures is also equivalent to the Radon-Nikodyn derivative, $p_1(\bm m)/p_0(\bm m) \propto \rho(\bm y-\mathcal{G}(\bm m))$ cf. Theorem 6.2 in~\cite{stuart_2010}. 

The relation~\eqref{BayesPost} characterizes the Bayesian posterior $p_1(\bm m)$ in a generic setting. In the case where $ M,  Y$ are both finite-dimensional, the noise $\eta$ is additive and Gaussian defined by $\eta=\mathcal{N}(0,\bm \Gamma_{noise})$, and the prior measure is Gaussian defined by $p_0 = \mathcal{N}(\bm m_0, \bm \Sigma_0)$,  the posterior distribution is given by
\begin{align}\label{bayesian_Gauss_post}
p_1(\bm m) \propto  \exp \Big( -\frac{1}{2}\|\bm y - \mathcal{G}( \bm m) \|^2_{\bm \Gamma_{noise}} - \frac{1}{2}\|\bm m - \bm m_0 \|^2_{\bm \Sigma_0} \Big) 
\end{align} 
where $\displaystyle \| \bm a \|^2_{\bm A}$ is the notation for the $\bm A$ weighted norm, i.e.  $ \| \bm a \|^2_{\bm A} =\bm a^T \bm A^{-1} \bm a$ and $\bm a^T$ is the transpose of $\bm a$.

In general, finding the posterior distribution in high dimensions is difficult. A suitable measure in such situations which provide useful information about the probabilistic content of the posterior is \emph{a maximum a posteriori} or MAP estimator. A MAP estimator is a point that maximizes posterior PDF  $p_1(\bm m)$. Given Equation~\eqref{bayesian_Gauss_post}, the MAP estimator is expressed by $\bm m_1 = \mathop{argmin}_{\bm m \in \mathbb{R}^k} ~\mathcal{P}(\bm m) $ where 
\begin{align}\label{opt_main}
\mathcal{P}(\bm m) =  \frac{1}{2}\|\bm y - \mathcal{G}(\bm m) \|^2_{\bm \Gamma_{noise}} + \frac{1}{2}\|\bm m - \bm m_0 \|^2_{\bm \Sigma_0}.
\end{align}

Considering Equation~\ref{bayesian_Gauss_post}, the posterior probability is not always Gaussian; it is only Gaussian when the map $\mathcal{G}$ is linear with the form of $\mathcal{G}(\bm m )= \bm A \bm m$ where $\bm A \in \mathbb{R}^{n \times d}$ and $\bm m \in \mathbb{R}^{d}$~\cite{stuart_2010}. In such scenarios, the following theorem from~\cite{stuart_2010}  provides analytical relations for the mean of the posterior (MAP point) and the covariance matrix. The derivations are based on completing the square process. For more details see the proofs of Theorem 6.20 and Lemma 6.21 in~\cite{stuart_2010}.

\begin{theorem}
Suppose the map $\mathcal{G}$ is linear with the form $\mathcal{G}(\bm m )= \bm A \bm m$ and the equation $\bm y = \bm A \bm m$ has unique solutions for every $\bm y \in \mathbb{R}^n$. Also further suppose that  $\eta=\mathcal{N}(0,\bm \Gamma_{noise})$, $p_0(\bm m) = \mathcal{N}(\bm m_0, \bm \Sigma_0)$ and that $\Gamma_{noise}$ and $\Sigma_0$ are invertible, then the posterior is Gaussian $p_1(\bm m) = \mathcal{N}(\bm m,\bm \Sigma)$ with 
\begin{equation}
\begin{array}{l l l}
\bm m &=& \bm m_0 + \bm \Sigma_0 \bm A^T (\bm \Gamma_{noise} + \bm A^T \bm \Sigma_0 \bm A)^{-1} (\bm y - \bm A \bm m_0)\\
\\
\bm  \Sigma &=& \bm \Sigma_0 - \bm \Sigma_0 \bm A^T (\bm \Gamma_{noise} + \bm A^T \bm \Sigma_0 \bm A)^{-1} (\bm A \bm \Sigma_0).
\end{array}
\end{equation}
\end{theorem}

In many practical situations, however, the map $\mathcal{G}$ is nonlinear or the map takes additional inputs (independent of $\bm m$) which we consider in this paper. In those cases one needs to solve the optimization problem~\eqref{opt_main} to find the MAP point $\bm m$. Subsequently assuming the posterior measure is linearized around the MAP, the local curvature of $\mathcal{P}$ at the MAP point i.e. the Hessian of $\mathcal{P}$ at the MAP denoted by $\bm H(\bm m_1)$ gives the inverse of the covariance operator~\cite{stuart_2010,CHEN2017147}
\begin{equation}\label{eq_basian_post}
\begin{array}{l l }
 \bm H(\bm m_1) =\displaystyle \frac{ \partial^2 \mathcal{P}(\bm m) }{\partial \bm m^2} = \bm H_{model} + \bm \Sigma_0^{-1},&  \bm \Sigma_1 \equiv \bm H^{-1}(\bm m_1)
 \end{array}
\end{equation}
where $\bm H_{model}$ is the Hessian associated with the first term which involves the forward model, i.e. $\bm H_{model} = 0.5( {\partial^2  \|\bm y - \mathcal{G}(\bm m) \|^2_{\Gamma_{noise}} }/{\partial \bm m^2} )$. In practice, the second derivative of $\mathcal{P}$ can be obtained from the last iteration of the unconstrained minimization of $\mathcal{P}$ using the BFGS method~\cite{Fletcher87}.  As part of numerical experiments, we compare our approach in this paper with the MAP-based approach described above in Section~\ref{Sec4}.

\subsection{Variational Bayesian Inference}\label{Sec2_2}
The goal in inference procedures and in general probabilistic models, is the evaluation of posterior probability given data $p_1(\bm m) \equiv p(\bm m | \bm D)$ where $\bm D \equiv \bm y$ and further computing the expectation of a function $f(\bm m)$ with respect to this distribution, i.e.  
\begin{equation*}
\mathbb{E}(f | \bm D) = \int_{\mathbb{R}^d} f(\bm m) p(\bm m | \bm D) d\bm m.
\end{equation*}
In variational inference, the main idea is to approximate the posterior distribution $p(\bm m | \bm D)$ with a variational distribution $q(\bm m)$ which admits a simple parameterization. A typical statistical quantity for optimization is the Kullback–Leibler (KL) divergence between the approximate and target density, i.e.
\begin{equation}
D_{KL}(q || p) \coloneqq - \int_{\mathbb{R}^d} q(\bm m) \log \frac{p(\bm m | \bm D)}{q(\bm m)} d\bm m. 
\end{equation}
It is \emph{supposed} that the direct optimization of KL divergence which involves the conditional probability $p(\bm m| \bm D)$ is difficult and the optimization with respect to joint probability $p(\bm m, \bm D)$ is significantly easier. To be precise, denoting the prior distribution by $p(\bm m) \equiv p_0(\bm m)$, the conditional probability $p(\bm m | \bm D)$ is 
\begin{equation}
 p(\bm m | \bm D) = \frac{p(\bm D | \bm m) p(\bm m)} {p(\bm D)}  =  \frac{p(\bm D | \bm m) p(\bm m)} {\int_{\mathbb{R}^d} p(\bm D, \bm m) d\bm m}
\end{equation} 
whose computation involves the marginalization over latent variable $\bm m$ (in the denominator). For high dimensional latent variables accurate computation of $\int_{\mathbb{R}^k} p(\bm D, \bm m) d\bm m$ is challenging. Accordingly, it is common to optimize a similar quantity, evidence lower bound (ELBO) involving the joint probability between latent variable and data. For any choice of $q(\bm m)$ the following decomposition holds (by simple algebraic manipulations): $\log~p(\bm D) = \mathcal{L}(q) + D_{KL}(q || p)$ where 
\begin{equation}
 \mathcal{L}(q) = \int_{\mathbb{R}^d} q(\bm m) \log \frac{p(\bm m , \bm D)}{q(\bm m)} d\bm m
\end{equation} 
is the lower bound for the evidence $\log~p(\bm D)$, ELBO. It is apparent that $\mathcal{L}(q) \leq \log~p(\bm D)$ since $D_{KL}(q || p) \geq 0$ with equality when $p(\bm m | \bm D) = q(\bm m)$. Maximizing $\mathcal{L}(q)$ is equivalent to minimizing KL divergence as $\log~p(\bm D)$ is fixed with respect to approximated distribution $q(\bm m)$. Using the relation $p(\bm m, \bm D) = p(\bm D | \bm m) p(\bm m)$, the negative of ELBO which should be minimized (it is a positive cost function), is expressed as
\begin{equation}\label{neg_elbo}
-\mathcal{L}(q)=\mathbb{E}_q(\log q(\bm m) - \log p(\bm m)) - \mathbb{E}_q(\log p(\bm D|\bm m)),
\end{equation} 
where $\mathbb{E}_q$ is the expectation with respect to the variational distribution, the first term, denoted by $C_1$ is the complexity cost (the cost of discrepancy between the variational posterior and the prior) and the second term, denoted by $C_2$ is the likelihood cost.

 Equation~\ref{neg_elbo} is the heart of variational Bayesian inference and training of a Bayesian neural net. A typical approach to minimize the cost function given by Equation~\eqref{neg_elbo} is (stochastic) gradient descent. In reality, the prior information is often insignificant and the most important part of the identification is the one associated with the observation data. As an example, in~\cite{Blundell15} the authors consider an adaptive weight on the complexity cost within the total cost $C = w_i C_1 + C_2$, namely $w_i = 2^{(M-i)}/2^M-1$ where $i$ is the mini-batch index and $M$ is the total number of mini-batches, which decreases as more mini-batches of data are observed in the training process. 
 
 In this paper, we follow the same logic and consider less emphasis on the complexity cost. However, the significant difference between our approach and the usual Bayesian inference procedures is the computation of likelihood. Neglecting the constant normalizing factors in multivariate Gaussian distributions, the standard way of computing log likelihood term for an instance of the latent parameter $\bm m^{(i)}$ is 
 \begin{equation}\label{log_lkl}
-\log p(\bm D | \bm m^{(i)})=\sum_{j=1}^{n_{data}}  \left(\bm y^{(j)}-\mathcal{G}(\bm x^{(j)},\bm m^{(i)})\right)^T \Gamma^{-1}_{noise} \left(\bm y^{(j)}-\mathcal{G}(\bm x^{(j)},\bm m^{(i)})\right)
\end{equation}
where $n_{data}$ is the number of observations. In the above equation, we include another set of inputs $\bm x \in \mathbb{R}^{l}$ which has one-to-one correspondence to observation $\bm y \in \mathbb{R}^n$. This is the setting that we consider in this paper for building a probabilistic surrogate between $\bm x$ and $\bm y$, i.e. we are given the data points as pairs  $\bm D \equiv \{\bm x^{(i)},\bm y^{(i)}\}_{i=1}^{n_{data}}$ and the goal is to build a probabilistic map between $\bm x$ and $\bm y$.  The expression~\eqref{log_lkl} allows for incorporation of noise which is discussed in Equation~\ref{eq_inverse1_prob}; however, as will be shown in all numerical examples the summation over all data points precludes the identification of more localized latent parameter distributions. Following loosely the ELBO formulation, we propose an alternative optimization problem to find the distribution of latent parameters $\bm m$:
\begin{equation}\label{our_VI}
  \begin{array}{r l l l}
    \displaystyle \mathop{\min}_{\bm m, \sigma} &  C_1&=&  \mathbb{E}_q(\log q(\bm m) - \log p(\bm m)) \\
    \\
    \text{subject to} & C_{2,i} &=&\bm y^{(i)}-\mathcal{G}(\bm x^{(i)},\bm m^{(i)}) = 0, \quad 1\leq i \leq n_{data} 
   \end{array}
\end{equation}
where we note that all variables are indexed by $i$ in the equality constraint. Some remarks are in order: 1) Based on Gibbs inequality the objective function $C_1$ is always positive, i.e. it is a KL distance. The objective function depends on the variational distribution which we parameterize as a Gaussian mixture model (GMM) with centers $\bm m^{(i)}$ and covariances that depend on a scalar parameter $\sigma$, i.e. $\bm \Sigma^{(i)}=\sigma \bm I$ where $\bm I \in \mathbb{R}^d$ is an identity matrix. We assume uniform weights for the GMM. 
2) The GMM centers are found from solving the equality constraints which is done in a sample-wise manner and described in the next section. The most important part of  our simulation-based inference is indeed solving these nonlinear constraints for $\bm m^{(i)}$. 3) Once the optimal centers are found, they could be plugged into the objective function to minimize the complexity cost with respect to the scalar $\sigma$. The result $(\bm m^{\ast}, \sigma^{\ast})$ will be a local minima which satisfies the optimality conditions i.e. $\partial C_1/\partial \sigma = 0$ and $C_{2,i}= 0,~\forall i$. 
   
\begin{remark}
We provide the details for computation of the scalar variable $\sigma$ for the sake of a complete presentation and for occasions when respecting the prior knowledge is crucial. As mentioned several times, the minimization of $C_1$ is often insignificant and could be misleading in cases where prior knowledge is different from reality. To generate large number of samples of posterior, we train a neural net which only uses the training data on the optimal $\bm m^{(i)}$ centers where themselves are informed by/obtained from the forward model. 
\end{remark}

As described, our inference procedure mainly revolves around solving $\mathcal{G}(\bm x^{(i)}, \bm m^{(i)})=\bm y^{(i)}$. This in essence is done in a deterministic manner. To consider the effect of noise $\eta$, we propose solving the noisy version of $C_{2,i}$ i.e. we solve  $\mathcal{G}(\bm x^{(i)}, \bm m^{(i)}) + \eta = \bm y^{(i)}$ for different realizations of $\eta$. The ensemble of solutions for $\bm m^{(i)}$ are then used as a larger (not necessarily richer) set for neural net training. We empirically study the effect of noise on our inference procedure in the first numerical example.

\section{Computational Framework}\label{Sec3}

\subsection{Inversion of the Forward Model}\label{Sec3_1}

The training samples for the transport map is obtained by inverting the forward model or solving the forward model for the latent parameters, i.e. we solve $\mathcal{G}(\bm x^{(i)}, \bm m^{\ast}) =  \bm y^{(i)}$. A common approach, widely-used in training neural nets, is to minimize the mean squared error or solve an unconstrained optimization problem 
\begin{equation}
 \bm m^{\ast} = \mathop{\textrm{argmin}}_{\bm m } \| \mathcal{G}(\bm x, \bm m) - \bm y \|_2^2
\end{equation}
where we omitted the dependence on every sample $i$. In a basic gradient descent approach, the optimization is done by updating the estimates of $\bm m$ with $\bm m \gets \bm m - \beta\Delta_{GD}$ where $\beta$ is a learning rate and $\Delta_{GD}$ is the derivative of loss function which is iteratively computed as $\Delta_{GD} \gets \partial \| \mathcal{G}(\bm x, \bm m) - \bm y \|_2^2/ \partial \bm m$. The popularity of gradient descent is due to its simplicity and its straight-forward application of its stochastic version with parallel processing (i.e. mini-batch computations); however, it is well-known that this approach has linear convergence as opposed to the Newton method which enjoys quadratic convergence. 

In this paper, we use  the Newton-Raphson method to solve nonlinear equations in both cases of solving $\mathcal{G}(\bm x^{(i)}, \bm m^{\ast}) =  \bm y^{(i)}$ and training the NNK in Section~\ref{Sec3_3}. The iterations are similar to gradient descent iterations; the only difference is the way that we compute (Newton) updates (or for rectangular nonlinear systems more commonly referred to as Gauss-Newton) $\Delta_{NR}$ in $\bm m \gets \bm m - \beta \Delta_{NR}$. In a multivariate setting, the Newton updates are obtained from $\Delta_{NR} = \bm J^{\dagger} \bm R$ where $\bm J^{\dagger}$ is the pseudo-inverse of the Jacobian $\bm J = [\partial  \mathcal{G}/\partial \bm m]$ and $\bm R = \mathcal{G}-\bm y$. To take advantage of fast built-in inversion tools, instead of computing the pseudo-inverse, we find the Newton's updates from a square linear system  $\Delta_{NR} = (\bm J^T \bm J + \delta_{Tikh} \bm I)^{-1} \bm J^T \bm R$ where $\bm I$ is an identity matrix and $\delta_{Tikh}$ is a user-specified  Tikhonov regularization parameter similar to the learning rate.

\begin{remark}
In general, it is possible that the NR approach does not converge, i.e. it is possible that it does not achieve a sufficiently small residual $\bm R$. As an example applying the NR approach for solving the forward model with noise, i.e. $\mathcal{G}(\bm x^{(i)}, \bm m^{(i)}) + \eta = \bm y^{(i)}$ may not always yield a small residual cf. Section~\ref{infer_noise}. The convergence of NR can also be slow for particularly highly nonlinear functions. As a high-level remedy for the convergence issue, one could start with a GD approach and perform several GD iterations until a suitable initial guess is found for the NR approach. The suitability of the initial guess however is not well defined for any generic function. In problems that we have solved in this paper, we have only used the NR approach, and we deem the obtained results to be satisfactory.
\end{remark}

%\begin{remark}
%A most significant part of the Newton-Raphson computation is the evaluation of $\bm J$. It should be noted that computing the Jacobian matrix is a more involved process compared to computing the derivative of loss function which involves summing over all data points and results in a vector. In popular machine learning platforms such as Tensorflow, the gradient operator yields derivatives of sum of the output $\bm y$ with respect to input $\bm x$. According to the Tensorflow documentation\footnote{~\href{https://www.tensorflow.org/api_docs/python/tf/gradients}{\textrm{https://www.tensorflow.org/api\_docs/python/tf/gradients}}}, denoting tensors or a list of of tensors with $\bm X$ and $\bm Y$, the operator \texttt{tf.gradients} returns a list of tensors of length equal to the length of $\bm X$ where each tensor is the sum of $d \bm y/d \bm x$ for $\bm y \in \bm Y$ and $\bm x \in \bm X$. According to this documentation, computing the Jacobian is not straight-forward and its proper computation may need several calls to the gradient operator which could be very time consuming. 
%\end{remark}

We provide the details of computation for the Jacobian $\bm J =[\partial \mathcal{G}/\partial \bm m]$ and the derivative of the scalar $\mathcal{P}(\bm m)$ cf. Equation~\eqref{opt_main} in~\ref{comp_J}. Algorithm~\ref{alg:opt_fwd} lists the main steps for optimization of the forward model using the Newton-Raphson approach.  
\begin{algorithm}
  \caption{ Inversion of the forward model}  
\flushleft \textbf{Input:}  $\bm x, \bm y, \beta,\delta_{Tikh},\delta_R$\quad \textbf{Output:}  $\bm m^{\ast}$ 
 \begin{algorithmic}[1]
 \State $\bm R \gets \bm 1 \in \mathbb{R}^{n_{data}n}$ where $n$ is the dimensionality of observation data $\bm y$
 \While{$\|\bm R\| > \delta_R$ }
 \State{$\mathcal{G} \gets \mathcal{G} (\bm x, \bm m),~ \bm J \gets \nabla_{\bm m} \mathcal{G}(\bm x,\bm m)$} 
 \State $\bm R \gets \mathcal{G} - \bm y$
 \State $\Delta_{NR} \gets (\bm J^T \bm J + \delta_{Tikh} \bm I)^{-1} (\bm J^T \bm R)$ \Comment{See~\ref{comp_J}}
 \State{$\bm m \gets \bm m - \beta \Delta_{NR}$} 
\EndWhile
\end{algorithmic}
\label{alg:opt_fwd}
\end{algorithm}

\subsection{Permutation of Prior Samples}\label{Sec3_2}

The transport map as shown in Figure~\ref{schematic} takes randomly distributed prior samples. These unstructured samples are mapped to the posterior samples that are supposed to follow the true distribution of the underlying parameter. This mapping in a general setting is expected to be highly nonlinear and utilization of deep networks alone may not be sufficient for its approximation. 

Performing the optimization explained in the previous subsection yields samples of the latent parameter $\bm m$. To facilitate the training of the neural net map, we permute the randomly distributed prior samples according to the formation of the optimized samples. To this end, we first scale the randomly distributed samples which we always consider to be uniformly distributed i.e. $\bm m_0 \sim U[0,1]^d$ with the information that we gain in the previous step from the range of the list of optimized samples denoted by $\bm m_{opt}$. The list of scaled prior samples are obtained as $\hat{\bm m}_{0} = \min(\bm m_{opt}) \bm m_0 + (\max(\bm m_{opt})-\min(\bm m_{opt}))$. Within the list $\hat{\bm m}_{0}$, we then find the closest samples (with respect to the Euclidean distance) to the optimized samples denoted by $\bm m_{opt}$ in a greedy procedure, i.e. for every sample in the optimized list we find the closest sample in $\hat{\bm m}_{0}$, add it to a new list which includes the permuted scaled random samples and remove it from $\hat{\bm m}_{0}$. The pseudocode associated with this procedure is provided in Algorithm~\ref{alg:perm}.

\begin{algorithm}
  \caption{ Permutation of the optimized samples }  
\flushleft \textbf{Input:}  A list of randomly distributed prior samples $\bm m_0$ and a list of optimized samples $\bm m_{opt}$
\vspace{-0.2cm}
\flushleft \textbf{Output:}  A list of permuted prior samples $\tilde{\bm m}_{0}$
 \begin{algorithmic}[1]
 %\State rand\_nodes\_dynamic $\gets$ rand\_nodes\_scaled
 \State $\hat{\bm m}_{0} \gets \min(\bm m_{opt}) + (\max(\bm m_{opt})-\min(\bm m_{opt})) \bm m_0$
 \State $\tilde{\bm m}_{0} \gets \{\}$
 \For{$i = 1, \ldots, n_{data}$ }
 \State{$\bm r \gets \| \bm m_{opt}^{(i)} - \hat{\bm m}_{0}\|_2$ where e.g. $\bm r \in \mathbb{R}^{n_{data}}$ for $i=1$, and  $\bm r \in \mathbb{R}$ for $i=n_{data}$}
 \State Find index $i_{min}$ where $\hat{\bm m}_{0}^{(i_{min})} = \min(\bm r)$
 %\Else
 \State{$\tilde{\bm m}_{0} \gets \textrm{append}(\tilde{\bm m}_{0}, \hat{\bm m}_{0}^{(i_{min})})$} 
  \State $\hat{\bm m}_{0}^{(i_{min})} \gets \{\}$
\EndFor
\end{algorithmic}
\label{alg:perm}
\end{algorithm}

\subsection{Neural Net Kernel (NNK)}\label{Sec3_3}

Kernel methods such as support vector machine or GP regressors have been extensively used for various machine learning tasks. The predictive power of kernels could be enhanced by combining them with a neural net architecture. The idea of using deep kernels (or many other similar names in the literature) has been gaining more attention in the past several years mainly due to the fact that they enjoy both elegant features of kernels such as describing the relationship between two data points via tangible measures i.e. some form of their distance and approximation power of deep/hierarchical architectures~\cite{Ingo16,pmlr_deepGP,manifold_GP,PANG2019270,lee2018deep,OWHADI201922,GIROLAMI2021}.    

One of the best known kernels is the Gaussian RBF or squared exponential kernel
\begin{equation}
 k (\bm m_{\ast}, \bm m_{\circ}) = \exp(-\|\bm m_{\ast}-\bm m_{\circ}\|_2^2/2\ell^2), \quad \bm m_{\ast},\bm m_{\circ} \in X
\end{equation}
where $X \subset \mathbb{R}^d$ and $\ell$ is often treated as an optimizable hyperaparameter in regression models. A typical approach in e.g. GP regression is to maximize the marginal likelihood which is parameterized with this hyperparameter. More sophisticated parameterizations are expected to yield more predictive regressors. To this end, we build a hierarchical kernel in the form of a neural network, depicted in Figure~\ref{schematic}. 

The input to the neural network is the absolute of difference between prior samples denoted by   $\bm m_{\ast_0}$ (equivalent to $\tilde{\bm m}_{0}$ in the previous subsection) and another set of randomly distributed points denoted by $\bm m_{\circ}$ in each dimension of samples, i.e. $|\bm m_{\ast_0,j}-\bm m_{\circ,j}|$. Note that the input features are matrices in our construction, i.e.  $|\bm m_{\ast,j}-\bm m_{\circ,j}| \in \mathbb{R}^{n_{\ast} \times n_{\circ}}$ where $n_{\ast}$ and $n_{\circ}$ are number of $\bm m_{\ast_0}$ and $\bm m_{\circ}$ samples. 

The hidden nodes are indexed by $ji$ where the first index denotes the layer number and the second one denotes the hidden node number in that layer which has a total of $n_i$ nodes. The hidden nodes in the first layer are evaluated as: 
\begin{equation*}
\begin{array}{l l}
k_{1i}(\bm m_{\ast_0}, \bm m_{\circ}) =\displaystyle \sum_{j=1}^{d} w_{ij} |\bm m_{\ast_0,j}-\bm m_{\circ,j}|+ b_i, & \hat{k}_{1i} = \tanh(k_{1i}), \qquad i=1,\ldots,n_{1}
\end{array}
\end{equation*}
Subsequently, the hidden nodes in the next layers $l=2,\ldots,L-1$ are 
\begin{equation*}
\begin{array}{l l}
 k_{li}(\bm m_{\ast_0}, \bm m_{\circ}) =\displaystyle \sum_{j=1}^{n_{l-1}} w_{ij} \hat{k}_{l-1j} + b_i, & \hat{k}_{li} = \tanh(k_{li}), \qquad i=1,\ldots,n_{l}
\end{array}
\end{equation*}
and finally in the last layer $l=L$, one unit/node is computed via
\begin{equation*}
\begin{array}{l l}
k_{L}(\bm m_{\ast}, \bm m_{\circ}) =\displaystyle \sum_{j=1}^{n_{L-1}}  w_{j} \hat{k}_{L-1j} + b, & k (\bm m_{\ast_0}, \bm m_{\circ})= (1+\tanh(k_{L}))/2
\end{array}
\end{equation*}
where we removed the index for one unit and the linear transformation ensures that $k (\bm m_{\ast_0}, \bm m_{\circ}) \in [0,1]$. In our implementation, we use $\tanh$ activation function as shown in above expressions. This choice is relatively generic in neural network computations. Other activation functions could be incorporated if there is particular knowledge about the problem at hand.  The output of the last node similarly to the input features is a matrix $ k(\bm m_{\ast_0}, \bm m_{\circ}) \in \mathbb{R}^{n_{\ast} \times n_{\circ}}$. The target data $\bm m_{\ast_1} \equiv \bm m_{opt}$ is predicted in a support vector machine format, i.e.  
\begin{equation}
\bm m_{\ast_1}= \sum_{i=1}^{n_{\circ}} \bm \alpha_i  k(\bm m_{\ast_0}, \bm m^{(i)}_{\circ}).
\end{equation}
The above equation can be written in the matrix form as $\bm m_{\ast_1} =  k(\bm m_{\ast_0}, \bm m_{\circ}) \bm \alpha$ where $\bm \alpha \in \mathbb{R}^{n_{\circ} \times d}$ are optimizable coefficients and $\bm m_{\ast_1} \in \mathbb{R}^{n_{\ast} \times d}$ are posterior samples. We recall that $d$ is the dimensionality of the prior and posterior samples. Training this neural network-based kernel entails optimization of hyperparameters in addition to coefficients i.e. $\{w_{ij}, b_i, \bm \alpha\}$. We train these parameters with Newton-Raphson approach similarly to what we described in Section~\ref{Sec3_1}, i.e. we solve nonlinear equations $ k(\bm m_{\ast_0}, \bm m_{\circ}) \bm \alpha -\bm m_{\ast_1} = \bm 0$ by forming the Jacobian that now includes gradients with respect to $\bm \alpha$ in addition to hyperparameters of $ k(\bm m_{\ast_0}, \bm m_{\circ})$.

Similarly to the discussion in Section~\ref{Sec3_1}, the most significant part of our neural network training is the computation of Jacobian within a neural network architecture. The derivative computation involves the well-known chain rule and the idea of backpropagation which evaluates the derivative of last layer as a product of derivatives obtained from each two successive layers from right to left. We have implemented the neural net kernel and the backpropagation with basic matrix operations using mainly fast vectorized MATLAB computations. An excerpt of the overall code, i.e. one function script which involves backpropagation is provided in~\ref{many_layer}. The codes for this paper are available upon request. We will publish more extensive version of these codes in our future research where we plan to investigate the possibility of solving generic PDEs such as linear elasticity problems with our neural net kernel.

\begin{remark}
Another approach for training this deep kernel is to maximize the marginal likelihood similar to GP regression. Based on our experience, optimization of marginal likelihood, i.e. minimization of a scalar parametrized via a neural net can be challenging as the objective function is highly non-convex. In addition, the optimized hyperparameters and the resulting kernel may not always result in more accurate regression of data. We note that the standard marginal likelihood as an objective function is derived from a statistical perspective. Training with these statistical quantities can be less effective compared to directly solving the residual equations $\bm R =  k(\bm m_{\ast_0}, \bm m_{\circ})\bm \alpha -\bm m_{\ast_1}$ which we consider in this paper.    
\end{remark}

\section{Numerical Examples}\label{Sec4}

\subsection{Illustrative Numerical Example with $d=2$}\label{S4_1_1}
\begin{figure}[!h]
\centering
\includegraphics[width=2in]{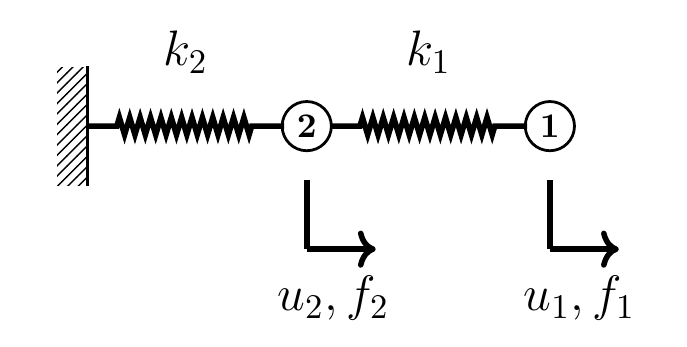}
\caption{\small{A system of two linear static springs.  }}\label{linear_sp}
\end{figure}

In the first example in this section we consider a simplistic system to elucidate our inference procedure. This system comprised of two linear static springs connected in series is shown in Figure~\ref{linear_sp}. 

The spring stiffnesses are parameterized according to  
\begin{equation}\label{stf_sp}
\begin{array}{ll}
k_1=x_1 g(m_1) +0.1, & k_2=x_2 g(m_2) +0.5, 
\end{array}
\end{equation}
where $x_1$ and $x_2$ are assumed to be the input to the system, $m_1$ and $m_2$ are the latent parameters, $g(m)=\exp(m)$ and the forces are assumed to be fixed $f_1=f_2=1$. In one of the examples where we investigate non-unique maps we consider $g(m)=m^2$. The displacements $u_1,~u_2$ are found from the following linear system
\begin{equation}\label{linear_FWD}
\left [ 
\begin{array}{ll}
k_1 & -k_1 \\
-k_1 & k_1+k_2
\end{array}
 \right]
 \left[ \begin{array}{ll}
u_1 \\
u_2
\end{array} \right] = \left[ \begin{array}{ll}
f_1 \\
f_2
\end{array} \right].
\end{equation}

The variables $x_1$ and $x_2$ are often considered as design parameters for optimizing the stiffness (compliance) in illustrative design optimization problems~\cite{KESHAVARZZADEH201647}. In this paper, we consider an uncertain form for $\bm x=[x_1, x_2]$ namely $\bm x = [0.5, 0.5] + \delta_x\hat{\bm x}$ where $\hat{\bm x} \sim U[-1,1]^2$ is a two dimensional uniform random variable. The parameter $\delta_x$ controls the scatter in the input data. In our numerical experiments, we mainly consider $\delta_x=0.005$.  

\begin{figure}[!h]
\centering
\includegraphics[width=1.58in]{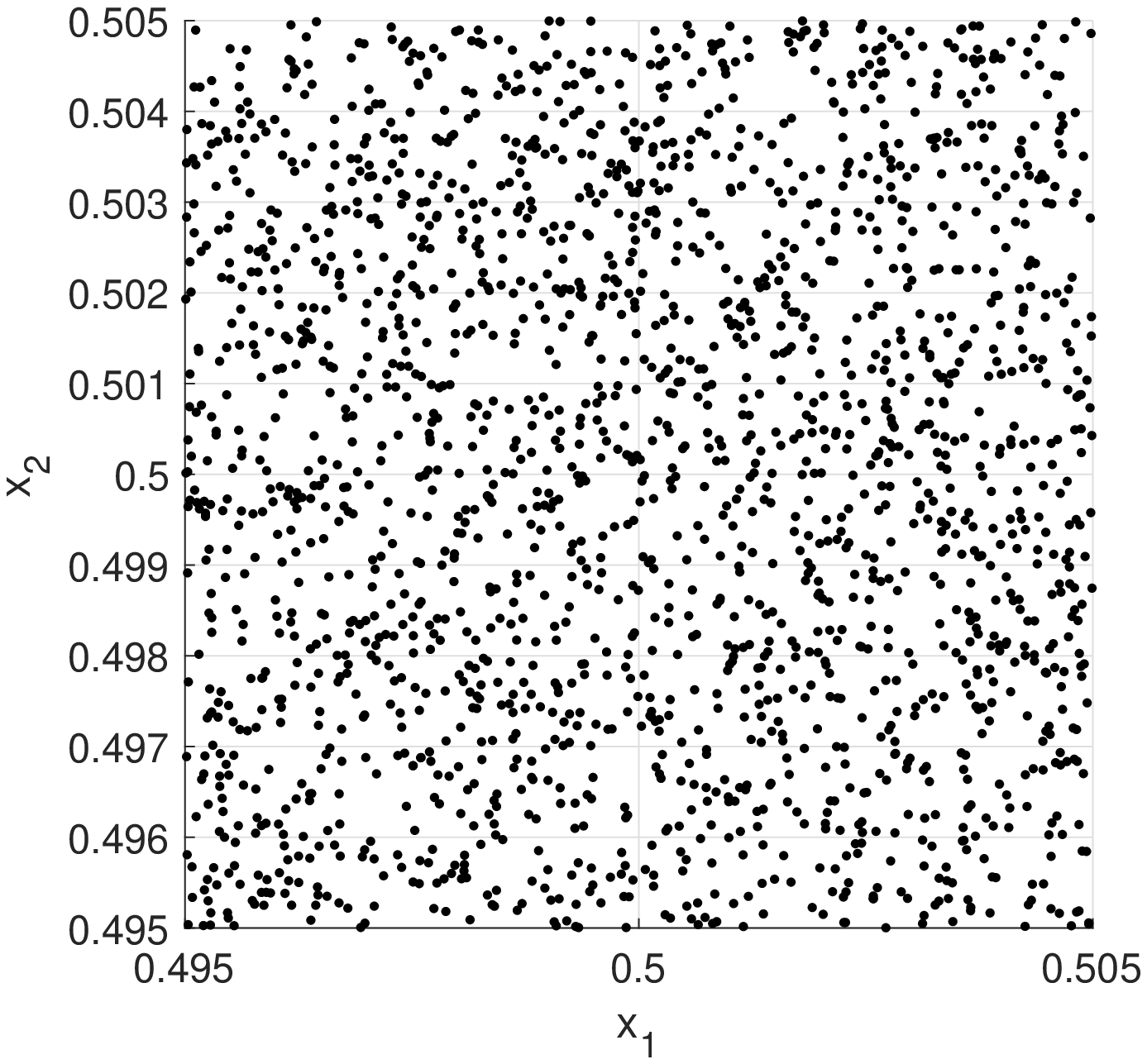}
\includegraphics[width=1.46in]{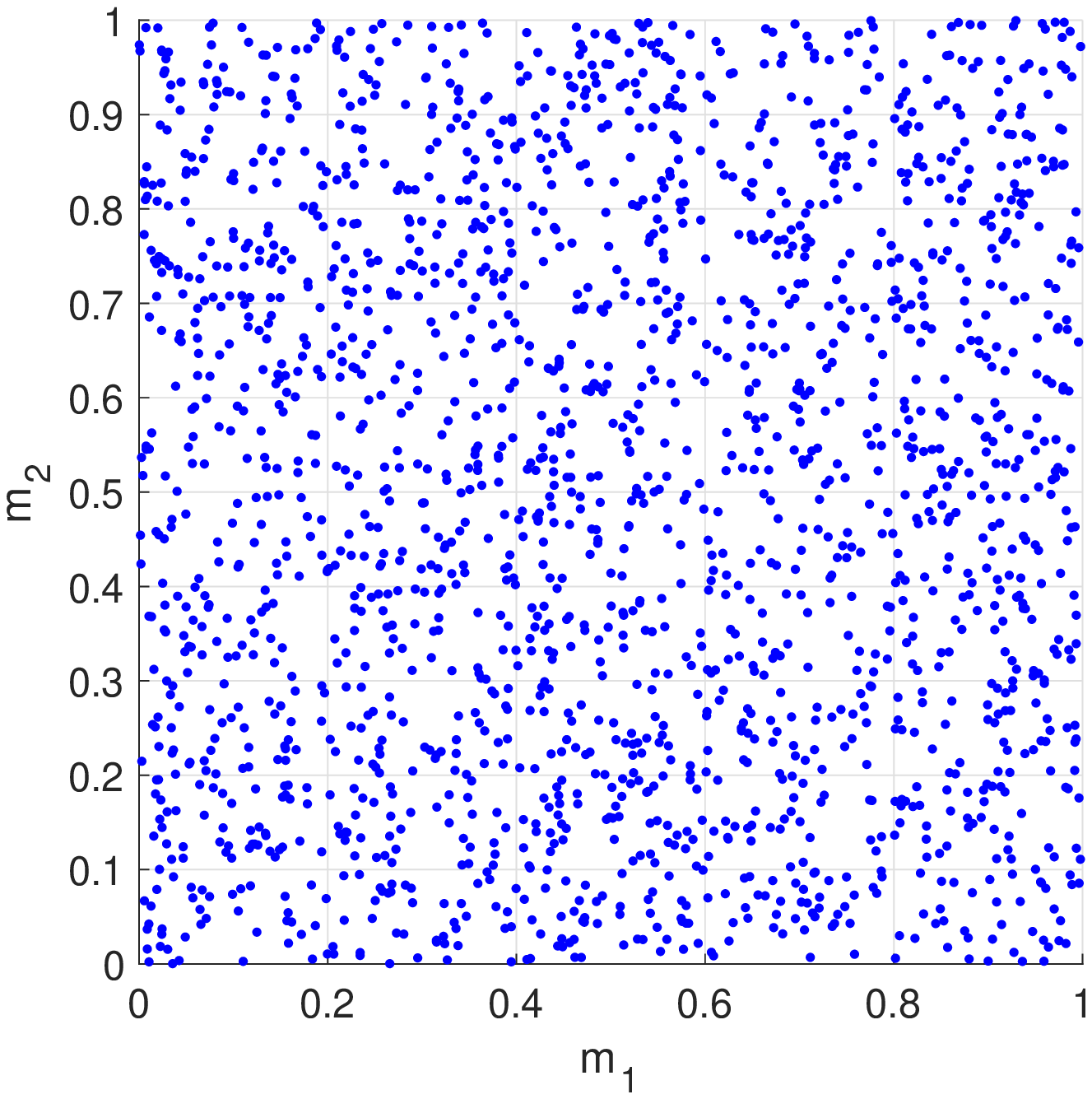}\\
\includegraphics[width=1.5in]{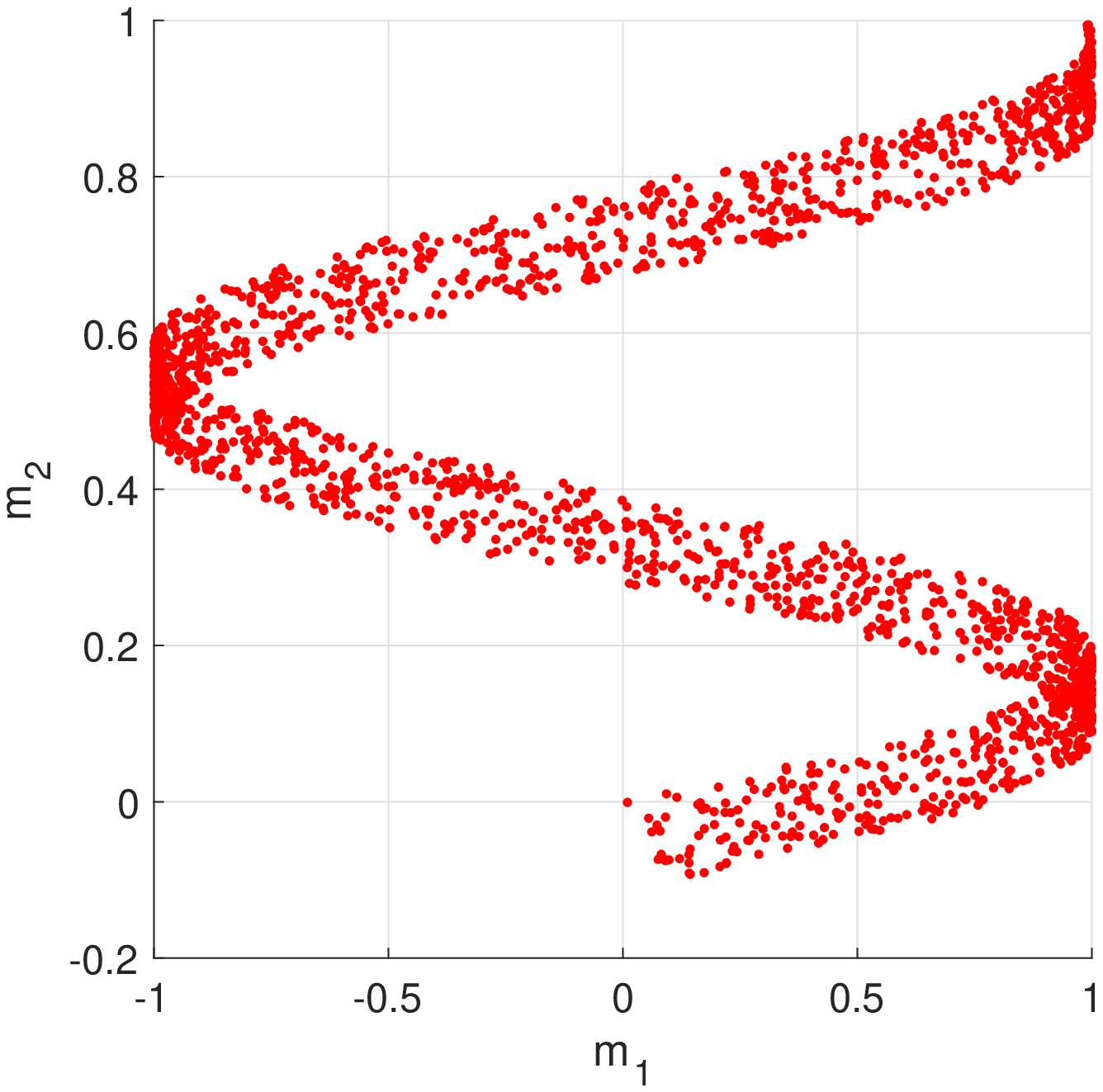}
\includegraphics[width=1.5in]{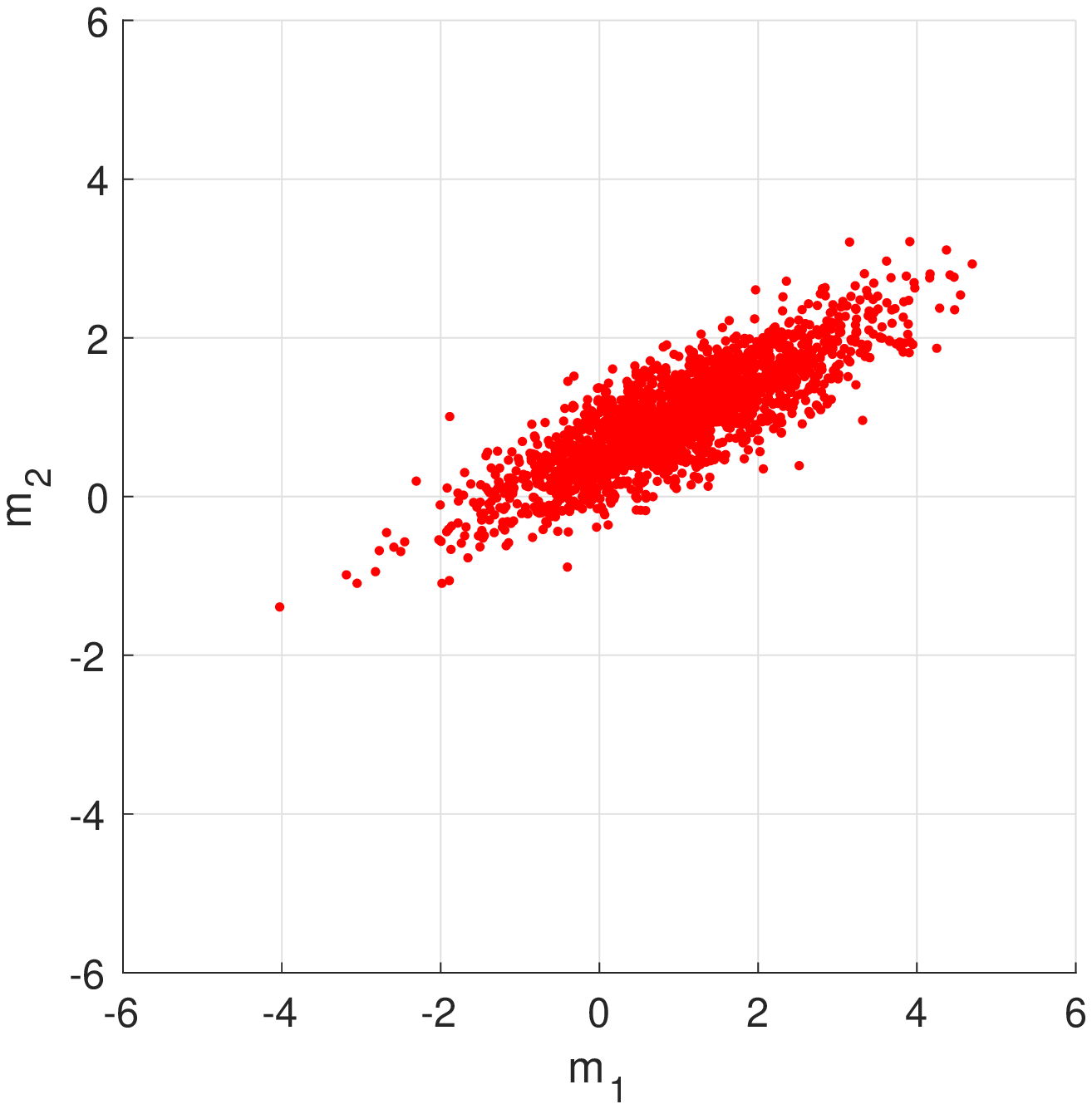}
\includegraphics[width=1.5in]{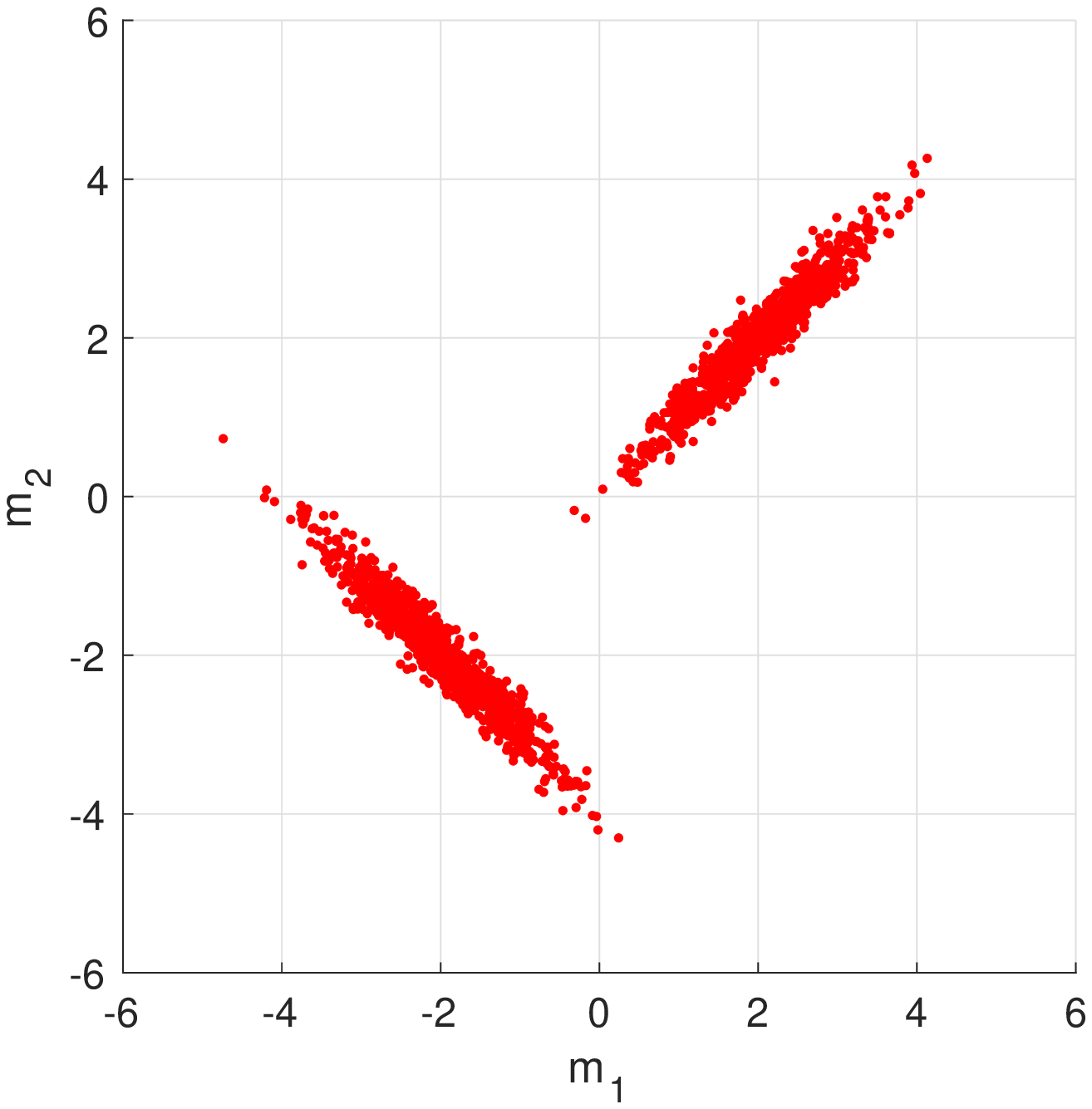}
\includegraphics[width=1.5in]{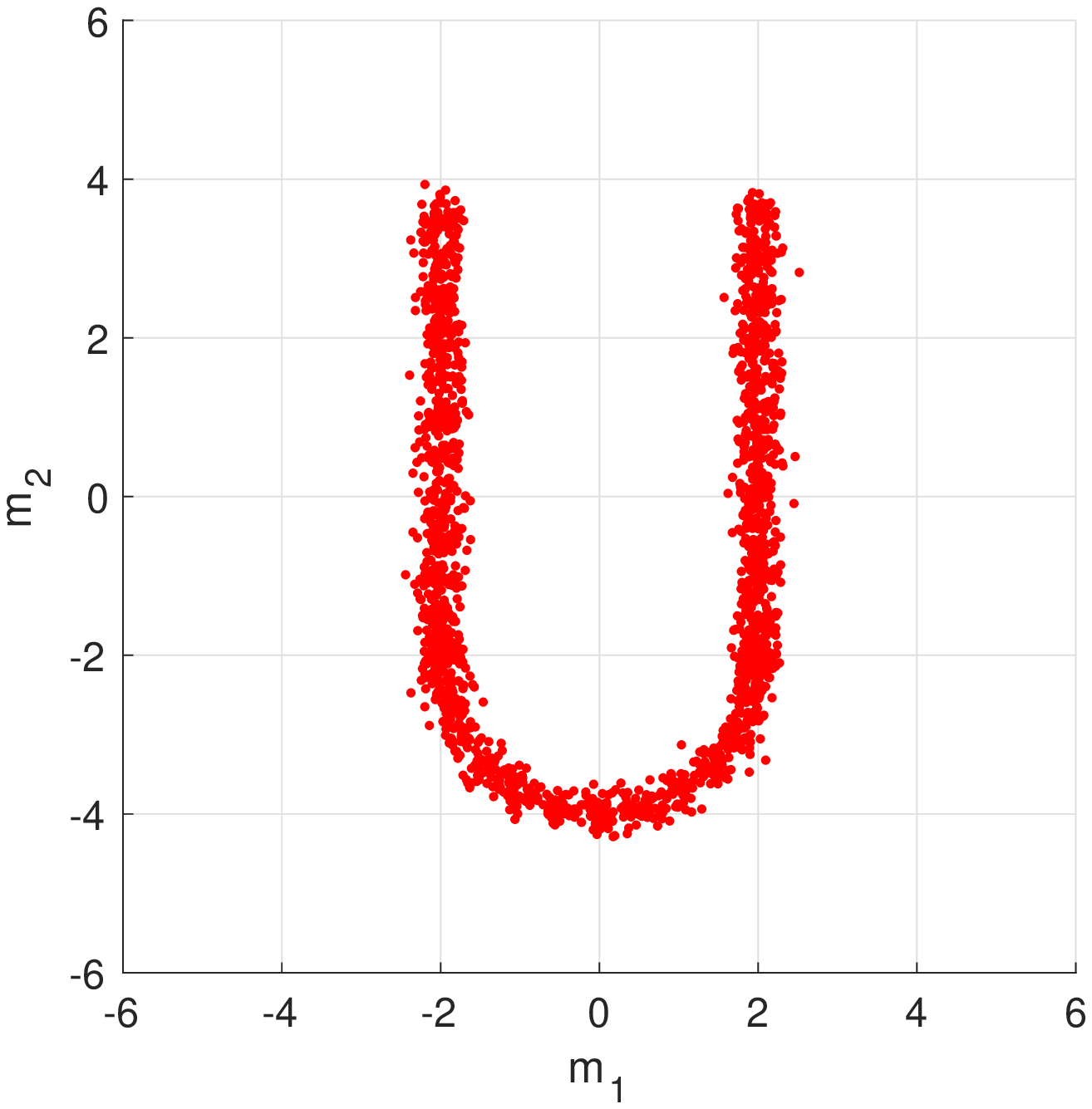}
\caption{\small{Top row: Distribution of input variables $\bm x$ for $\delta=0.005$ (left) and prior distribution of latent variables $\bm m$ (right); Bottom row: The true distribution of latent variables $\bm m$ in the spring problem; analytical (left pane), unimodal (second pane), bimodal (third pane), and $U$ shape (right pane).}}\label{fig_dist_true}
\end{figure}

In this example, we investigate the performance of the approach on four different true distributions on $\bm m$. The distributions of $\bm x$, prior $\bm m$ which is ${\bm m}_0 \sim U[0,1]^2$ and four different true distributions, namely an analytical, unimodal, bimodal and U shape on $\bm m$ are shown in Figure~\ref{fig_dist_true}.

In all examples, the number of training and test samples (denoted by $n_{\ast}$) are $200$ and $1000$. The number of $\bm m_{\circ}$ used within the kernels is variable in different examples. The number of data points $n_{data}$ cf. Algorithm~\ref{alg:opt_fwd} is equal to the number of training data points $n_{\ast}$.

\subsubsection{An Analytical Map}
We start the experiments with the analytical map. In this example, we only consider the performance of the NNK in predicting an analytical distribution. There is no observation or simulation data in this example. The expression for the analytical distribution is
\begin{equation}
\bm m=[\sin(8x_1+0.1x_2),~x_1-0.1x_2]
\end{equation} 
where $\bm x \in U[0,1]^2$ only in this example. The goal in this example is to train an NNK which maps $\bm x$ to $\bm m$ (unlike the original setting which maps permuted prior samples $\tilde{\bm m}_0$ to posterior samples $\bm m_1$). For the sake of comparison, we consider two approaches for training NNK, the Newton-Raphson (NR) and Gradient-Descent (GD). The network architecture is comprised of four layers with  ${[2,7,4,1]}$ nodes and $\bm \alpha \in \mathbb{R}^{20 \times 2}$, i.e. the NNK has two hidden layers with $7, 4$ nodes, the output layer (always) with $1$ node in addition to the input layer which takes two features $|x_{\ast,1}-x_{\circ,1}|$ and $|x_{\ast,2}-x_{\circ,2}|$, i.e. two nodes.

The learning rate for the Newton-Raphson is set to $\beta=5 \times 10^{-3}$. This learning rate is too steep for the GD approach and results in divergence, therefore we decrease the learning rate to $\beta=5 \times 10^{-4}$ in the case of GD. Also, we set a fixed (Tikhonov) regularization parameter in inversion of $\bm J^T \bm J$ cf. Algorithm~\ref{alg:opt_fwd} as following
\begin{equation}\label{tikh_numerics}
\delta_{tikh} = 
\begin{cases}
10^{-5}, \qquad & \| R \|_2 \geq 0.01 \\
10^{-6}, \qquad  & otherwise \\
\end{cases}
\end{equation}
This regularization yields robust results throughout the examples that we consider in this paper.  The training can be accelerated by employing more sophisticated approaches for solving the linear system $\bm J \Delta_{NR} = \bm R$ which is an idea for our future research. Also, the learning rate could be estimated by a formal line search method which we also consider as further developments for the NNK tool.
 
\begin{figure}[!h]
\centering
\includegraphics[width=2.1in]{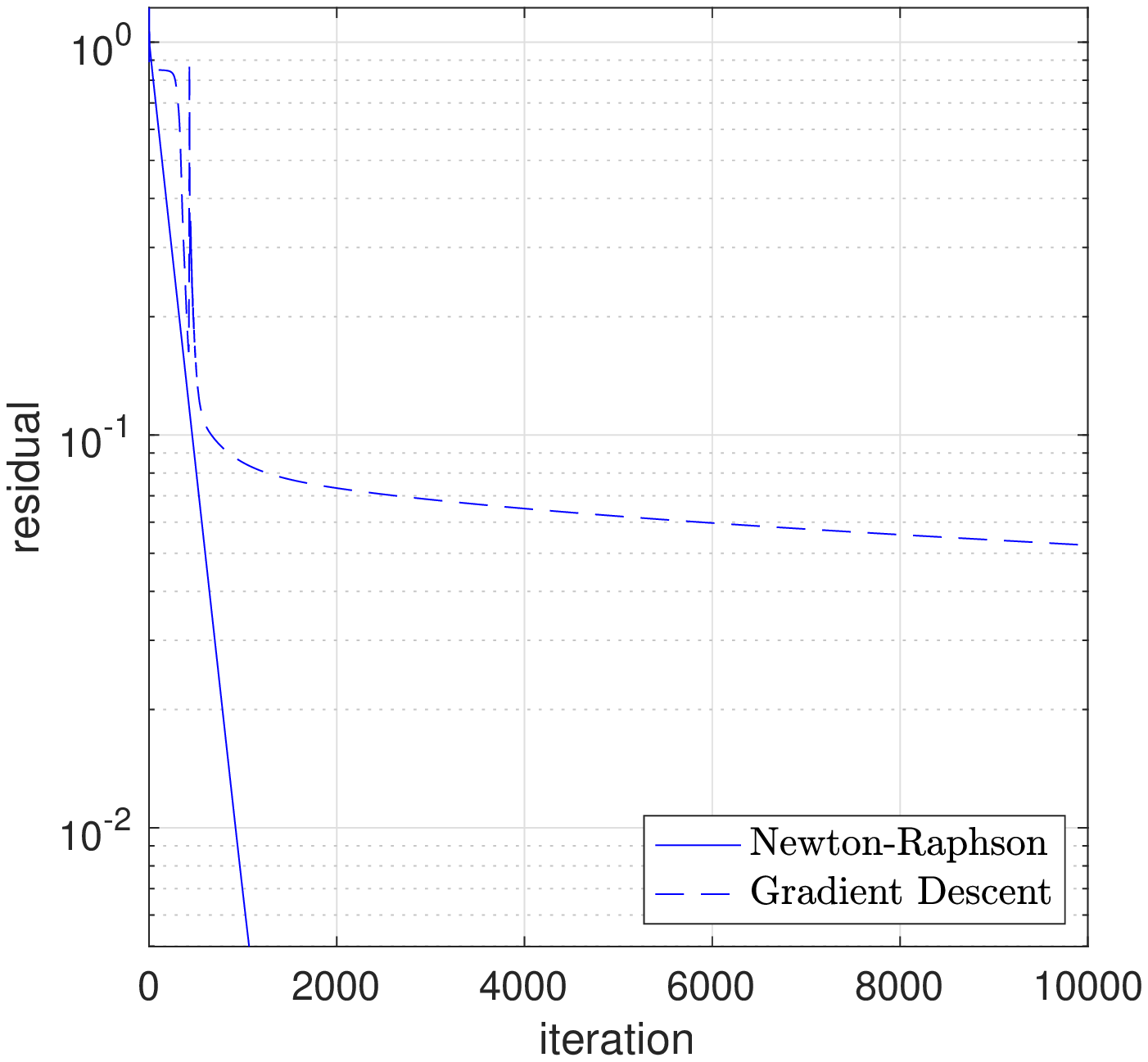}
\includegraphics[width=2.0in]{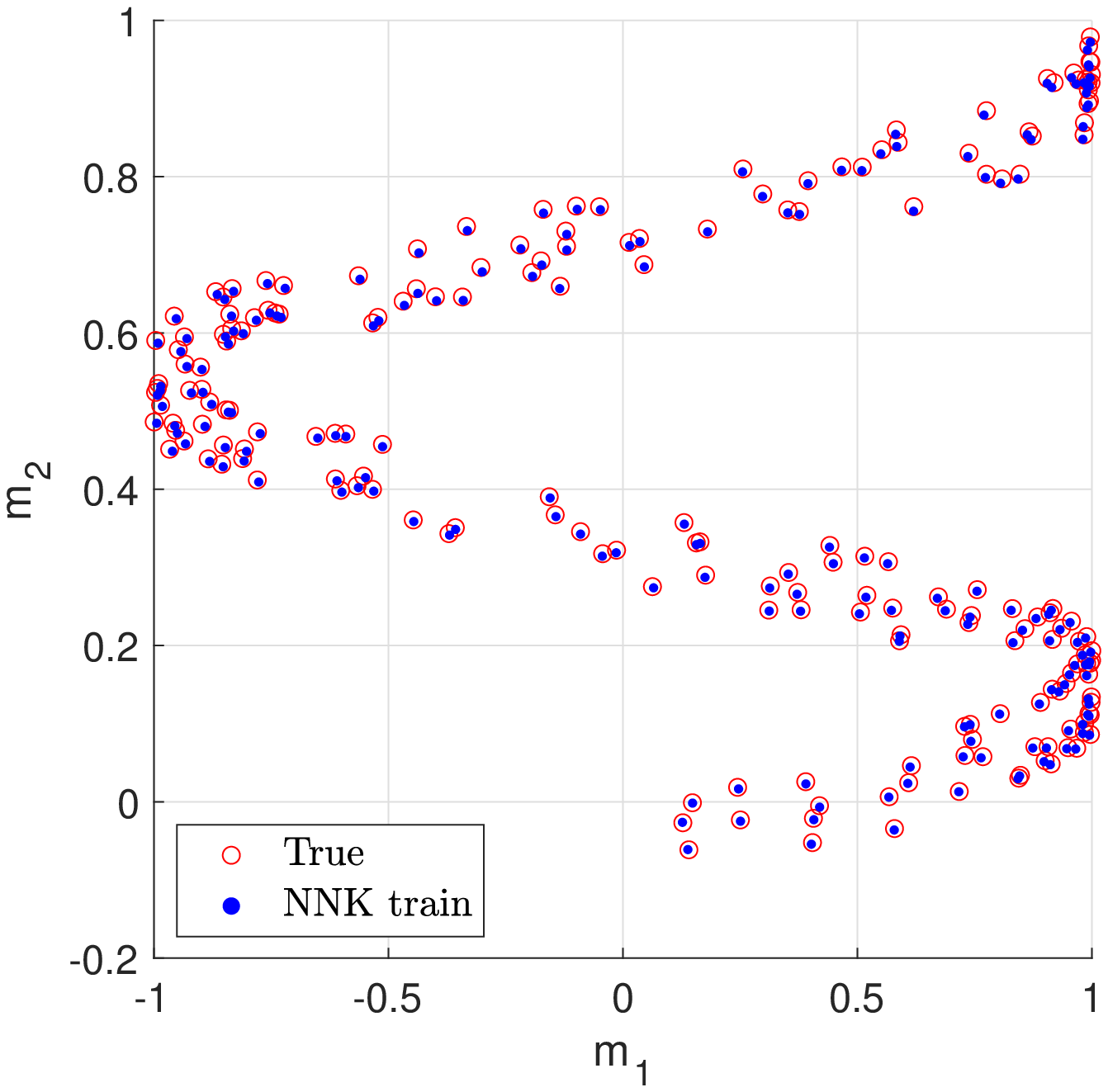}
\includegraphics[width=2.0in]{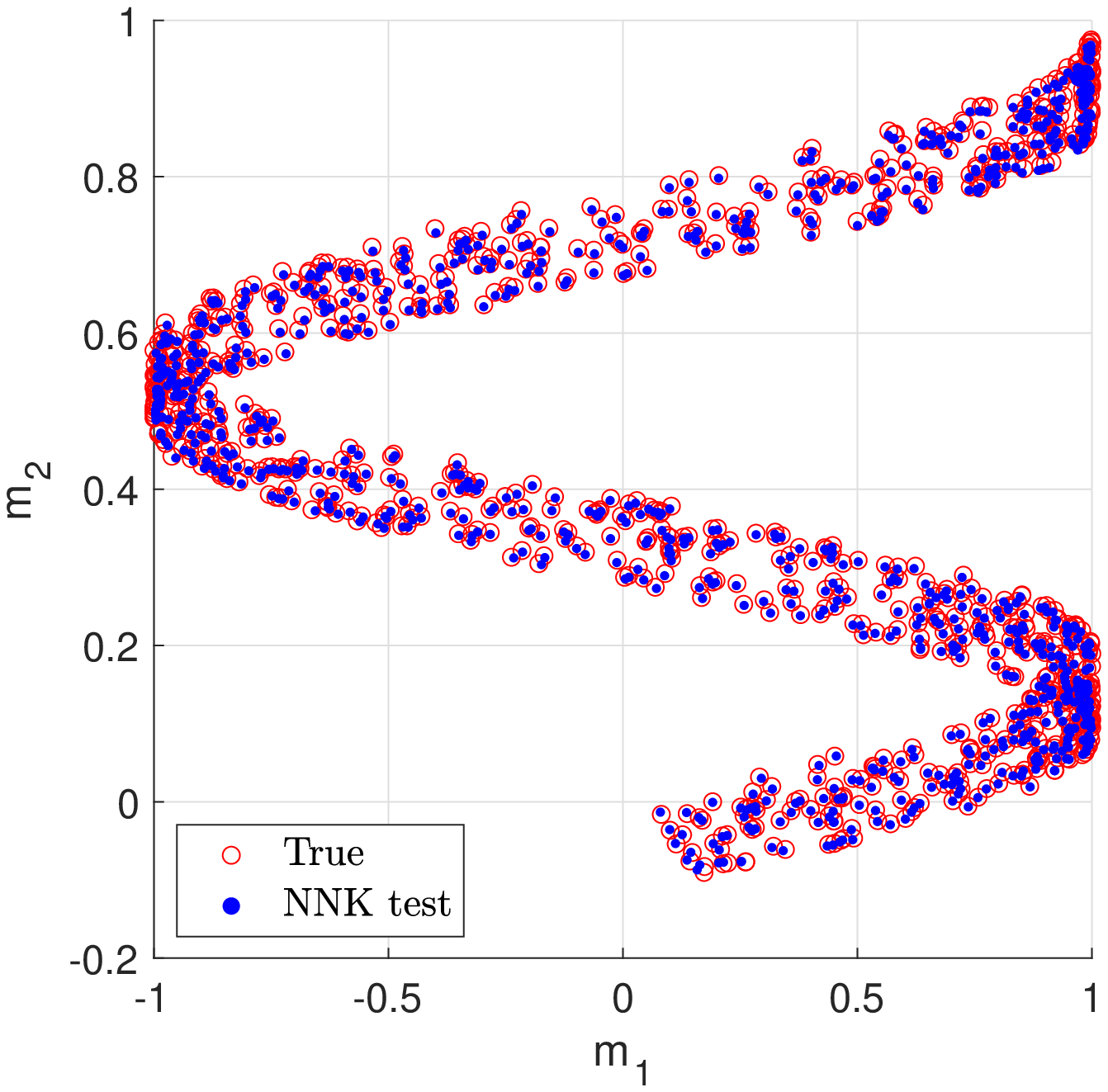}\\
\caption{\small{Iterations of the NNK training with NR and GD optimization (left), NNK training with NR associated with the analytical map (middle), NNK prediction (test) with randomly distributed prior nodes (right). Based on these results, we deem NNK is capable of predicting the analytical map accurately. More sophisticated maps are considered in the next examples.}}\label{fig_NR_GD}
\end{figure}

Figure~\ref{fig_NR_GD} (first pane) shows the convergence of both NR and GD approaches for the training of the analytical map. The second and third panes show training and testing for the NR approach. To investigate the accuracy, we compute the normalized error between the predicted samples $\bm m$ and exact samples $\bm m_{exc}$ via $e=\|\bm m - \bm m_{exc}\|/\|\bm m_{exc} \|$ in both training and testing. We also measure the training time for both NR and GD. The results are presented in Table~\ref{tab_GD_NR}.  It is apparent that the NR approach is more accurate and significantly faster. Both approaches are done on the same computer on a single processor. From the first pane of Figure~\ref{fig_NR_GD}, note that the GD approach exhibits poor convergence; we indeed manually limit the number of iterations to $10000$. Getting a better performance from GD for this neural network architecture might require significant hand engineering which is less desirable for a robust numerical tool. On the other hand, the NR approach requires proper computation of $\bm J$ and the Newton step $\Delta_{NR}$ where both involve linear solves that could be further accelerated for utilization on large scale problems. 

We again mention that the conclusion drawn is solely pertinent to this particular small scale example. For larger scale problems, a combination of both approaches i.e. starting with a cheap GD approach for the sake of finding a suitable starting point for the NR approach may result in less computational time. 

 %%\delta_R=0.005

\begin{table}[!h]
\caption{Accuracy and speed of two optimization approaches for NNK hyperparameters}
\normalsize
\centering
\begin{tabular}{c c c c c }
\hline\hline
   & $e_{train}$  & $e_{test}$    &  $n_{iter}$ & time(sec) \\
\hline
Newton-Raphson & $5.00 \times 10^{-3}$ & $5.20 \times 10^{-3}$ & $1066$ & $20.39$	 \\
Gradient Descent & $5.24 \times 10^{-2}$ & $5.77 \times 10^{-2}$ & $10000$ &$191.51$  \\
\hline
\end{tabular}
\label{tab_GD_NR}
\end{table}

\subsubsection{Effect of Permutation of Prior Samples} 

In this example, we consider the unimodal distribution for $\bm m$ and investigate the effect of the second step i.e. permutation of prior samples in our procedure. The unimodal distribution is a bivariate Gaussian distribution $\bm m \sim \mathcal{N}(\bm \mu, \bm \Sigma)$  where $\bm \mu =[1,1]$ and $\bm \Sigma = \begin{bmatrix}
1.4 & 0.63\\
0.63 & 0.41
\end{bmatrix}$.

To begin the procedure, we generate $200$ observation data points $\bm y$ and solve the $\mathcal{G}(\bm x, \bm m) = \bm y$. To solve these nonlinear equations associated with the simulation model, we use the same Tikhonov parameter cf. Equation~\eqref{tikh_numerics}  to regularize $\bm J^T \bm J$. We also consider a fixed learning rate for solving the forward throughout all examples in this paper, i.e. we set $\beta=0.1$.   

\begin{figure}[!h]
\centering
\includegraphics[width=1.5in]{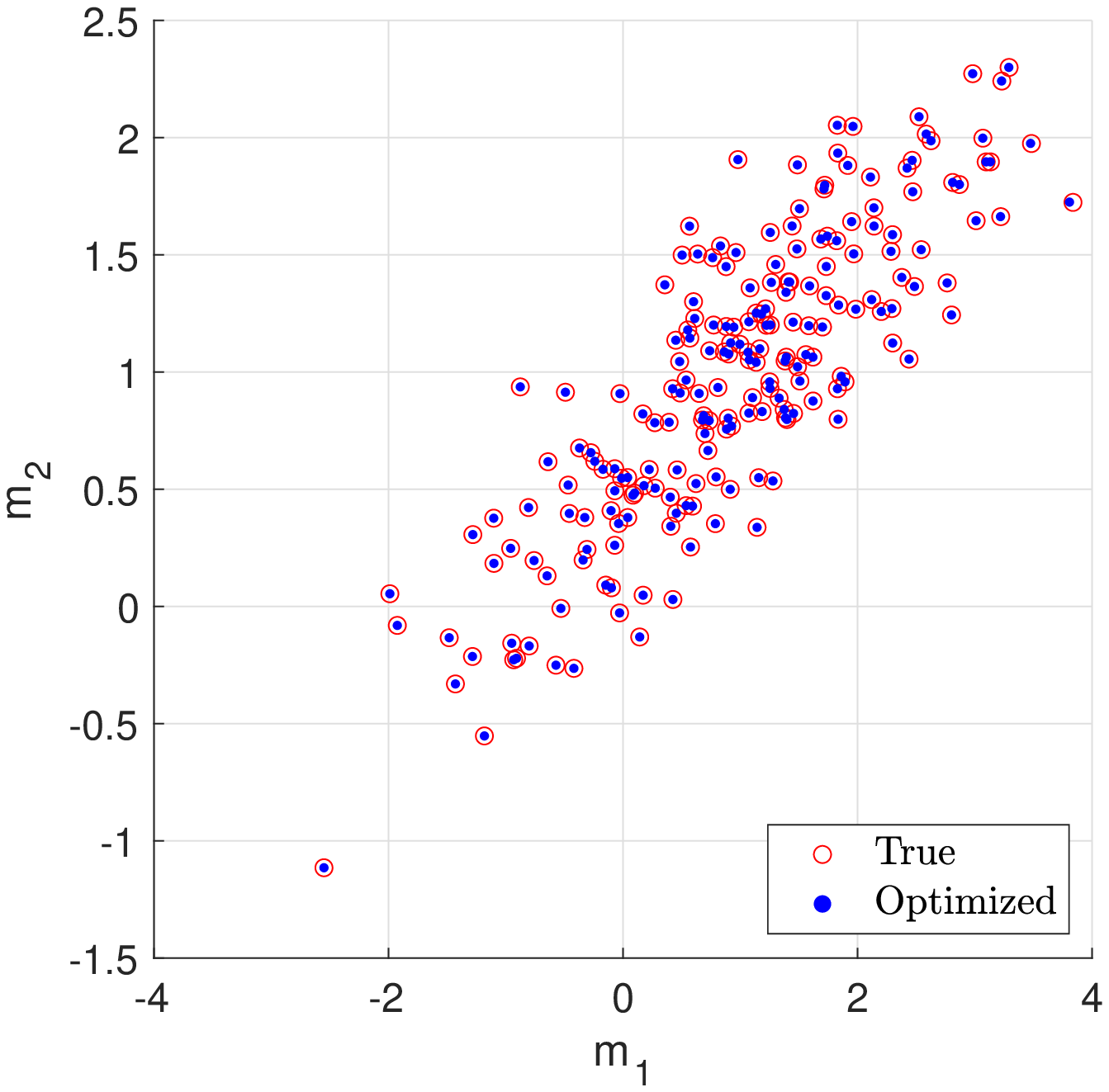}
\includegraphics[width=1.5in]{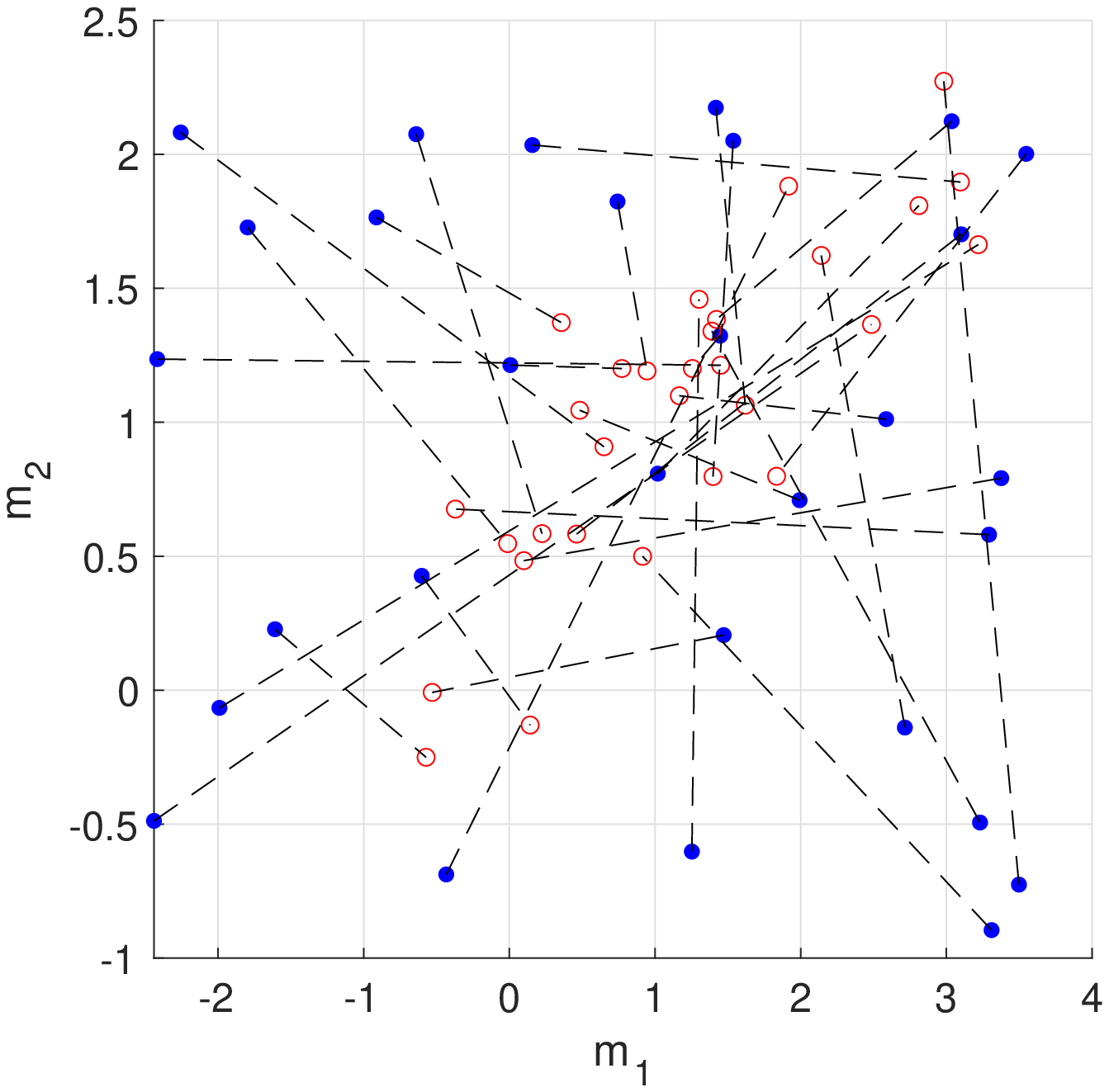}
\includegraphics[width=1.5in]{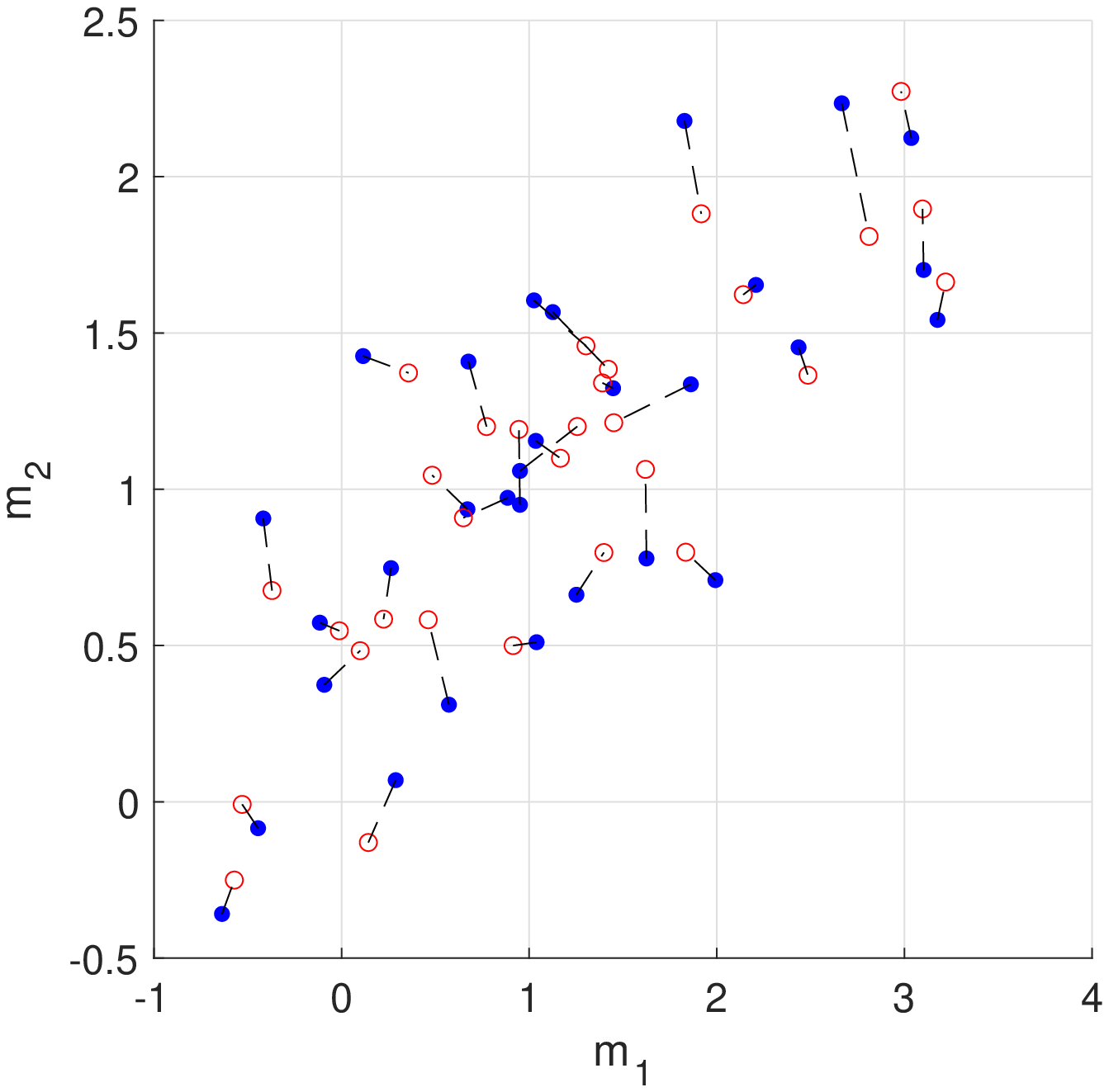}\\
\includegraphics[width=1.5in]{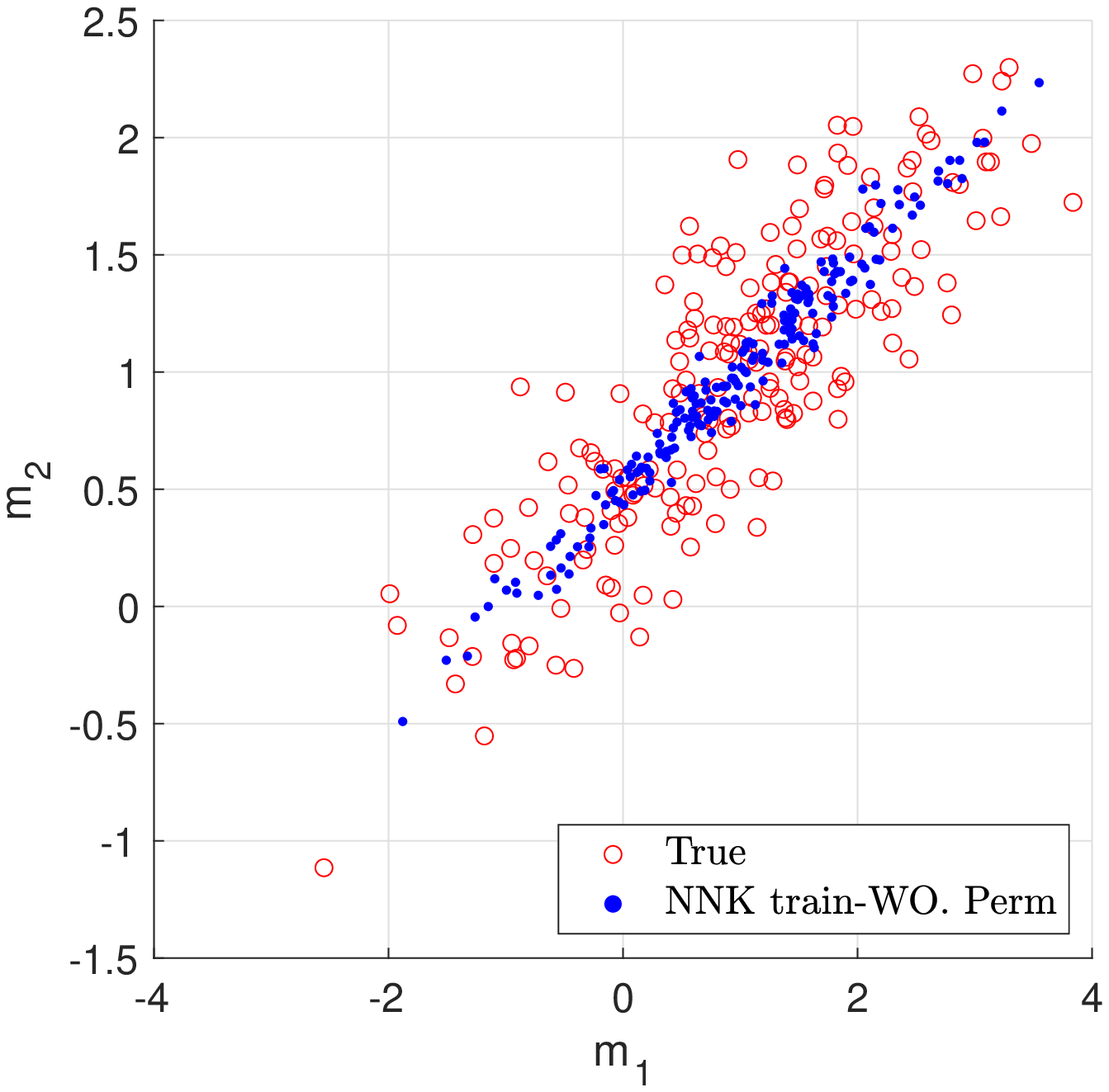}
\includegraphics[width=1.5in]{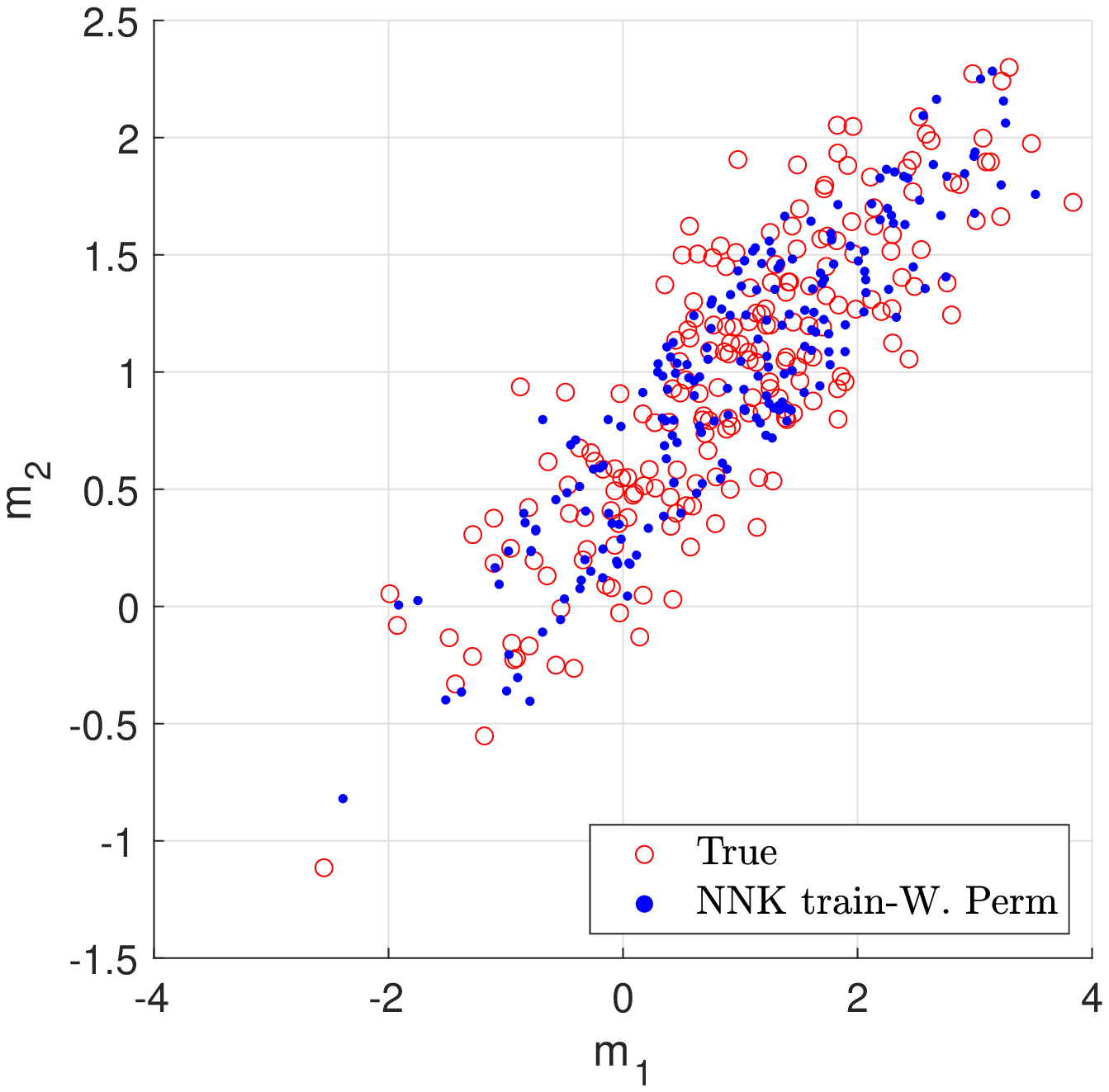}
\includegraphics[width=1.5in]{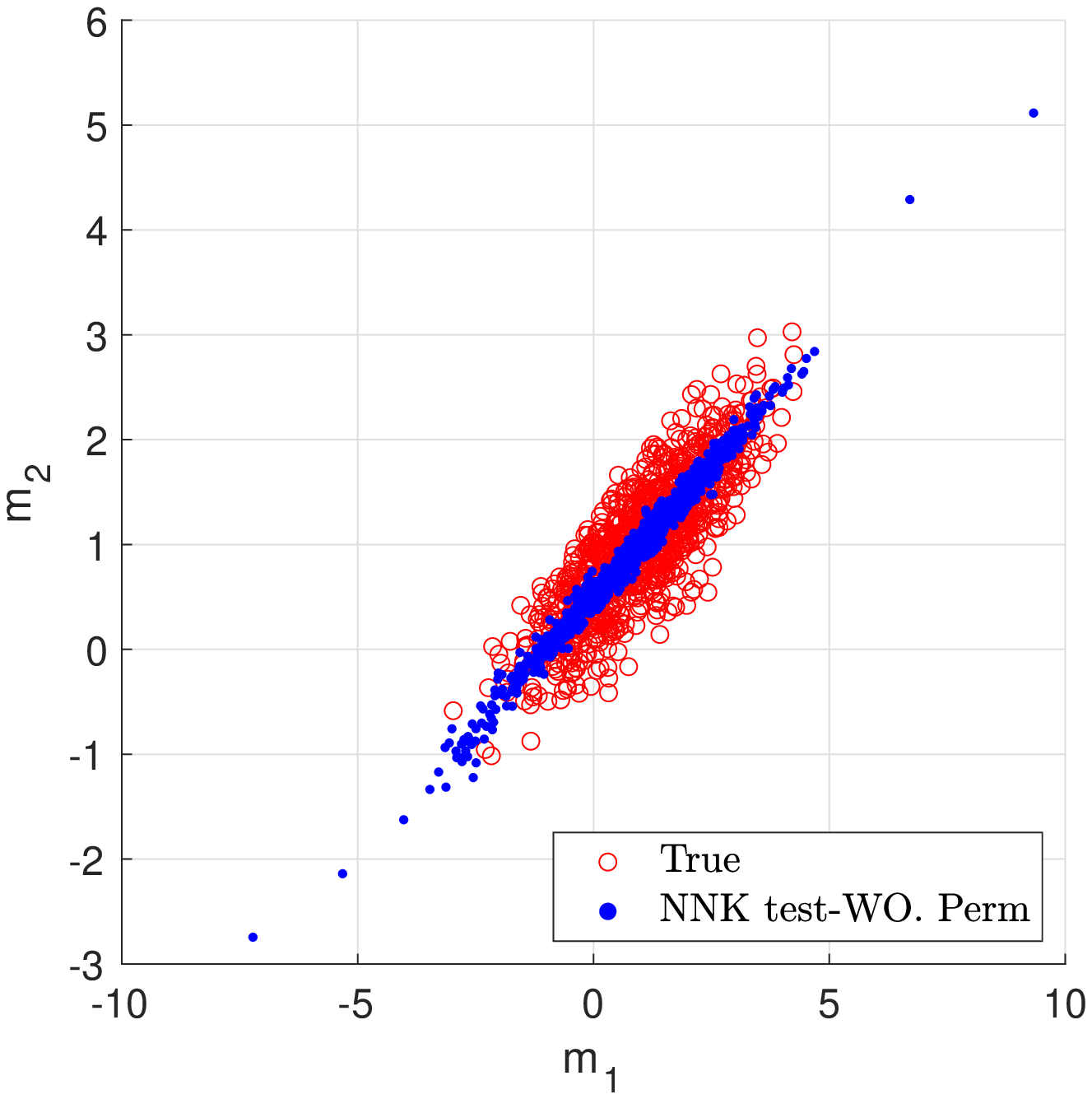}
\includegraphics[width=1.5in]{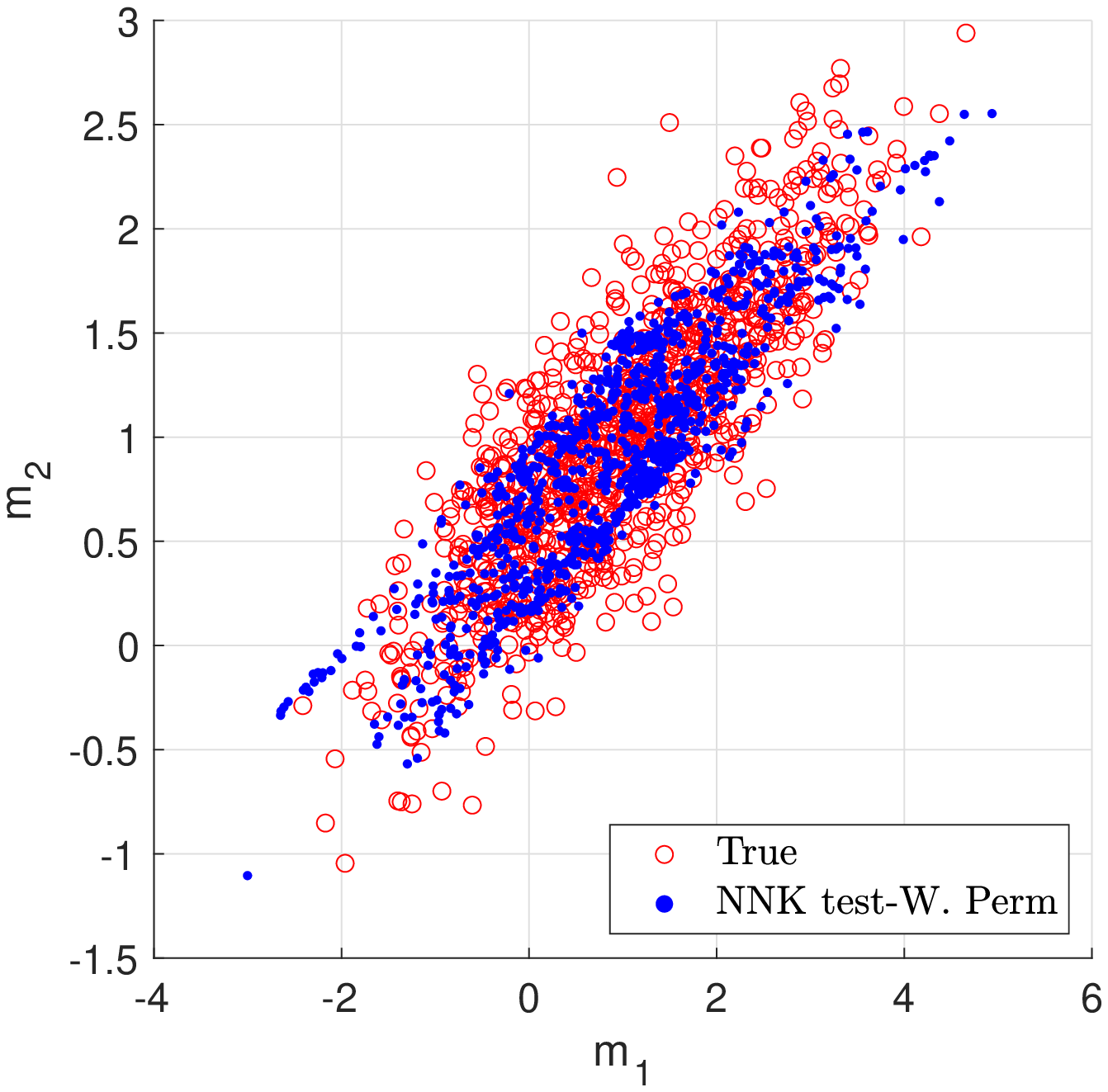}
\caption{\small{Different steps of the inference procedure: Inversion of the forward model (top row - first pane), visualization of the prior samples permutation (top row - second and third panes) and training and testing with two scenarios of with and without permutation of prior samples (bottom row). It is apparent that permutation of prior samples results in more accurate prediction in the training and testing datasets. See the numerical results in Table~\ref{Tab_perm}.}}\label{fig_permute}
\end{figure}

To test the robustness of this approach to different levels of scatter in $\bm x$, we consider a larger $\delta_x$, i.e. $\delta_x=0.05$. We also set the tolerance for residual of nonlinear equations to $\delta_R=10^{-3}$ cf. Algorithm~\ref{alg:opt_fwd}. In both cases i.e. $\delta_x=0.005$ and $\delta_x=0.05$ the NR approach converges successfully and satisfies the residual tolerance. The result of this optimization for samples $\bm m$ are shown in the first pane of top row in Figure~\ref{fig_permute}. Note that we only show one scenario as there is no difference between two cases. 

After finding optimized samples $\bm m$ we consider two scenarios for prior samples: 1) we only scale the uniformly distributed samples according to the optimized samples and do not  permute, 2) we permute the uniformly distributed samples using Algorithm~\ref{alg:perm}. To visualized both cases, we connect the prior samples to the optimized samples. The results are shown in the second and third panes of the top row in Figure~\ref{fig_permute}. For better visualization we only show a small number of points in these figures. It is apparent that  the map on the left is more erratic and its training is expected to be more challenging.

We use both sets of prior samples for training the NNK. The network architecture in this example is $[2,10,4,1]$ with $\bm \alpha \in \mathbb{R}^{20 \times 2}$ and $\beta=5 \times 10^{-3}$. As these are not analytical maps, they do not achieve similar small residuals for training as the previous experiment. The training residual for without and with permutation is $\|\bm R\|=0.39,~0.16$ after $n_{iter} =4197,~ 2073$ iterations. The results for training and testing for both cases are shown in the bottom row of Figure~\ref{fig_permute}.
It is observed that the training and testing results associated with the permuted samples are more effective in predicting the scatter in the underlying parameter. To numerically investigate the predictions, we compute normalized error (for prediction of $\bm m$) in mean and standard deviation for training and testing with both scenarios of with and without permutation. The results are provided in Table~\ref{Tab_perm}. Based on the numerical results, the permutation improves the results in both mean and standard deviation for training and testing.

%err in mean and std, [dir1], [dir2]
%with perm
%train $[1.005 \times 10^{-5}, 0.0189 ], [3.97\times 10^{-4}, 0.0551]$
%test [0.0729 0.0429],[0.0779 0.0644]
%without perm
%train$[1.2 \times 10^{-3}, 0.1564], [2.8\times 10^{-3}, 0.2196]$
%test [0.2641 0.2237],[0.1319 0.1324]

\begin{table}[!h]
\caption{Accuracy of mean and standard deviation of $\bm m$ in training and testing with NNK}
\normalsize
\centering
\begin{tabular}{l c c c c }
\hline\hline
   & $e_{\mu_1}$  & $e_{\mu_2}$  & $e_{\sigma_1}$  & $e_{\sigma_2}$   \\
\hline
NNK train-WO. Perm & $1.20 \times 10^{-3}$ & $2.80\times 10^{-3}$ & $1.56 \times 10^{-1}$ & $2.19 \times 10^{-1}$	 \\
NNK train-W. Perm & $1.00 \times 10^{-5}$ & $3.97\times 10^{-4}$ & $1.89 \times 10^{-2}$ & $5.51 \times 10^{-2}$	 \\
NNK test-WO. Perm & $2.64 \times 10^{-1}$ & $1.31 \times 10^{-1}$ & $2.23 \times 10^{-1}$ & $1.32 \times 10^{-1}$	 \\
NNK test-W. Perm & $7.29 \times 10^{-2}$ & $7.79 \times 10^{-2}$ & $4.29 \times 10^{-2}$ & $6.44 \times 10^{-2}$	 \\
\hline
\end{tabular}
\label{Tab_perm}
\end{table}

\begin{figure}[!h]
\centering
\includegraphics[width=1.5in]{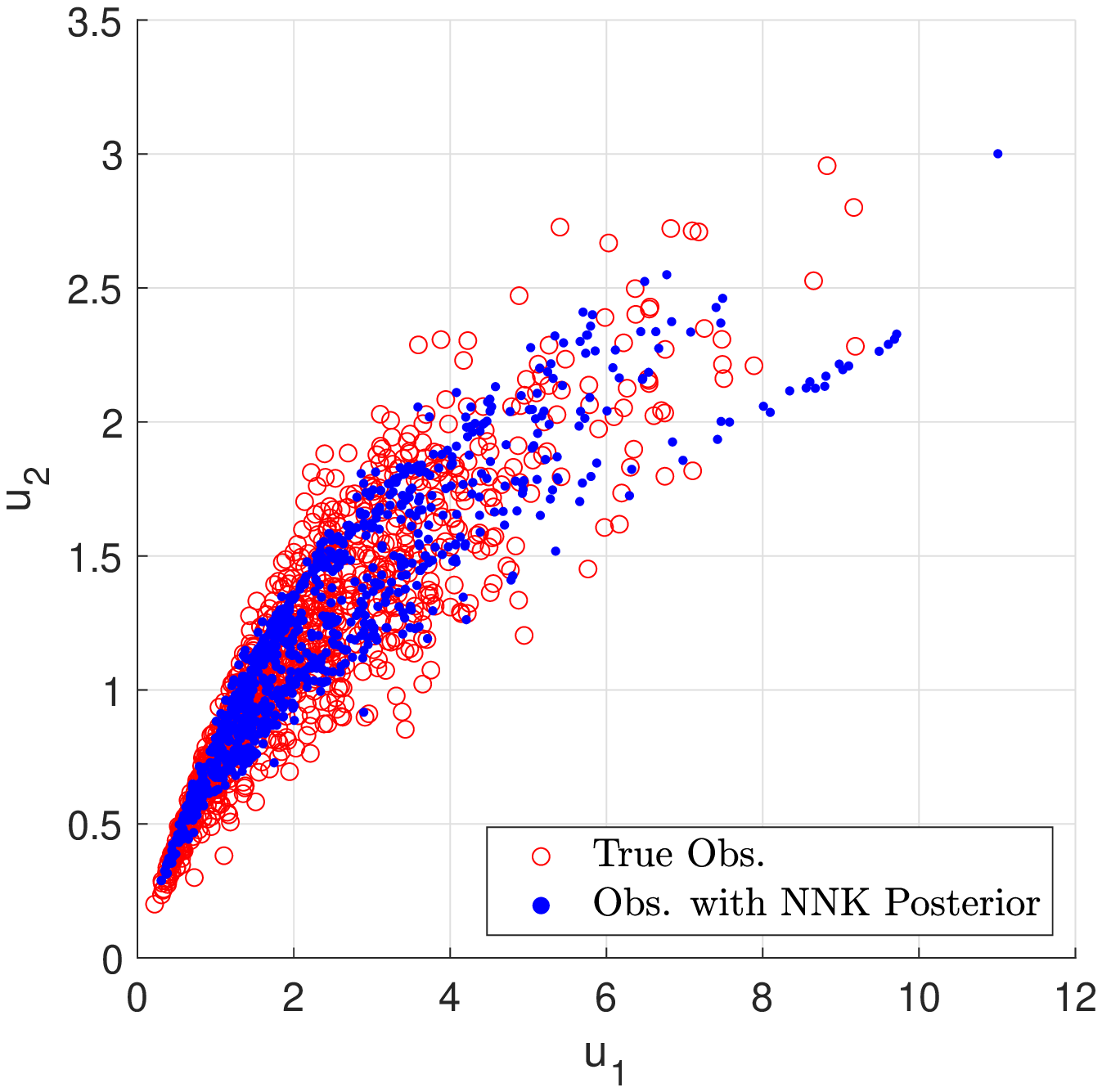}
\includegraphics[width=1.5in]{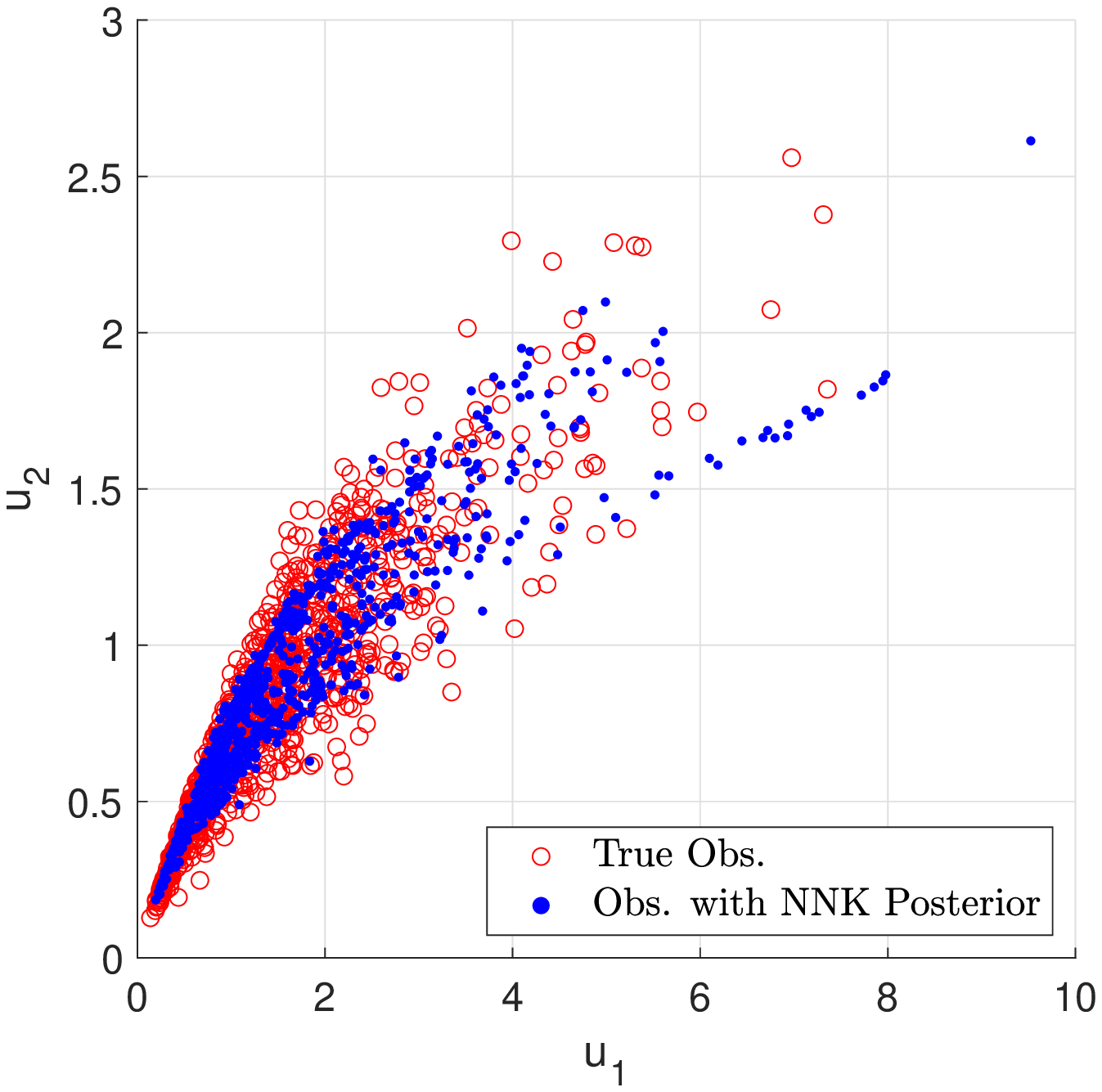}
\caption{\small{Distribution of observations (spring displacements) for two input variables $\bm x$, inside $\bm x = [0.501,0.502]$ (left) and outside $\bm x= [0.9,0.8]$ (right) the training dataset. }}\label{fig_sim_unimodal}
\end{figure}

In the final part of this example, we use the predicted samples of NNK, i.e. test samples associated with the second scenario (with permutation) to generate displacements $u_1$ and $u_2$ and make comparison with true observations. We consider two individual sample inputs, one within the range of training inputs i.e. $\bm x = [0.501,0.502]$ cf. Figure~\ref{fig_dist_true} and one without the range i.e. $\bm x= [0.9,0.8]$. The results are shown in Figure~\ref{fig_sim_unimodal}. For numerical investigation, we also compute the normalized error in the mean of $u_1$ and find $e_{\mu_{u_1}} = 6.37 \times 10^{-2},~7.51 \times 10^{-2}$ for both cases which are almost similar. 

Based on the results in this experiment, we deem our inference procedure yields accurate enough estimates for underlying parameters which could be used for prediction of simulation outputs. In the next examples, we directly compare the predictions obtained from our inference procedure and the ones obtained with a number of standard approaches.

%validation in two cases $[0.501,0.502]$ and $[0.9,0.8]$
%in the first case the mean of validation in the first direction: $6.37 \times 10^{-2}$
%in the second case the mean of validation in the first direction: $7.51 \times 10^{-2}$
%which are almost similar. 

\subsubsection{Bimodal and Irregular Distributions and Non-uniqueness of the Inverse Map}\label{Bi_Irr_nonunique}
In this example, we apply the inference procedure to a bimodal and an irregular distribution with the shape of U and investigate the performance of the procedure on a non-bijective map by considering $g(m)=m^2$ in Equation~\eqref{stf_sp}. 

\begin{figure}[!h]
\centering
\includegraphics[width=1.3in]{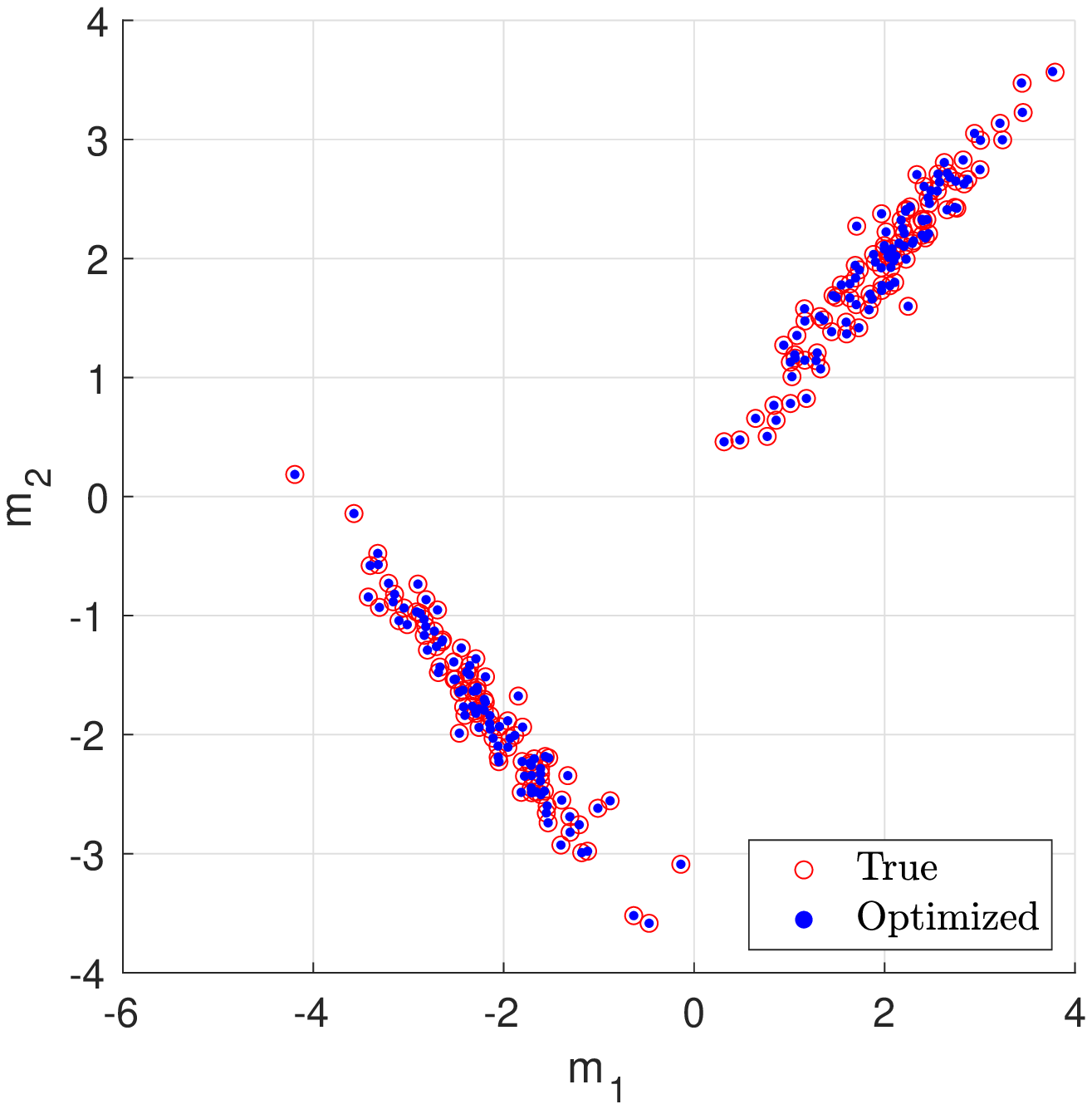}
\includegraphics[width=1.3in]{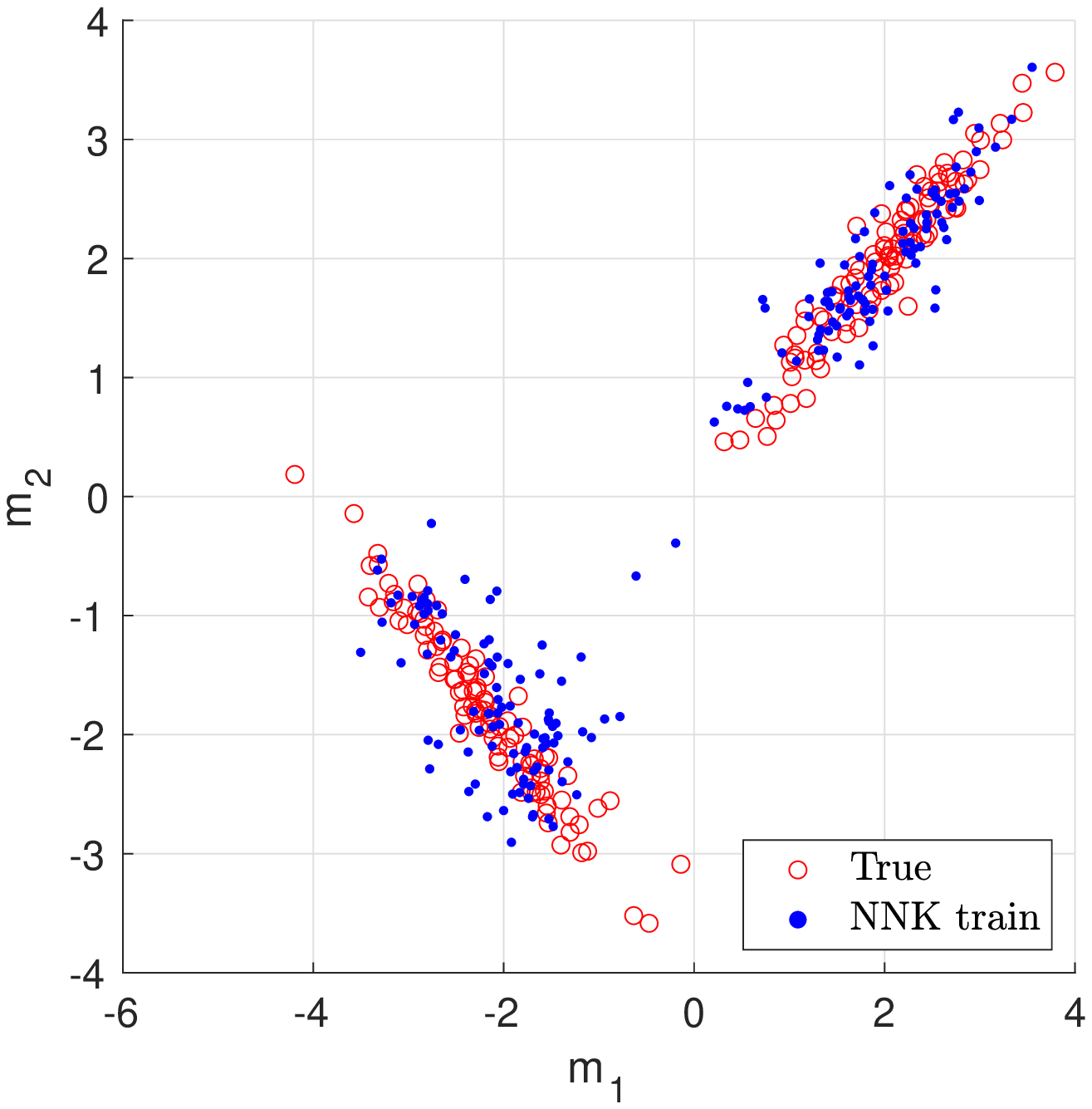}
\includegraphics[width=1.3in]{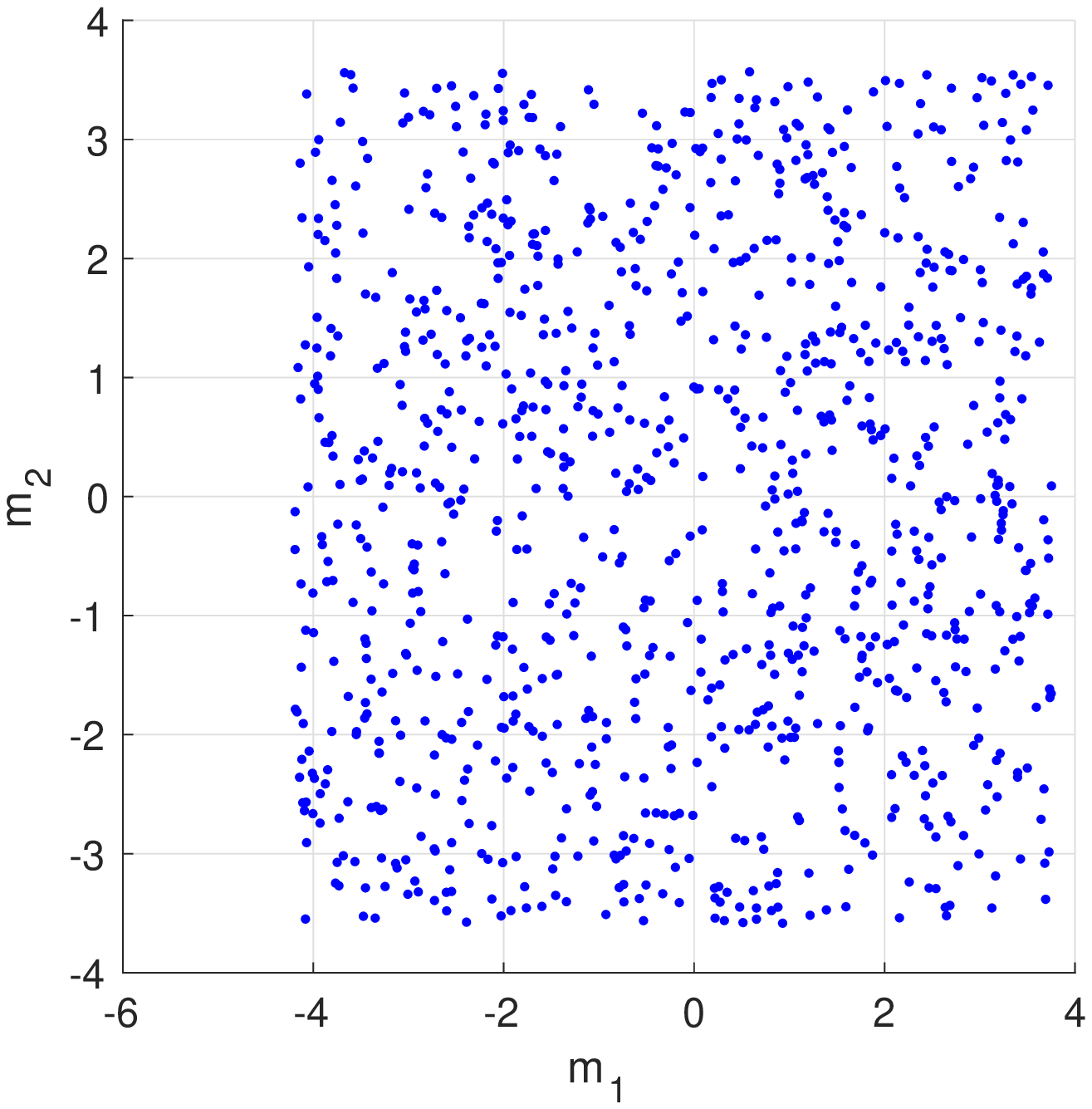}
\includegraphics[width=1.3in]{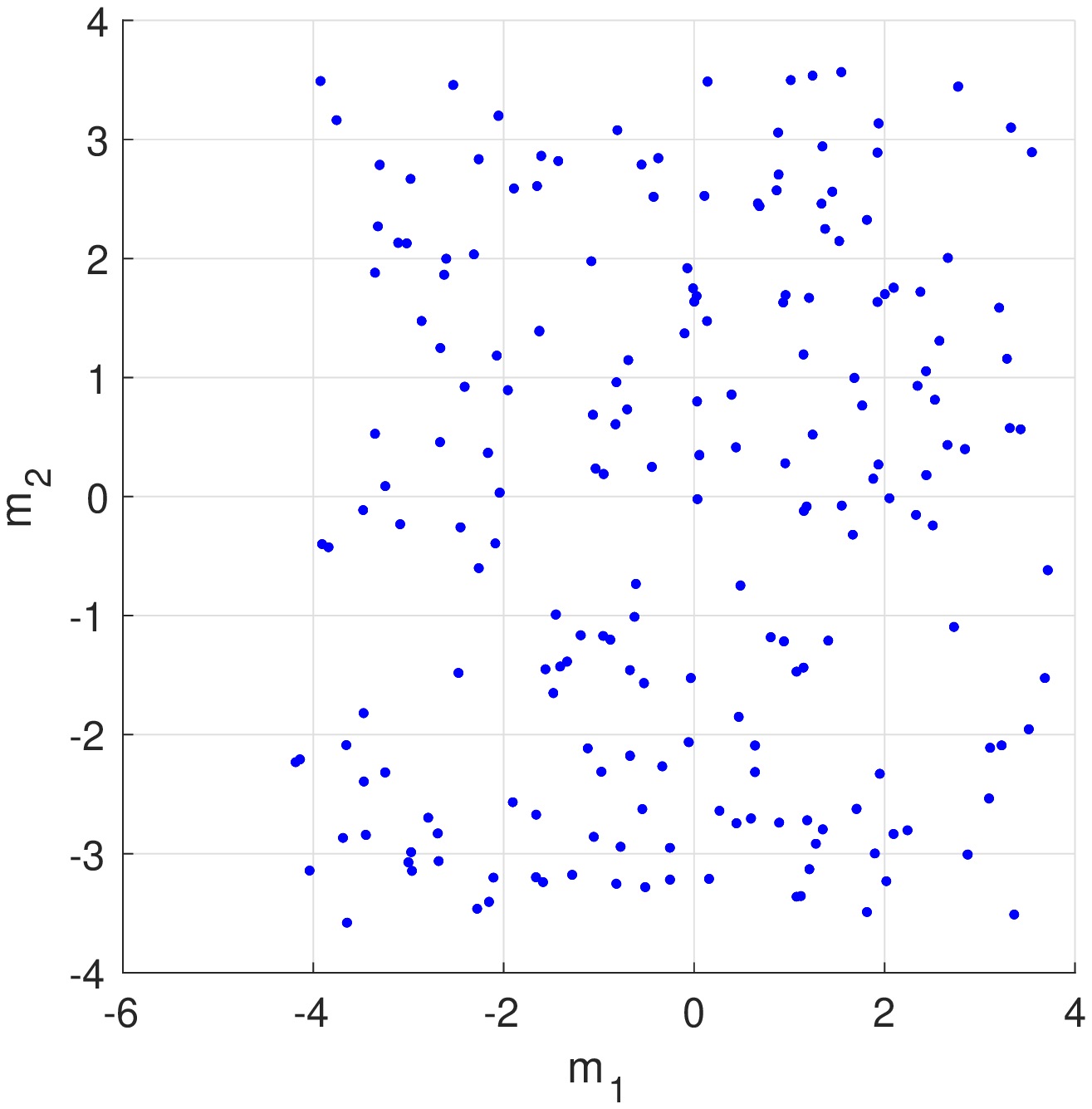}\\
\includegraphics[width=1.3in]{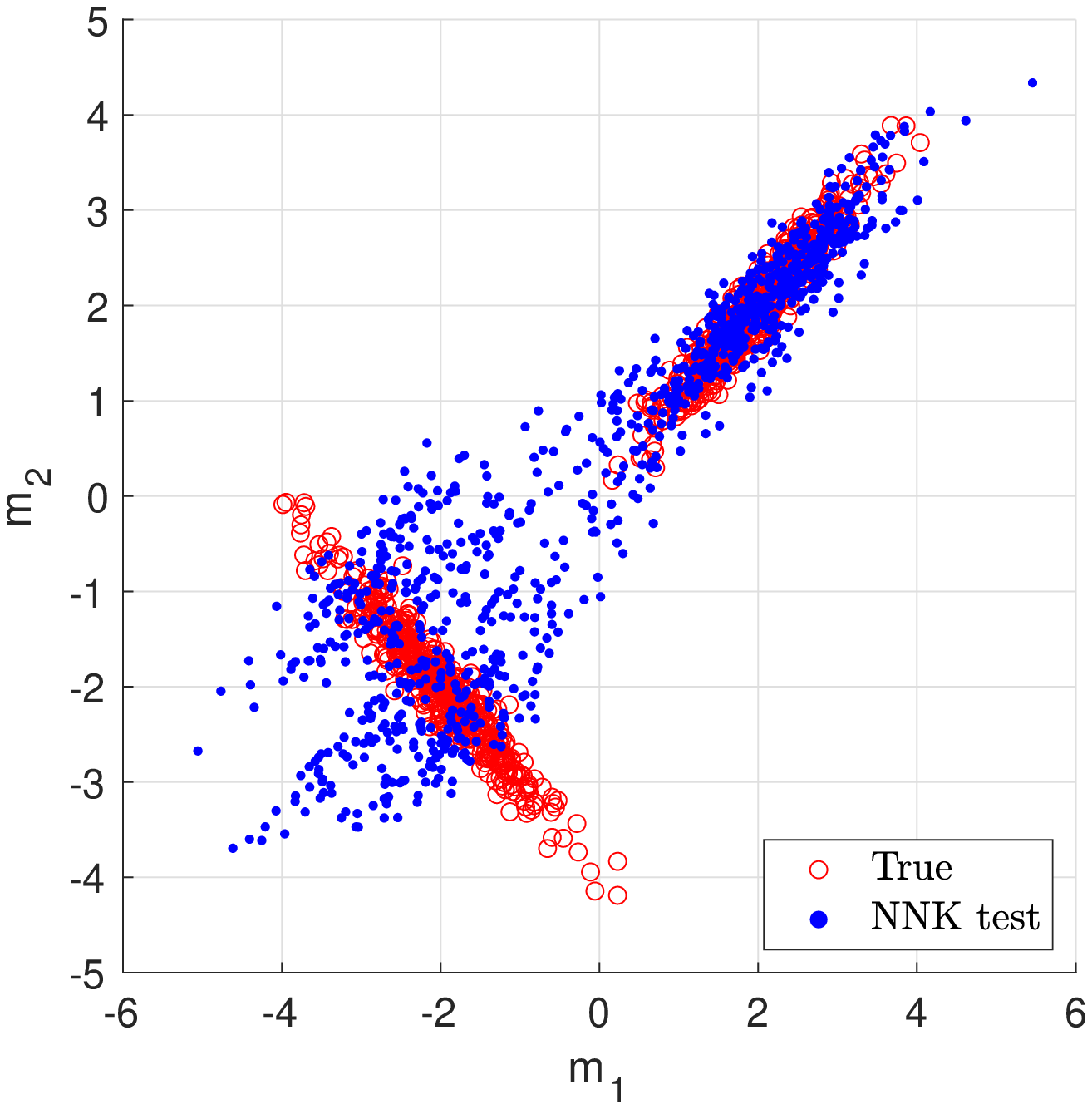}
\includegraphics[width=1.3in]{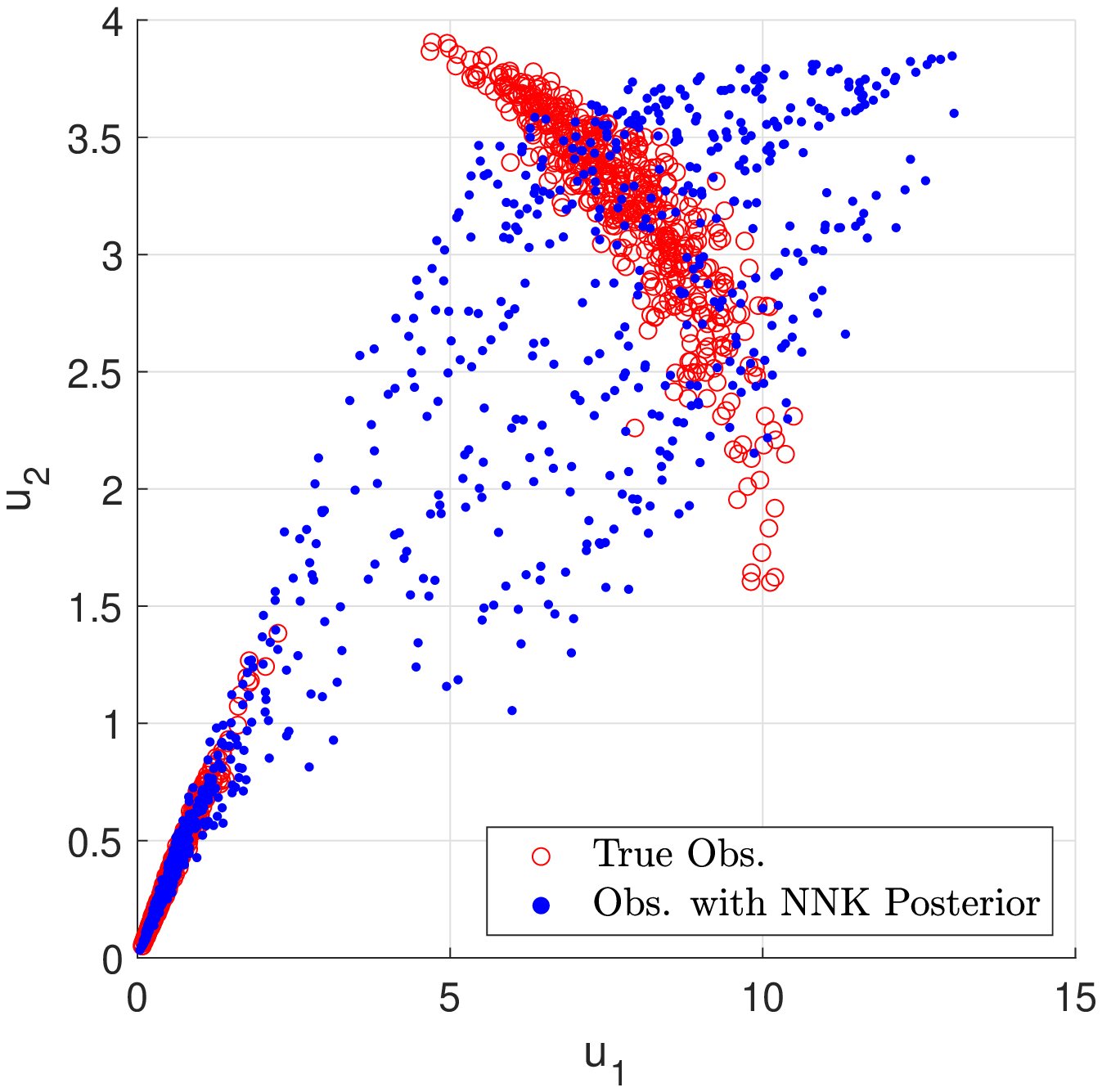}
\includegraphics[width=1.3in]{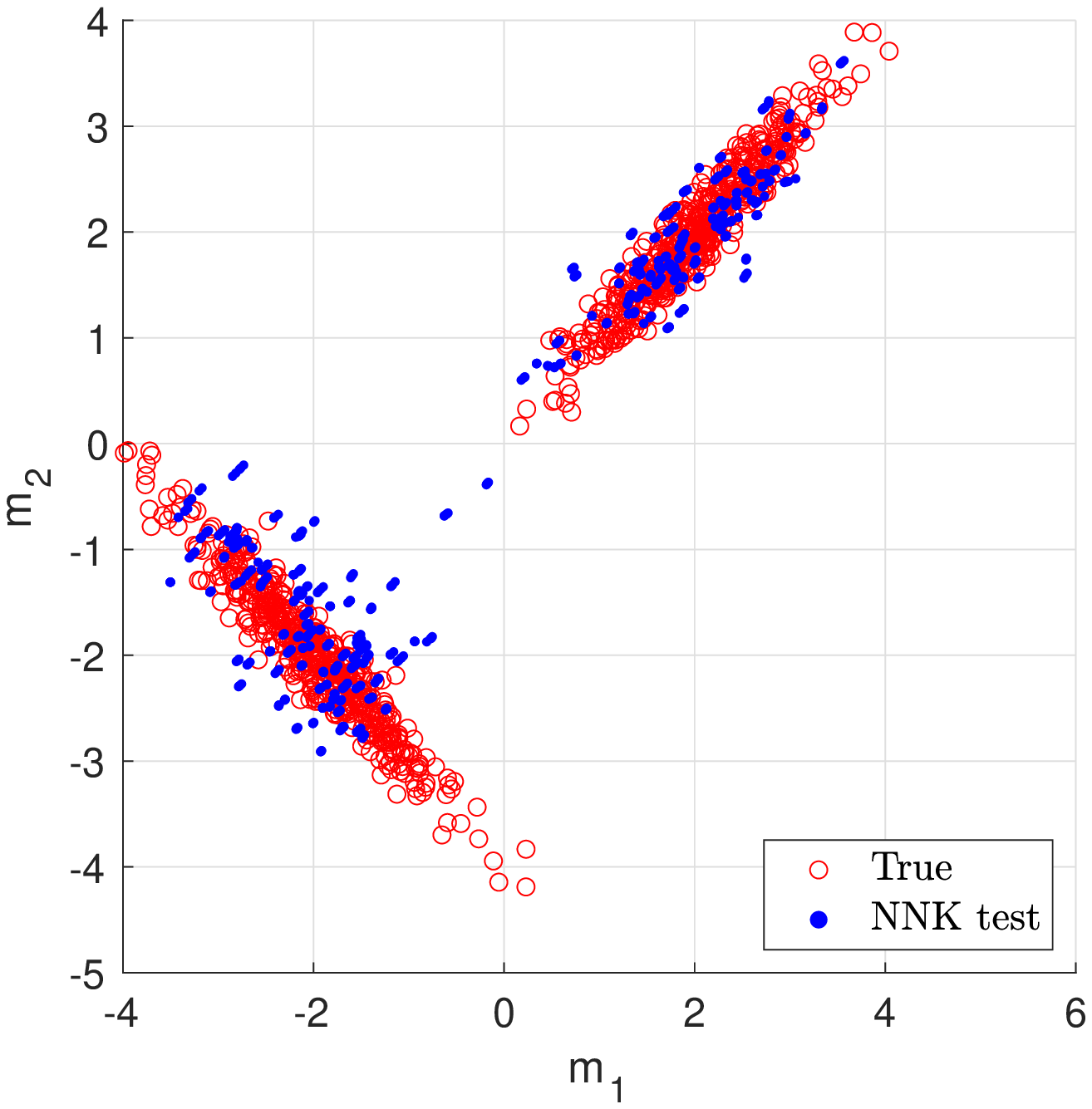}
\includegraphics[width=1.3in]{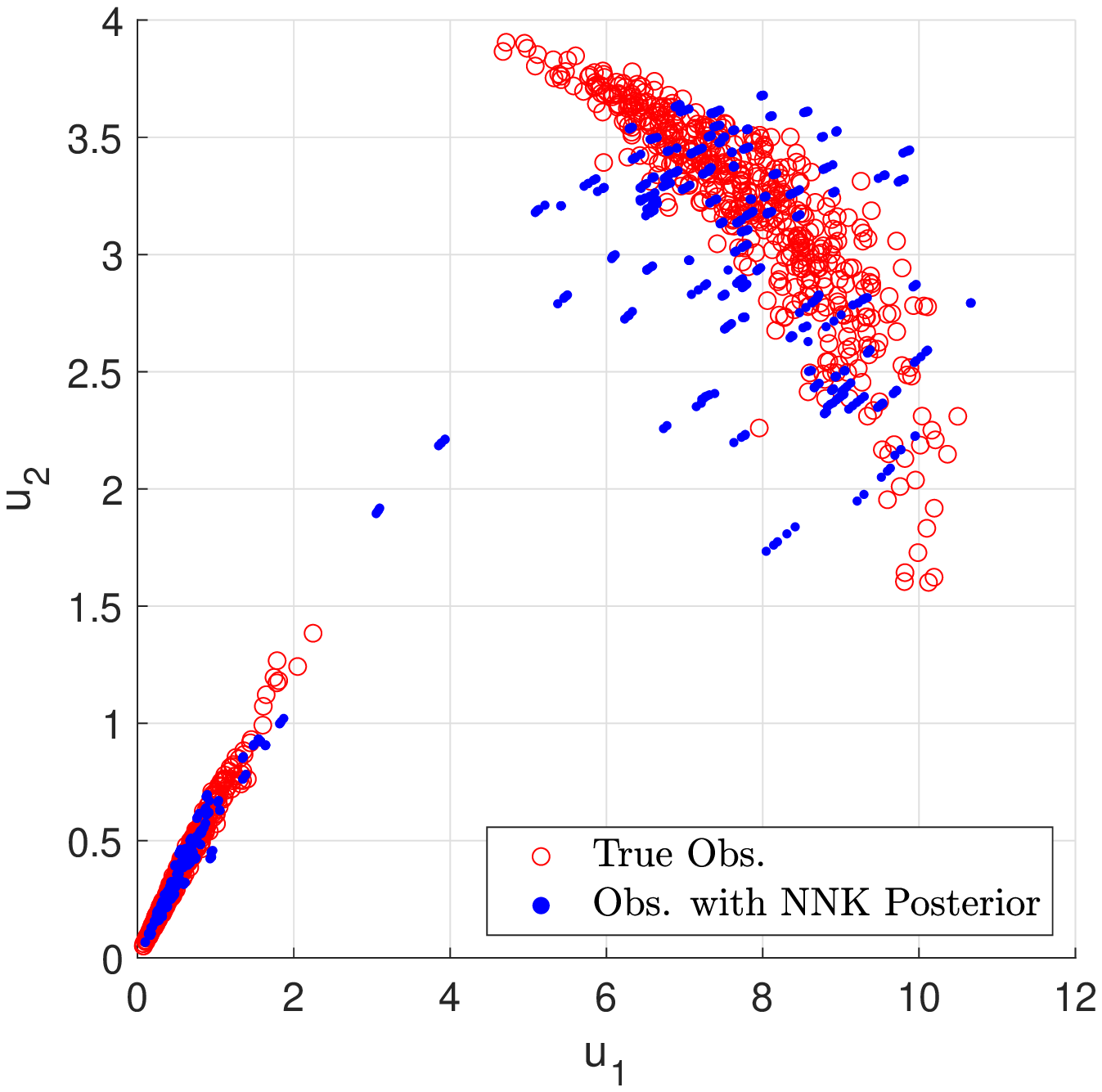}
\caption{\small{Inference procedure for a bimodal distribution: Top row: Inversion of the forward model (first), network training (second), randomly distributed prior samples (third), augmented training samples as prior (fourth); Bottom row: NNK prediction with randomly distributed prior samples and augmented training samples (first and third), distribution of observation using the latent parameter samples in the first and third panes (second and fourth). The result of training and testing in this plot will be compared with the result of MCMC sampling on the same bimodal dataset in Section~\ref{MCMC_numerics}. The NNK approach outperforms MCMC sampling in predicting bimodal distributions.}}\label{fig_bim_iden}
\end{figure}

The bimodal distribution is comprised of two equally weighted bivariate Gaussian distributions
i.e.
\begin{equation}\label{bim_dist}
\bm m \sim 0.5\mathcal{N}(\bm \mu_1, \bm \Sigma_1)+0.5\mathcal{N}(\bm \mu_2, \bm \Sigma_2)
\end{equation}
where
\begin{equation}\label{bim_dist_details}
\begin{array}{ll}
\bm \mu_1 =[2,2], & \bm \mu_2=[-2,-2] \\
\\
\bm \Sigma_1 = \begin{bmatrix}
0.51 & 0.49\\
0.49 & 0.51
\end{bmatrix}, & \bm \Sigma_2 = \begin{bmatrix}
0.51 & -0.49\\
-0.49 & 0.51
\end{bmatrix}
\end{array}
\end{equation}
The results of inference procedure, including the optimization in the first step are shown in Figure~\ref{fig_bim_iden}. The numerical parameters and the NNK architecture are similar to the previous example. Similarly to the previous example, the training residual saturates at some level as this map is also non-analytical, i.e. it converges to $\|\bm R\|=0.209$ after $n_{iter} =755$ iterations. 

The result of training is shown in the second pane of first row which appears to be promising for a challenging bimodal map. To generate test samples, we first generate $1000$ uniformly distributed samples according to  the scaled samples of prior training samples (top row, third pane). The NNK test samples for this case are shown in the first pane of the bottom row. The prediction of displacements associated with these samples with the input $\bm x = [0.9,0.8]$ is shown in the second pane. As can be seen, the test samples using a uniformly distributed prior exhibit a large scatter as seen in both first and second panes of the bottom row. To restrict these large scatters we consider a set of prior samples that are close to the training samples. To this end we add a small Gaussian noise to the prior training samples according to
\begin{equation}
\tilde{\bm m}_0\gets \bm m_0 + 0.002 \mathcal{N}(0,1)
\end{equation} 
where we refer to $\tilde{\bm m}_0$  as augmented training samples. Note that we add five noise values to each $\bm m_0$, therefore we generate $1000$ samples for testing from the original $200$ training samples. These augmented training samples are shown in the fourth pane of the first row. The results associated with this set of training samples are shown in the third and fourth pane of the bottom row. The samples in these cases are apparently less scattered and provide better representations for the bimodality. 
 
As a numerical investigation, similarly to the previous example, we compute the normalized error in the mean of $u_1$ and find $e_{\mu_{u_1}} = 4.53 \times 10^{-2},~7.61 \times 10^{-3}$ for both cases of uniformly distributed prior and augmented prior samples.  We also compute the error in standard deviation and find $e_{\sigma_{u_1}} = 7.13 \times 10^{-2},~3.80 \times 10^{-3}$ which we deem to be accurate for this very challenging distribution. The less accurate case of randomly distributed prior can still predict bimodality to some extent. For example, in a reliability analysis if we need to find the probability of $u_1 >4$ the obtained predictions could still provide a viable estimate. In many other approaches such as BNN (as we will see in next examples), the output is only predicted as a unimodal distribution in a narrow range. Therefore in those cases, it is possible that the probability of  $u_1 >4$ is erroneously estimated as zero.

%We do the bimodal: the tolerance reaches $0.2092$. We perform the optimization, then permute, then we do train shown and we do test by performing xtest and augmented xtest with noise 0.002. %the error of mean of validation in first direction $0.0453$ and $0.0076$

%but the error in standard for samples is significant $u1>4$ for reliability analysis
%$0.8073$ $0.0139$
%In the next examples we directly compare the performance of the approach on this example with other approaches. 

%We then perform the same procedure on a U shape distribution using both maps.
%We do the U: the tolerance reaches $0.2307$ for exp
%the square is more challenging since the mapping should work in both directions
%we consider a larger net $\{[2,20],[20,7],[7,4],[4,1]\}$  
% and larger $\alpha$ i.e. $\alpha \in \mathbb{R}^{40 \times 2}$...the residual reaches $0.2513$

\begin{figure}[!h]
\centering
\includegraphics[width=1.3in]{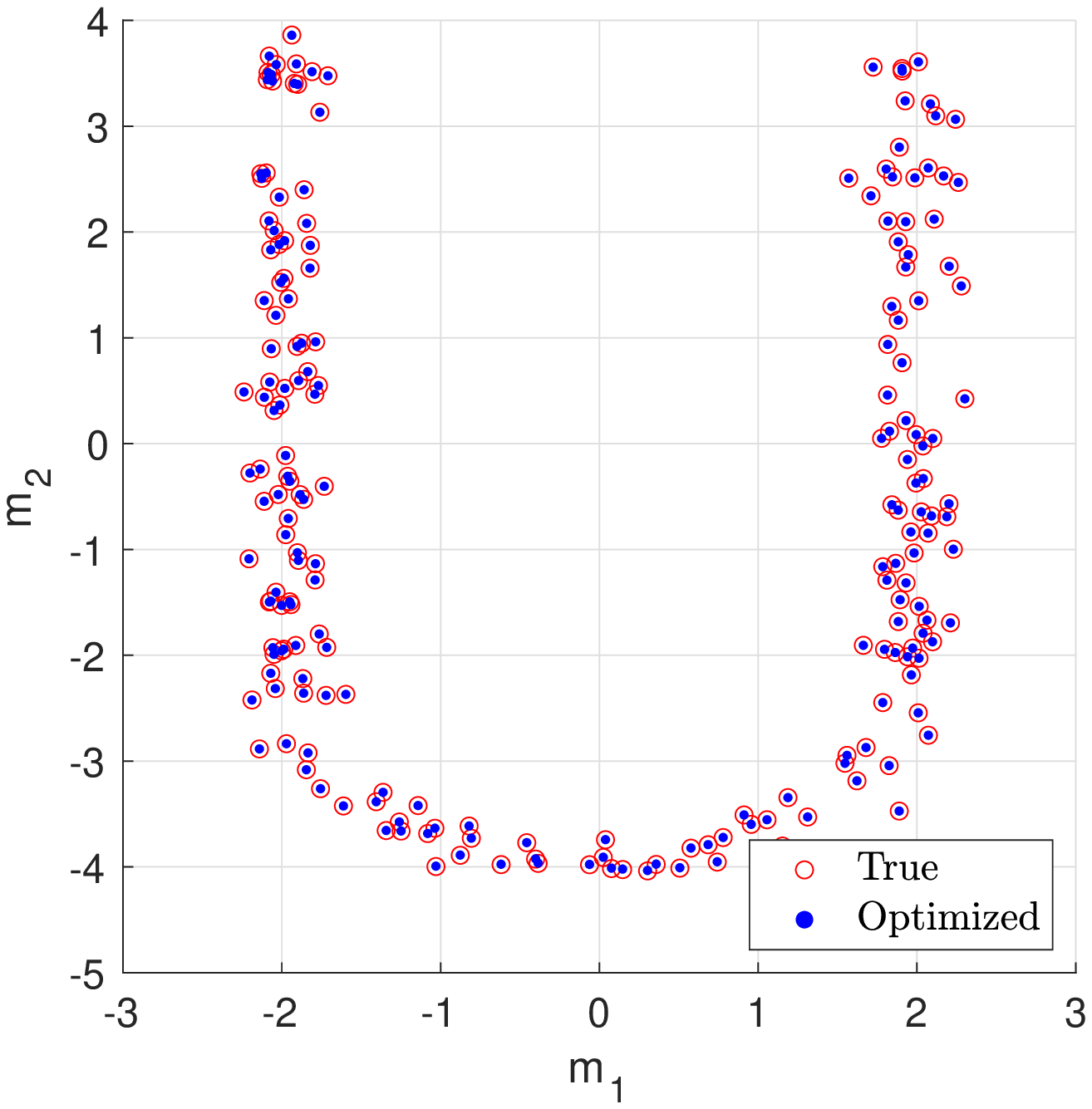}
\includegraphics[width=1.3in]{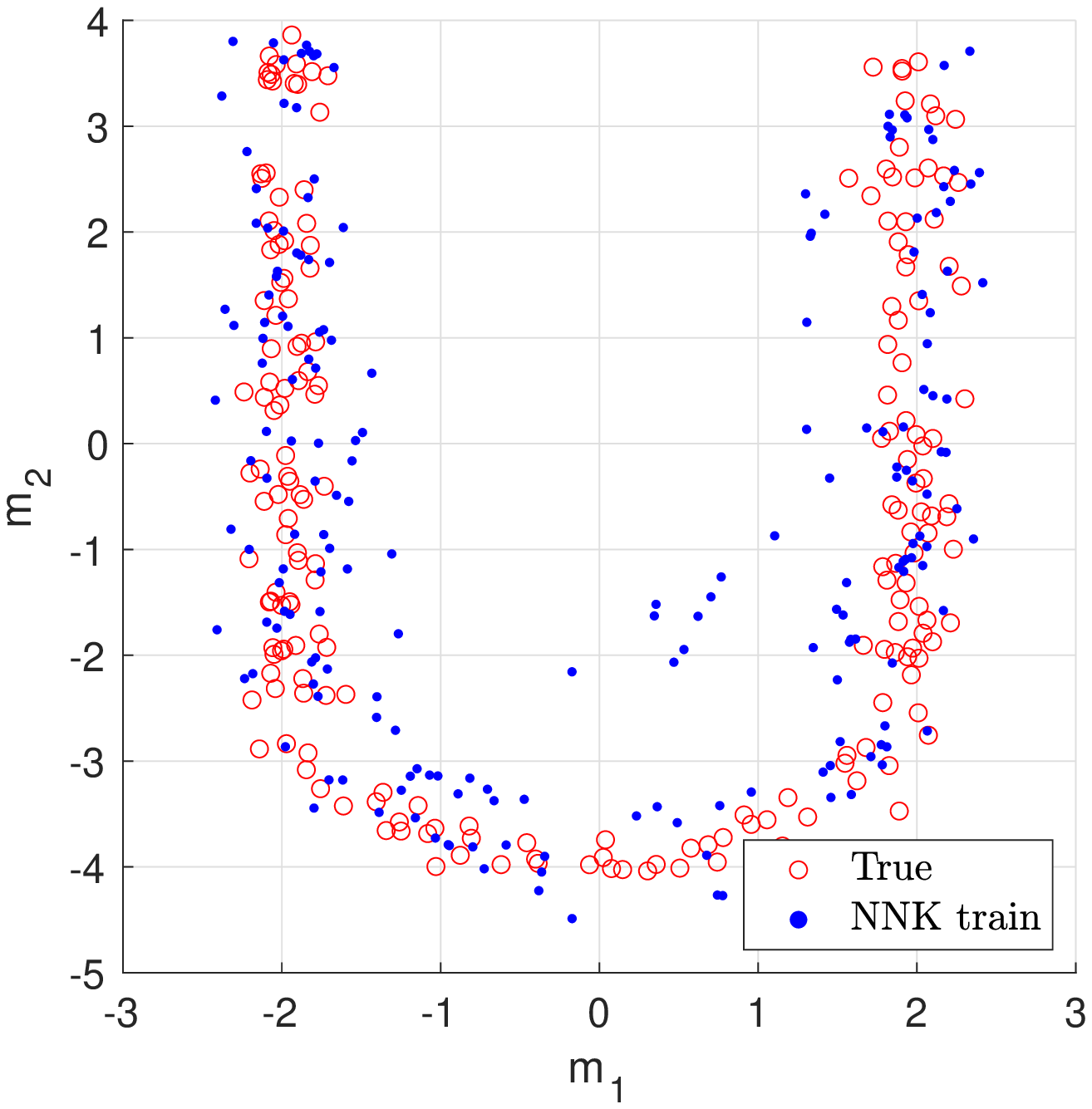}
\includegraphics[width=1.3in]{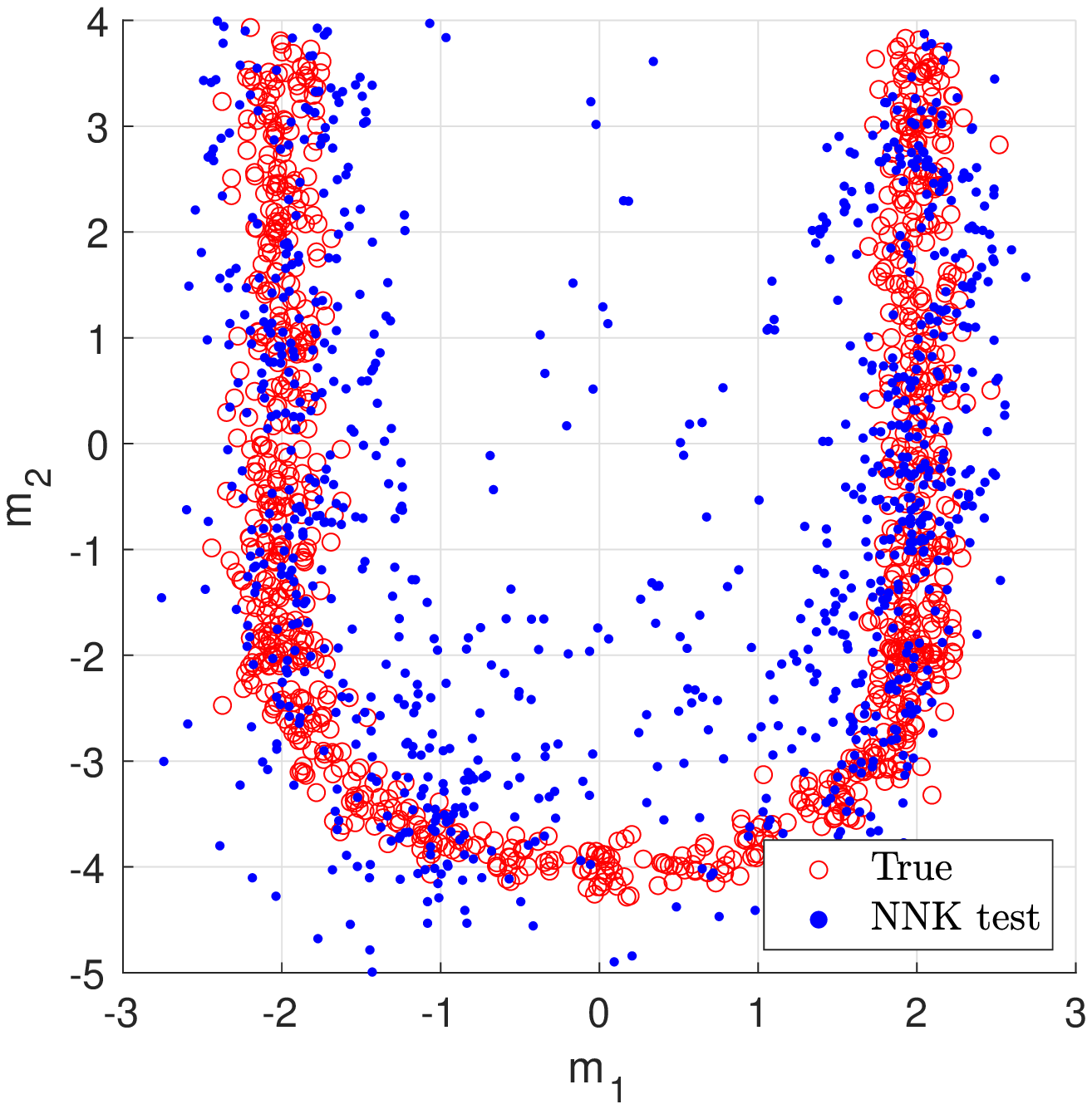}
\includegraphics[width=1.3in]{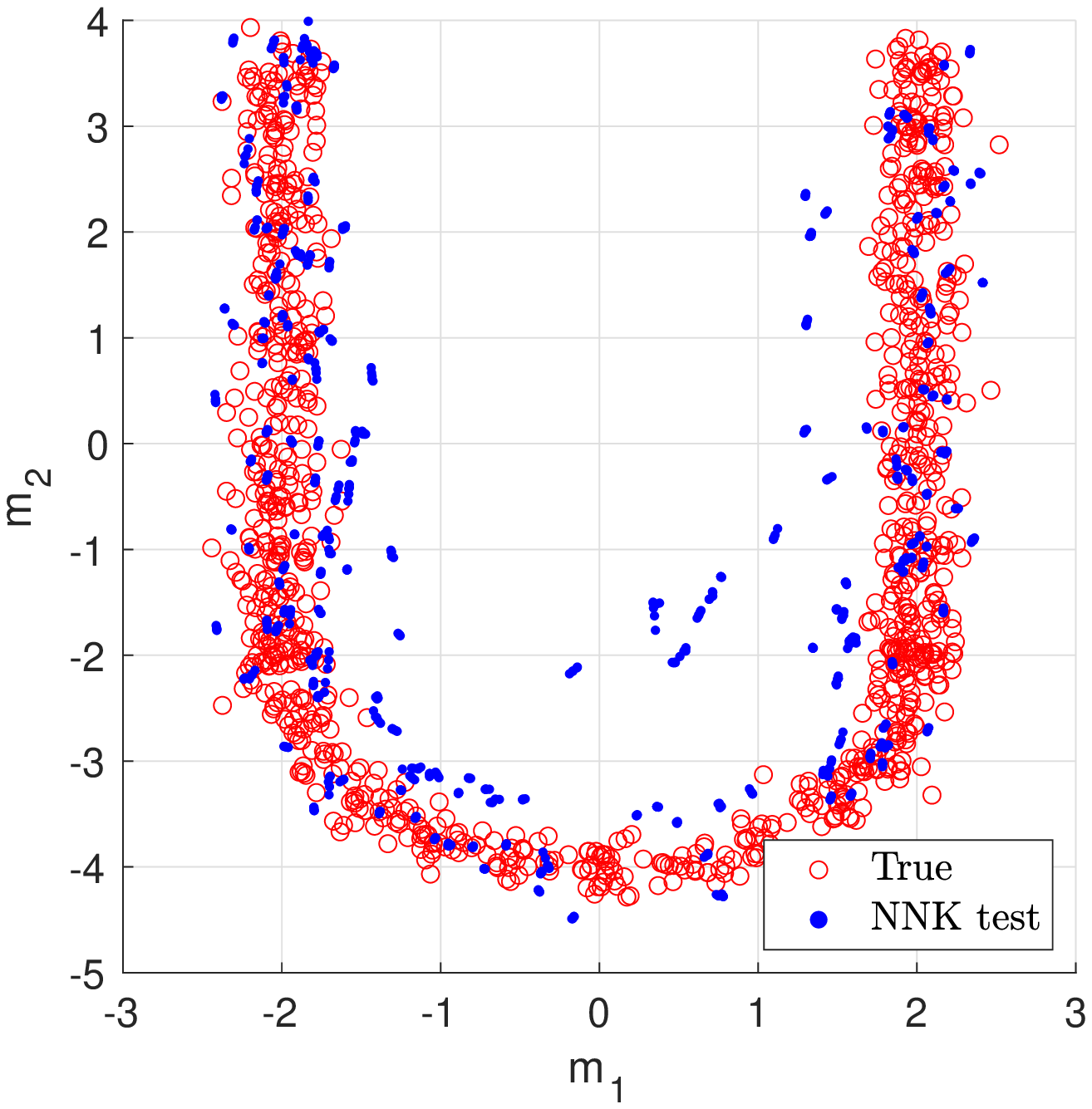}\\
\includegraphics[width=1.3in]{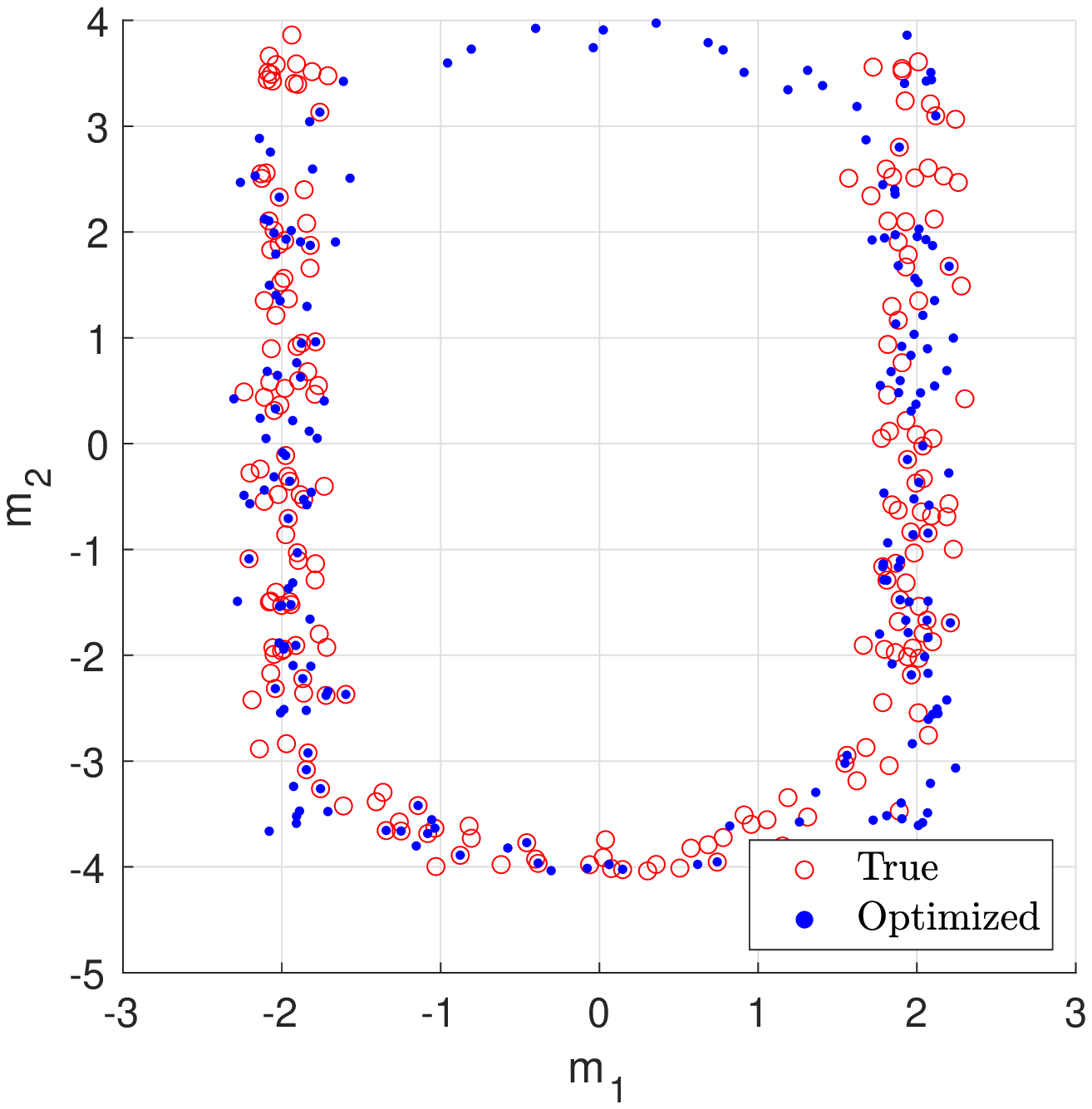}
\includegraphics[width=1.3in]{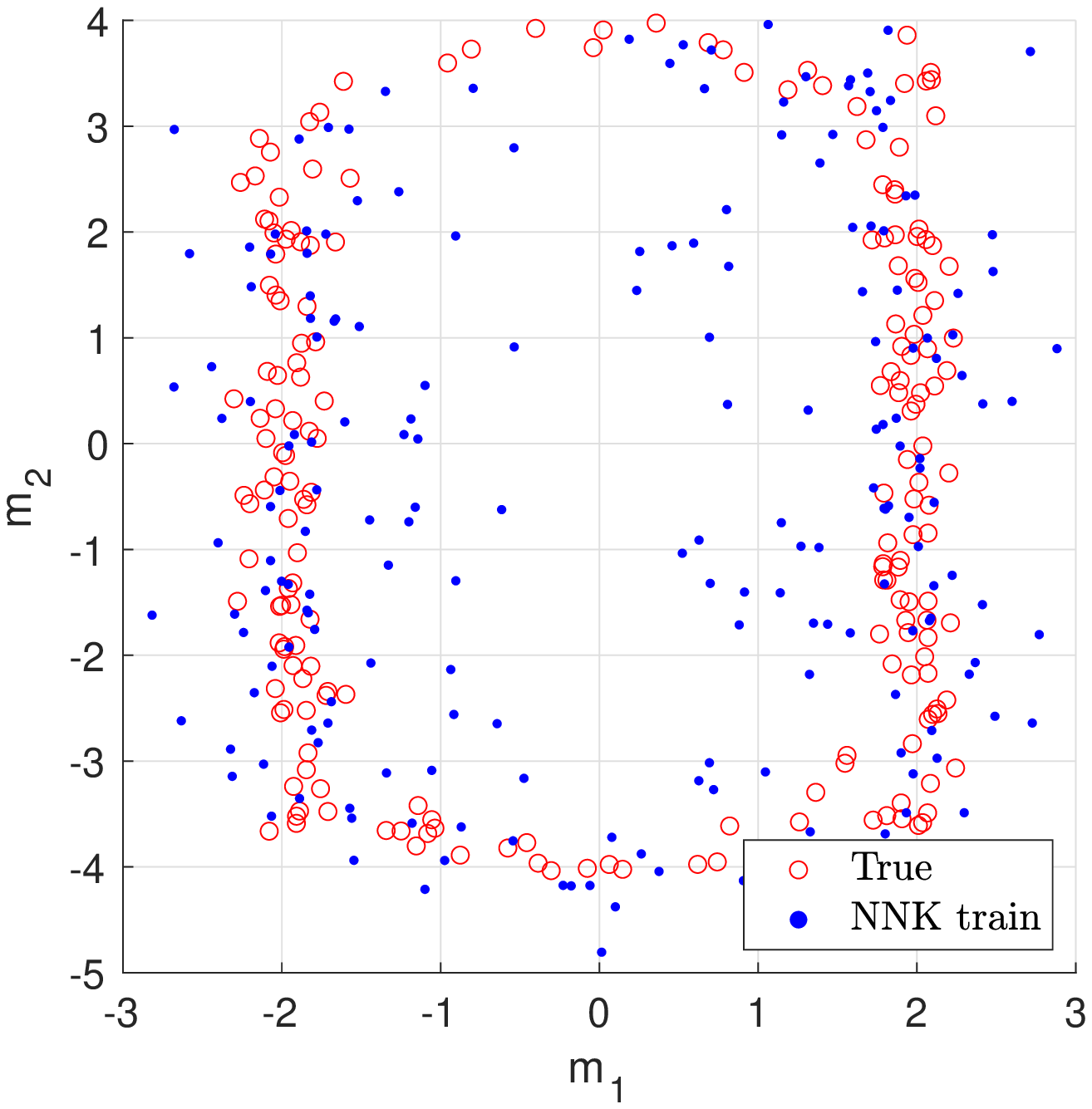}
\includegraphics[width=1.3in]{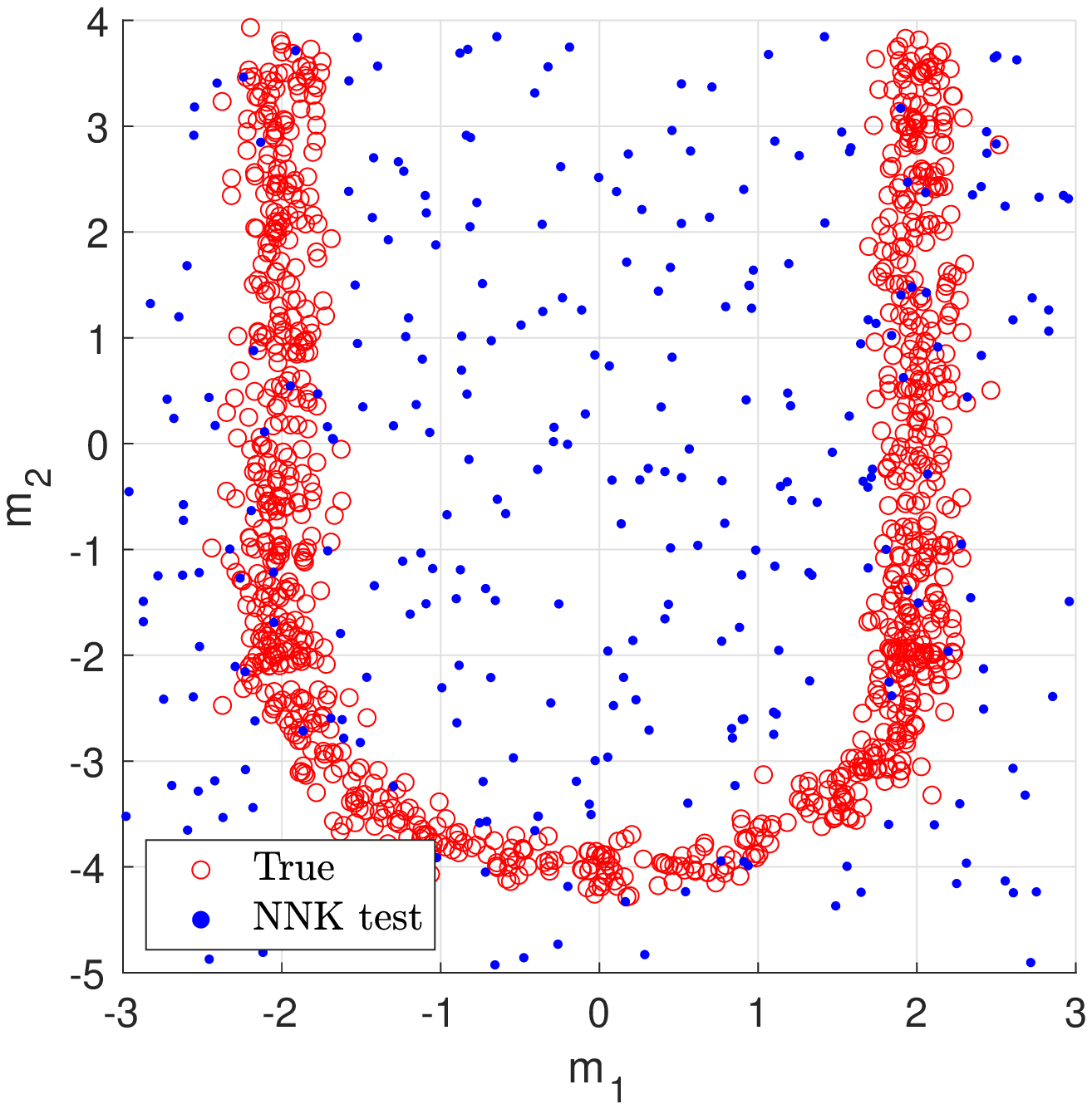}
\includegraphics[width=1.3in]{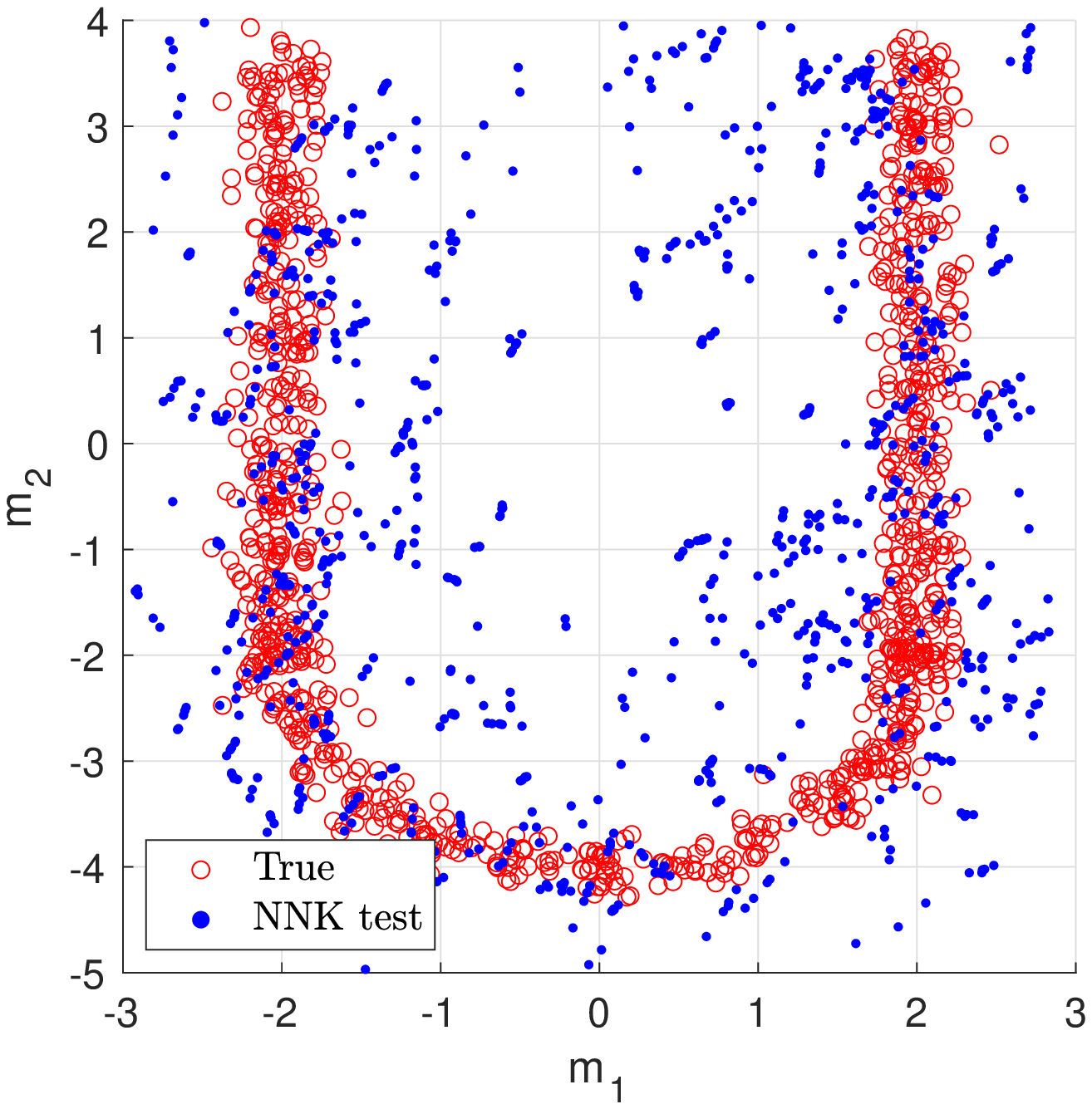}
\caption{\small{Inference procedure for an irregular distribution $U$ with two forward maps $g(m)=\exp(m)$ (top row) and $g(m)= m^2$ (bottom row): Inversion of observation (first), NNK training (second), NNK prediction with randomly distributed prior samples and augmented training samples (third and fourth). As expected, the predictions with a one-to-one map i.e. $g(m)=\exp(m)$ is more accurate than the many-to-one map i.e. $g(m)=m^2$. See the numerical results associated with Figure~\ref{fig_U_obs} below.}}\label{fig_U_iden}
\end{figure}

We now apply the same procedure to the distribution with the shape of $U$. The samples are generated from a Gaussian mixture model with several centers that are located on a shape of U. For this example we also consider a larger network, i.e. we build an NNK with five layers and $[2,20,7,4,1]$ nodes with $\bm \alpha \in \mathbb{R}^{40 \times 2}$. We perform the inference procedure on both forward models with $g(m)=\exp(m)$ and $g(m)=m^2$. The results are shown in Figure~\ref{fig_U_iden}: The top and bottom rows show results for $g(m)=\exp(m)$ and $g(m)=m^2$ respectively. The main difference is in the first step where the non-uniqueness of the forward map $g(m)=m^2$ results in emergence of samples on top of the U shape. 

The NNK appears to be relatively successful in learning the U shape. The test samples for both 
uniformly distributed prior samples and augmented training samples exhibit the shape of U. However in the case of $g(m)=m^2$, since the NNK training is done with non-true samples the resulting performance in identification of the true underlying parameter is inferior which could be seen visually. 
 
\begin{figure}[!h]
\centering
\includegraphics[width=1.3in]{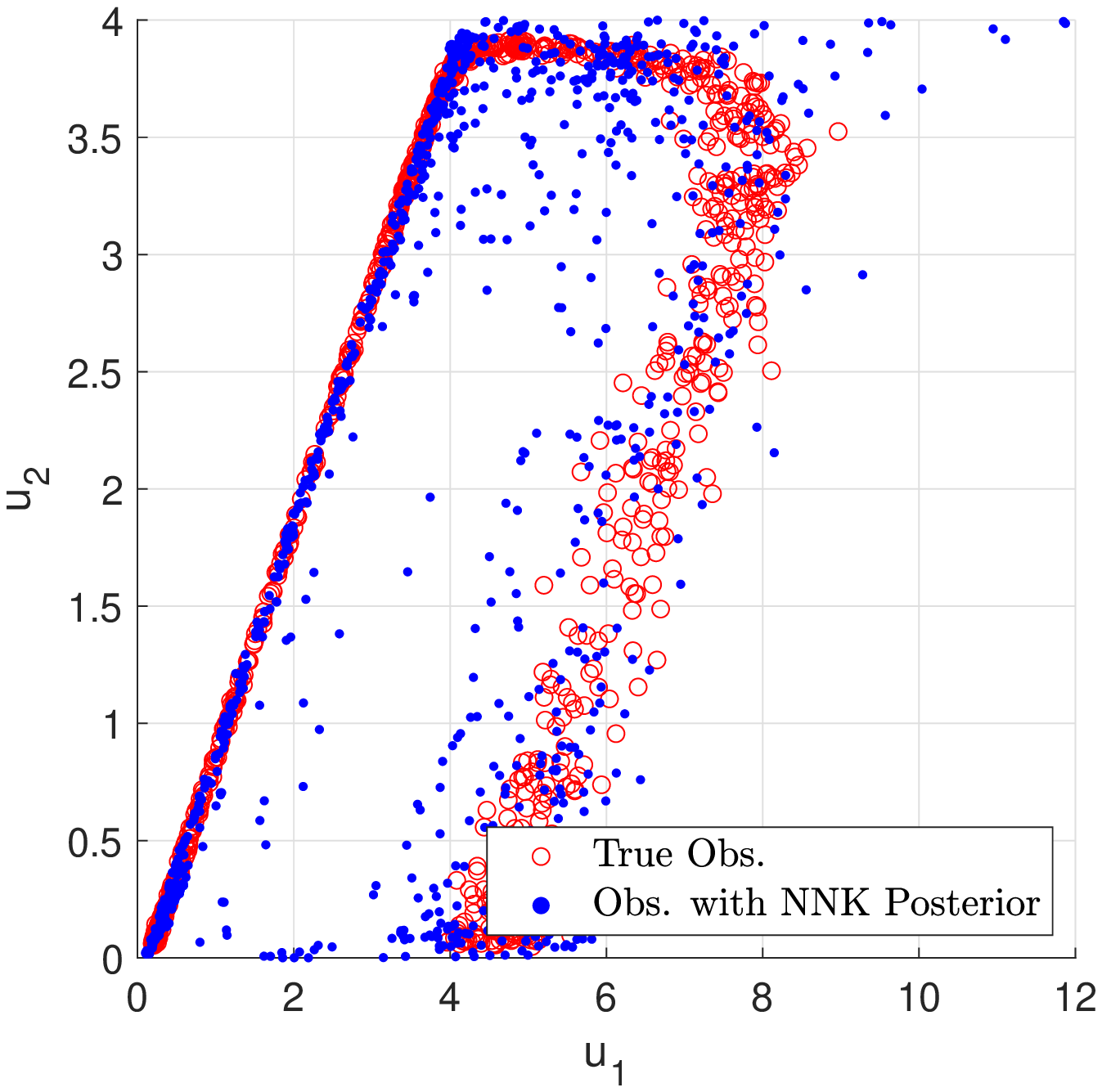}
\includegraphics[width=1.3in]{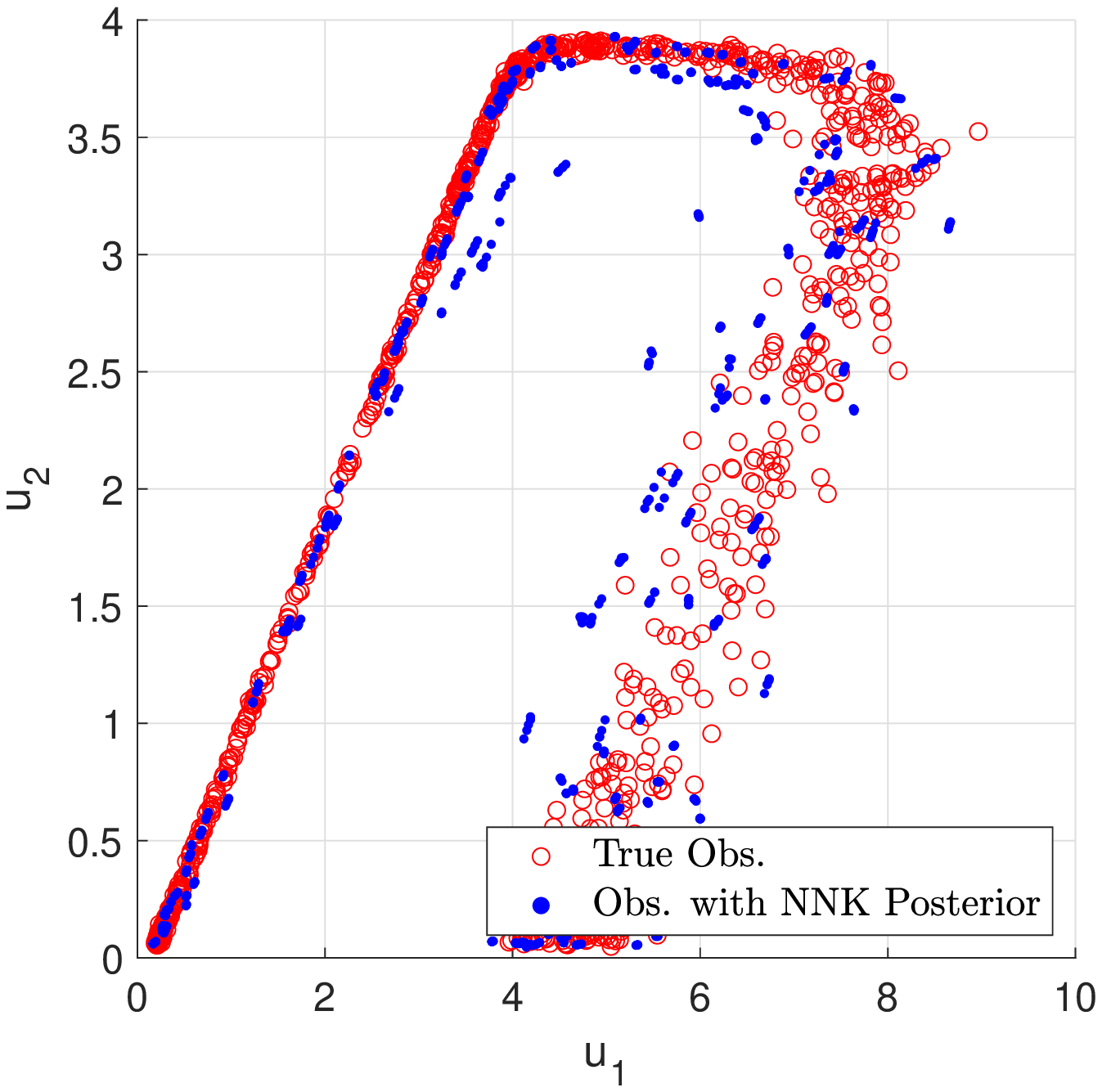}
\includegraphics[width=1.3in]{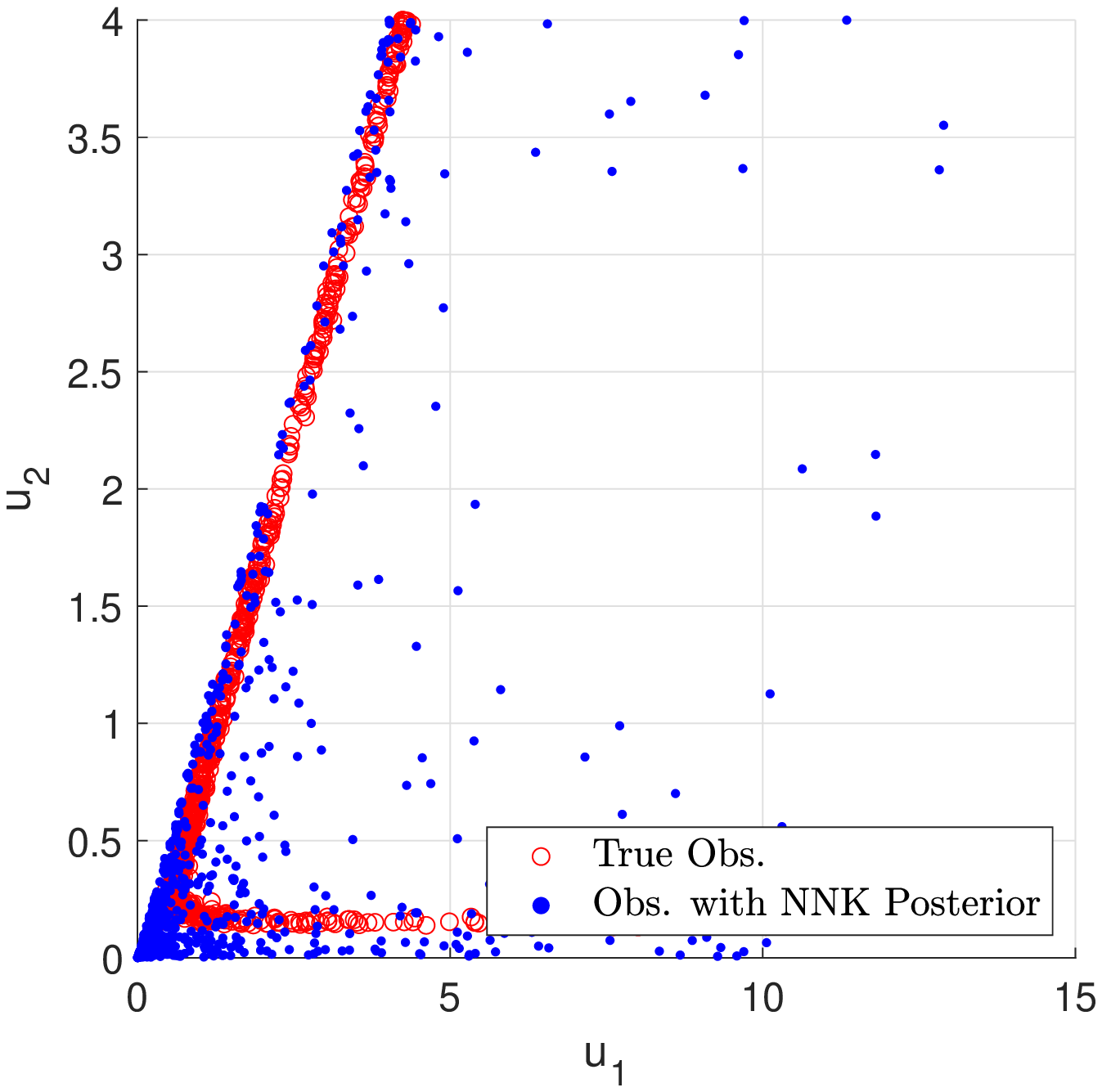}
\includegraphics[width=1.3in]{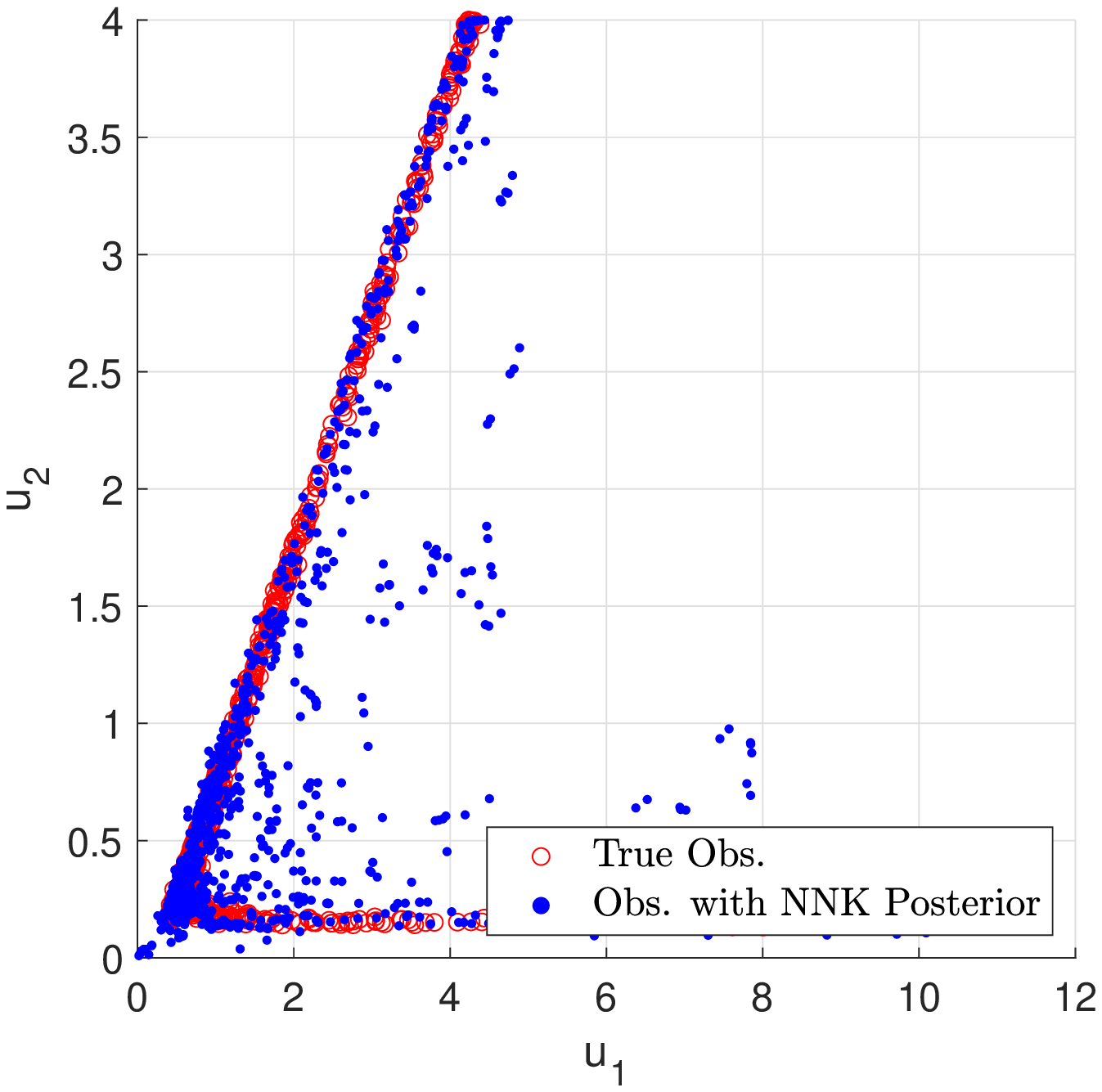}
\caption{\small{Forward model outputs computed with $\exp(\bm m)$ (first two panes) and $\bm m^2$ (second two panes). The left and right panes in each pair are associated with uniformly distributed prior samples and augmented training samples. The result of output prediction in this plot (for $g(m)=\exp(\bm m)$) will be compared with the result of prediction from a Bayesian neural network in Section~\ref{BNN_numerics}. The NNK approach outperforms BNN in predicting the output.}}\label{fig_U_obs}  %The error in mean of first variable is $0.0799,~0.0132,~0.1775$ and $0.3058$ and in the second direction is $0.0432, 0.0454,~0.4634$ and $0.1120$.}}\label{fig_U_obs}
\end{figure}

We use the test samples for both cases of $g(m)=\exp(m)$ and $g(m)=m^2$ and both cases of uniformly distributed prior samples and augmented training samples to generate forward model outputs on $\bm x=[0.9, 0.8]$. The results are shown in Figure~\ref{fig_U_obs}. The normalized errors in the mean of $u_1$ and $u_2$ associated with the prediction in the four panes of  Figure~\ref{fig_U_obs} are $e_{\mu_{u_1}}=7.99 \times 10^{-2},~1.32 \times 10^{-2},~1.77 \times 10^{-1},~3.05 \times 10^{-1}$ and $e_{\mu_{u_2}}=4.32 \times 10^{-2},~4.54 \times 10^{-2},~4.63 \times 10^{-1},~1.12 \times 10^{-1}$. Based on these results, the predictions associated with $g(m)=\exp(m)$ are more accurate than the ones with $g(m)=m^2$. Also, there are cases that the uniformly distributed training samples yield more accurate estimates than the augmented training samples. For example, the mean of $u_1$ associated with $g(m)=m^2$ is predicted more accurately using the uniformly distributed training samples.

\subsubsection{Comparison with a MAP-based Inference} In this example, we investigate the MAP-based formulation as described in Section~\ref{Sec2_1}. To this end, we consider the dataset associated with the unimodal distribution. The MAP-based approach involves a deterministic optimization cf. Equation~\eqref{opt_main}. To perform the optimization, we need to assign values to $\Gamma_{noise}$ and $\Sigma_0$. Specifying these values is one of the inherent challenges in Bayesian inference as they are usually unknown. As we have no prior knowledge we assign identity matrix to these quantities. 

\begin{figure}[!h]
\centering
\includegraphics[width=1.3in]{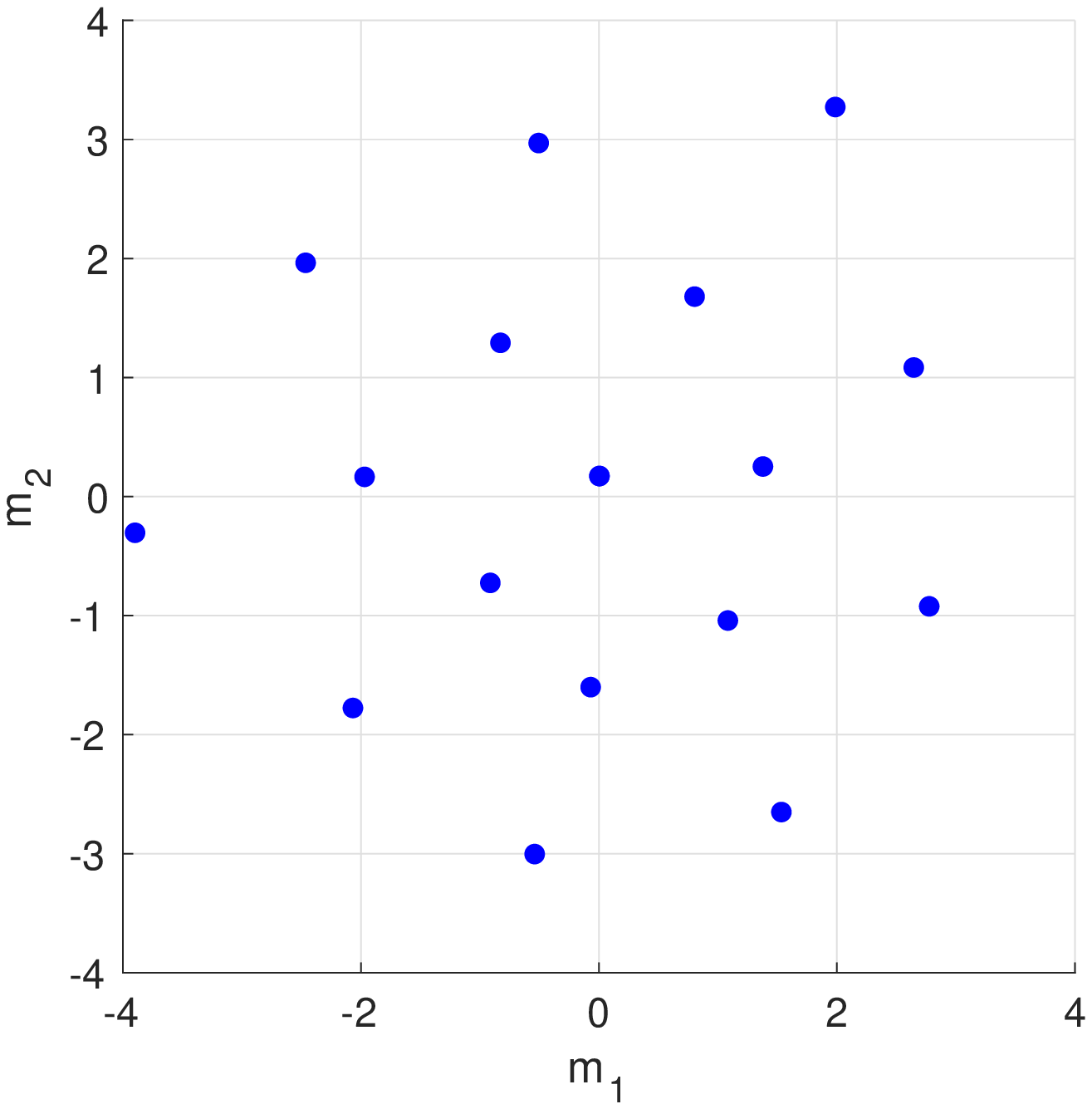}
\includegraphics[width=1.4in]{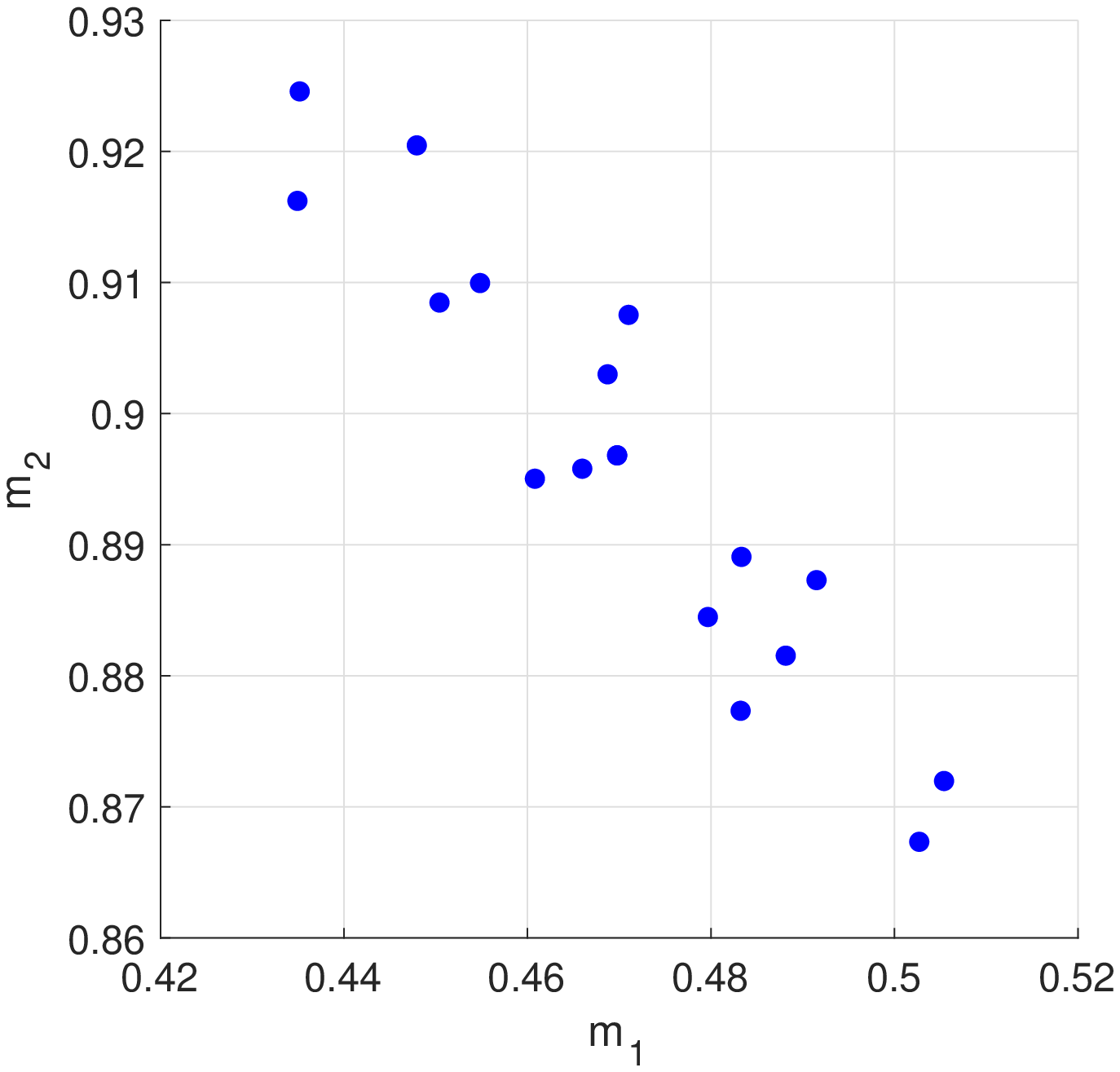}
\includegraphics[width=1.35in]{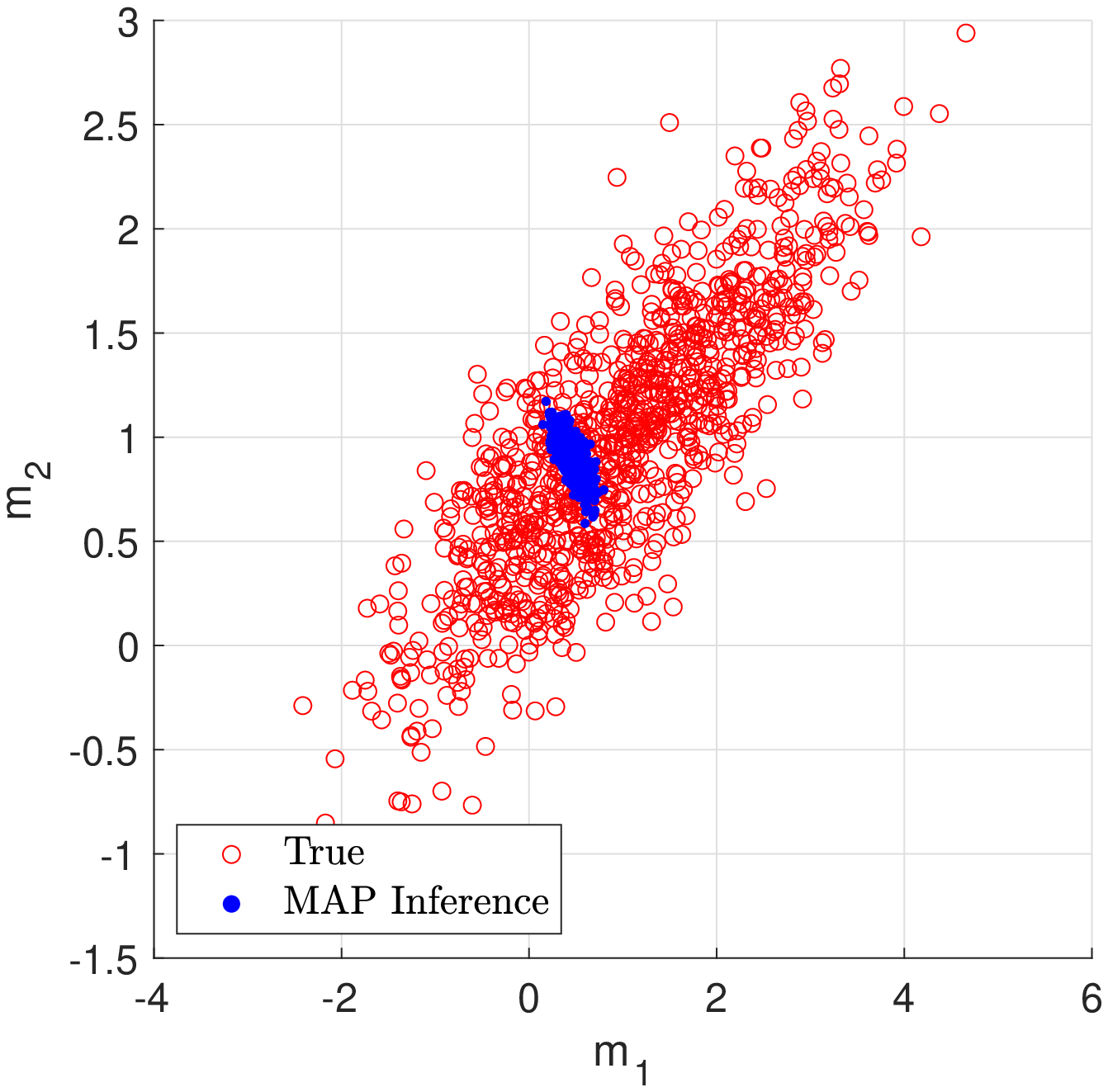}
\includegraphics[width=1.35in]{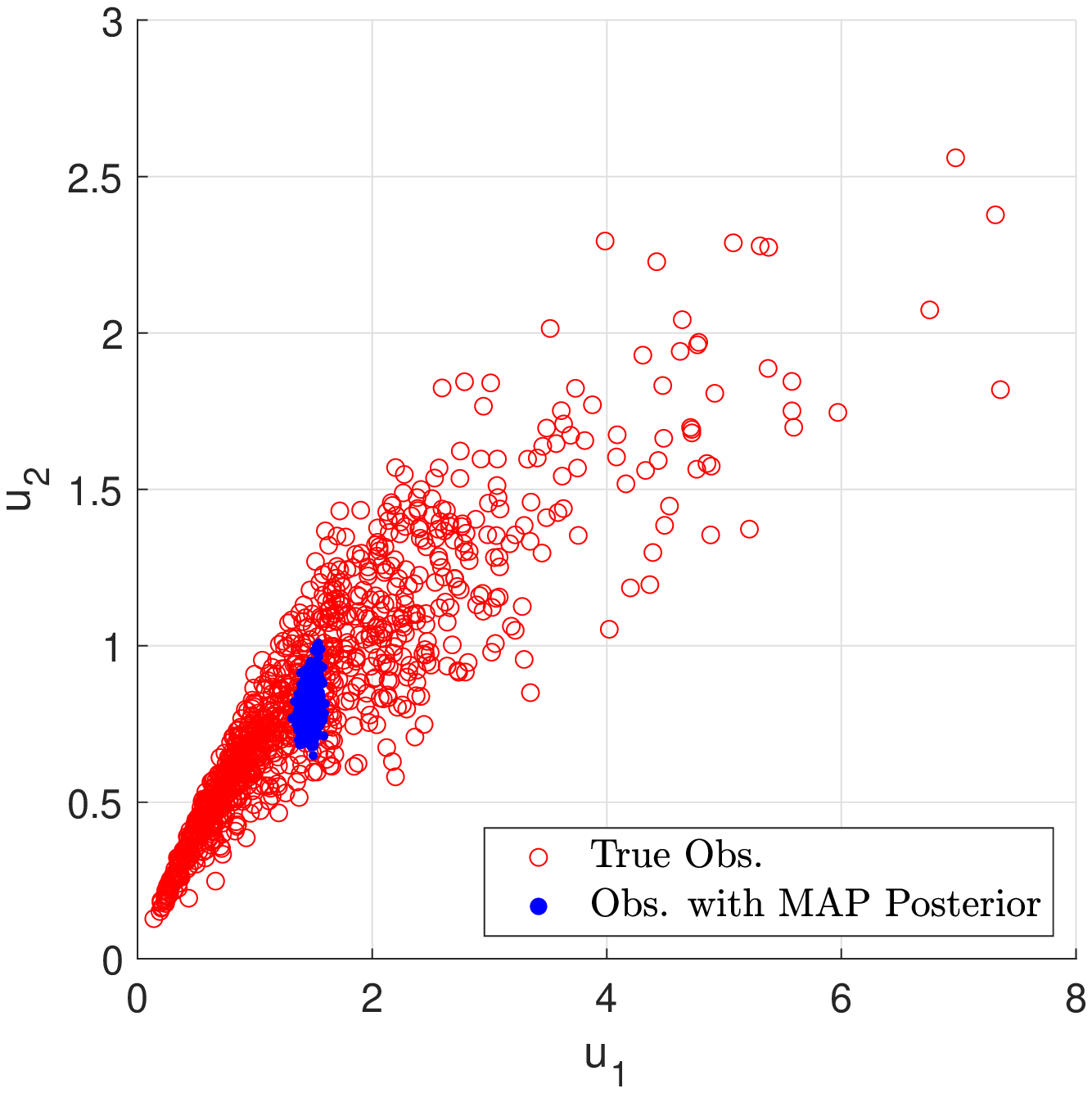}
\caption{\small{MAP-based inference: prior and posterior samples (first and second panes), distribution of the latent parameter $\bm m$ (third), and output (observations) obtained by the true distribution and the one inferred by the MAP-based approach.}}\label{fig_MAP}
\end{figure}

Assuming a prior sample for $\bm m_0$ we can compute the posterior mean  $\bm m_1$ by solving Equation~\eqref{opt_main} which consequently yields the hessian of $\mathcal{P}$ using the BFGS approach. To get a richer approximation for the posterior we assume several prior samples and perform the MAP-based approach to find associated mean $\bm m_1$ and the covariance $\bm \Sigma_1$. To this end, we consider designed quadrature nodes (obtained for Gaussian weight)~\cite{Keshavarzzadeh_DQ} as prior samples and compute the MAP point and the covariance associated with them individually. Once we find the MAP points and the covariances, we estimate the posterior probability as the Gaussian mixture model with uniform weights, i.e.
\begin{equation}\label{MAP_pdf}
p(\bm m_1) = \sum_{i=1}^{n_{dq}} (1/n_{dq}) \mathcal{N}(\bm m_1^{(i)},\bm \Sigma_1^{(i)}). 
\end{equation}
The quadrature rule in this example is comprised of $n_{dq}=17$ points. The prior samples (designed quadrature nodes) and optimized MAP points  are shown in the first and second panes of Figure~\ref{fig_MAP}. The posterior samples drawn from the distribution~\ref{MAP_pdf} are shown in the third pane. Comparing this result with the result shown in the fourth pane of Figure~\ref{fig_permute}, it is seen that our inference procedure provides an obviously closer distribution to the underlying true distribution. As we mentioned before, the typical Bayesian approaches yield only averaged-type/like distributions which is apparent in these results. We also further use the identified samples into the forward model and compute the displacements for the input $\bm x= [0.9,0.8]$. The result is shown in the fourth pane. We compute the normalized errors in mean and standard deviation of $u_1$. The normalized error $e_{\sigma_{u1}}=2.57$ is significant for the MAP-based approach. Similar calculation on the right pane of Figure~\ref{fig_sim_unimodal} yields $e_{\sigma_{u1}}=4.36 \times 10^{-1}$ which is one order of magnitude more accurate than the MAP-based approach. In terms of the mean, we reported $e_{\mu_{u1}}=7.51 \times 10^{-2}$ for our inference procedure. The MAP-based approach yields slightly better estimate for the mean, i.e. $e_{\mu_{u1}}=2.17 \times 10^{-2}$ which might be due to the very fact that we just mentioned. The Bayesian approaches might be suitable for estimating mean behavior however they are incapable of revealing the localized behavior of the underlying parameters. We also expect that the success of the Bayesian approach in estimating the mean is only pertinent to this less sophisticated spring problem. For more challenging forward maps our sample-wise approach may yield better estimates for the mean.

%The error in the standard deviation of first $2.57$...this error for our approach in plot 5 right is $0.4368$
%In terms of mean: we reported before $0.0751$ but the MAP-based is slightly better: 0.0217...this might be due to the fact that MAP approaches and in general Bayesian inference provides an estimate for mean behavior however for reliability based approaches such inference can be very misleading. Also different forward maps may not always work. 

\subsubsection{Comparison with MCMC Sampling: Metropolis-Hastings and Hamiltonian Monte Carlo}\label{MCMC_numerics} In this example we use the bimodal dataset and compare the results we obtained in the first part of Example~\ref{Bi_Irr_nonunique} with MCMC samplings, particularly two approaches: Metropolis Hastings and Hamiltonian Monte Carlo.

A key part of MCMC sampling is the evaluation of likelihood as well as the process of acceptance/rejection of samples. In Metropolis-Hastings the new samples are generated by a random realization from a proposal distribution which is typically a multivariate Gaussian distribution centered on the previous accepted sample. The covariance of this proposal distribution is user-defined. The new sample is accepted if it satisfies the criterion which mainly depends on the likelihood function. If the new sample has a higher likelihood it is more likely to be accepted. For details of Metropolis-Hastings implementation see~\cite{Navarro11}.

\begin{figure}[!h]
\centering
\includegraphics[width=1.4in]{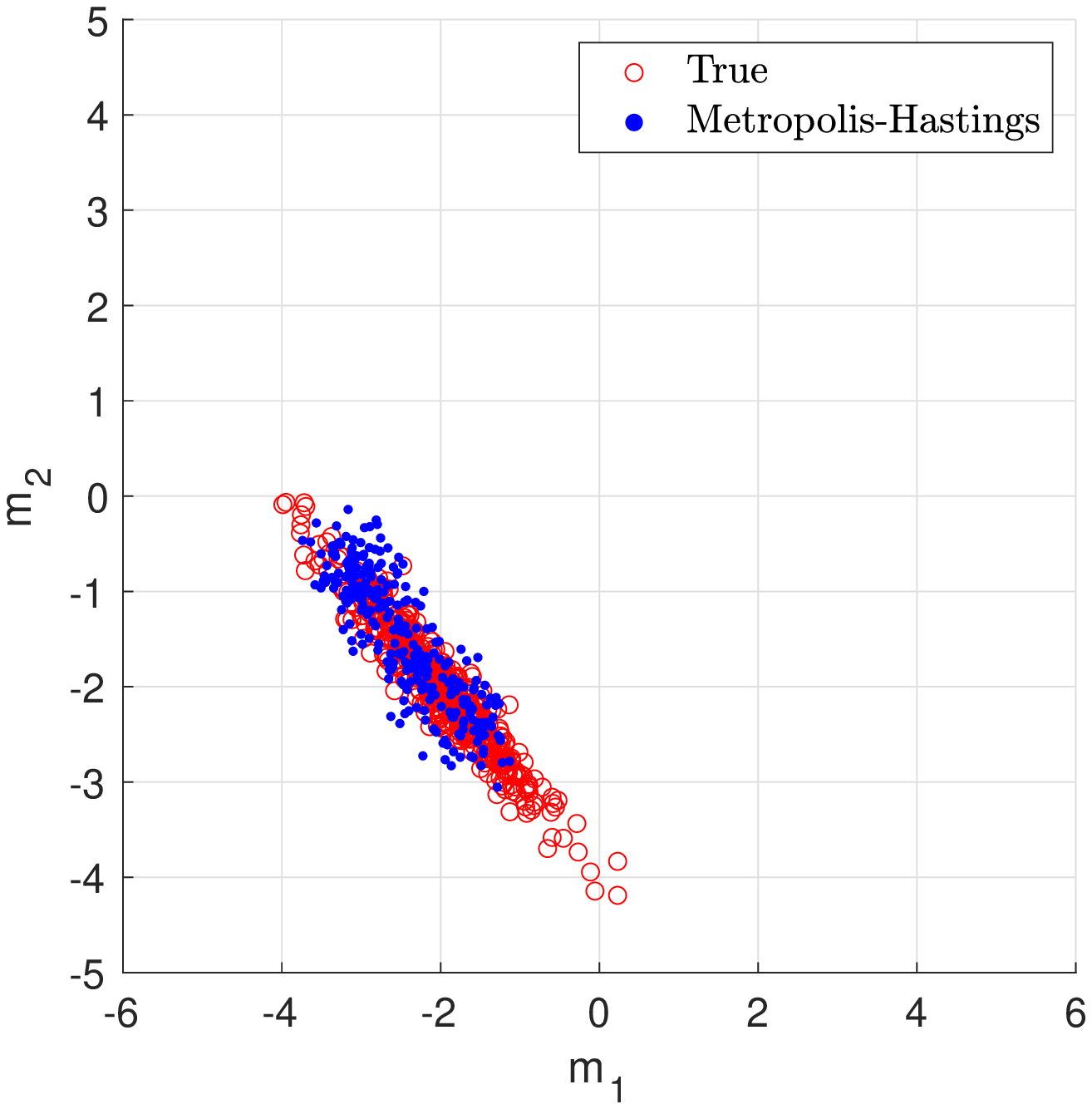}
\includegraphics[width=1.4in]{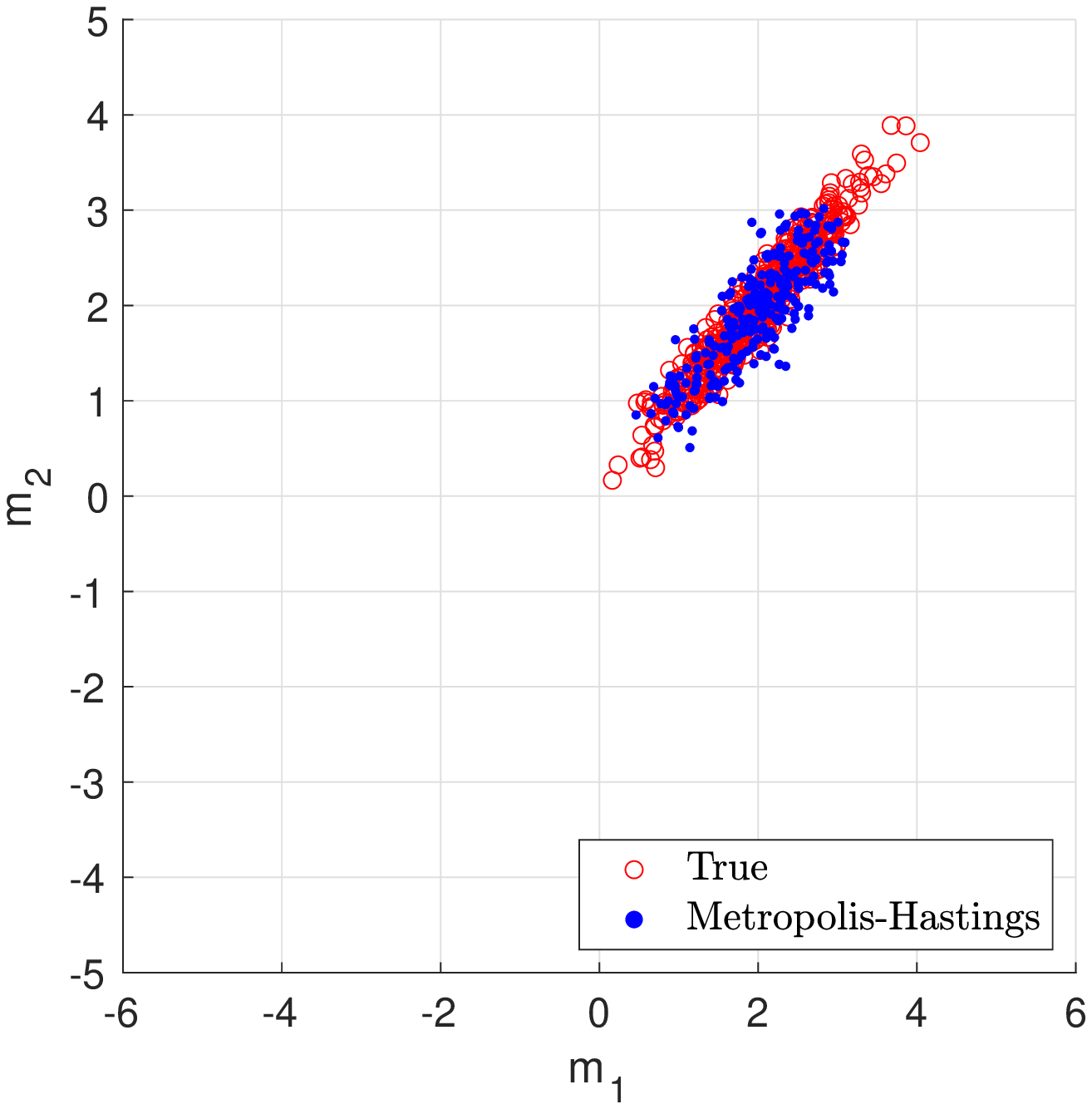}
\includegraphics[width=1.4in]{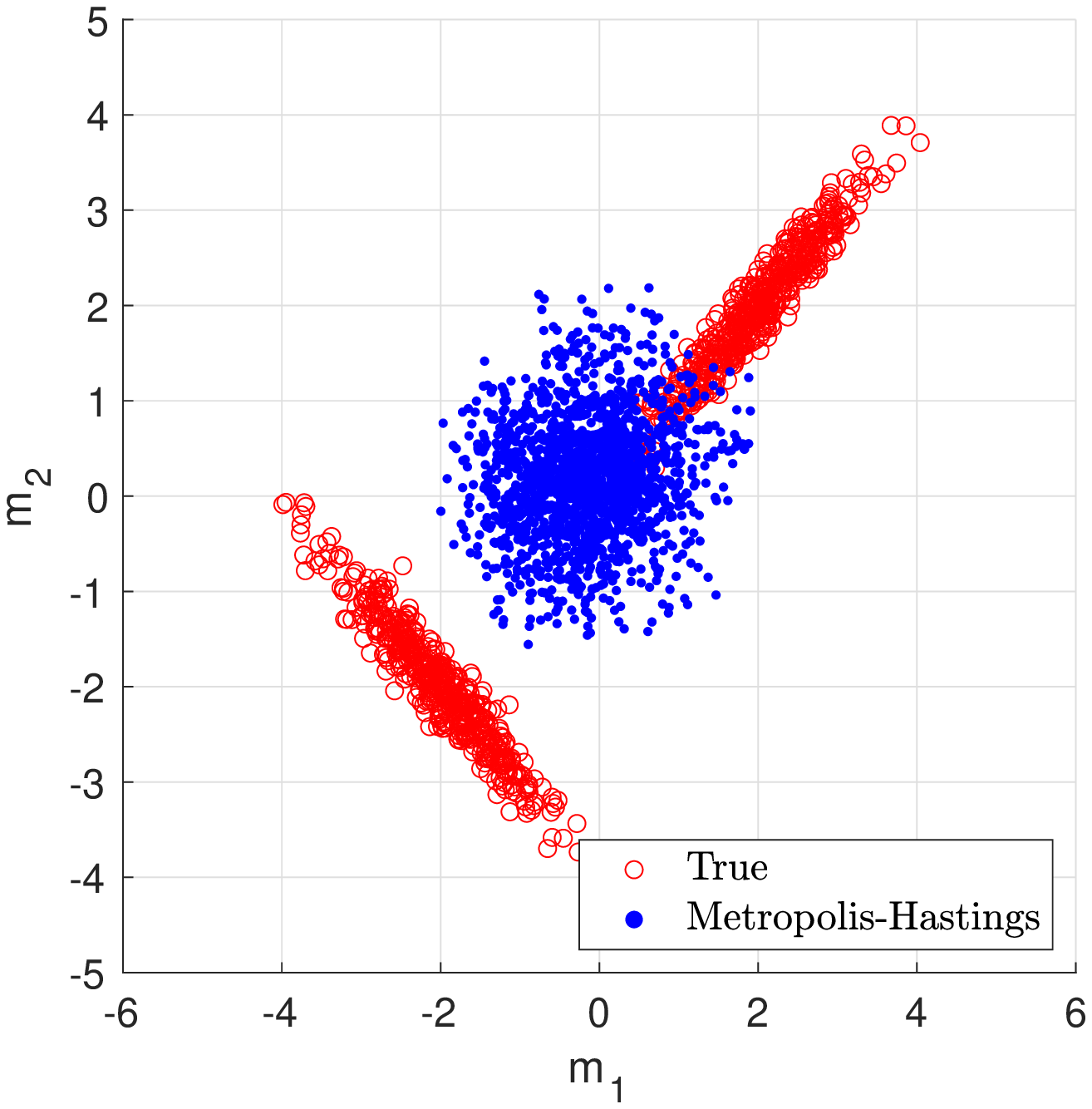}
\includegraphics[width=1.4in]{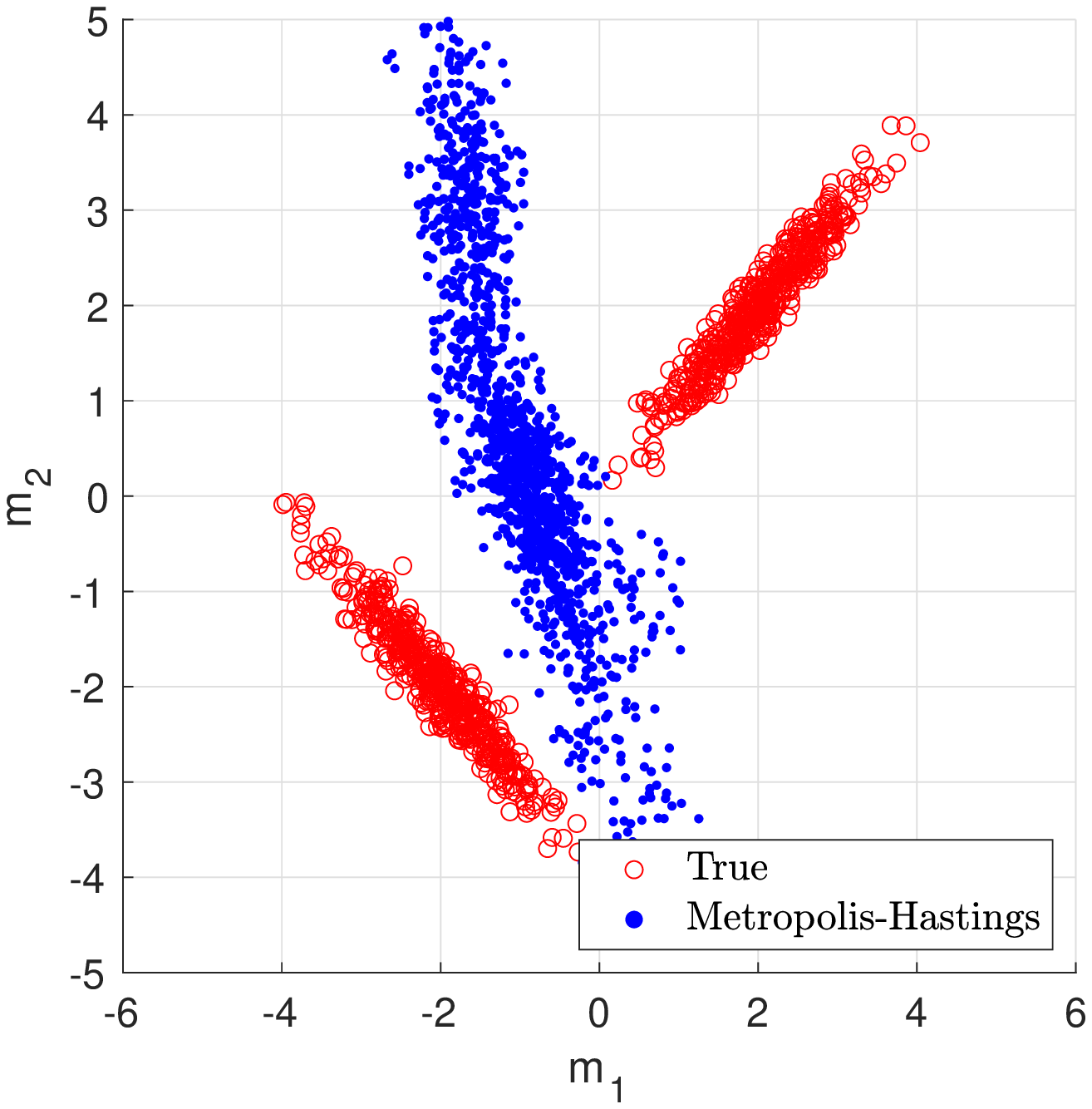}
\caption{\small{MCMC sampling - Metropolis-Hastings: Evaluating likelihood as the distance directly on latent parameters and considering one mode (first two panes), evaluating likelihood as the distance directly on latent parameters and considering both modes (third pane), sampling from the Bayesian posterior (right pane).}}\label{fig_MH}
\end{figure}

To test the correctness of our Metropolis-Hastings (MH) algorithm we first consider a unimodal distribution, i.e. only one of two modes shown in the bimodal distribution cf. Figure~\ref{fig_dist_true} (bottom row-third pane). We also compute the likelihood slightly differently from a typical evaluation in Bayesian inferences. We compute the likelihood directly based on the distance between the underlying parameter samples and the proposed samples using a small number of small distances. To this end we first compute the list of distances between the proposed sample and true underlying samples 
\begin{equation}\label{rs}
\bm r^{(i)} = \| \bm m_{\ast} - \bm m^{(i)}\|_2^2, \qquad i=1,\ldots,n_{data}
\end{equation}
then sort this list $\hat{\bm r} \gets \textrm{sort}(\bm r,``ascend")$ and finally compute the (log) likelihood associated with $\bm m_{\ast}$ as
\begin{equation}\label{log_lkl_MH}
-\log p(\bm D|\bm m_{\ast}) = \sum_{i=1}^{n_{lkl}} \bm r^{(i)} 
\end{equation}
with a small number of sample, i.e. we assume $n_{lkl}=20$. We consider small number of samples for this particular experiment to avoid the averaged behavior that we have mentioned many times throughout the paper. The results for both lower left and top right modes of bimodal distribution are shown in the first and second panes of Figure~\ref{fig_MH}. The identification appears to be successful with Metropolis-Hastings algorithm, considering a unimodal distribution, computing the likelihood directly as distance between the proposed sample and the underlying true samples, and considering a limited number of data points with high likelihood.

In practice, however, we utilize the MH algorithm differently. We consider both modes, we assume $n_{lkl}=200$ (i.e. all data points) and we set $\Gamma_{noise}=0.01 \bm I_{2 \times 2}$ (larger values of $\Gamma_{noise}$ will result in narrower distributions for the realized MH samples) and subsequently compute $\bm r^{(i)} = \| \bm m_{\ast} - \bm m^{(i)}\|_{\Gamma_{noise}}^2$. The result is shown in the third pane which clearly indicates that the computed likelihood and the resulting MH algorithm is incapable of identifying both modes. Similarly to the previous example, a unimodal distribution is identified. 

In the result of third pane, we still compute the likelihood directly based on the samples distances. In Bayesian inference, the likelihood is computed based on the simulation output. Therefore in the fourth pane we compute the likelihood in the standard way i.e. we use Equation~\eqref{log_lkl}. We assume a uniform prior which does not make any impact on the ratio between two likelihoods in the MCMC procedure. The result is shown in the fourth pane which is similar to the third pane in terms of not identifying double modes. 

One issue with these statistical samplings is their slow convergence compared to optimization-based or variational approaches. For example, in the first and second panes of Figure~\ref{fig_MH}, out of $500$ samples only $272$ and $286$ samples are accepted, slightly over $50 \%$ success. In the third and fourth panes we assume longer chains and consider $2000$ samples. Out of this number of proposed samples, $1587$ and $1348$ samples are accepted which in terms of percentage are more successful than the first two panes. This might be due to the fact that, the likelihood is less localized in the last two panes as we consider all data points for calculation of the likelihood. We expect that the MCMC convergence is slower in higher dimensions.

%To test the correctness of our MCMC sampling we 
%
%To show that MCMC can work in some cases we consider unimodal and only consider likelihood with the samples itself and also consider only first 20 samples...
%
%Then we consider both of them; in this case we consider all samples..and we generate more samples
%
%We consider the noise 0.01 for third and fourth panes...
%
%The reality however is that we should consider likelihood which is shown in the end.... 
%
%
%We consider $n=200$ burnin and run $500$ times..for UR in total we find $272$ unique samples for the lower left we get 286
%
%from 2000 run for both we get 1587 accept and for sim we get 1348

A more effective approach in MCMC sampling is the Hamiltonian Monte Carlo (HMC) approach which replaces the random walk step in MH with a deterministic walk, consequently resulting in more effective traverse of the probability distribution. The deterministic walk is performed by simulating a dynamical system, specifically evolving Hamiltonian dynamics. The dynamics simulation requires the gradient of the target distribution modeled as position state; however, the approach in essence is statistical since another key part of the simulation, the momentum state  is sampled randomly. This approach is discussed with practical details in~\cite{Neal11}. 

\begin{figure}[!h]
\centering
\includegraphics[width=1.7in]{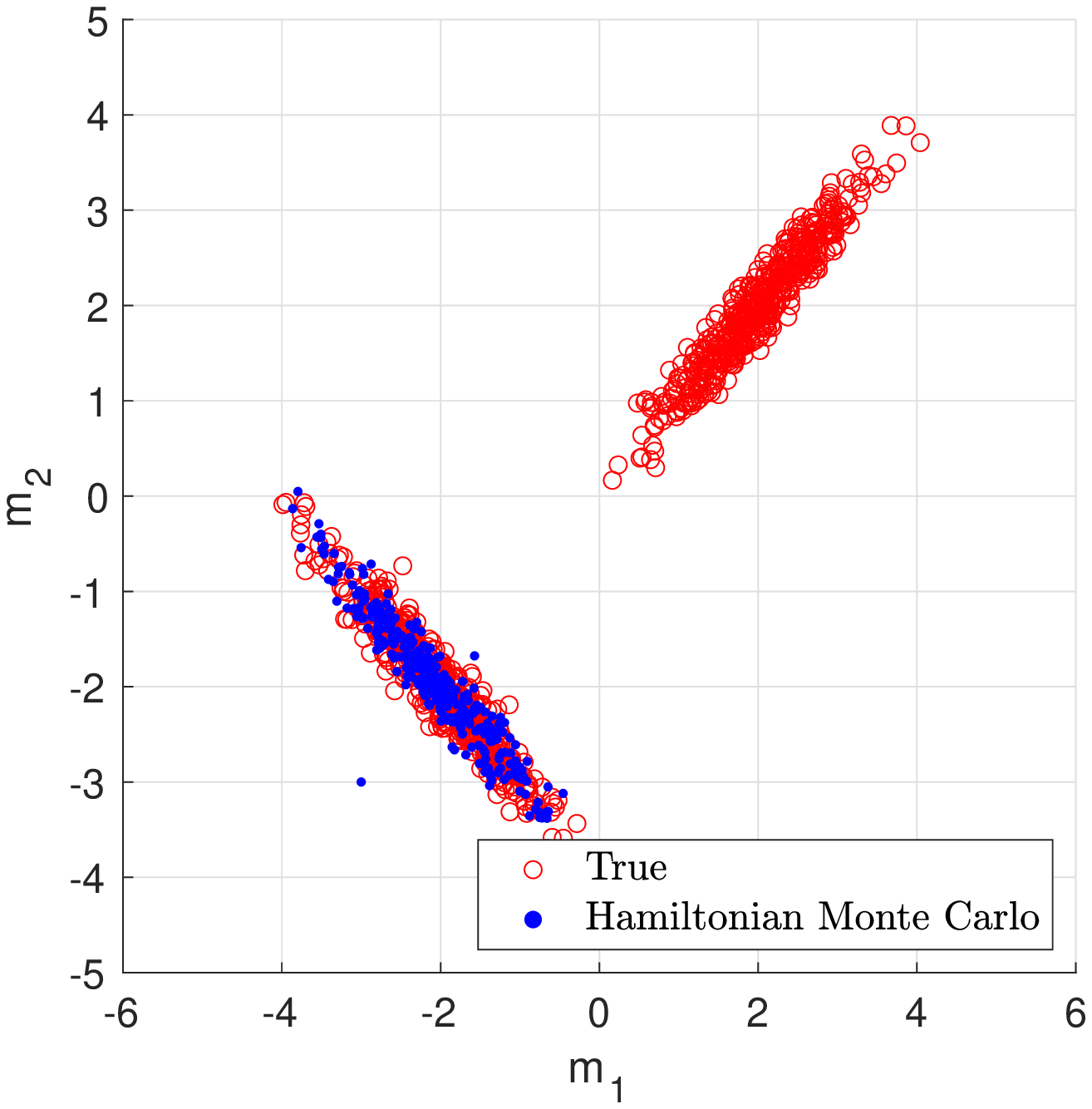}
\includegraphics[width=1.7in]{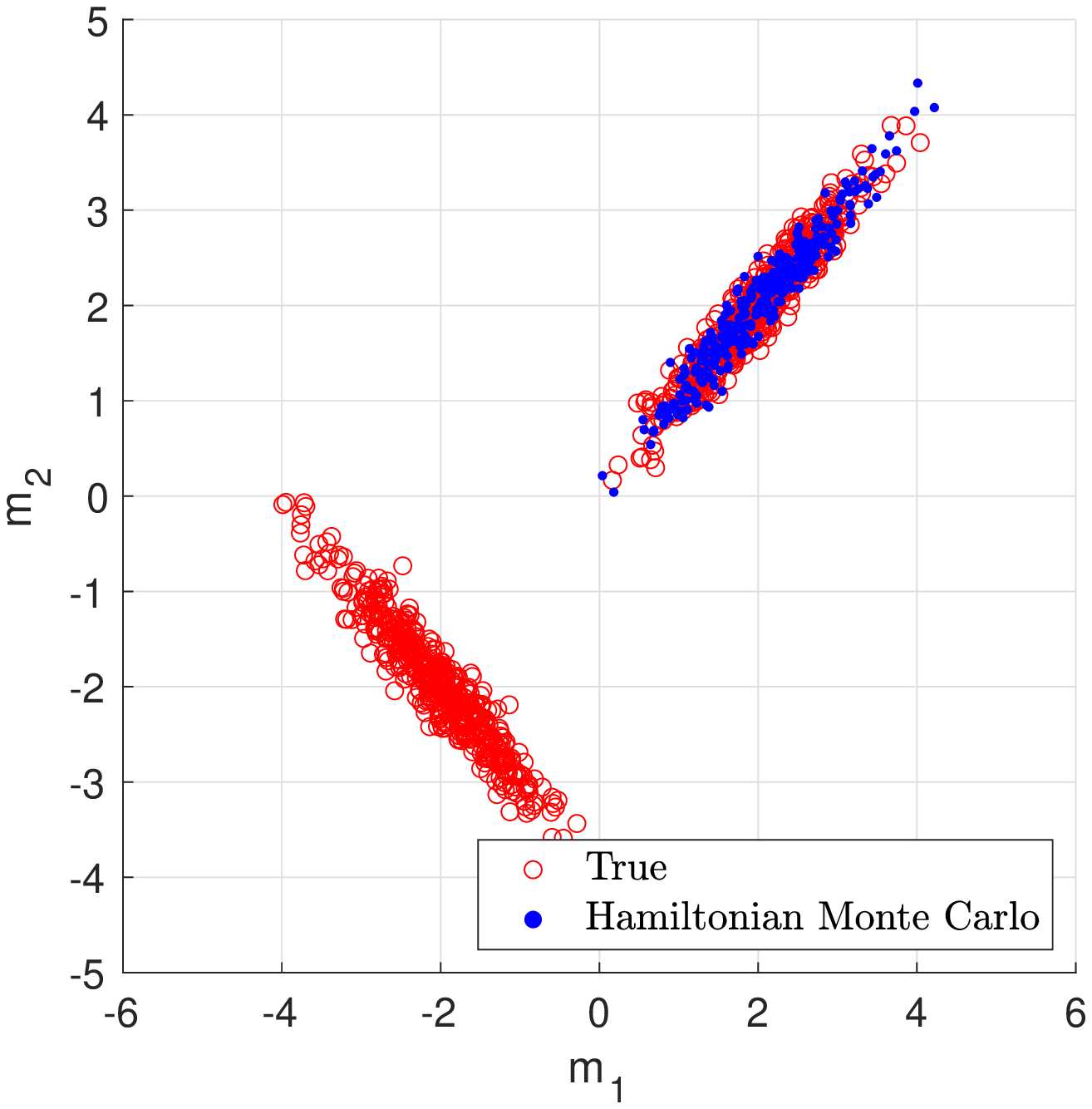}
\includegraphics[width=1.7in]{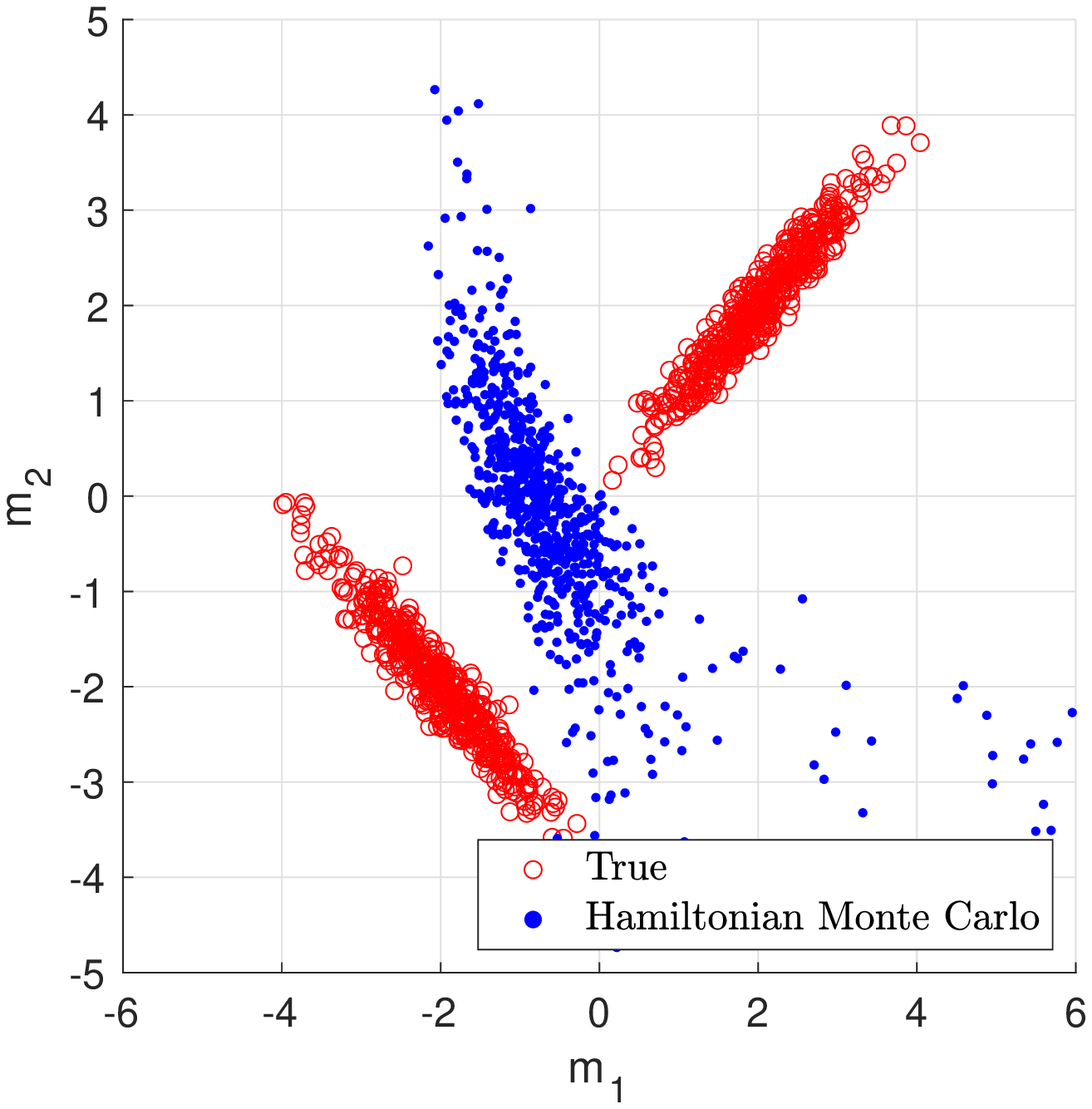}
\caption{\small{MCMC sampling - Hamiltonian Monte Carlo: Evaluating likelihood (and its derivative) as the log pdf of the known multivariate normal associated with bimodal latent parameters,  considering both modes and starting the chain from $[-3,-3]$ (left pane) and $[3,3]$ (middle pane); sampling from the Bayesian posterior (right pane).}}\label{fig_HMC}
\end{figure}

Similarly to the previous part of the example, we first consider HMC for the true distribution of the underlying parameter. To this end, we consider the target distribution according to~\eqref{bim_dist} where the computation of derivative is straight-forward. We start the chains from initial points $\bm m=[-3,-3]$ and $\bm m = [3,3]$ and consider $500$ iteration of the Markov chain. The results are shown in the first and second panes of Figure~\ref{fig_HMC}. It is again observed that the HMC approach is incapable of identifying both modes and it converges to only one of two modes. We also start the chain from $\bm m = [0, 0]$ a number of times and find that the algorithm mainly identifies the one on the lower left. To test the approach for a practical situation, we again consider the likelihood of Equation~\eqref{log_lkl}. To run the HMC algorithm, we need to provide the derivative which is already discussed in Section~\ref{Sec2_2}, see~\ref{comp_J}. The result of HMC sampling using the standard likelihood is shown in the third pane which again appears to be representing only the middle of two modes. The success rates for the HMC algorithm is slightly higher, i.e. in the first two panes and running the chain for $500$ iterations we find $391$ and $291$ samples and In the third pane from $2000$ iterations, we find $1602$ samples which indicate that incorporating the derivative and performing HMC algorithm makes the MCMC sampling more effective/successful.

%Results for HMC start the chain from -3,-3 generate 500 samples also again the likelihood is the distance
%
%HMC UR acc 291 from 500
%HMC LL acc 311 from 500
%
%HMC sim acc 1602 from 2000

\begin{figure}[!h]
\centering
\includegraphics[width=1.5in]{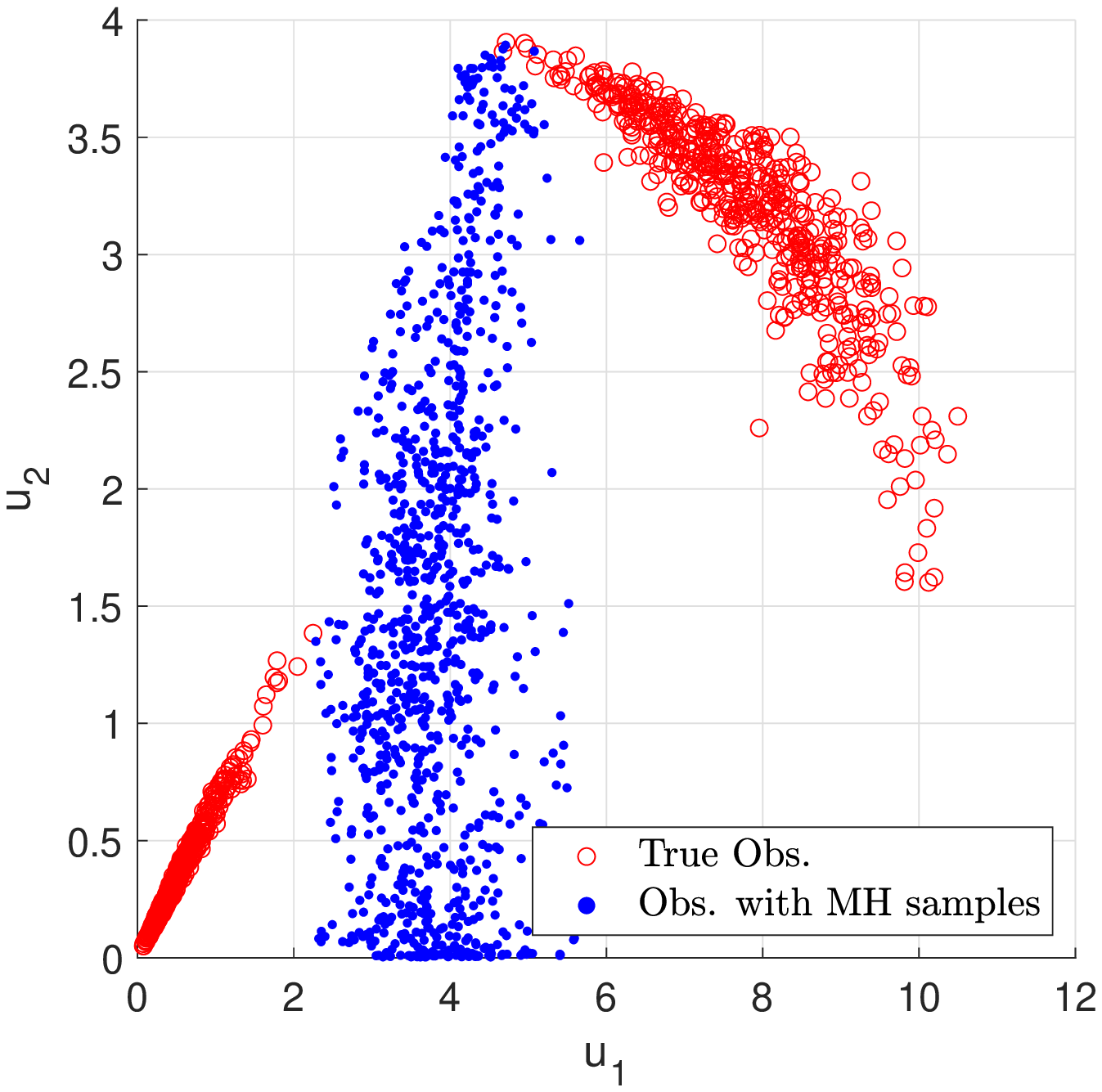}
\includegraphics[width=1.5in]{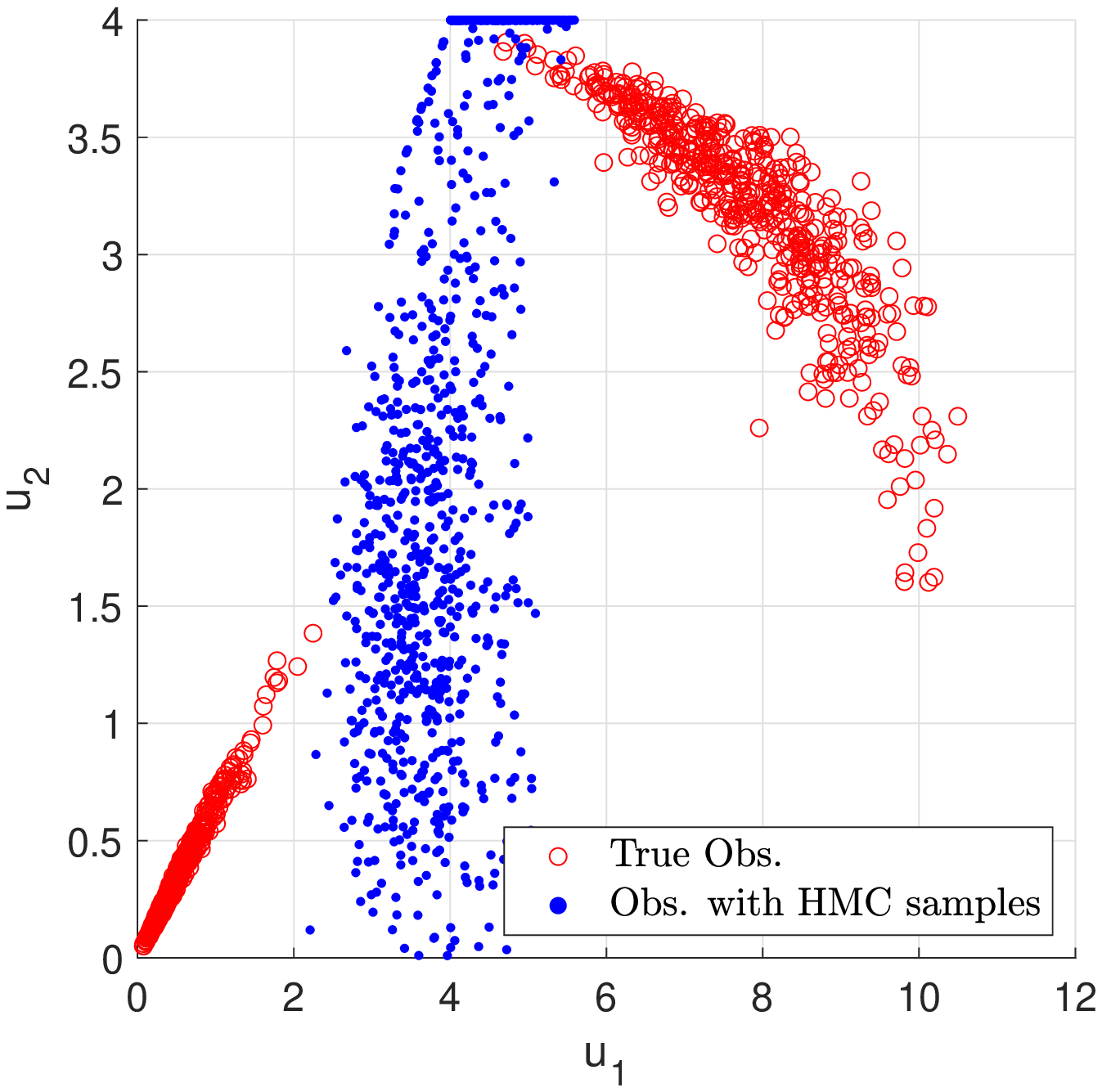}
\caption{\small{Predictions with MCMC samples: Metropolis-Hastings (left) and Hamiltonian Monte Carlo (right). }}\label{fig_sim_MCMC}
\end{figure}

Similarly to previous examples, we use the identified samples in the last panes of Figures~\ref{fig_MH} and~\ref{fig_HMC} in the forward model with the input $\bm x = [0.9, 0.8]$ and compute the simulation outputs $u_1$ and $u_2$ cf. Figure~\ref{fig_sim_MCMC}. The normalized errors in mean and standard deviation of $u_1$ are $e_{\mu_{u_1}}=7.98 \times 10^{-2} ,~4.84 \times 10^{-2}$ and $e_{\sigma_{u_1}}=8.30 \times 10^{-1},~8.24 \times 10^{-1}$  for the left and right panes respectively.  Comparing these results with the results of the second and fourth panes of the bottom row in Figure~\ref{fig_bim_iden}, i.e. $e_{\mu_{u_1}} = 4.53 \times 10^{-2},~7.61 \times 10^{-3}$,~$e_{\sigma_{u_1}} = 7.13 \times 10^{-2},~3.80 \times 10^{-3}$, it is apparent that the sample-wise inference procedure outperforms the MCMC sampling approaches especially in prediction of the standard deviation.

%The error in mean and standard deviation of first variable is [0.0798, 0.8303] and [0.0484, 0.8245] for the left and right panes.
%Compare these results with second and fourth panes in figure 6..Second pane [0.0453, 0.0713]
%fourth pane: [0.0076, 0.0038]

\subsubsection{Comparison with Bayesian Neural Network}\label{BNN_numerics}
In this example, we use the dataset associated with the $U$ shape distribution and consider  $g(m)=\exp(m)$. The main goal in this example is to compare the statistical prediction of the simulation output obtained from a trained BNN with the prediction of our inference procedure we presented in Section~\ref{Bi_Irr_nonunique}. 

The BNN implementation is based on the Bayes by Backdrop algorithm~\cite{Blundell15} which is explained with implementation details based on the Keras API (API: Application Programming Interface) of Tensorflow in~\cite{McAteer}. The main idea behind the Bayes by Backdrop approach is consideration of the mean field approximation for the variational distribution of neural net weights. More precisely, the approach  assumes independent one dimensional Gaussian random variables (with two optimizable parameters: mean and standard deviation) for each neural net weight which facilitates the variational inference using the standard backpropagation approaches. 

In our implementation of BNN, we assume $M=4$ batches where each batch includes $n_{data}=50$ samples. We assume standard normal prior for the variational inference procedure and we consider the recommendation in~\cite{Blundell15} for weighting the complexity cost based on the number of batches cf. Section~\ref{Sec2_2}. For our inference procedure, we consider the same network as the one in Section~\ref{Bi_Irr_nonunique}. 

\begin{figure}[!h]
\centering
\includegraphics[width=1.5in]{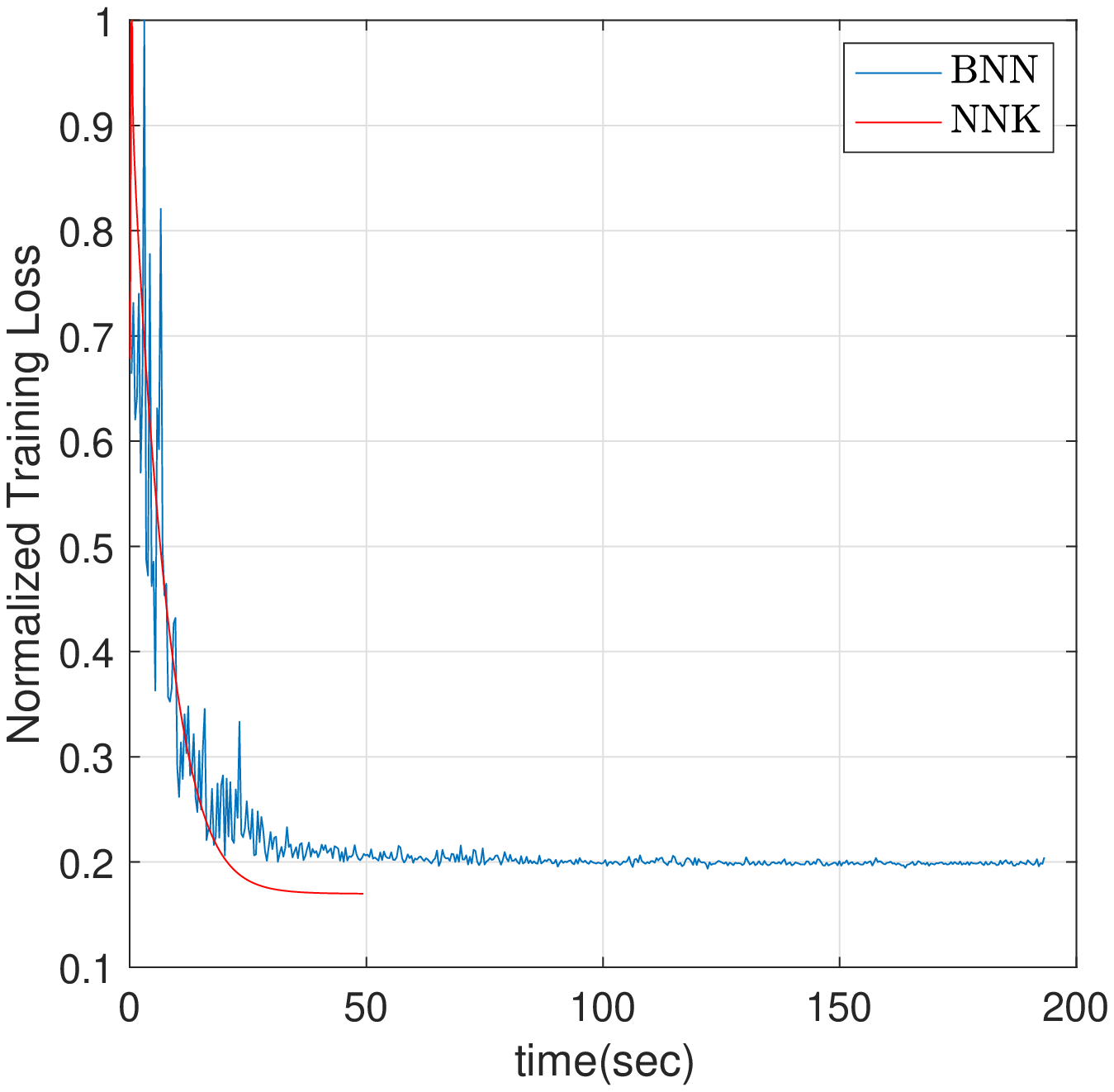}
\includegraphics[width=1.5in]{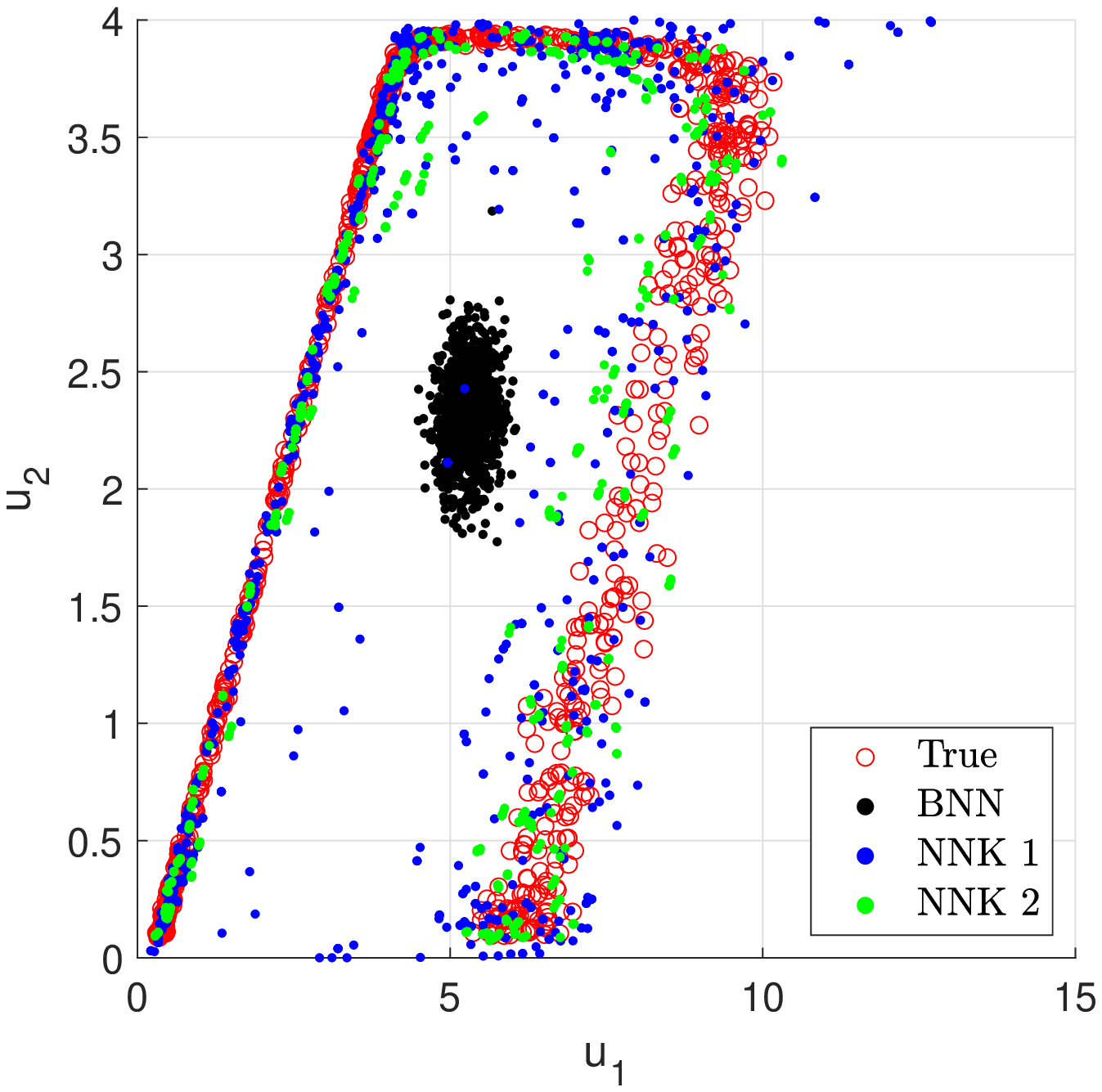}
\includegraphics[width=1.5in]{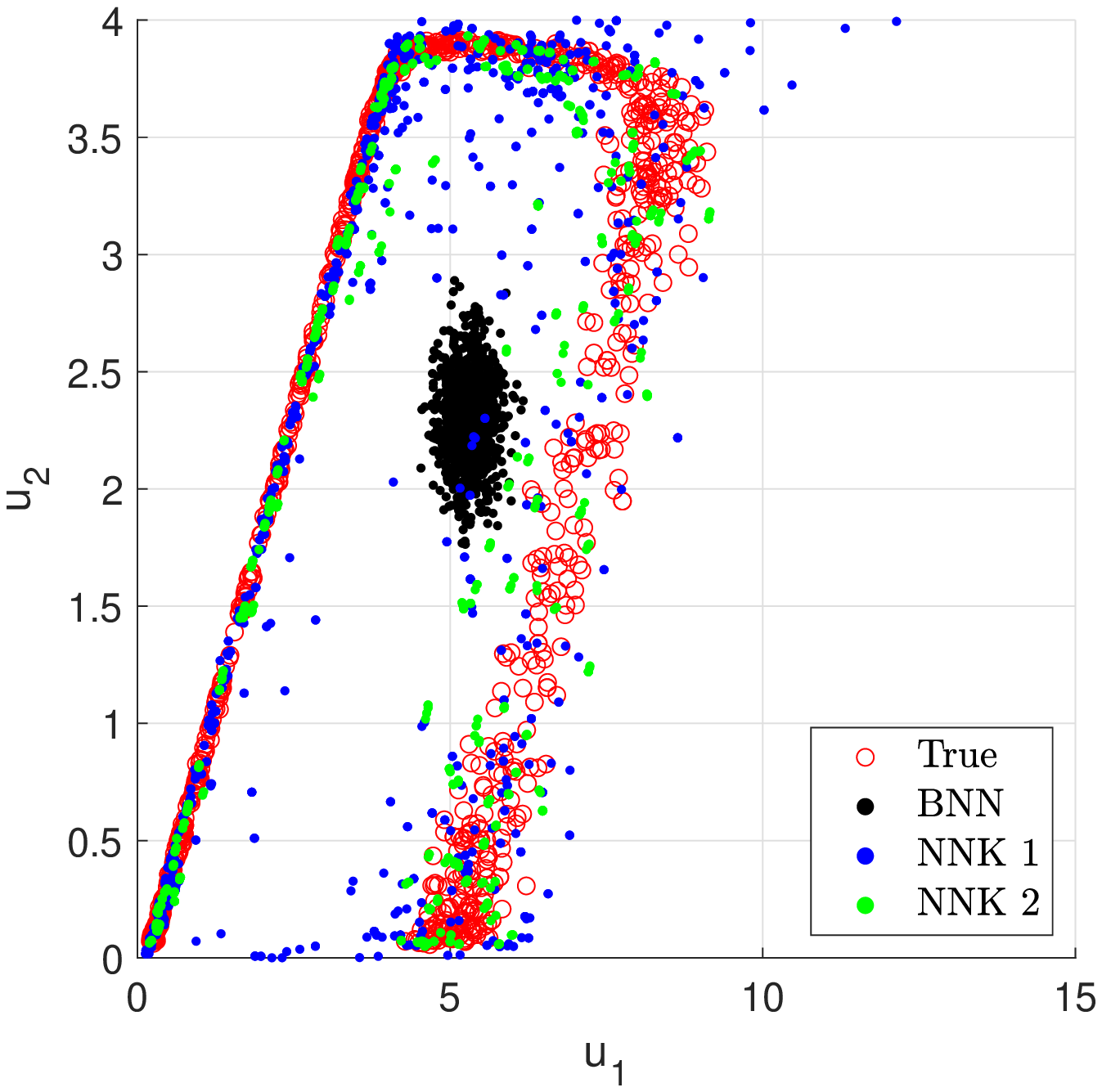}
\caption{\small{Predictions with BNN and NNK on a point inside the training $x=[0.5,0.5]$ (left) and outside the training $\bm x = [0.75,0.75]$. NNK 1 and 2 refer to randomly distributed and augmented training samples for latent parameter prior.}}\label{fig_BNN_iden}
\end{figure}

Figure~\ref{fig_BNN_iden} (first pane) shows the normalized residual versus timing for the BNN and NNK where we normalized the residuals to the maximum value in order to provide an insight about the convergence of both approaches which have different implementations. The final time $t_{BNN}=193.22$~(sec) for the BNN is associated with $500$ iterations. The final time $t_{NNK}=49.26$ is associated with $1076$ iterations. While the BNN approach could be stopped ealier since it has achieved the final residula in about $50$ seconds similar to NNK, it is clear that each iteration of NNK, $t_{iter,NNK}=0.0458$~(sec) is much faster than the iterations of BNN, $t_{iter,BNN}=0.3864$; almost one order of magnitude faster. This is a promising result which motivates further development of NNK for large scale problems.  To be fair in our comparison, we mention that our inference procedure requires inversion of the forward model which in this example takes $17.22$ seconds and permutation of prior samples which takes $0.00776$ seconds. Therefore in total the BNN approach might be slightly faster than our inference approach (only if smaller number of iterations is assumed, however this is also after the fact) but we expect that for larger scale problems our total time of inference will be less than the total time of the BNN inference.

The second and third panes show the distribution of displacements for two inputs $\bm x=[0.5, 0.5]$ and $\bm x=[0.75, 0.75]$. NNK 1 and 2 refer to considering randomly distributed and augmented training samples for the prior parameters in our inference procedure. It is again visually obvious that, the BNN is suitable for representing the mean behavior of displacement whereas both NNK approaches are able to reveal localized behavior of the output. 

%for bnn we consider num batch=4 and each bach with 50 samples 
%1/num batch is the weight for kl loss
%also we do the same net 20,7,4
%For implementation with Keras, the blog post~\cite{McAteer} is useful

\begin{figure}[!h]
\centering
\includegraphics[width=1.5in]{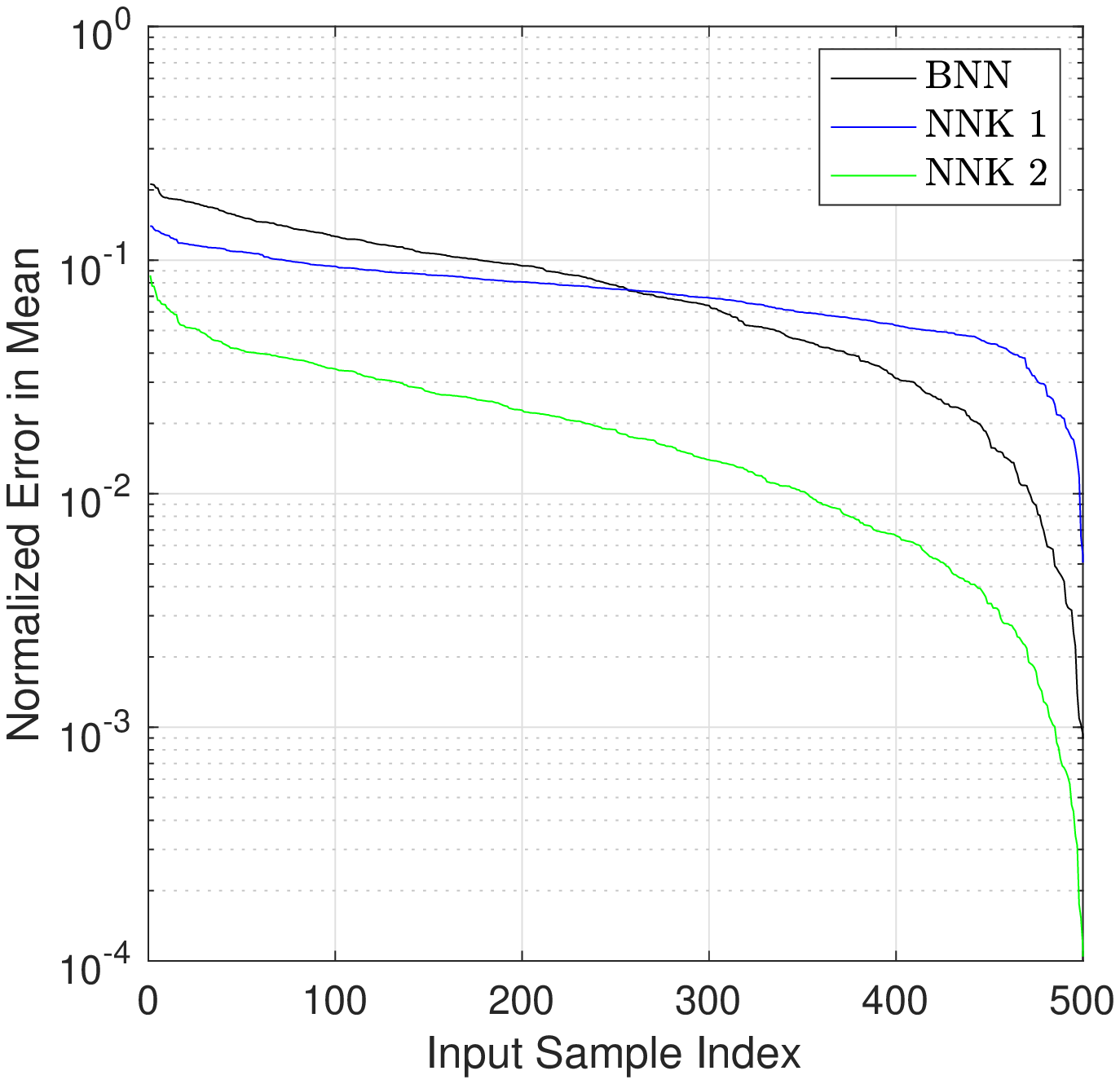}
\includegraphics[width=1.5in]{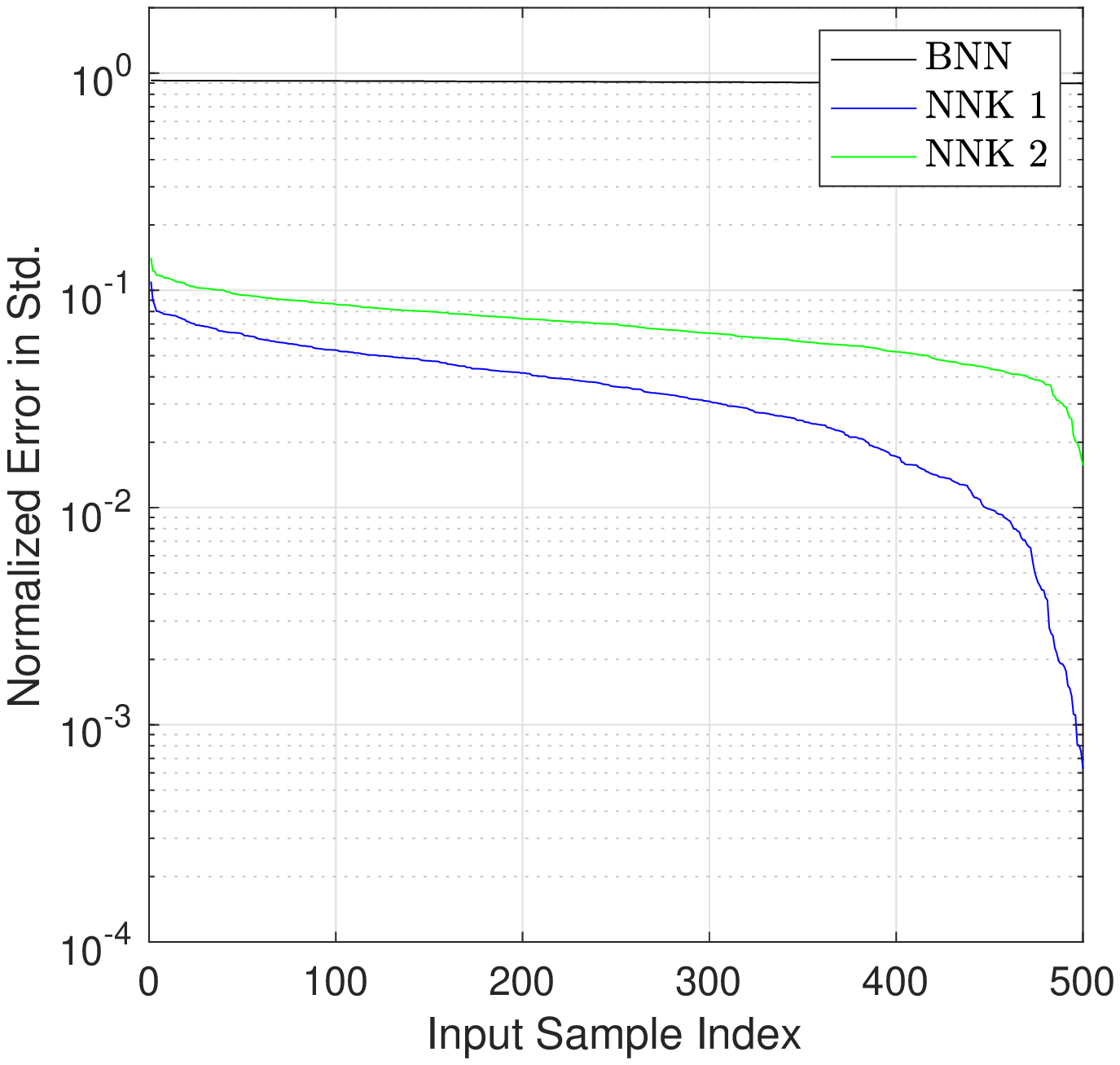}
\caption{\small{Normalized errors in mean (left) and standard deviation (right) of $u_1$ evaluated on input points outside the training range, i.e. $x_1\in [0.25, 0.75], x_2 = 0.75$.}}\label{fig_BNN_norm_err}
\end{figure}

\begin{figure}[!h]
\centering
\includegraphics[width=2in]{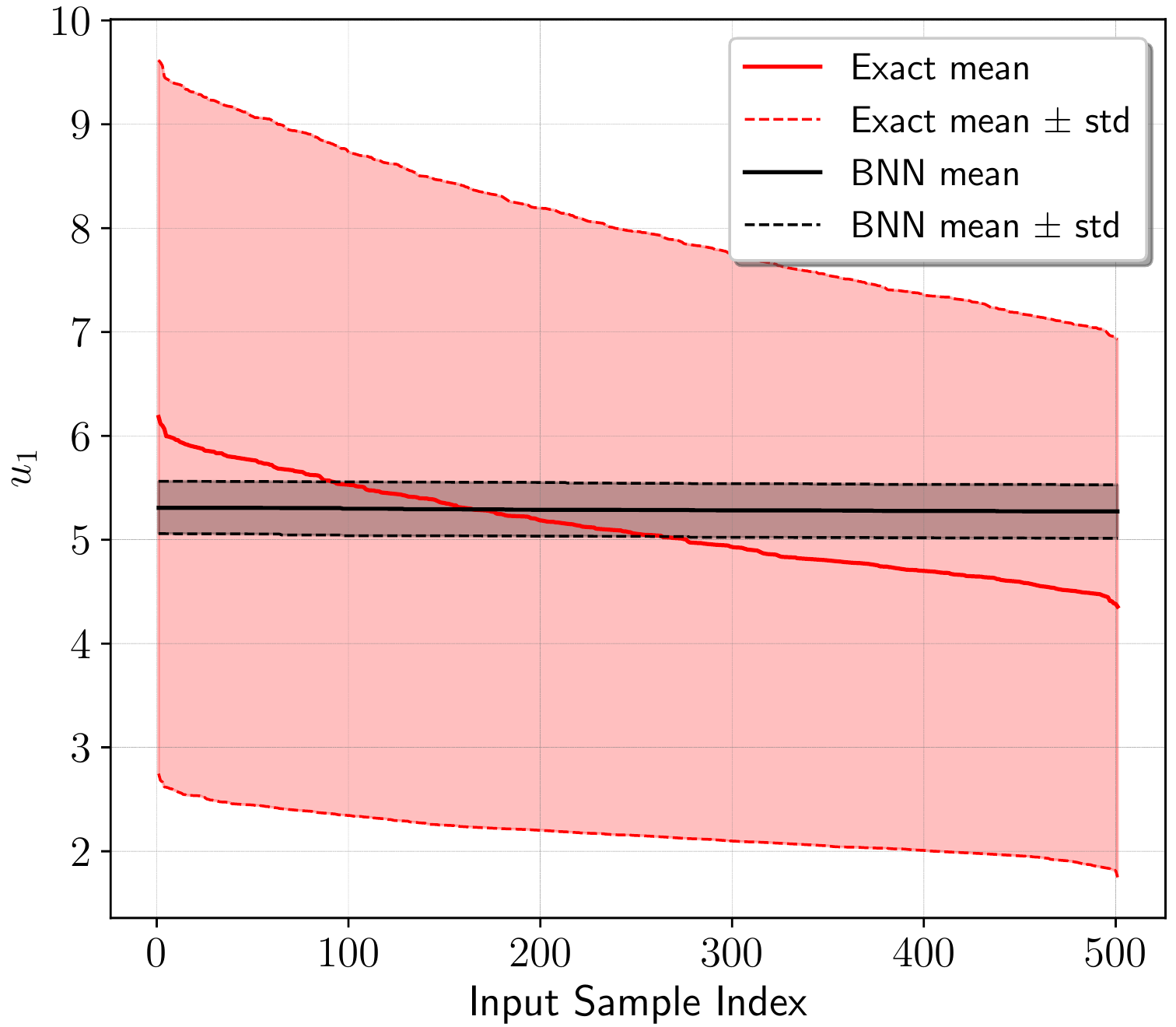}
\includegraphics[width=2in]{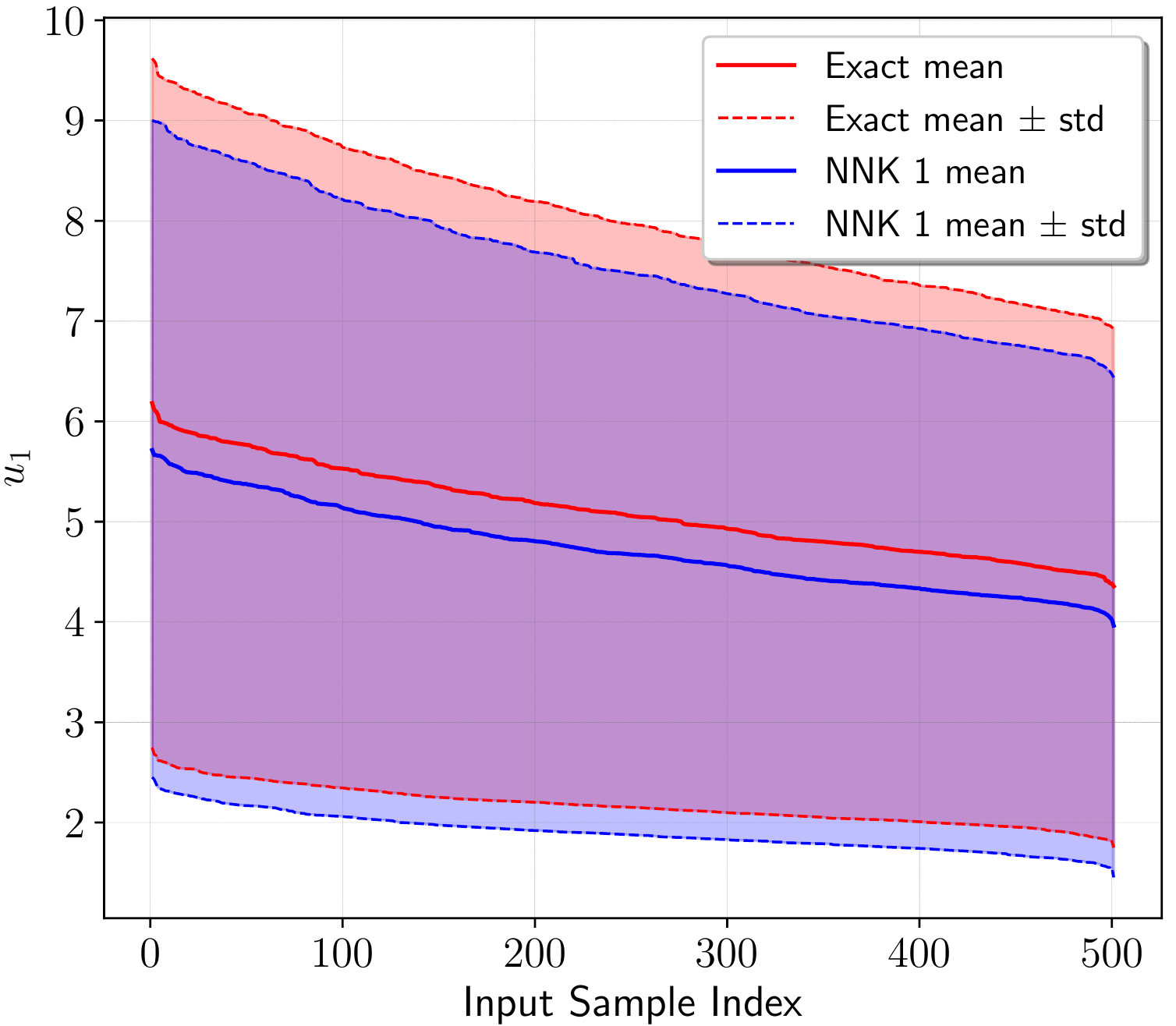}
\includegraphics[width=2in]{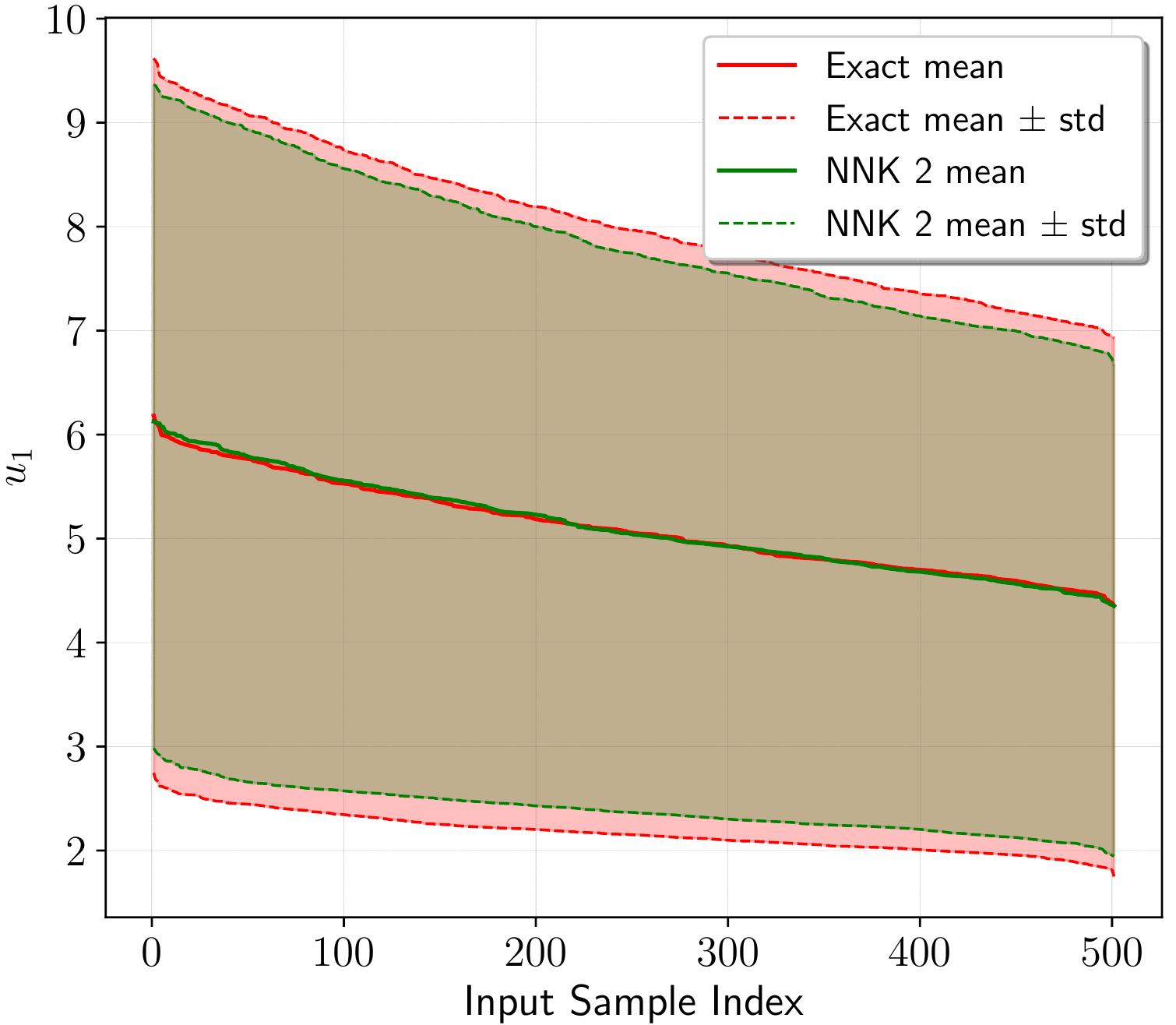}
\caption{\small{Mean and standard deviation along the line $x_1\in [0.25, 0.75], x_2 = 0.75$.}}\label{fig_BNN_std}
\end{figure}

Focusing on the output $u_1$, we evaluate the normalized errors in its mean and standard deviation for a larger input set. In particular we consider a line outside the training range  $x_1\in [0.25, 0.75], x_2 = 0.75$ and compute normalized errors for $500$ equally spaced samples on this line. The sorted values associated with these errors are shown in Figure~\ref{fig_BNN_norm_err}. We notice that for some samples BNN yields more accurate estimates in terms of mean compared to NNK 1. However, the difference are less significant compared to the difference in standard deviation which clearly indicate inadequacy of BNN for approximating localized behaviors.  We also visualize the mean values and standard deviations around mean values for these statistical results on $u_1$ in Figure~\ref{fig_BNN_std}. Poor performance of BNN and appreciable performance of NNK 2 in approximating the standard deviation are evident.

%The timing for 500 iterations of bnn is 193.22, i.e. each iteration takes 0.3864
%The timing for NNK for 1076 is 49.26 i.e. each iteration takes 0.0458..NNK also has previous stages: simulation opt takes 17.22 second and permutation takes $0.00776$ seconds. 

\subsubsection{Inference with Noise}\label{infer_noise}

As the last part of the examples in this section, we investigate the performance of our approach in the presence of noise. Specifically, the presence of noise impacts the identification of true samples in the first step of our procedure. 

\begin{figure}[!h]
\centering
\includegraphics[width=6in]{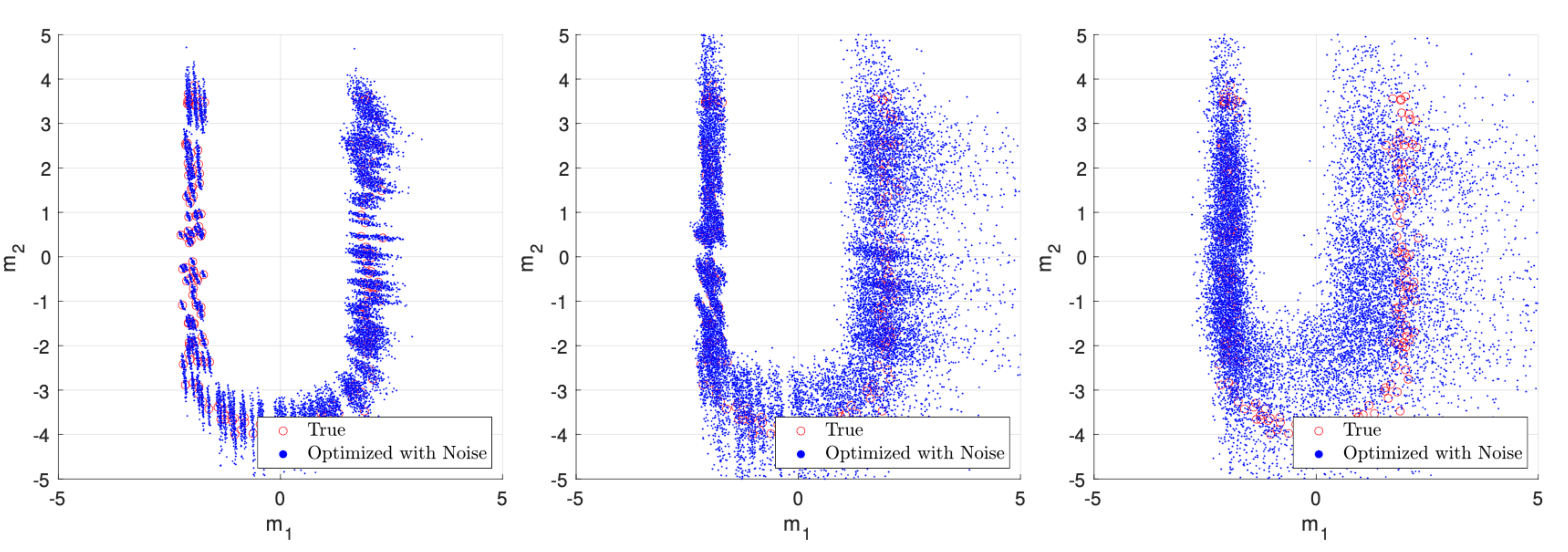}
\caption{\small{Inversion of the forward model for different levels of noise $\delta_{noise} = 10^{-3},~ 10^{-2},~ 10^{-1}$ from left to right. }}\label{fig_opt_noise}
\end{figure}

We consider the noise in the first step by solving $G(\bm x^{(i)}, \bm m^{(i)}) + \eta = \bm y^{(i)}$
follows the distribution $\eta~\delta_{noise} \mathcal{N}(0,1)$. For each training sample we assume $100$ noise values. Therefore, in total we find $2 \times 10^{4}$ optimized samples after inverting the forward model with noise.  Figure~\ref{fig_opt_noise} shows the optimized samples for three levels of noise $\delta_{noise} = 10^{-3},~ 10^{-2},~ 10^{-1}$. It should be noted that for the larger levels of noise, specifically $\delta_{noise} =10^{-2},~ 10^{-1}$ the convergence to the preset tolerance $\delta_R=0.01$ in the NR approach is not always achieved. We set the maximum number of iteration as $\max_{iter}=500$ and record the obtained sample as the optimized sample at the end of $500$th iteration if the tolerance is not met.

\begin{figure}[!h]
\centering
\includegraphics[width=1.75in]{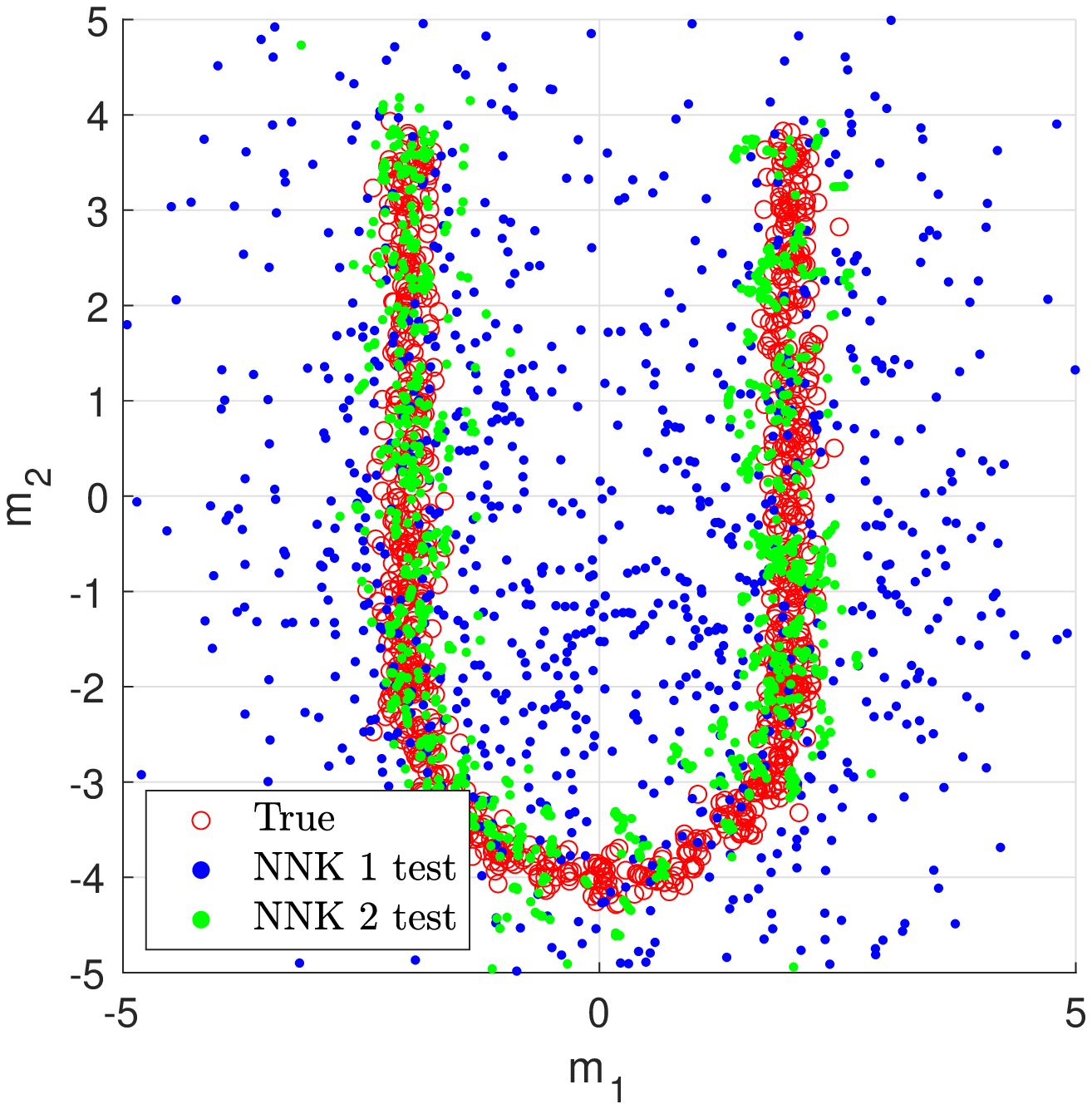}
\includegraphics[width=1.75in]{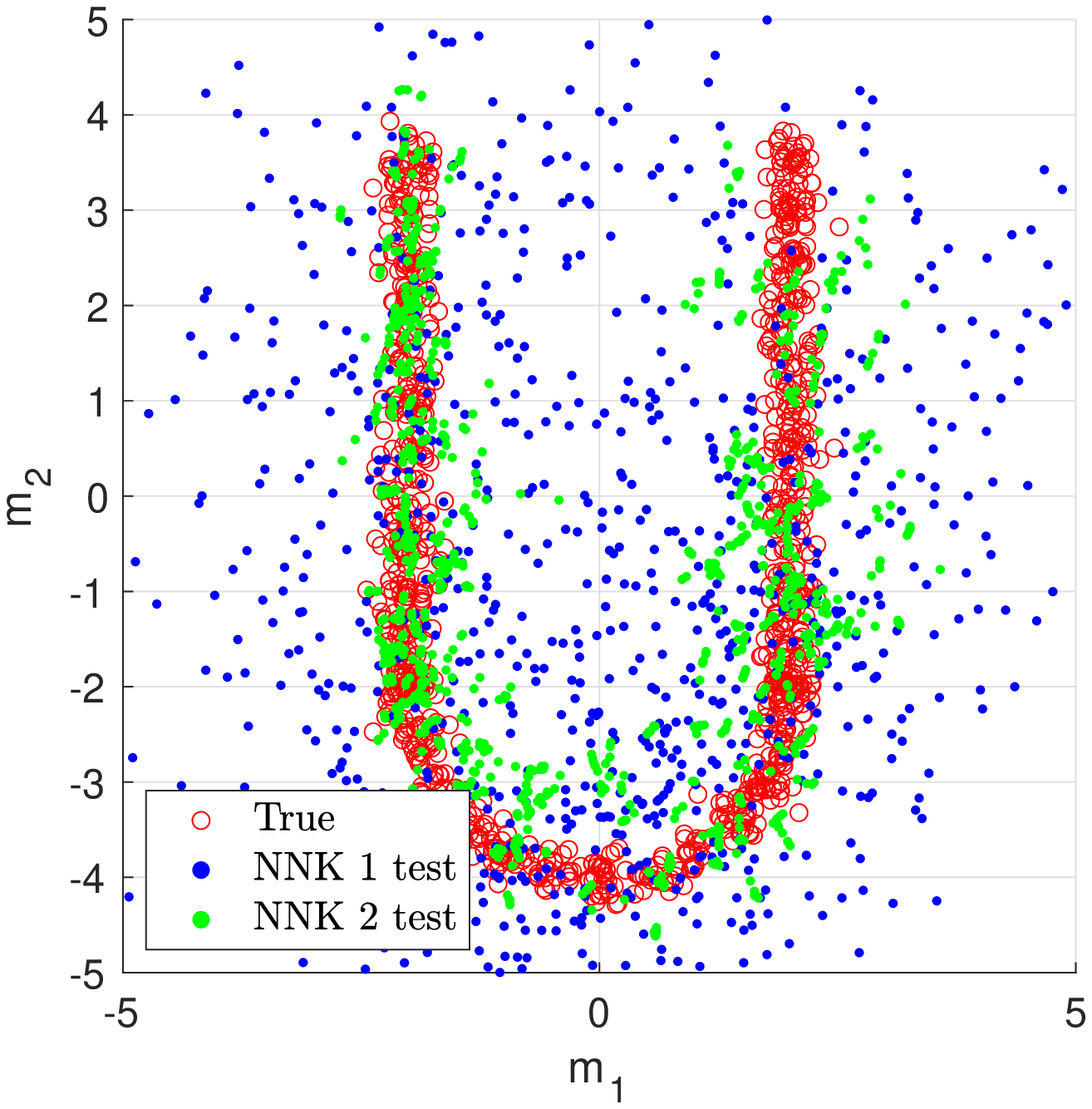}
\includegraphics[width=1.75in]{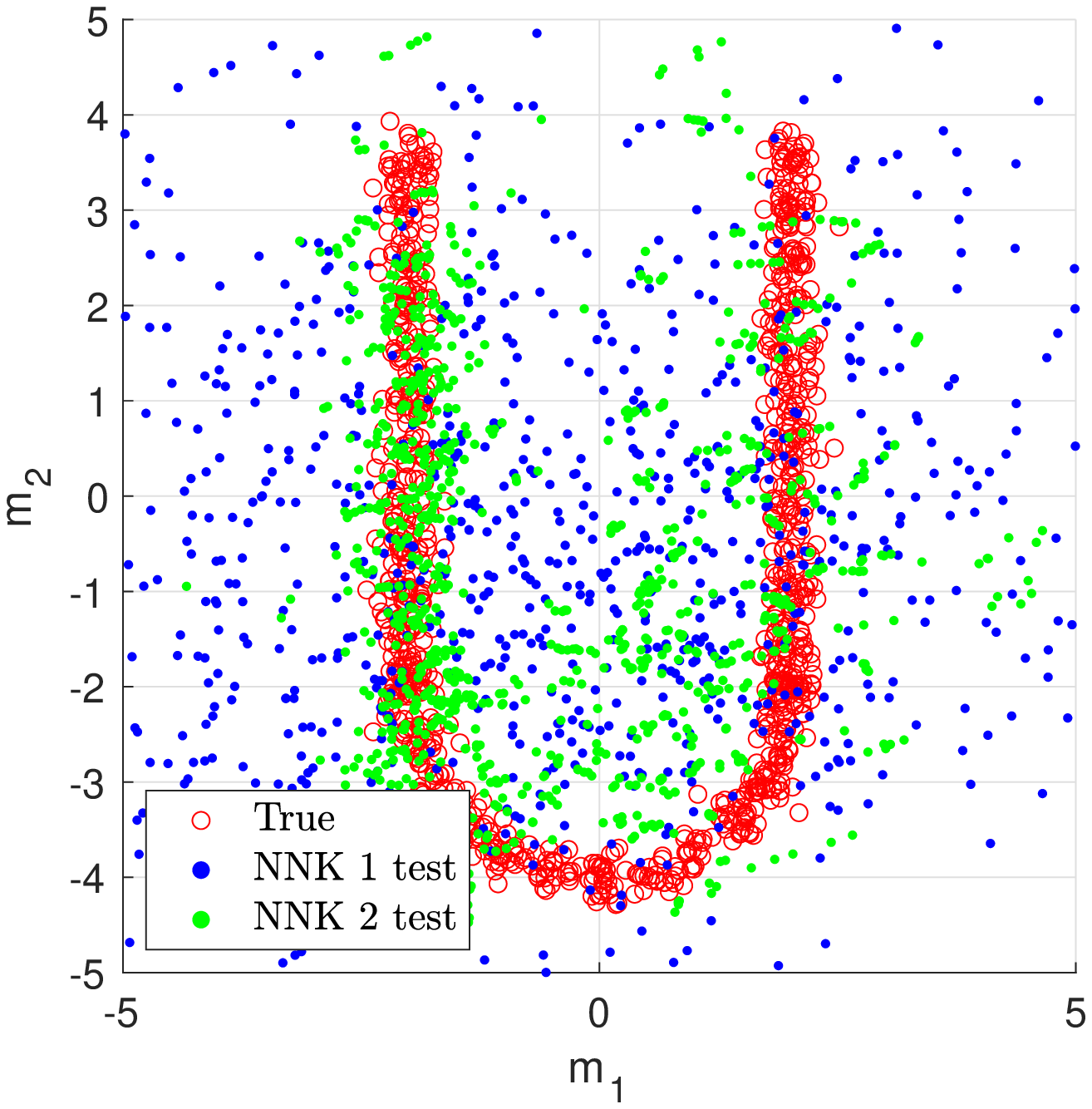}\\
\includegraphics[width=1.75in]{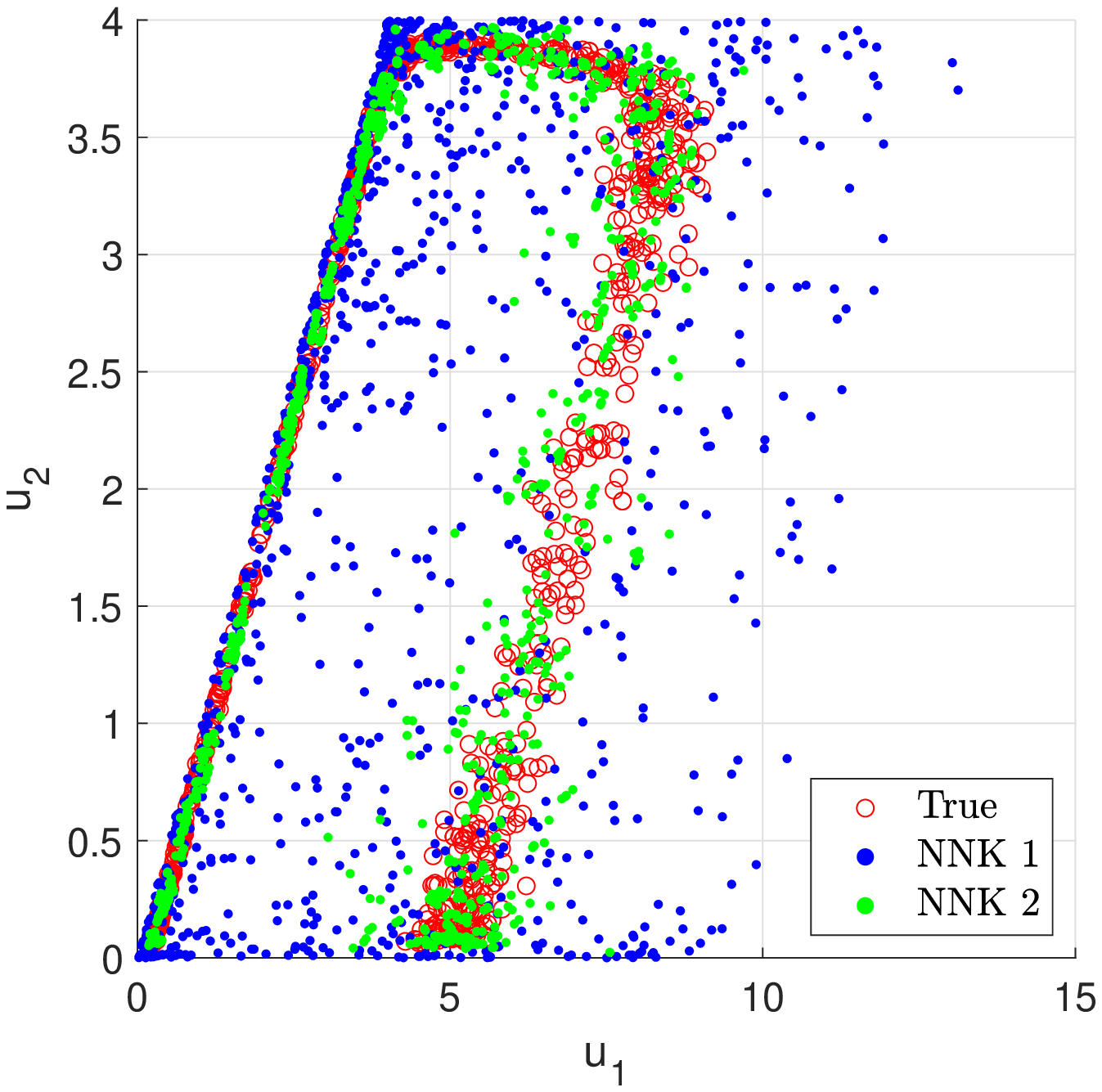}
\includegraphics[width=1.75in]{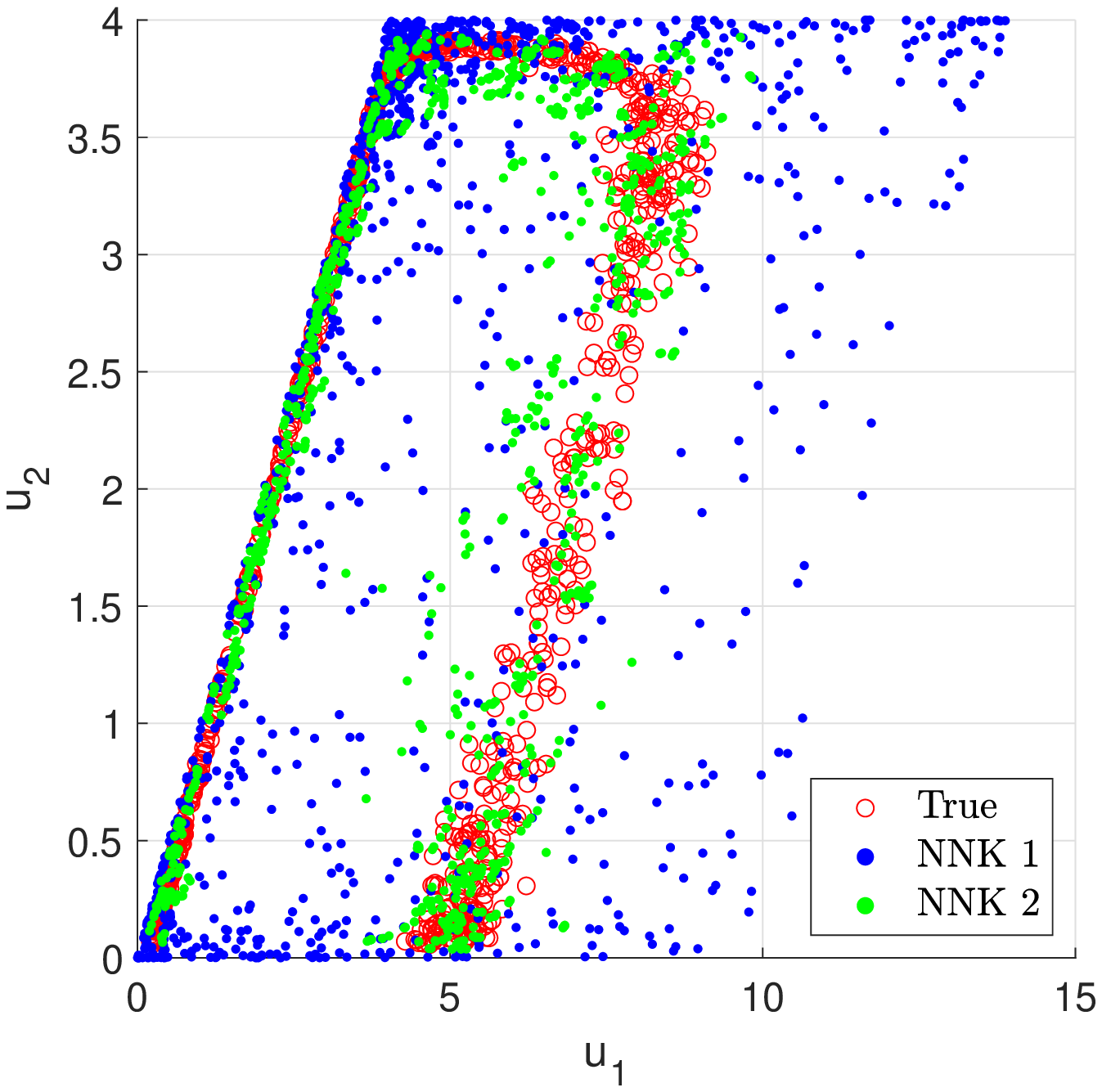}
\includegraphics[width=1.75in]{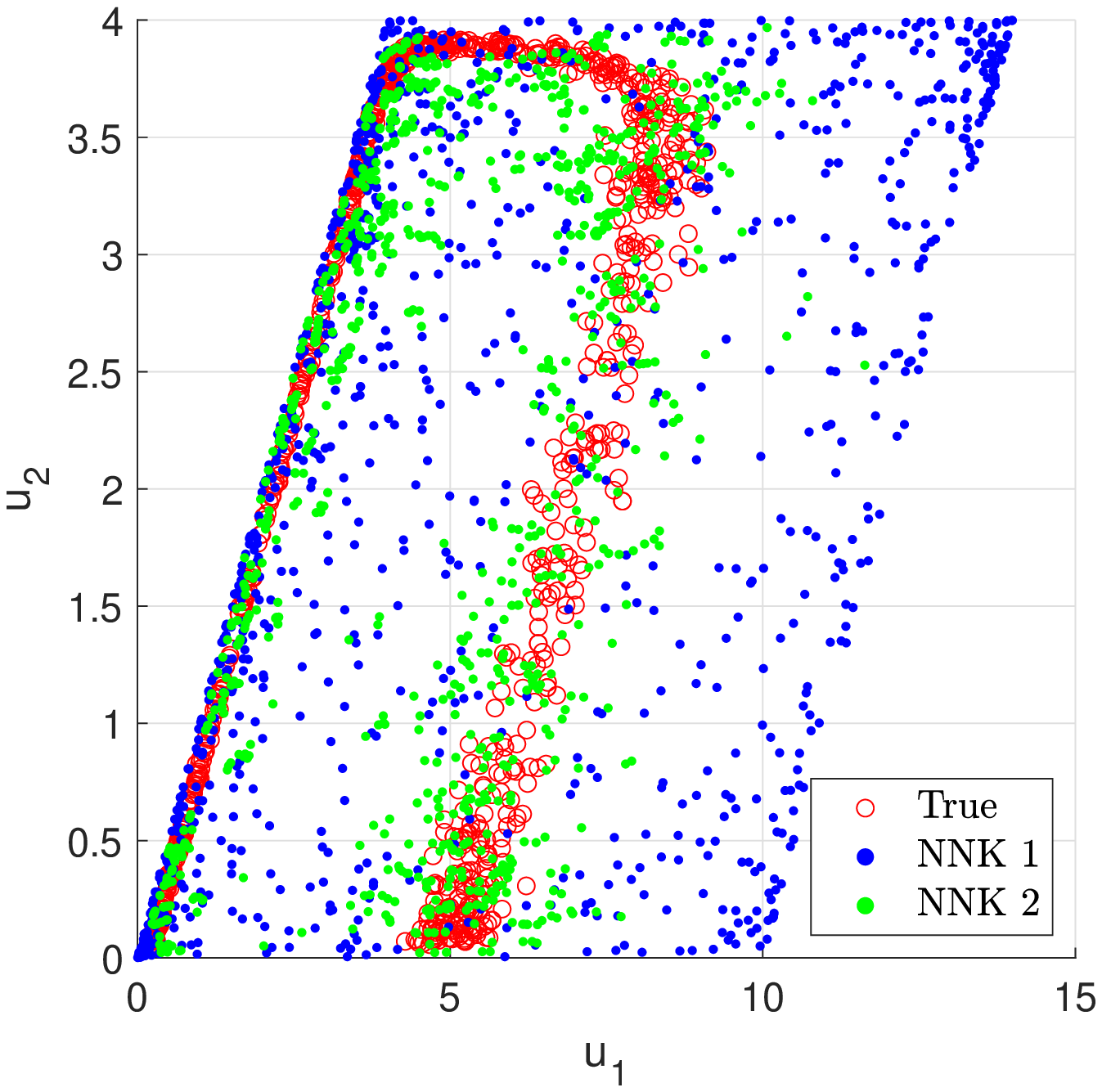}\\
\caption{\small{The inference of latent parameter (top row) and the prediction of displacement on the input $\bm x=[0.75,0.75]$ associated with three levels of noise $\delta_{noise} = 10^{-3},~ 10^{-2},~ 10^{-1}$ from left to right.}}\label{fig_noise_pred}
\end{figure}

Using the optimized samples we proceed with the next steps of the inference procedure. We specifically choose $200$ samples out of $2 \times 10^4$ optimized samples and perform NNK training similarly to the study in Section~\ref{Bi_Irr_nonunique}.  The results of latent parameter inference and prediction of displacements on the input $\bm x=[0.75, 0.75]$ are shown in Figure~\ref{fig_noise_pred}.  The normalized errors in mean and standard deviation of $u_1$ for these three predictions are listed in Table~\ref{Tab_noise}.

\begin{table}[!h]
\caption{Normalized errors in mean and standard deviation of $u_1$ with the noisy version of the sample-wise inference procedure}
\normalsize
\centering
\begin{tabular}{l c c c  }
\hline\hline
   & $\delta_{noise} = 10^{-3}$   & $\delta_{noise} = 10^{-2}$  & $\delta_{noise} = 10^{-1}$  \\
\hline
NNK 1 - $e_{\mu}$ & $7.80 \times 10^{-3}$ & $1.10\times 10^{-1}$ & $3.47 \times 10^{-1}$  \\
NNK 1 - $e_{\sigma}$ & $1.12 \times 10^{-1}$ & $2.7\times 10^{-1}$ & $5.94 \times 10^{-1}$ 	 \\
NNK 2 - $e_{\mu}$ & $1.58 \times 10^{-2}$ & $8.39\times 10^{-2}$ & $1.01 \times 10^{-1}$  \\
NNK 2 - $e_{\sigma}$ & $6.58 \times 10^{-2}$ & $1.02\times 10^{-1}$ & $7.84 \times 10^{-2}$  \\
\hline
\end{tabular}
\label{Tab_noise}
\end{table}
The result of this table can be compared with the BNN prediction: the normalized errors in mean and standard deviation of $u_1$ associated with the input $\bm x=[0.75, 0.75]$ are $e_{\mu_{u_1}}=1.85 \times 10^{-1}$ and $e_{\sigma_{u_1}}=9.02 \times 10^{-1}$. Even the least accurate NNK result in apprxoimation of standard deviation i.e. NNK 1 with rather significant noise $\delta_{noise} = 10^{-1}$ is more accurate than the BNN approximation. In terms of mean however BNN is more accurate compared to this particular scenario. 

We cannot draw a strong conclusion by making a comparison between the noisy version of our procedure and generic probabilistic surrogates such as BNN. However, as it is apparent visually, the sample-wise inference provides more insight in terms of the localized behavior of the underlying parameter and simulation output even in the presence of noise. In extreme cases of noise, the true underlying parameter will not be identifiable however in those challenging situations, Bayesian approaches yield narrow distributions (prior will have more impact in the inference) which are also distant from reality. 

\subsection{Identification of Elastic Modulus in 2D Topology Optimization}
The goal in this example is to demonstrate the applicability of our generic framework in identifying random parameters for standard simulation problems such as topology optimization. This example involves much larger number of degrees of freedom as well as larger dimensionality of the underlying parameter. The randomness is assumed on the elastic modulus which is modeled as a function of a Karhunen-Lo\'eve (KL) expansion with multiple modes. 

\begin{figure}[!h]
\centering
%\includegraphics[width=1.90in]{./mesh_2D.pdf}
%%\hspace{1.3cm}
%\includegraphics[width=2.1in]{./elas2D.pdf}
%\includegraphics[width=2.1in]{./elas2D_transformed.pdf}
\includegraphics[width=6.1in]{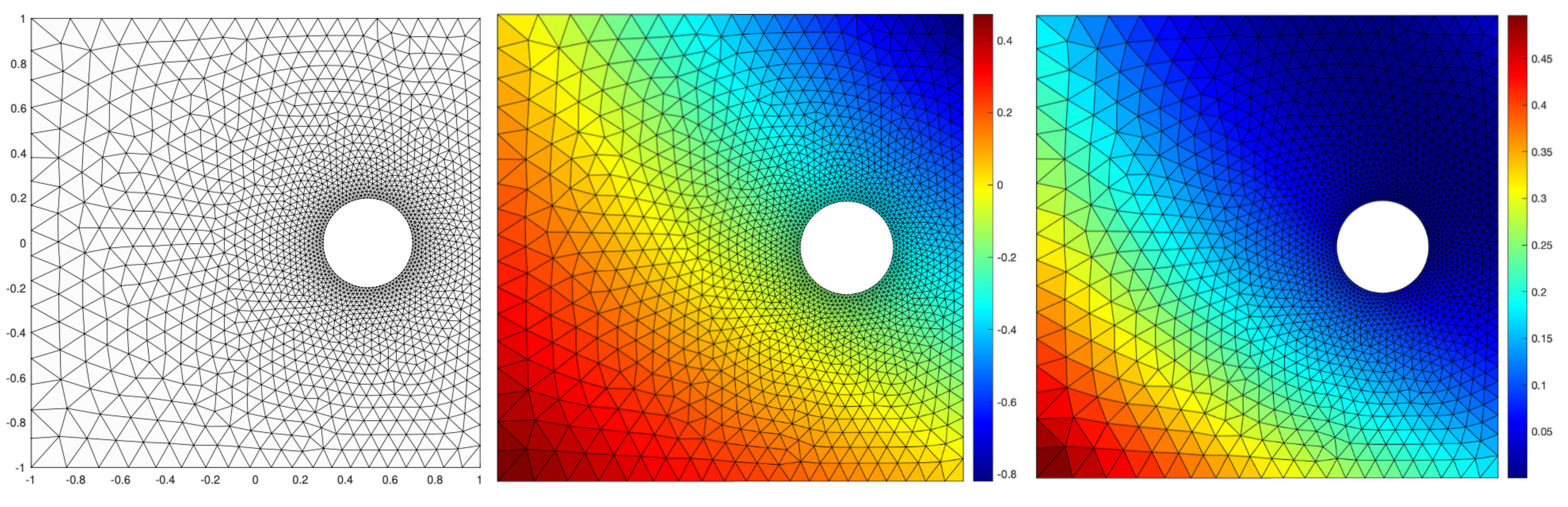}
\caption{\small{Geometry and discretization of a square domain with a circular hole (left) and visualization of a Karhunen-Lo\'{e}ve basis vector $\bm E_1$ (middle) and its transformed one, i.e. $\sin(\bm E_1)$.}}\label{mesh2D}
\end{figure}

Figure~\ref{mesh2D} (left) shows the geometry of the structure and its discretization which is achieved with triangular elements. The structure has fixed degrees of freedom in both $x$ and $y$ directions on the left side i.e. $x=-1$ and it is subject to a vertical point load on the right side at $x=1, y=0$. The mesh is generated using the \texttt{distmesh} code~\cite{Persson04,Persson05}. We developed the finite element and topology optimization solvers with triangular and tethrahedral meshes (next example) as part of the overall framework in this paper. We plan to extend the current codes in terms of incorporating fast linear algebra approaches as well as application to nonlinear elasticity problems.  

\begin{figure}[!h]
\centering
\includegraphics[width=2.0in]{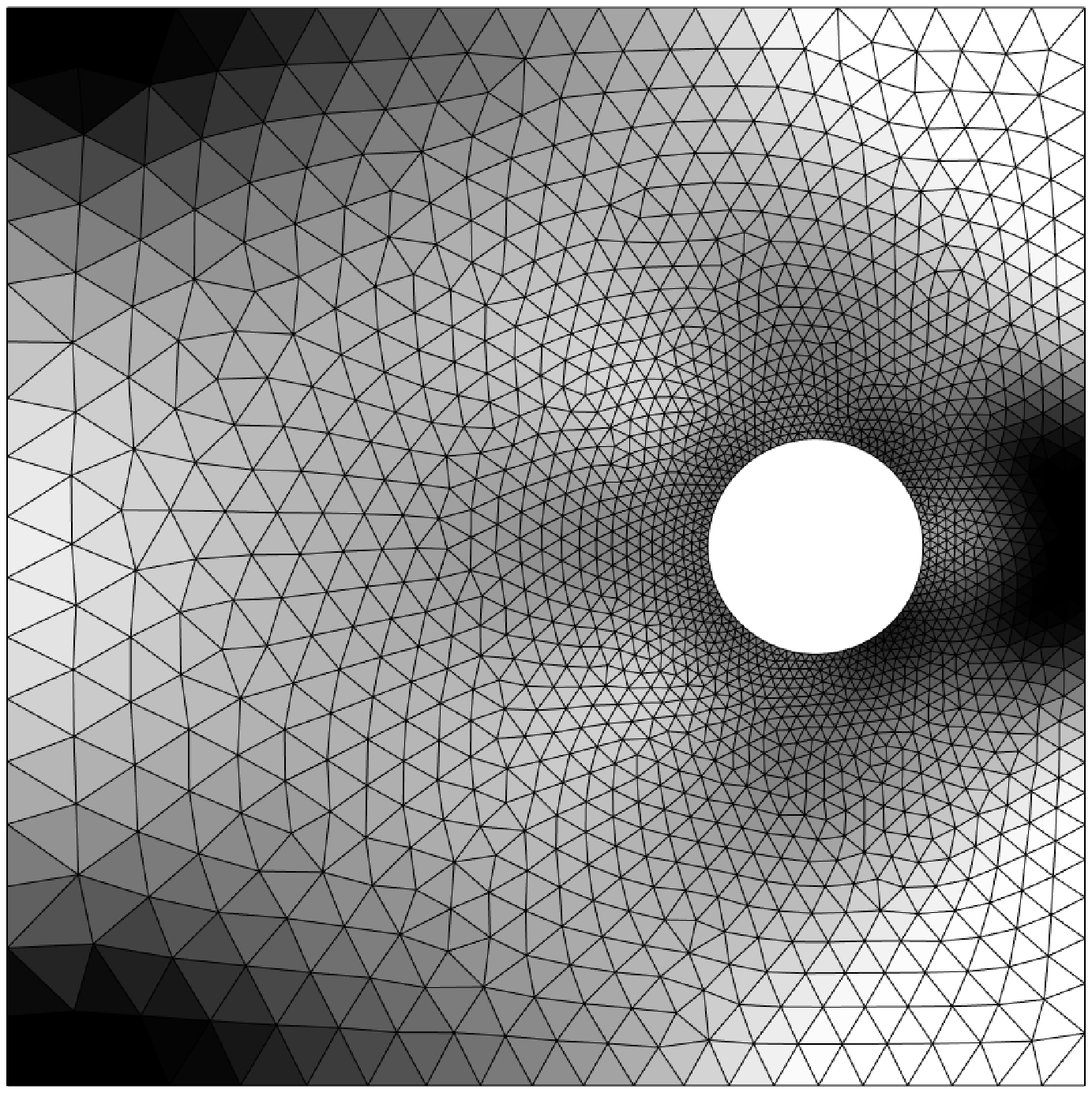}
\includegraphics[width=2.0in]{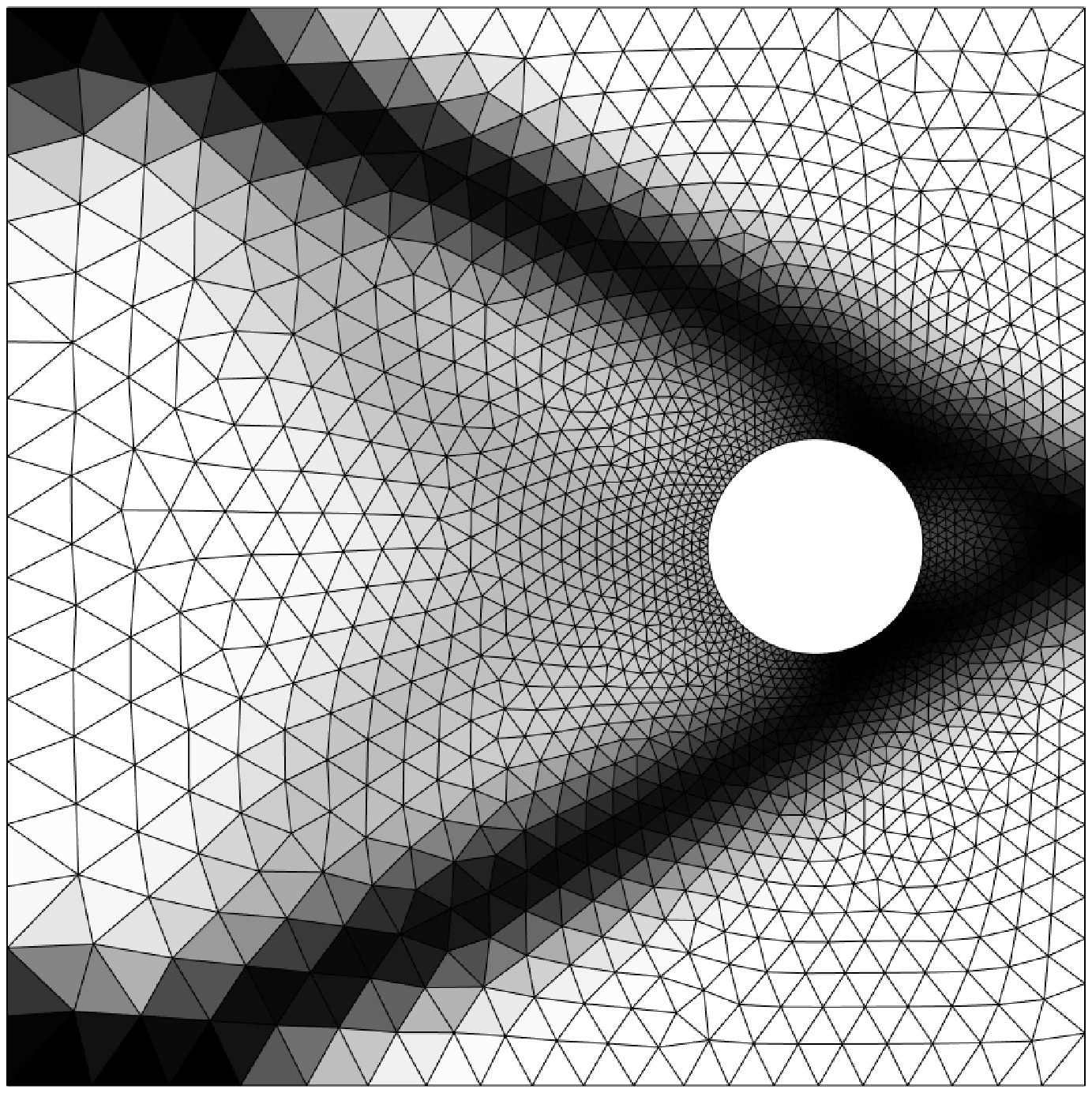}
\includegraphics[width=2.0in]{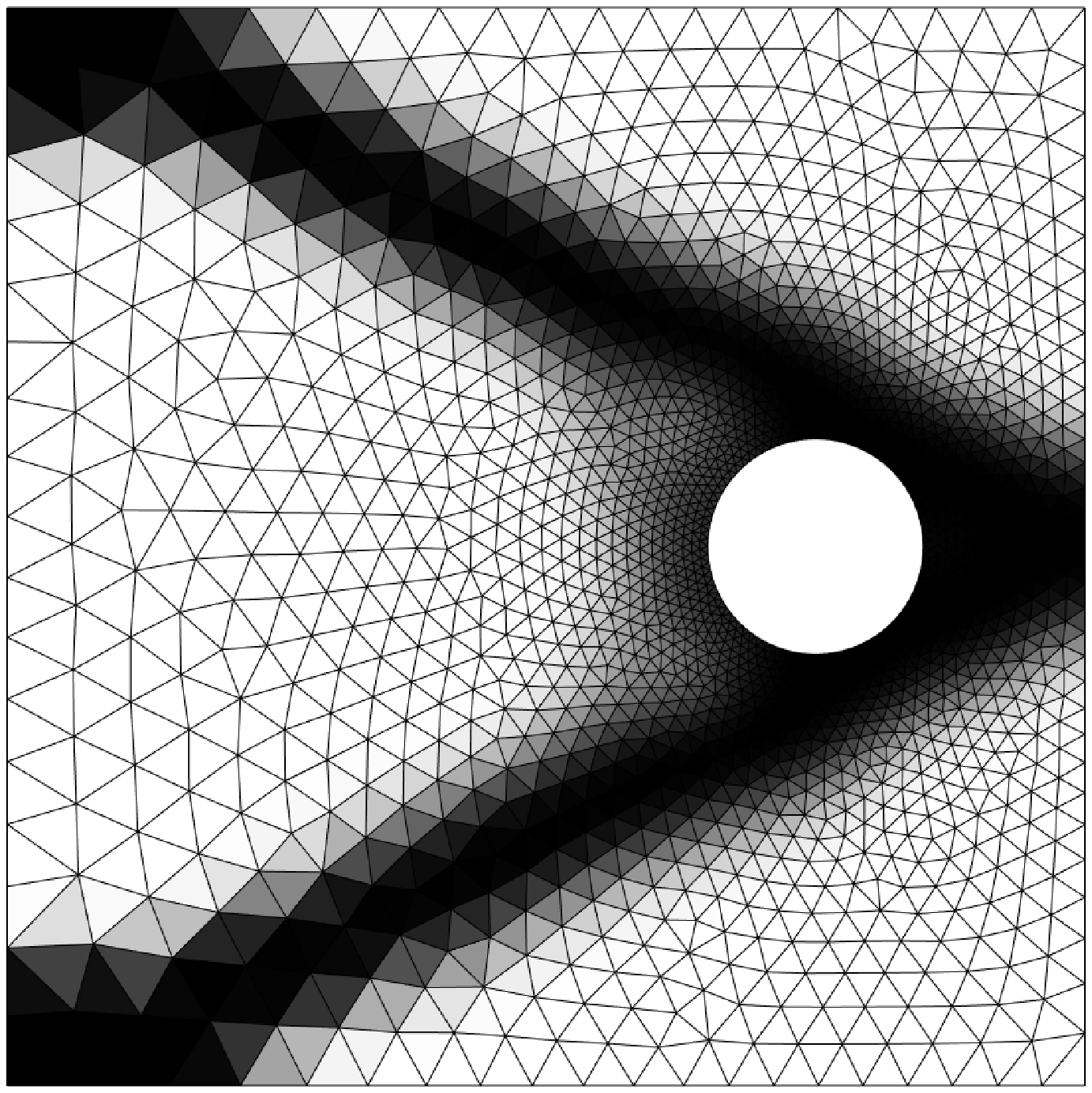}
\caption{\small{Different iterations (used as input $\bm x$) in a deterministic 2D topology optimization.}}\label{2D_xs}
\end{figure}

In this example, we seek to find the distribution for elastic modulus $E(\bm x, \bm m)$  given the observations on the nodal displacements $\bm u$ using a dataset which involves high-dimensional topology optimized designs $\bm x$. The equation governing the forward model is a linear elliptic PDE 
\begin{equation}
-\nabla \cdot (E(\bm x, \bm m) \nabla u ) = f
\end{equation}
where the elastic modulus $E(\bm x, \bm m) =  \exp(\tilde{E}(\bm x, \bm m))$ is expressed via the following expansion,
\begin{align}\label{KL_numerical}
\tilde{E}(\bm x, \bm m) = \sum_{i=1}^d \sqrt{\lambda_i} E_i(\bm x) m_i .
\end{align}
In this examples, the input $\bm x$ is the vector of design variables in topology optimization, in particular the volume fraction of each element. Three different input $\bm x$ are visualized in Figure~\ref{2D_xs}. These design instances are obtained from solving a similar deterministic topology optimization and they correspond to intermediate iterations within the overall topology optimization process. The number of iterations in the deterministic topology optimization however is not necessarily equal to the number of training points (in this example again we consider $n_{data}=200$). To generate $200$ individual input vectors $\bm x$ we select $20$ intermediate iterations from the deterministic topology optimization and add $10$ noise vector to each one of them to generate total of $200$ samples, i.e. we generate the input samples via
\begin{equation}
\bm x \gets \min(1,\max(0.001,\bm x + 0.01 \mathcal{N}(0,1)))
\end{equation}
where we ensure that the volume fractions are in the range $[10^{-3},1]$. 

The eigenvalues $\lambda_i$ and basis functions $E_i$ are taken as those for an exponential kernel $C(\bm x,\bm x')=\exp(-\|\bm x- \bm x' \|_1)$ on $D=[0,1]^2$. The details of KL expansion with this choice of kernel function is provided in~\cite{Teckentrup2015}. 

%In this example to ensure positivity of the elastic modulus we set $E_0=3$ and $\delta=0.02$.

The first step is to invert the forward model $\bm K(\bm x, \bm m) \bm u = \bm f$ and find samples of $\bm m$. We have noticed that using the original basis vectors $E_i$ result in a map that is many-to-one. To demonstrate this issue we first consider a smaller problem i.e. $d=6$ and apply the inference procedure to the dataset simulated with the original $E_i$. In this simpler case (in terms of underlying uncertainty) we assume a unimodal distribution. To make the map one-to-one (or perhaps less smooth) we apply $\sin$ and $\cos$ functions to the original basis functions and find the transformed basis vectors $\hat{E}_i$  according to
\begin{equation}\label{bss_fnc}
\hat{E}_i = 
\begin{cases}
\sin(E_i), \qquad & $i$~ is~ odd \\
\cos(E_i), \qquad  & $i$~is~ even. \\
\end{cases}
\end{equation}
The original basis vector $E_1$ and the the transformed vector are shown in Figure~\ref{mesh2D} (second and third panes). We then consider a larger dimension for the KL expansion $d=28$, use the transformed basis vectors and assume a bimodal distribution.

The true underlying parameter for the lower dimensional problem (with many-to-one map) is a multivariate Gaussian distribution $\bm m \sim \mathcal{N}(\bm \mu, \bm \Sigma)$ with
\begin{equation}
\bm \mu = [1.1:0.1:1.6]^T, \qquad 
\bm \Sigma = \bm I_{6 \times 6}
\end{equation}
where the notations $[a:i:b]$ denotes equidistant points between $a$ and $b$ with the distance $i$ including both ends $a$ and $b$.

\begin{figure}[!h]
\centering
\includegraphics[width=1.5in]{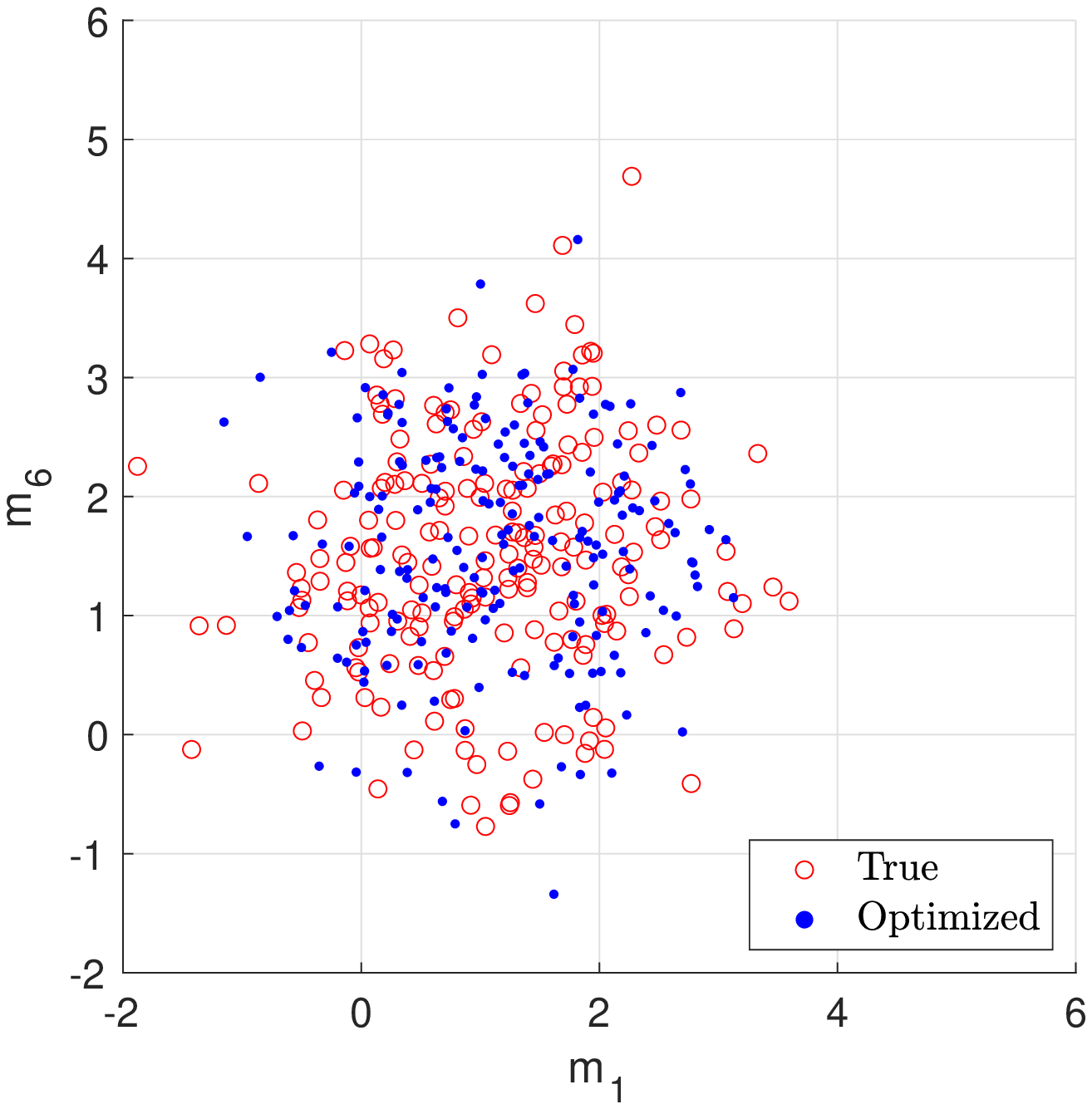}
\includegraphics[width=1.5in]{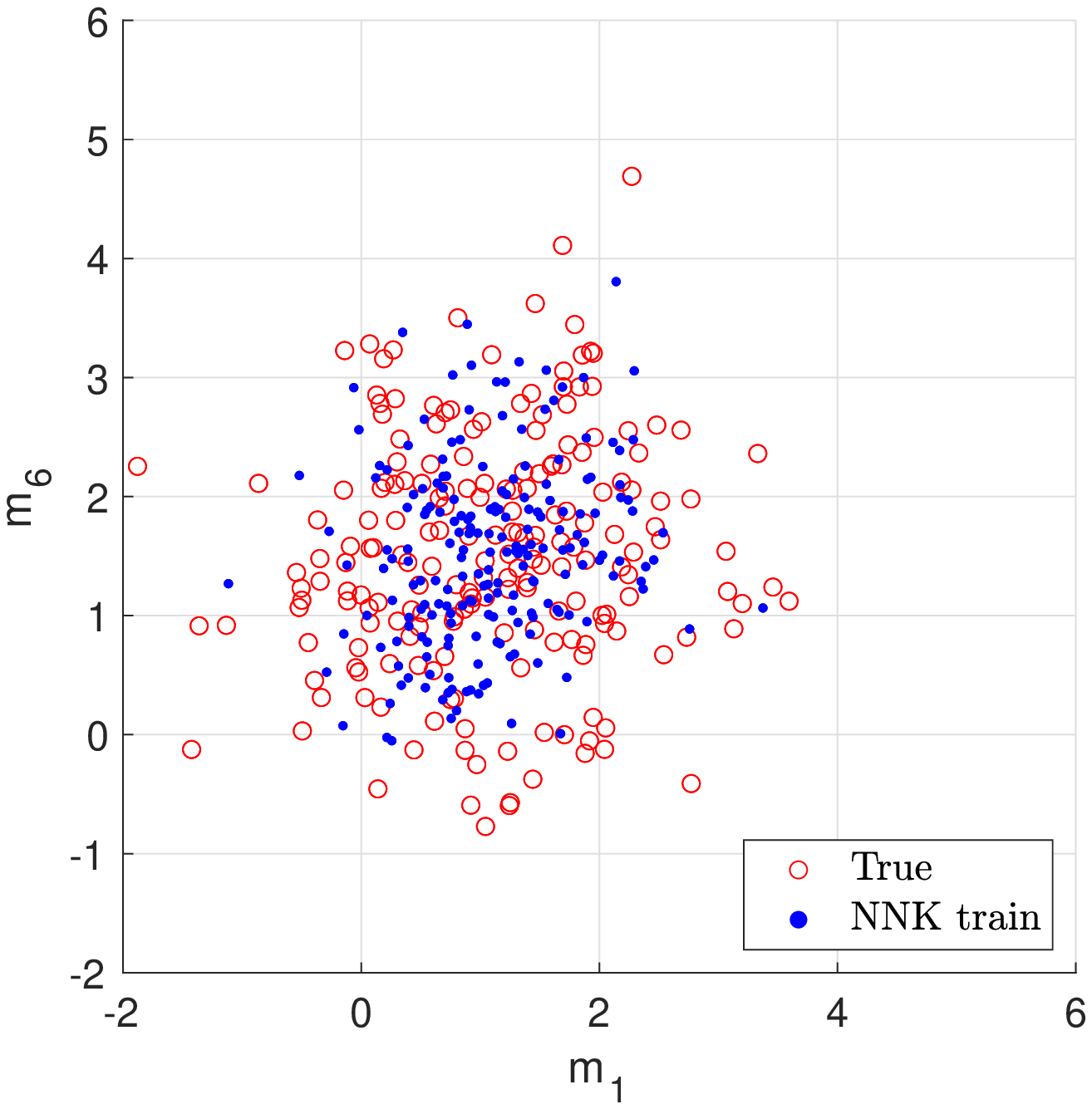}
\includegraphics[width=1.5in]{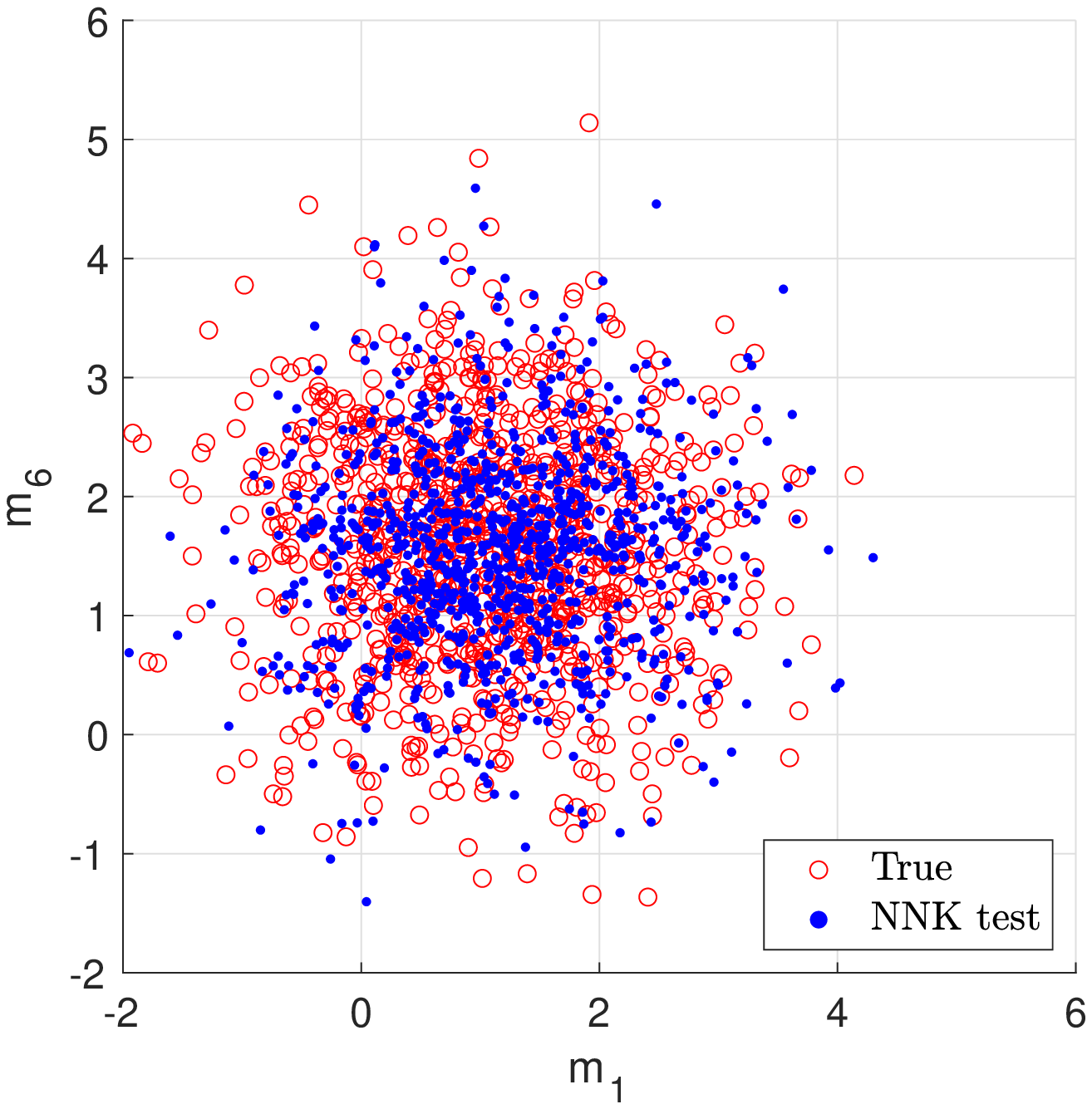}
\includegraphics[width=1.6in]{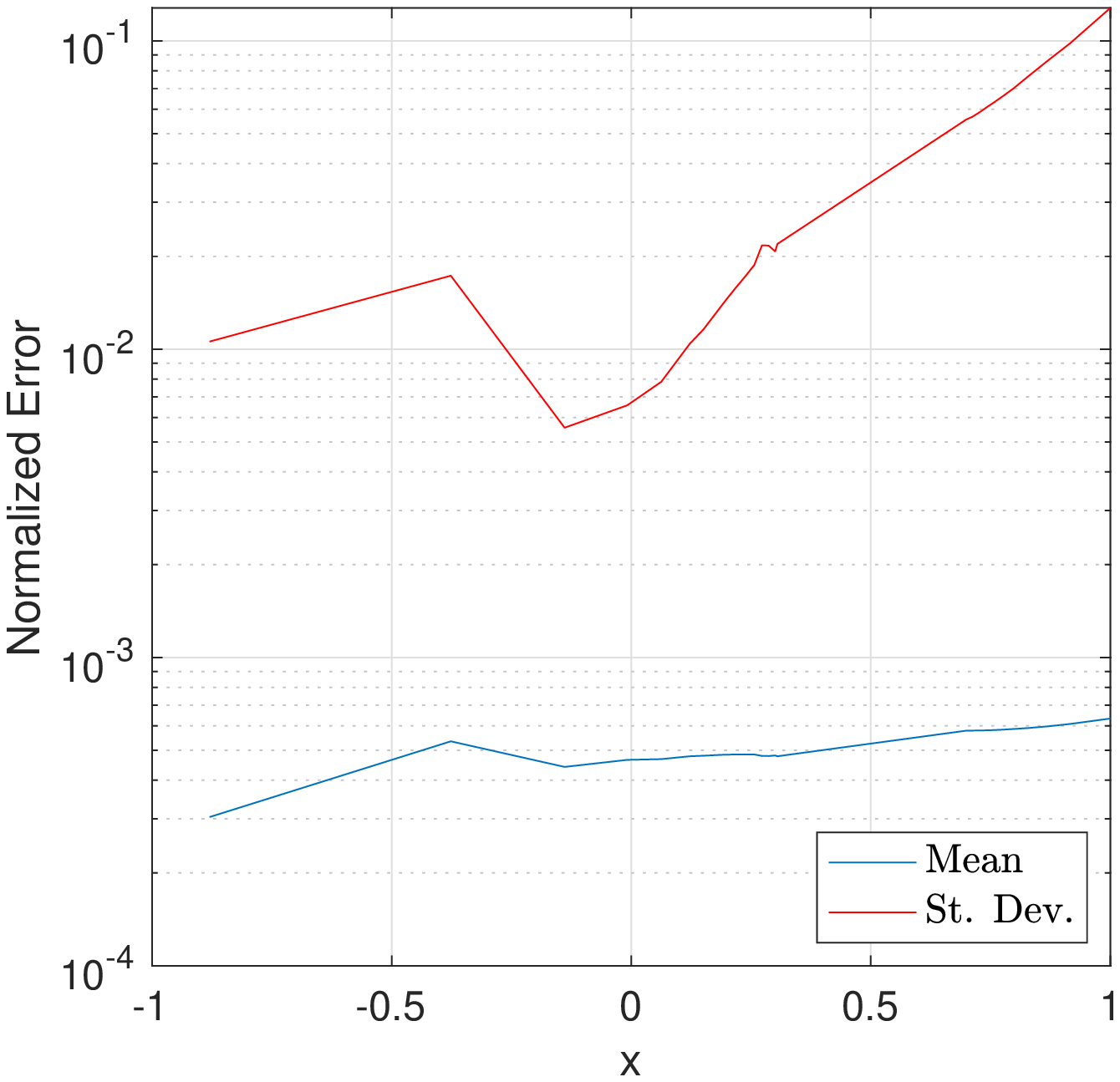}
\caption{\small{Finding samples of latent parameters via optimization of the forward model $\bm K(\bm x, \bm m) \bm u = \bm f$ (left), training and testing of NNK (second and third panes), and normalized errors in mean and standard deviation for particular degrees of freedom (right).}}\label{fig_unimodal_2D}
\end{figure}

The result of inversion of the forward map is shown in the left pane of Figure~\ref{fig_unimodal_2D} for the slice of $m_1,~m_6$. Note that there is a discrepancy between the true samples and the optimized ones. It should be noted that the NR approach in this case achieves the $\delta_R=0.01$ tolerance. The issue here is the non-uniqueness of the inverse map, which results in samples that are not equivalent to underlying parameter samples but they satisfy the forward model equation, i.e. $\mathcal{G}(\bm x^{(i)}, \bm m^{(i)}) = \bm y^{(i)}$. We also again mention that the NR approach in Algorithm~\ref{alg:opt_fwd}  involves  a large number of degrees of freedom. More precisely, there are $3380$ degrees of freedom, i.e. the vector of nodal displacements is $\bm y \in \mathbb{R}^{3380 \times 1}$. Considering only one sample of $\bm m$, the solution of the nonlinear equations take $t=23.08$ seconds. The total time for all dataset i.e. $200$ samples is $76$ minutes.  The Jacobian computation is more involved as it entails the computation of $\partial \bm K/\partial \bm m$ which itself requires another matrix assembly and later a linear solve in the form of $\bm A \bm x = \bm b$, cf.~\ref{comp_J}.

Finding the optimized samples, we proceed with the next steps of our inference procedure. We train a network with three hidden layers. The number of nodes in all layers is $[6,12,7,4,1]$ and the coefficients are $\bm \alpha \in \mathbb{R}^{25 \times 6}$. The results of training and testing are shown in the second and third panes of Figure~\ref{fig_unimodal_2D}. Similarly to previous examples, we compute the normalized error in mean and standard deviation of one direction of displacement (in this example $y$ direction) for particular degrees of freedom and a fixed design iterate (last iterate of the deterministic optimization). The particular degrees of freedom are those whose associated nodes admit $-0.05<y<0.05$, i.e. we consider (almost) a line in the middle of the structure (in the $y$ direction). For true solution we compute the displacements with $1000$ Monte Carlo samples drawn from the true distribution. Based on these results, it appears that the inference has been successful even considering the first step resulted in less accurate samples. The test dataset which is the prediction of the NNK seems almost similar to the distribution of the underlying parameter.

We now consider larger number of dimensions $d=28$ with the transformed basis vectors $\hat{E}_i$. For this part of the example, we assume the underlying parameter has a bimodal distribution which consists of two equally weighted multivariate Gaussian distributions in the form of $\bm m \sim 0.5\mathcal{N}(\bm \mu_1, \bm \Sigma_1) + 0.5\mathcal{N}(\bm \mu_2, \bm \Sigma_2)$
with $\bm \mu_1 = \bm 0_{28 \times 1},~\bm \mu_2 = 4 \times \bm 1_{28 \times 1}$ and $\bm \Sigma_1=[0.11:0.01:0.38]^T,~\bm \Sigma_2=[0.38:-0.01:0.11]^T$ where $\bm 0_{28 \times 1}$ and $\bm 1_{28 \times 1}$ are the vectors of zeros and ones with size $d=28$.

\begin{figure}[!h]
\centering
\includegraphics[width=2.0in]{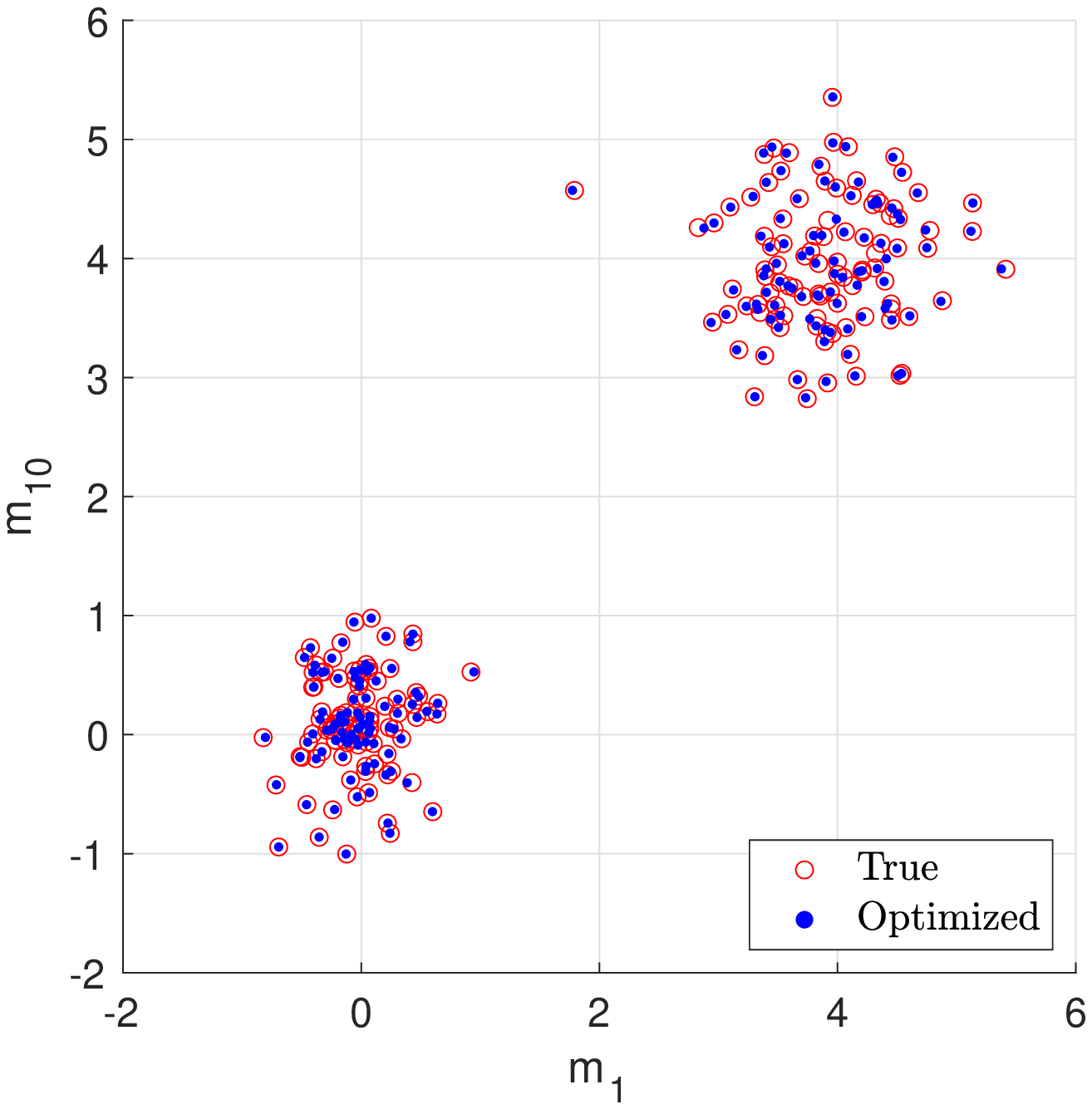}
\includegraphics[width=2.0in]{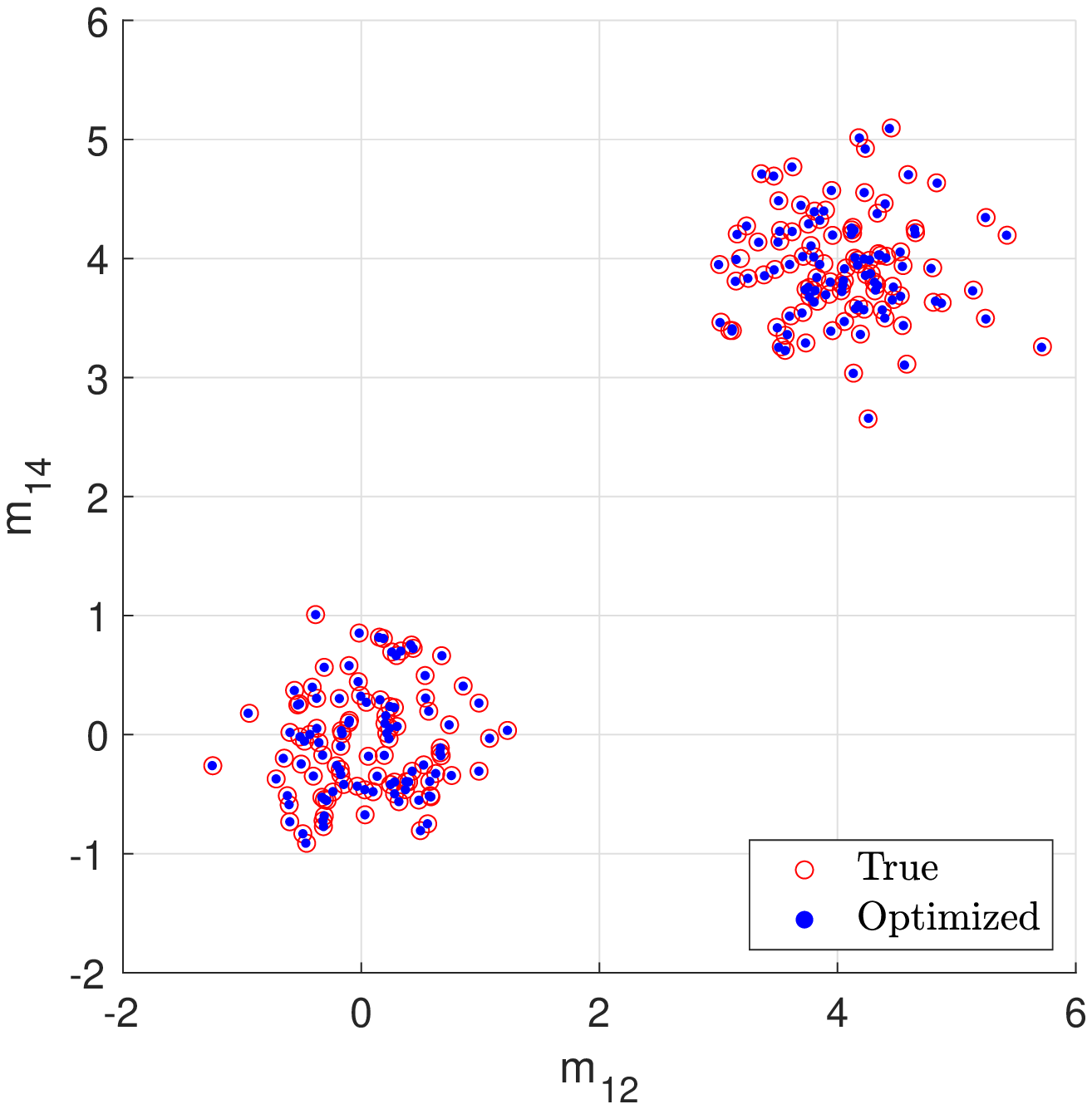}
\includegraphics[width=2.0in]{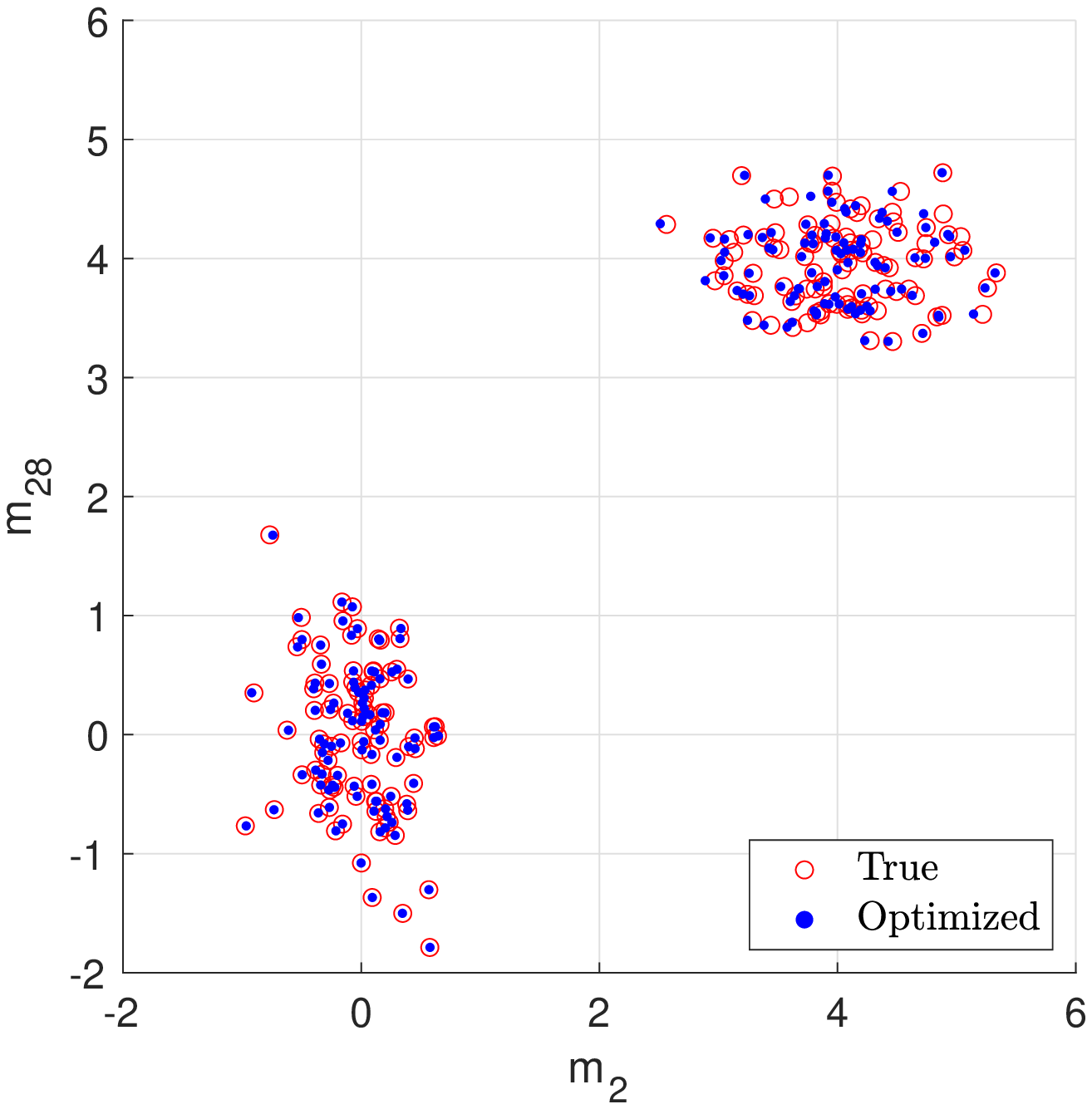}
\caption{\small{Finding samples of latent parameters via optimization of the forward model $\bm K(\bm x, \bm m) \bm u = \bm f$ for three slices of $m_1-m_{10}$, $m_{12}-m_{14}$ and $m_2-m_{28}$.}}\label{fig_bimodal_2D}
\end{figure}

The result of optimization for the samples is shown in Figure~\ref{fig_bimodal_2D}. As we had shown in the previous example, the step of forward model optimization/inversion can accurately find bimodal distributions so long as the forward map is one-to-one and noise-free.  

\begin{figure}[!h]
\centering
\includegraphics[width=2.0in]{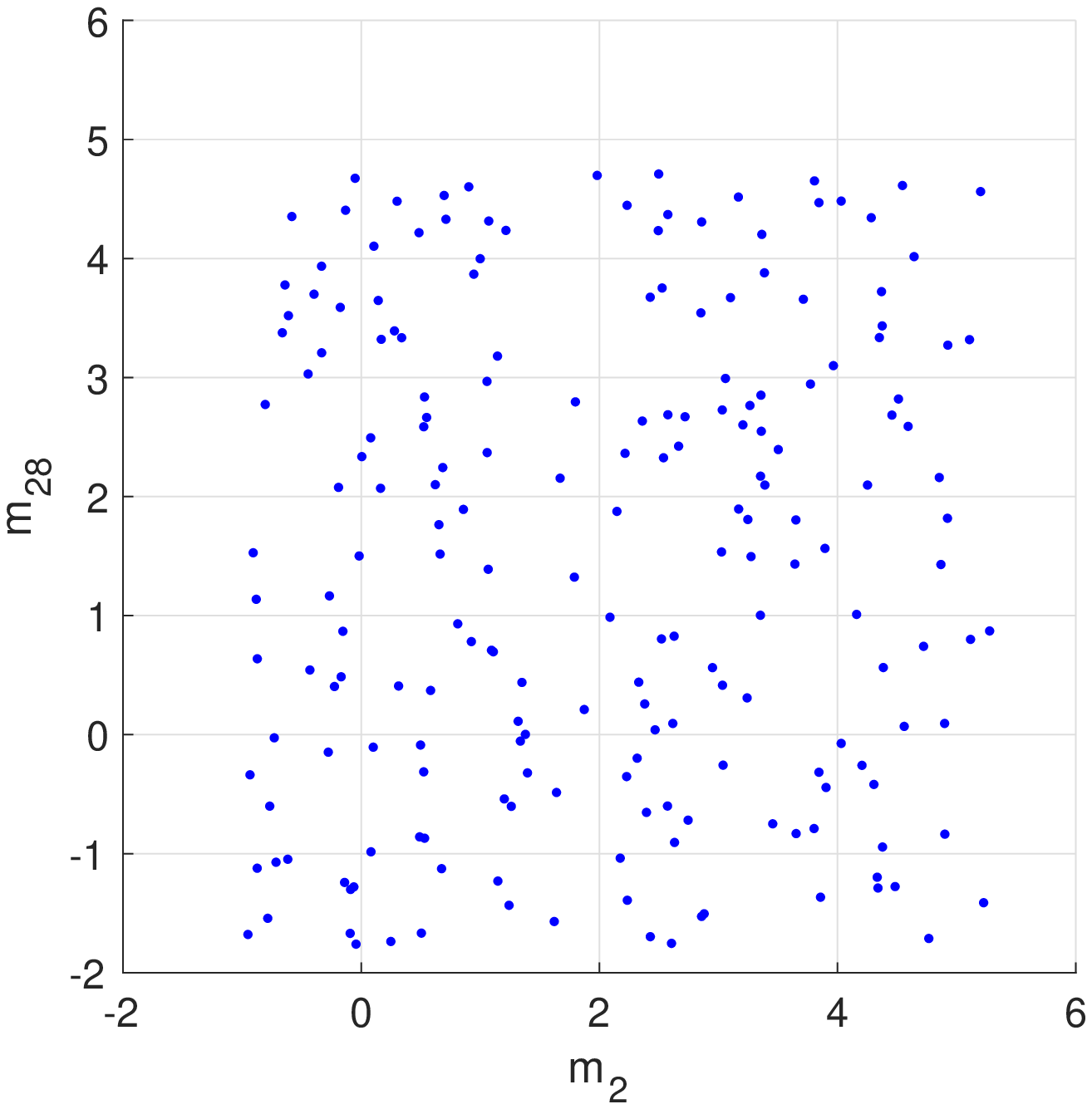}
\includegraphics[width=2.0in]{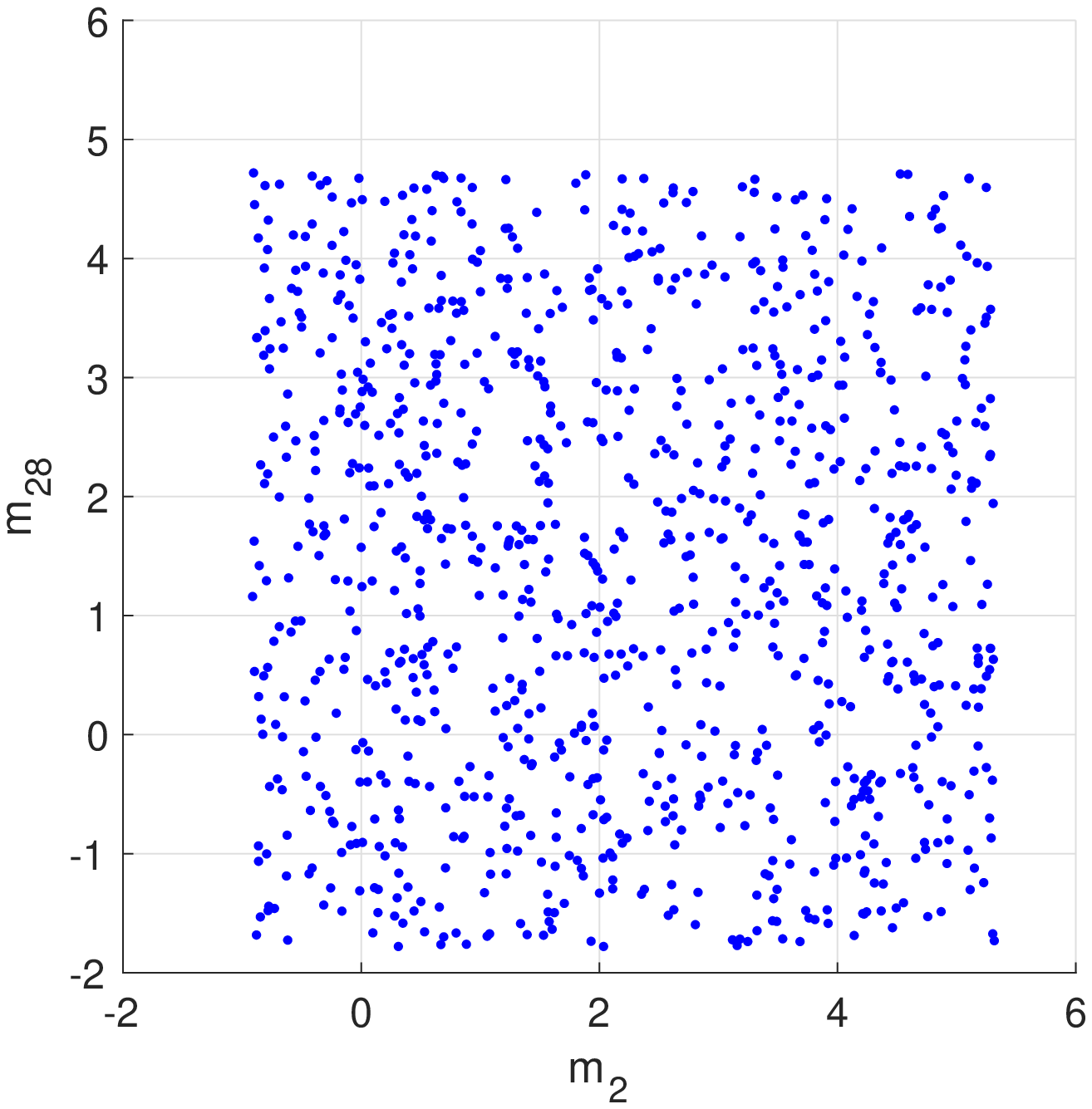}
\includegraphics[width=2.0in]{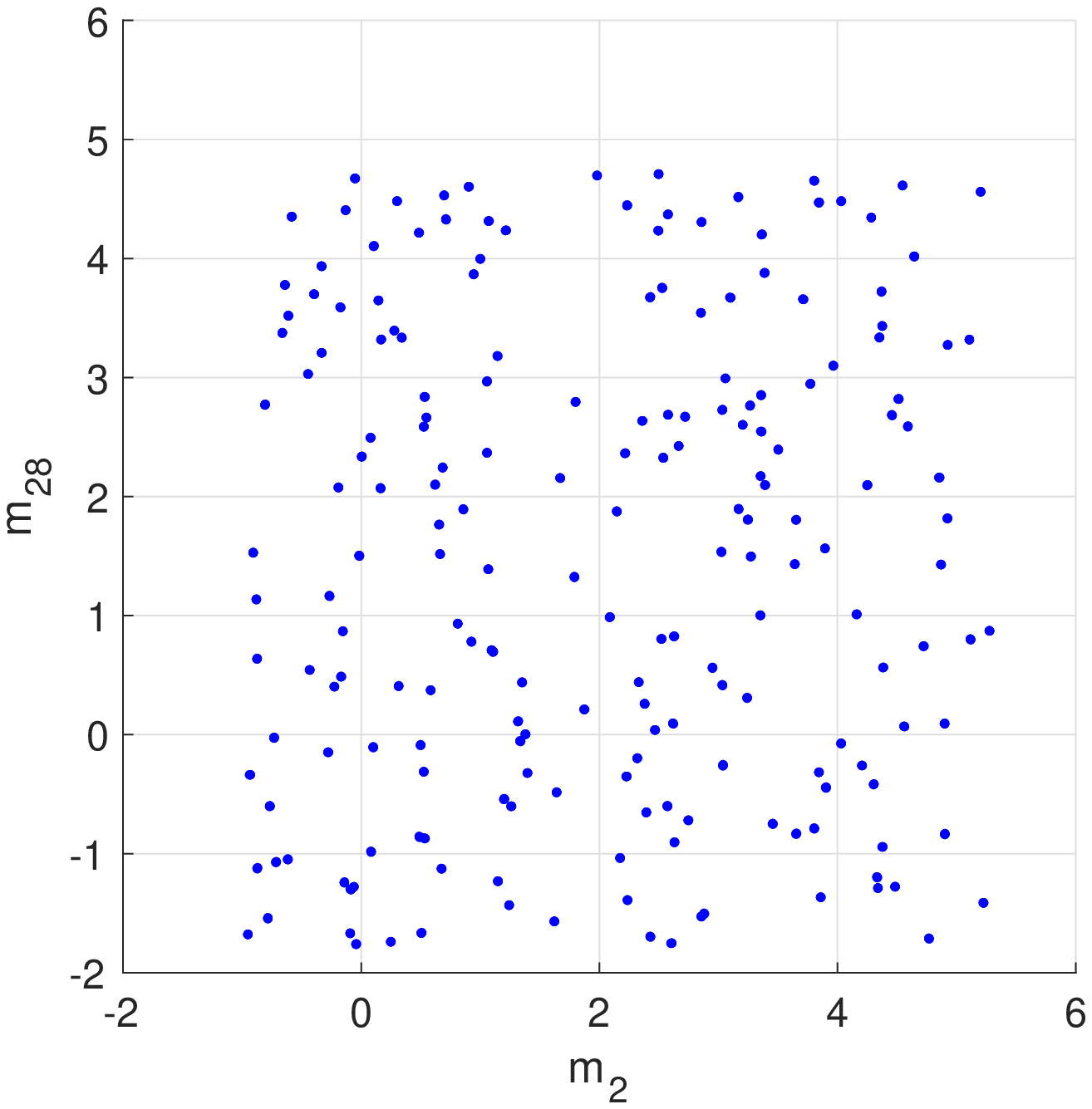}\\
\includegraphics[width=2.0in]{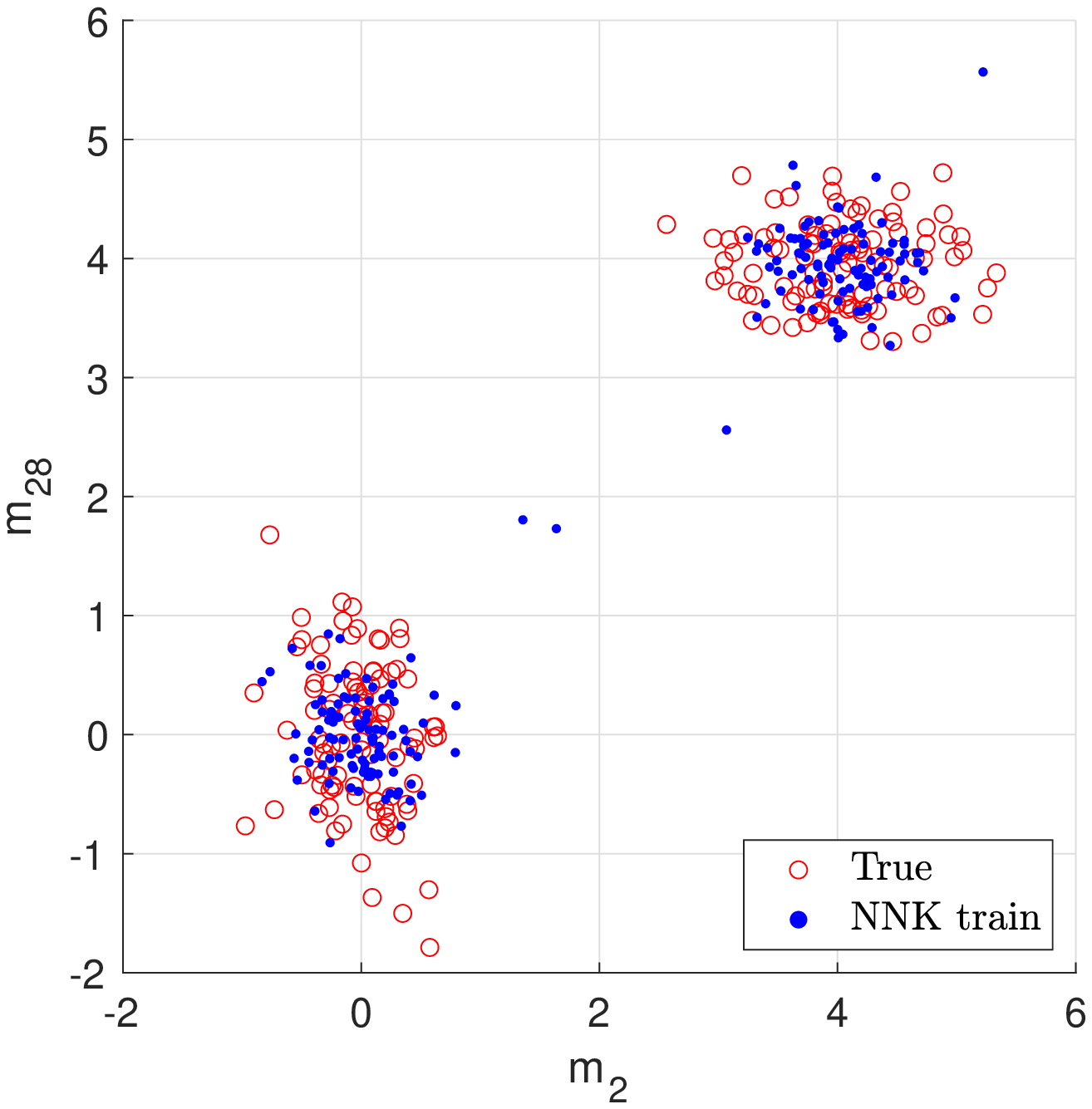}
\includegraphics[width=2.0in]{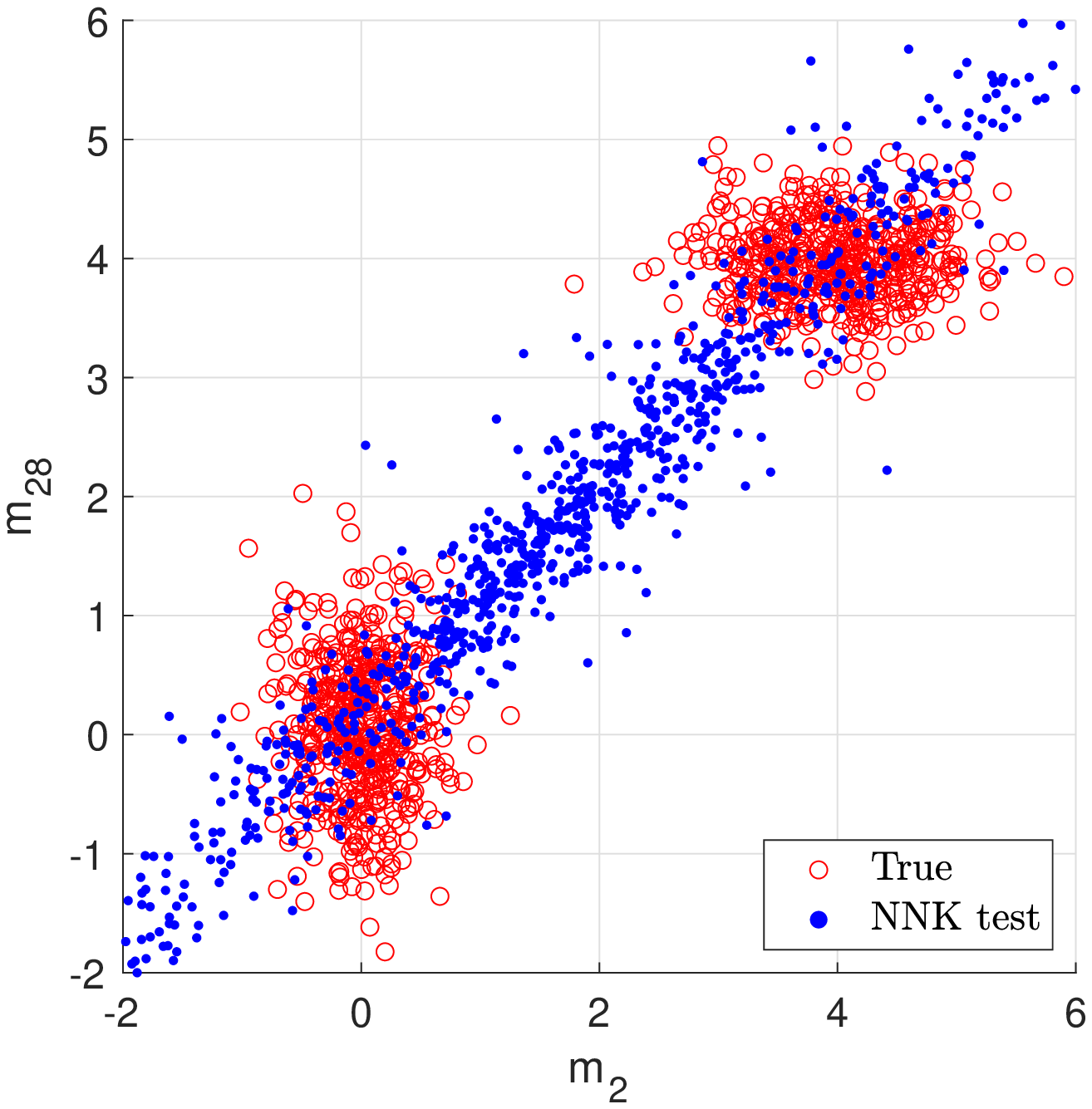}
\includegraphics[width=2.0in]{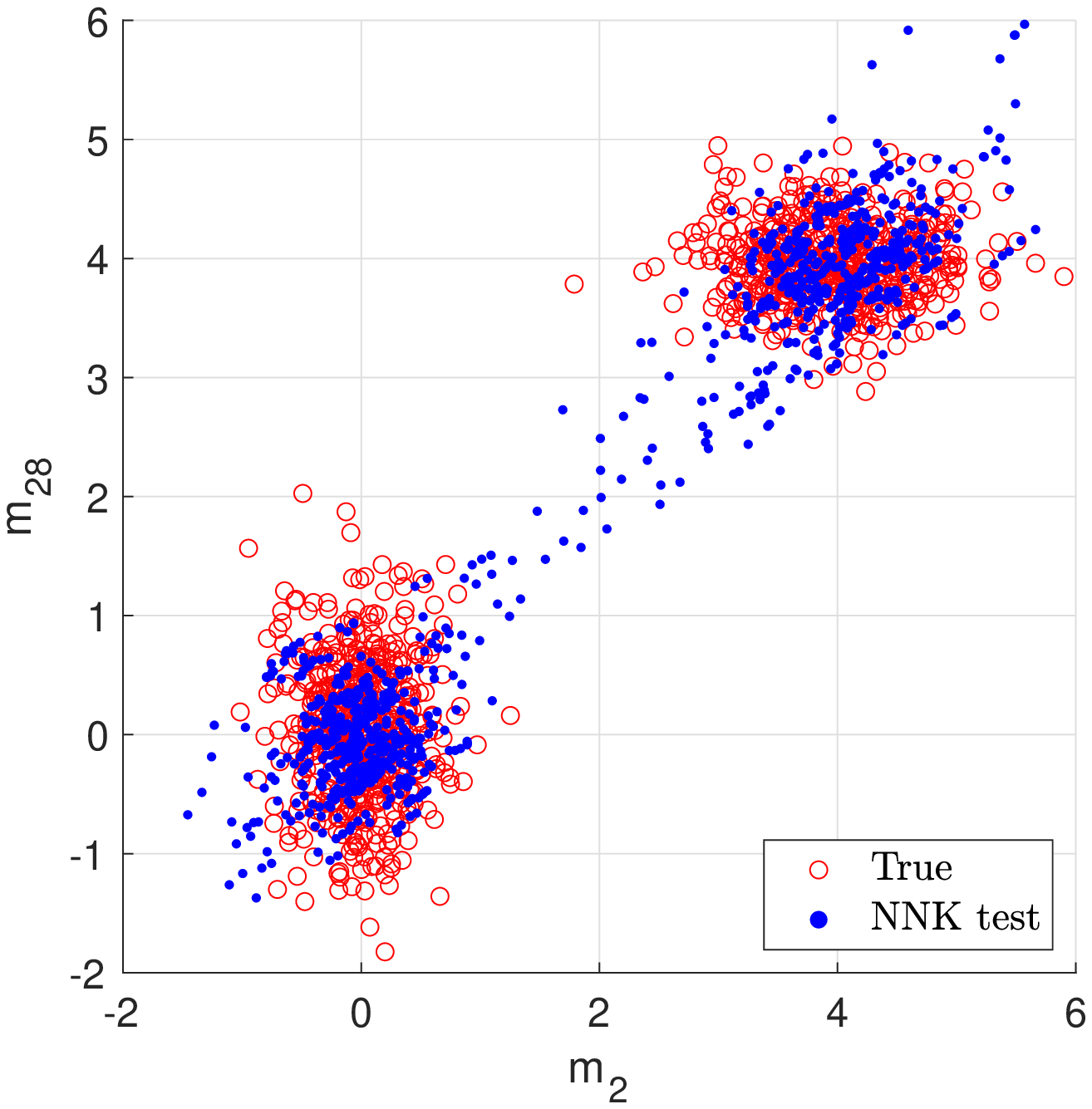}
\caption{\small{Inference in $28$ dimensions and bimodal uncertainty associated with the 2D topology optimization problem. Prior samples: training (top-left), randomly distributed samples for testing (top-middle), augmented training prior samples (top-right). Posterior samples, i.e. the prediction of the NNK map: training (bottom-left), prediction with randomly distributed prior samples (bottom-middle) and prediction with augmented training prior samples (bottom-right).}}\label{fig_2D_bimodal_iden}
\end{figure}

\begin{figure}[!h]
\centering
\includegraphics[width=2.0in]{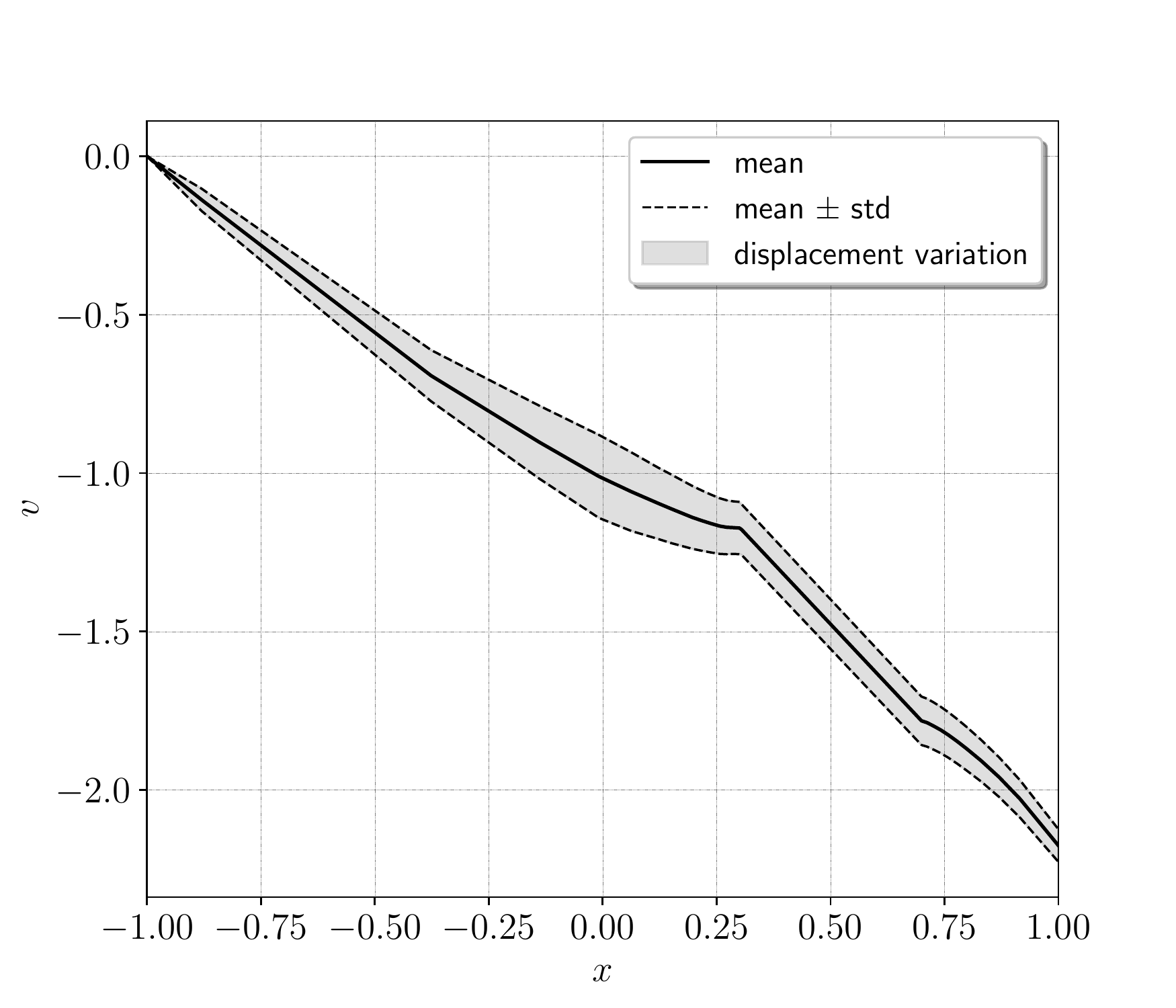}
\includegraphics[width=2.0in]{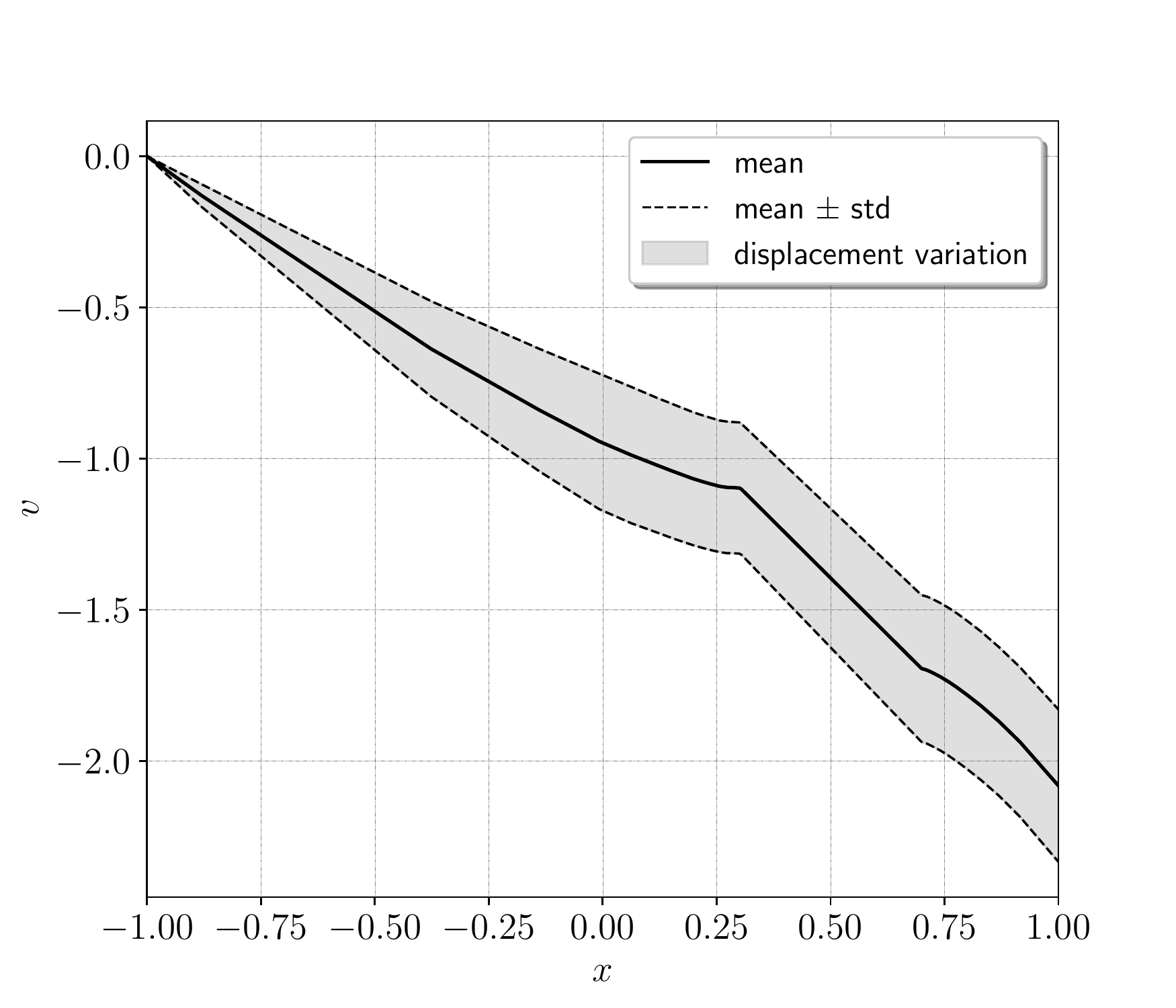}
\includegraphics[width=2.0in]{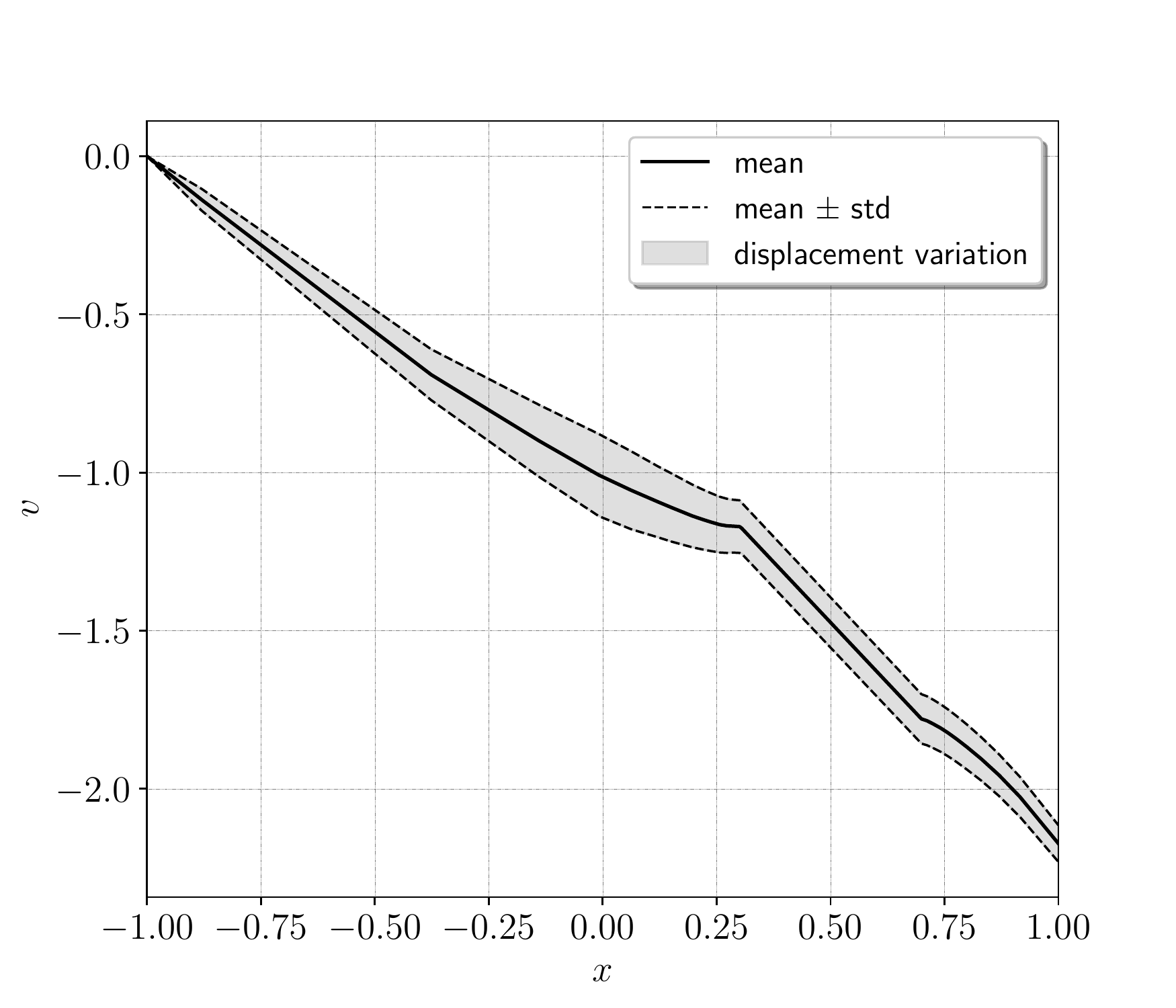}
\caption{\small{Statistical study on displacement in $y$ direction (denoted by $v$) using Monte Carlo samples of the latent variables, i.e. true solution (left), posterior samples from the neural net map with randomly distributed prior samples (middle), posterior samples from the neural net map with augmented training prior samples (right).}}\label{fig_2D_stat}
\end{figure}

Next, we perform training and testing with a similar NNK, the only difference is the number of nodes in the first layer, i.e. $[28,12,7,4,1]$ and the second dimension of $\bm \alpha$, i.e. $\bm \alpha \in \mathbb{R}^{25 \times 28}$. The results of training and testing for two scenarios of randmoly distributed prior samples and augmented training samples are shown in Figure~\ref{fig_2D_bimodal_iden}. As expected, the test dataset with augmented training samples yield closer distribution to the bimodal uncertain parameter.
To investigate the effectiveness of these predictions, similarly to the previous part we compute statistical displacements (in $y$ direction denoted by $v$) on the same finite element nodes considered previously. The result of this statistical study is shown in Figure~\ref{fig_2D_stat}. The prediction with uniform samples of prior exhibits some discrepancy with the true solution (Monte Carlo analysis) especially in terms of scatter, i.e. standard deviation, however the prediction on the third pane which is associated with the augmented training samples is very similar to the true solution.  

%We then predict the forward solution.. it is not too shabby for the second case. Note that this is a bimodal one. 

\subsection{Identification of Elastic Modulus in 3D Topology Optimization}

\begin{figure}[!h]
\centering
\includegraphics[width=2.5in]{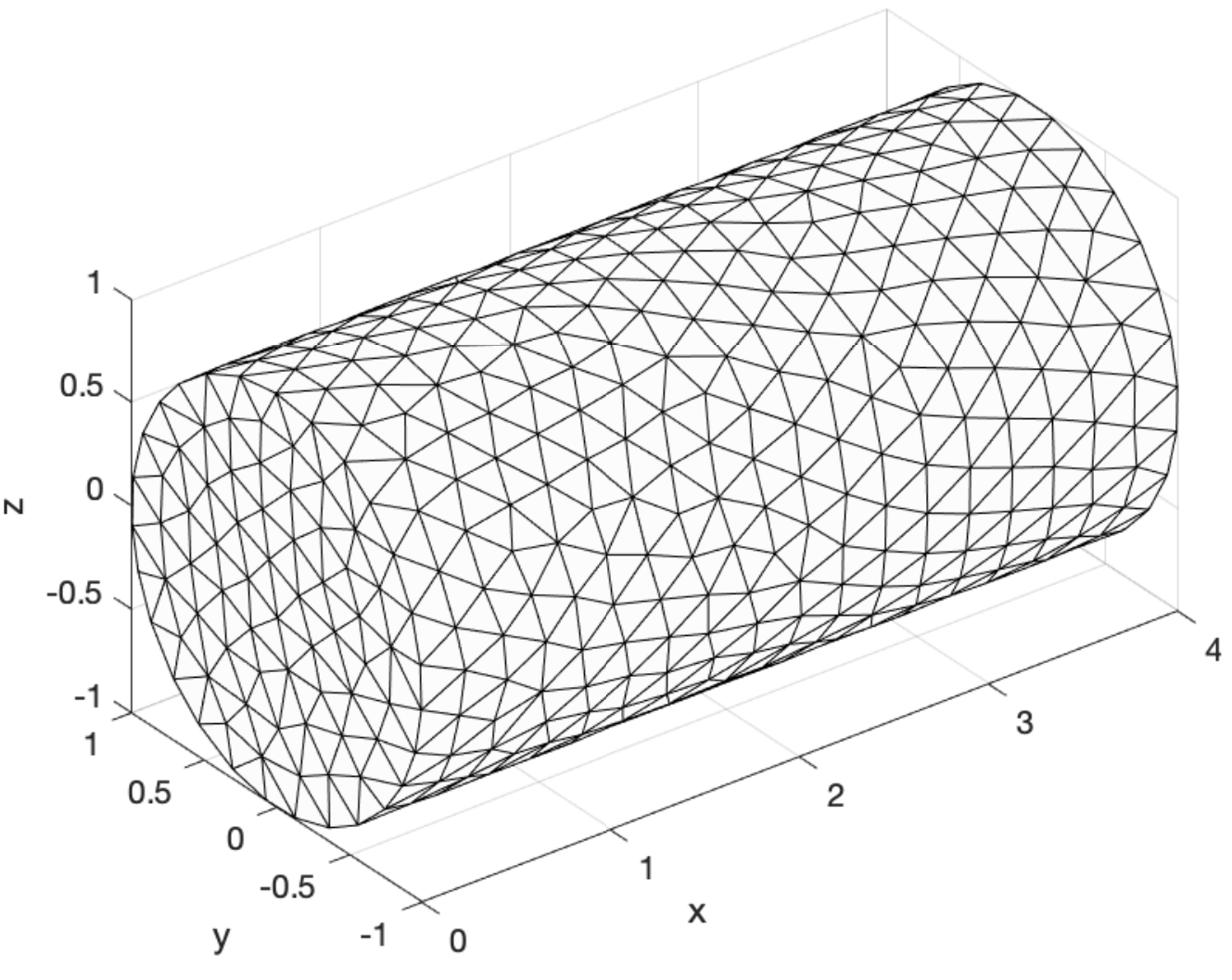}
\hspace{0.75cm}
\includegraphics[width=2.5in]{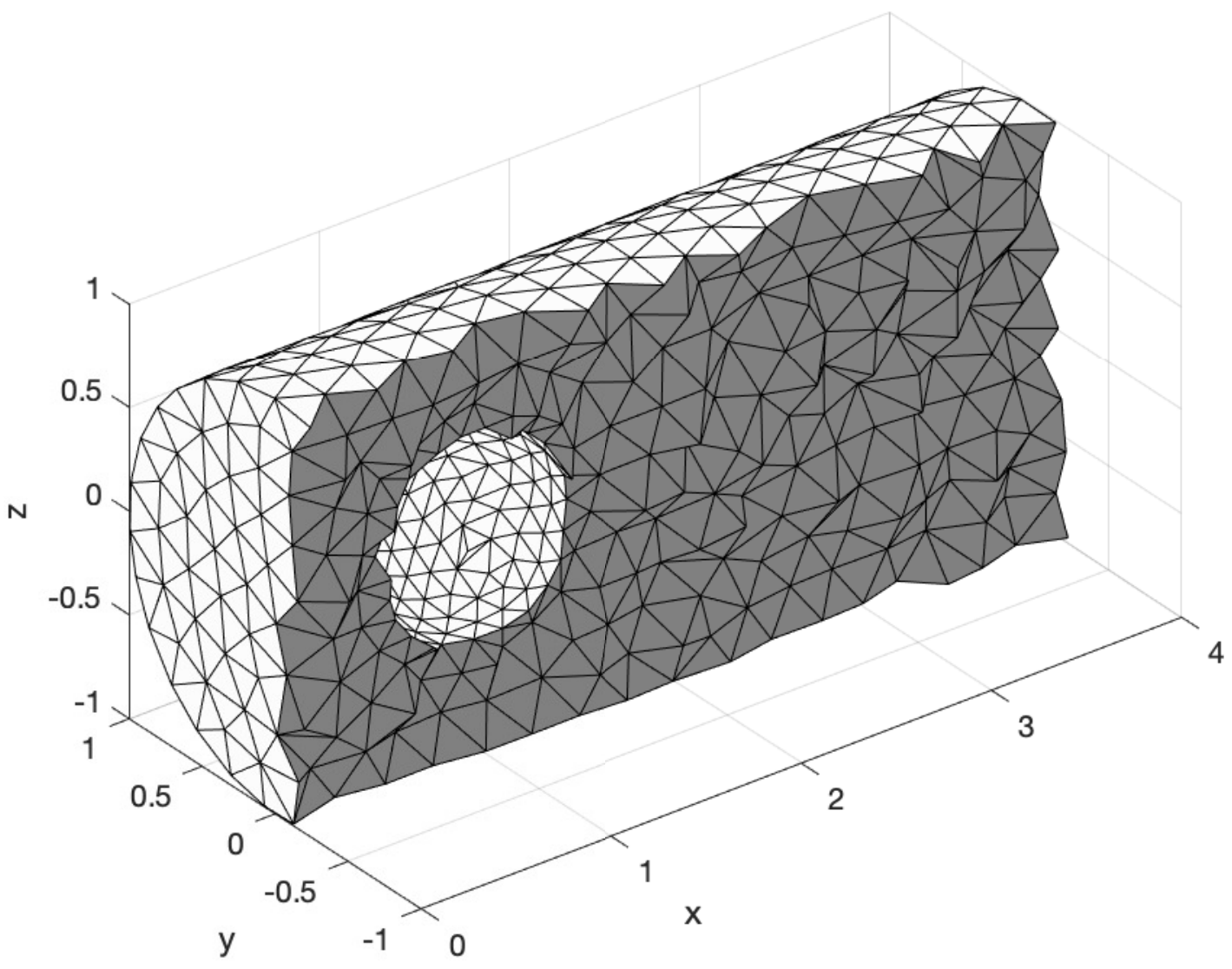}
\caption{\small{Geometry and discretization of a cylindrical domain with a spherical hole: full domain (left) half domain (right).}}\label{mesh3D}
\end{figure}

\begin{figure}[!h]
\centering
\includegraphics[width=2.5in]{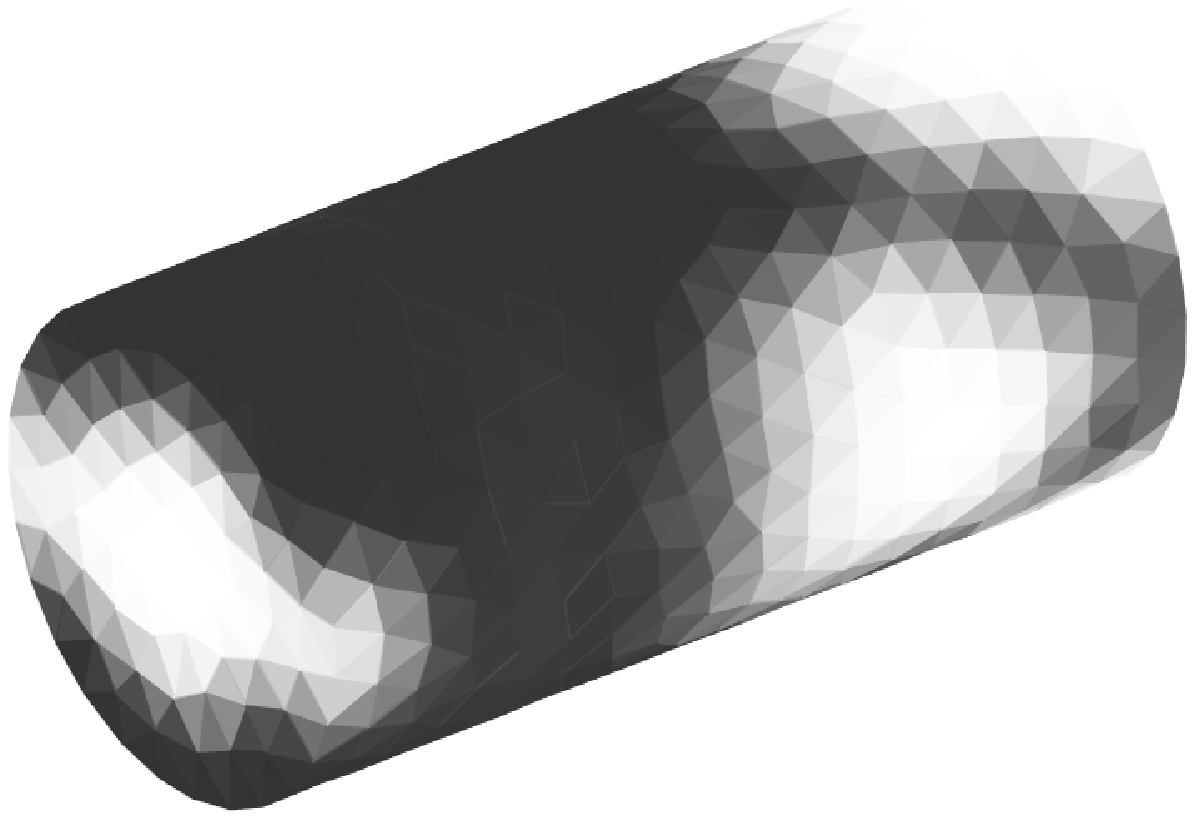}
\hspace{0.75cm}
\includegraphics[width=2.5in]{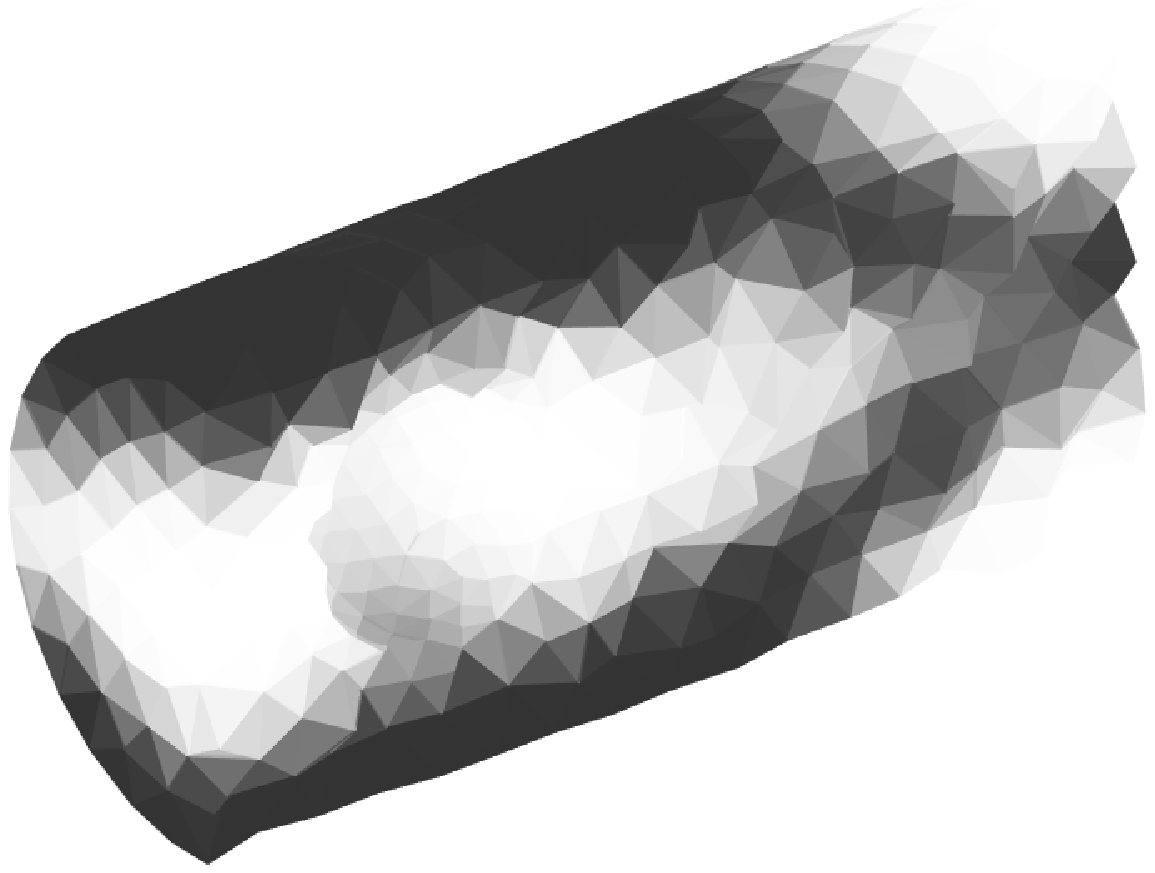}
\caption{\small{Final iteration (used as one input $\bm x$) in the 3D topology optimization: full domain (left) and half domain (right).}}\label{fig_3D_top}
\end{figure}

\begin{figure}[!h]
\centering
\includegraphics[width=2.0in]{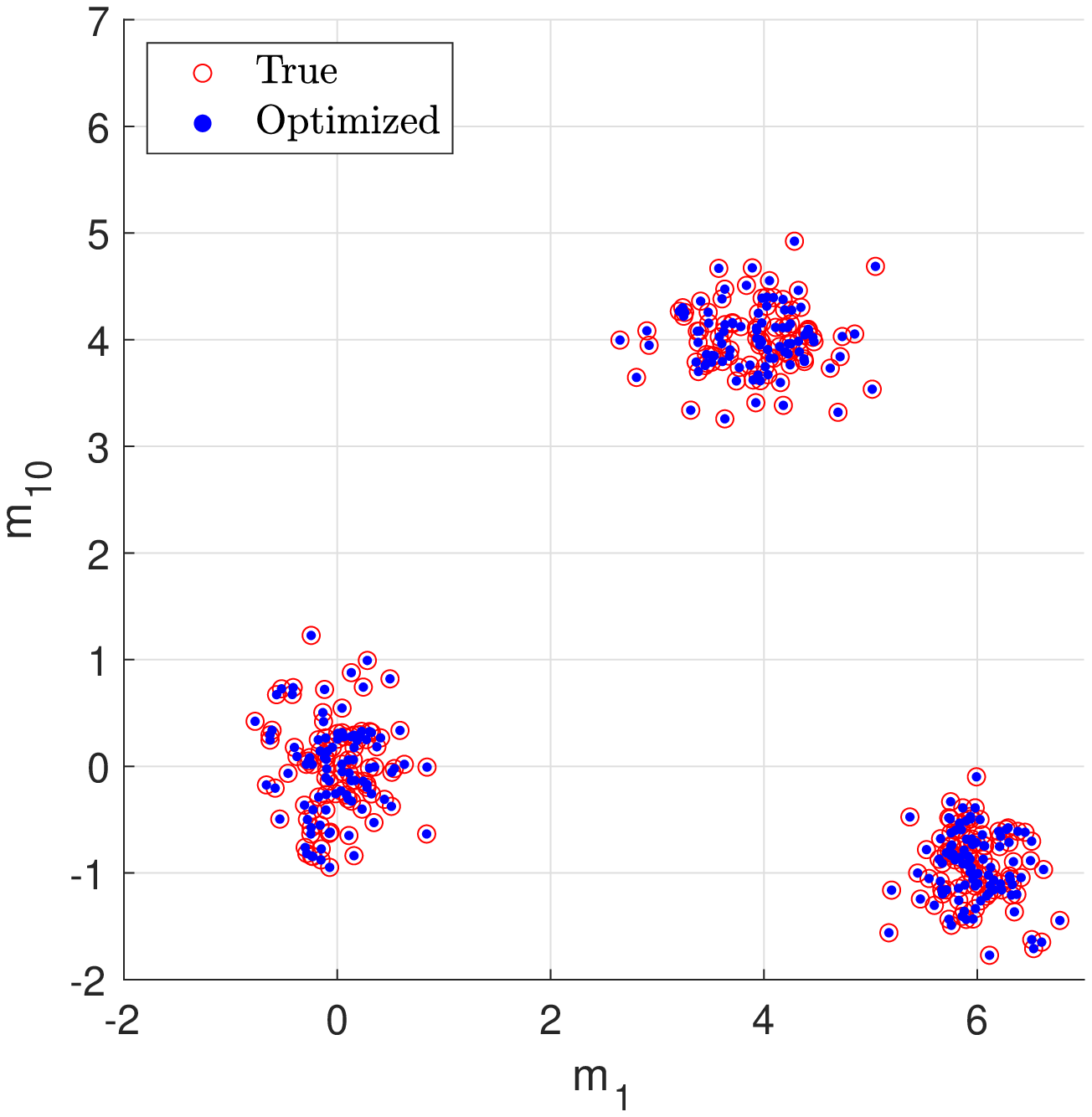}
\includegraphics[width=2.0in]{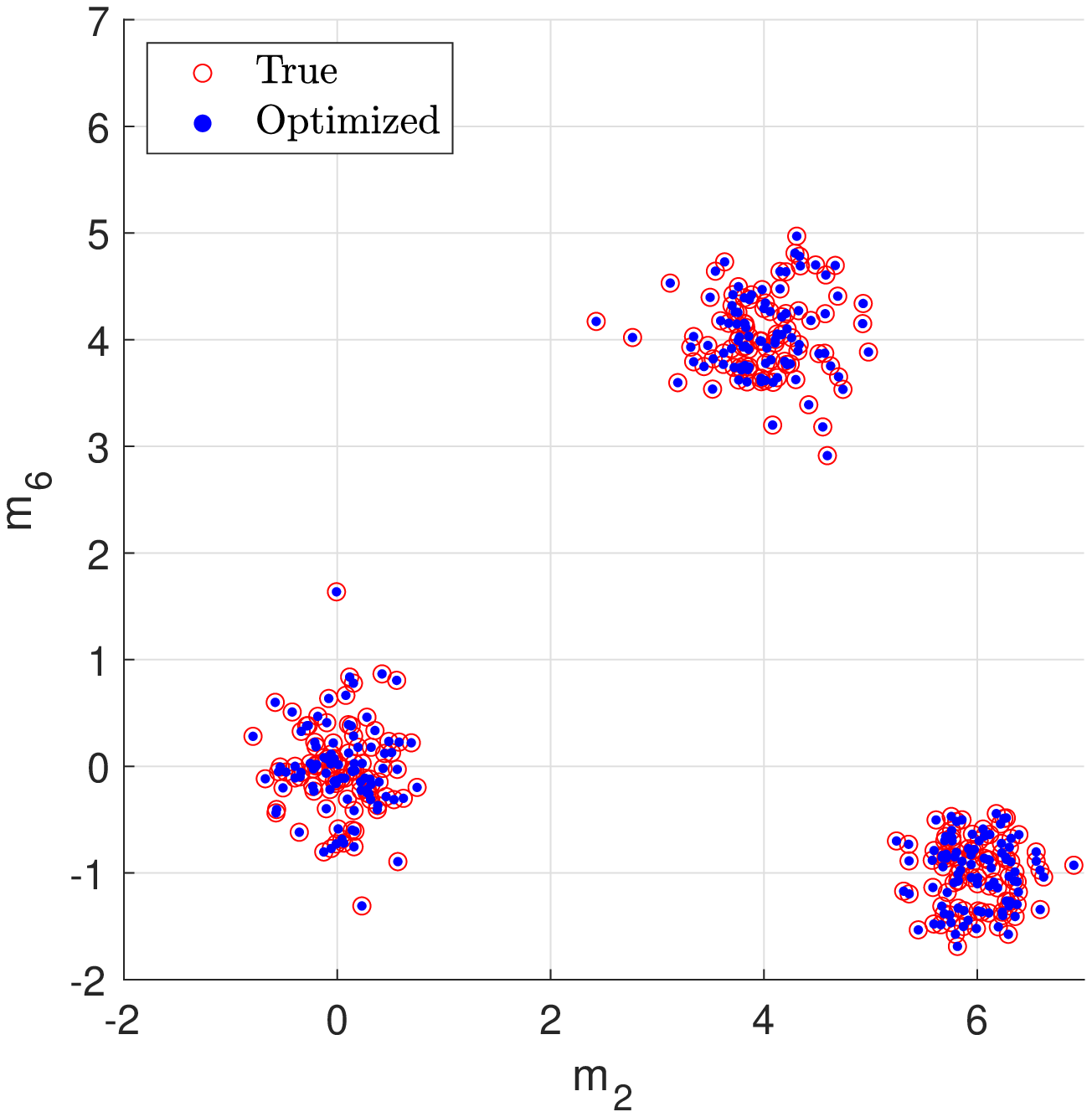}
\includegraphics[width=2.0in]{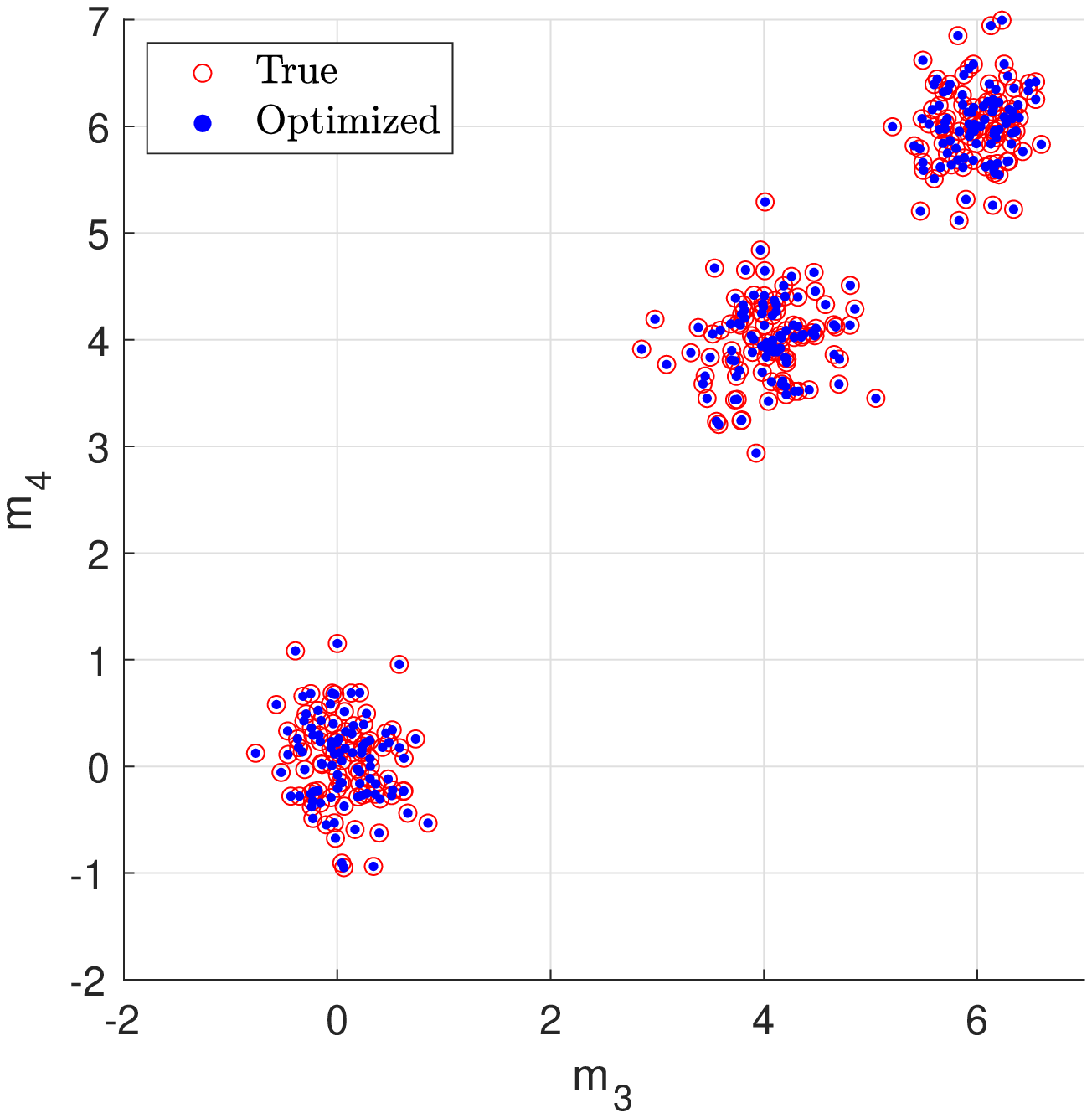}
\caption{\small{Finding samples of latent parameters via optimization of the forward model $\bm K(\bm x, \bm m) \bm u = \bm f$ for three slices of $m_1-m_{10}$,~$m_2-m_6$ and $m_3-m_4$.}}\label{fig_3D_opt}
\end{figure}

In this example we investigate the performance of our approach on a three dimensional topology optimization problem.  The geometry of the full and half structure and their discretization with tetrahedral elements are shown in Figure~\ref{mesh3D}
The structure is assumed as a cantilever beam which is clamped in $x=0$. Vertical point loads are applied in $z$ direction at the nodes with $x=4,~ -1\leq y \leq 1,~ -0.05 \leq z \leq 0.05$. The final result of the deterministic topology optimization is shown in Figure~\ref{fig_3D_top}.
The procedure is similar to the previous case, however the problem involves larger number of degrees of freedom, i.e. $\bm y \in \mathbb{R}^{5817 \times 1}$. We assume a KL expansion with $d=10$ basis vector and consider a more challenging trimodal distribution. In particular the true underlying parameter is comprised of three equally weighted multivariate Gaussian distributions i.e. $\bm m \sim  (\mathcal{N}(\mu_1, \Sigma_1) + \mathcal{N}(\mu_2, \Sigma_2) +\mathcal{N}(\mu_3, \Sigma_3) )/3$ where
 \begin{equation}
\begin{array}{lll}
\bm \mu_1 = \bm 0_{10 \times 1}, & \bm \mu_2 = 4 \times \bm 1_{10 \times 1}, & \bm \mu_3 = [6 \times \bm 1_{1 \times 5}, -\bm 1_{1 \times 5}]^T \\
\bm \Sigma_1 = [0.11 : 0.01 : 0.2]^T & \bm \Sigma_2 = [0.2 : -0.01 : 0.11]^T & \bm \Sigma_3 = 0.1 \times \bm 1_{10 \times 1}.
\end{array}
\end{equation}
The computation of KL basis vectors is similar to the previous case; the only difference is that the analytical results of the eigenvalues and eigenvectors of the exponential kernel are associated with a three dimensional domain, i.e. $D=[0,1]^3$~\cite{Teckentrup2015}. We also apply the same transformation cf. Equation~\eqref{bss_fnc} to ensure that the inversion of the forward map results in unique solutions. 
In this example we also consider larger number of training data, i.e. $n_{data}=300$. The test dataset is also generated with $1500$ samples. The results associated with the inversion of the forward model are shown in Figure~\ref{fig_3D_opt}. Again, it is observed that the challenging trimodal distribution can be accurately identified in the first step of the procedure. 
The NNK in this example is slightly larger; in the second layer we consider $18$ nodes instead of $12$ in the previous example. The quality of training prediction and testing predictions with randomly distributed prior samples and augmented training prior samples is similar to the previous example. Since there are three modes covering both dimensions of $m_1 - m_{10}$, the testing dataset for the randomly distributed prior samples shows scatter that cover both dimension. In the previous example, trained NNK produced an almost line that was going through three modes.  

\begin{figure}[!h]
\centering
\includegraphics[width=2.0in]{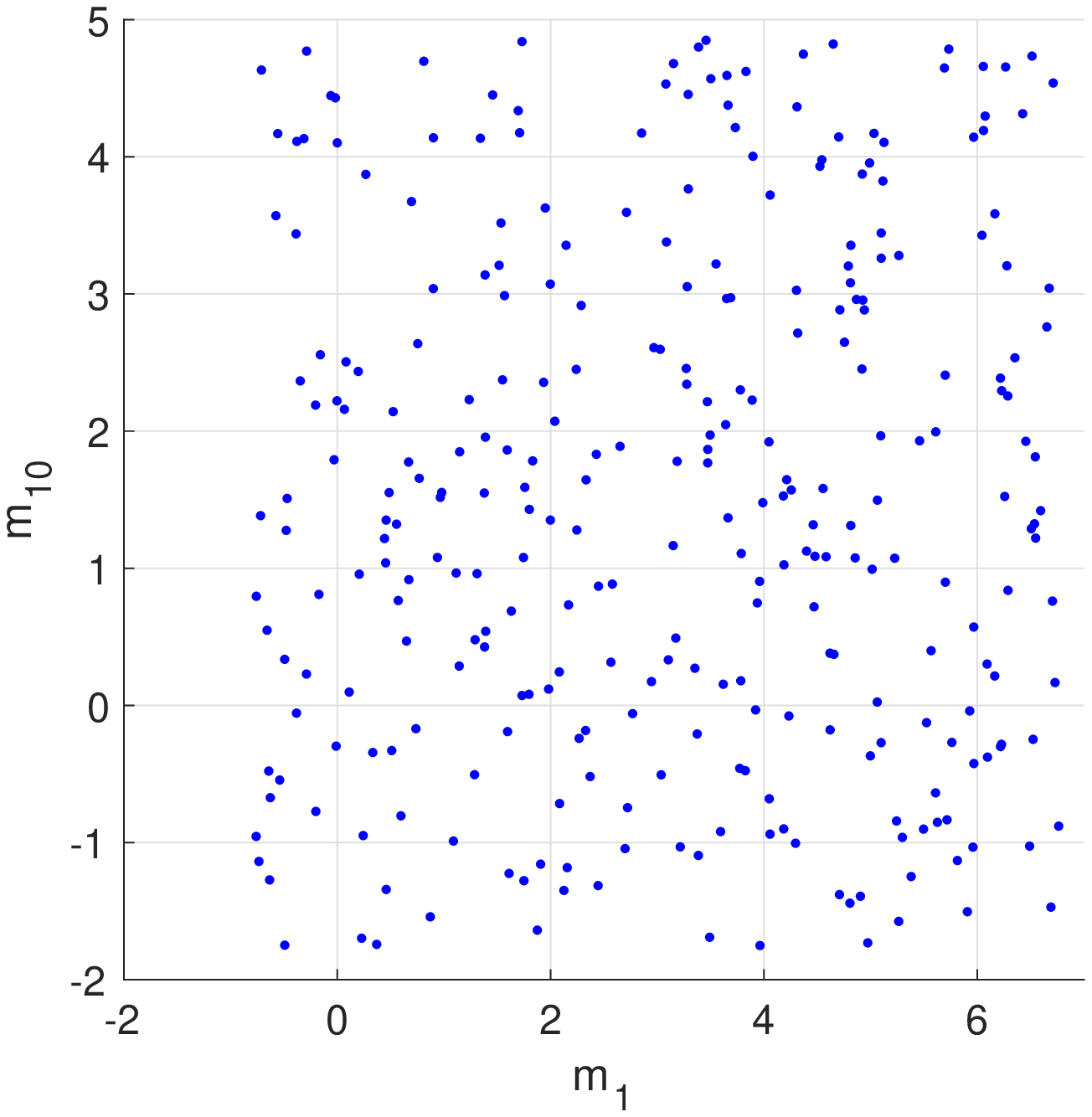}
\includegraphics[width=2.0in]{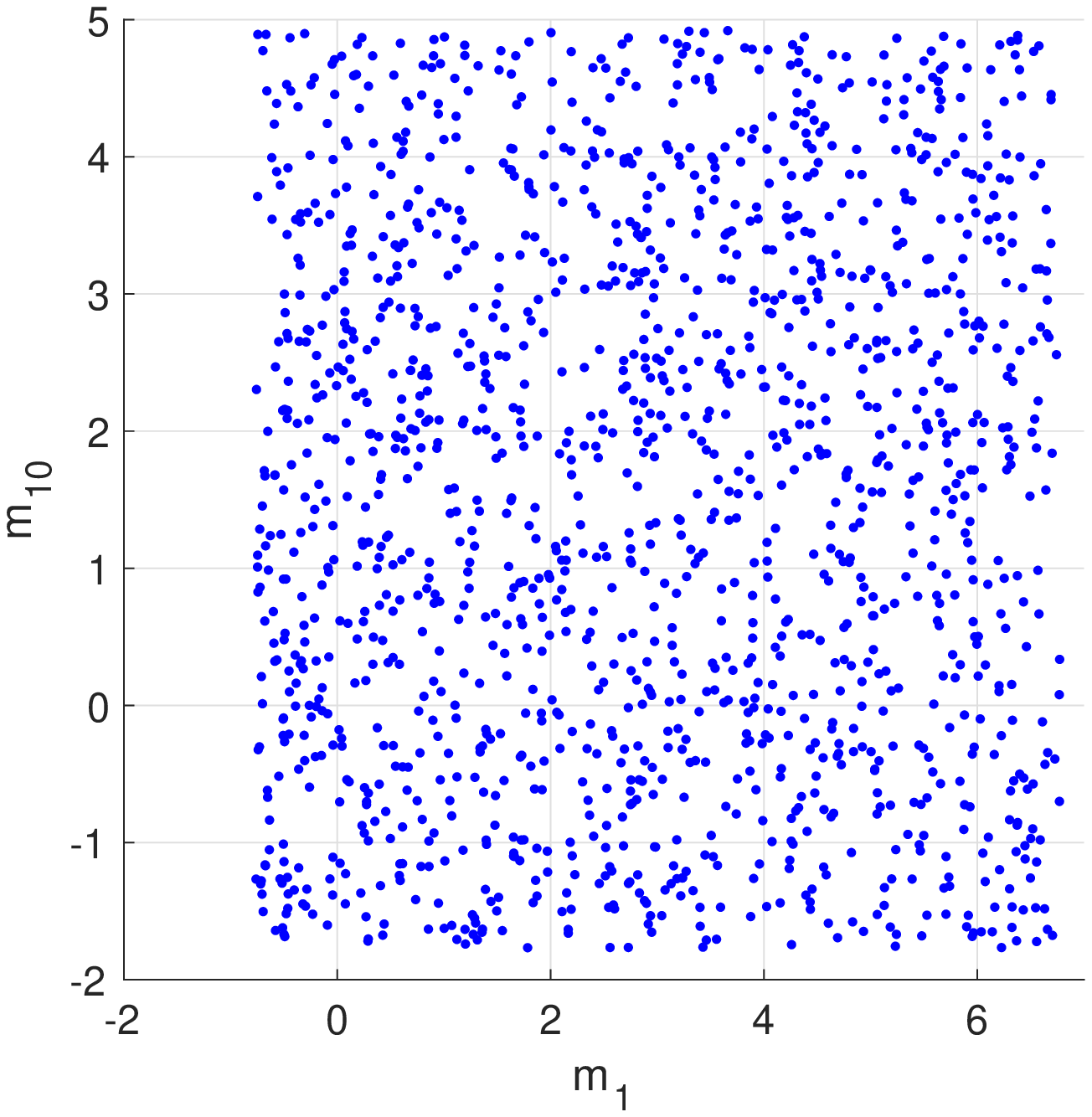}
\includegraphics[width=2.0in]{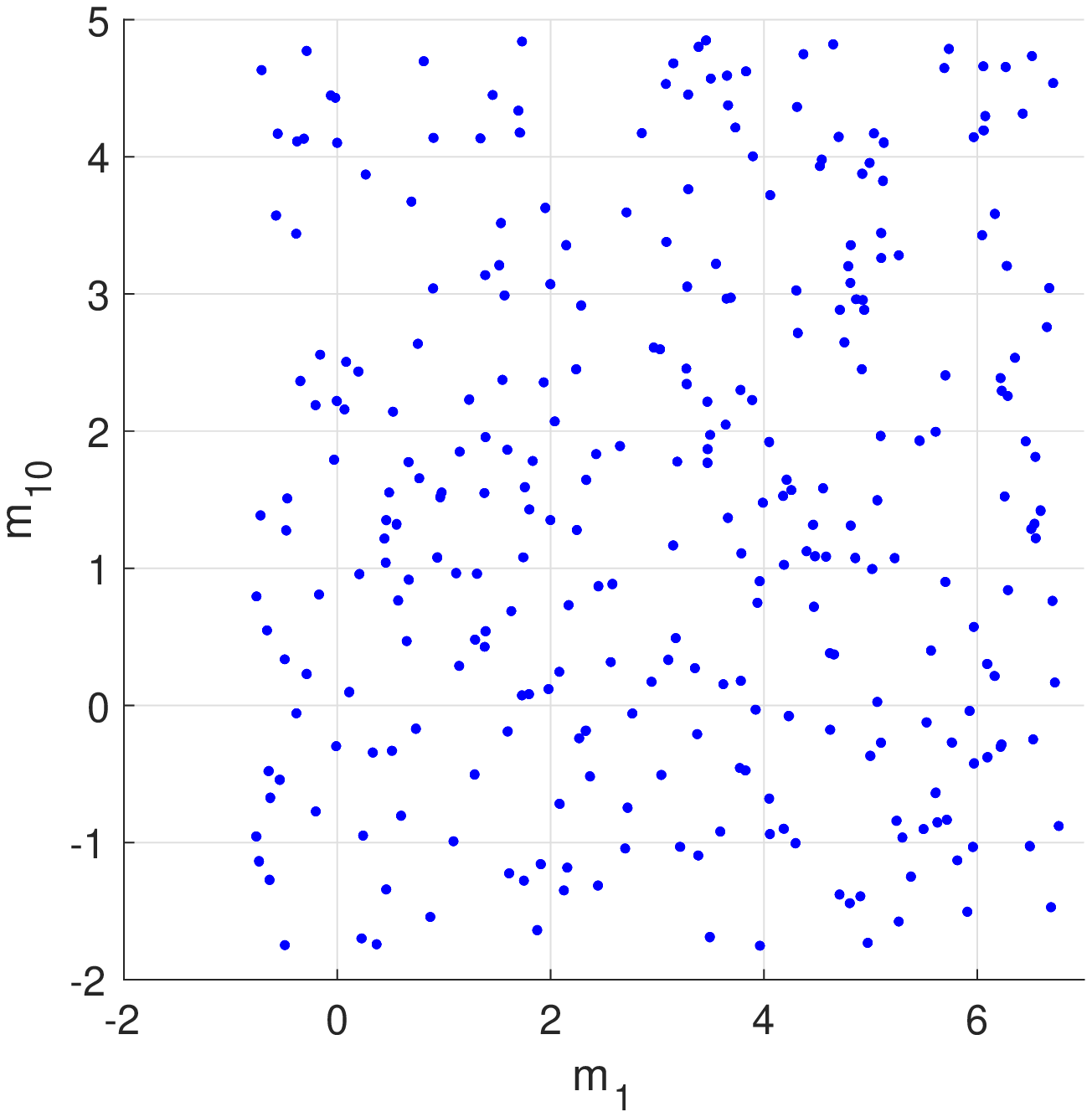}\\
\includegraphics[width=2.0in]{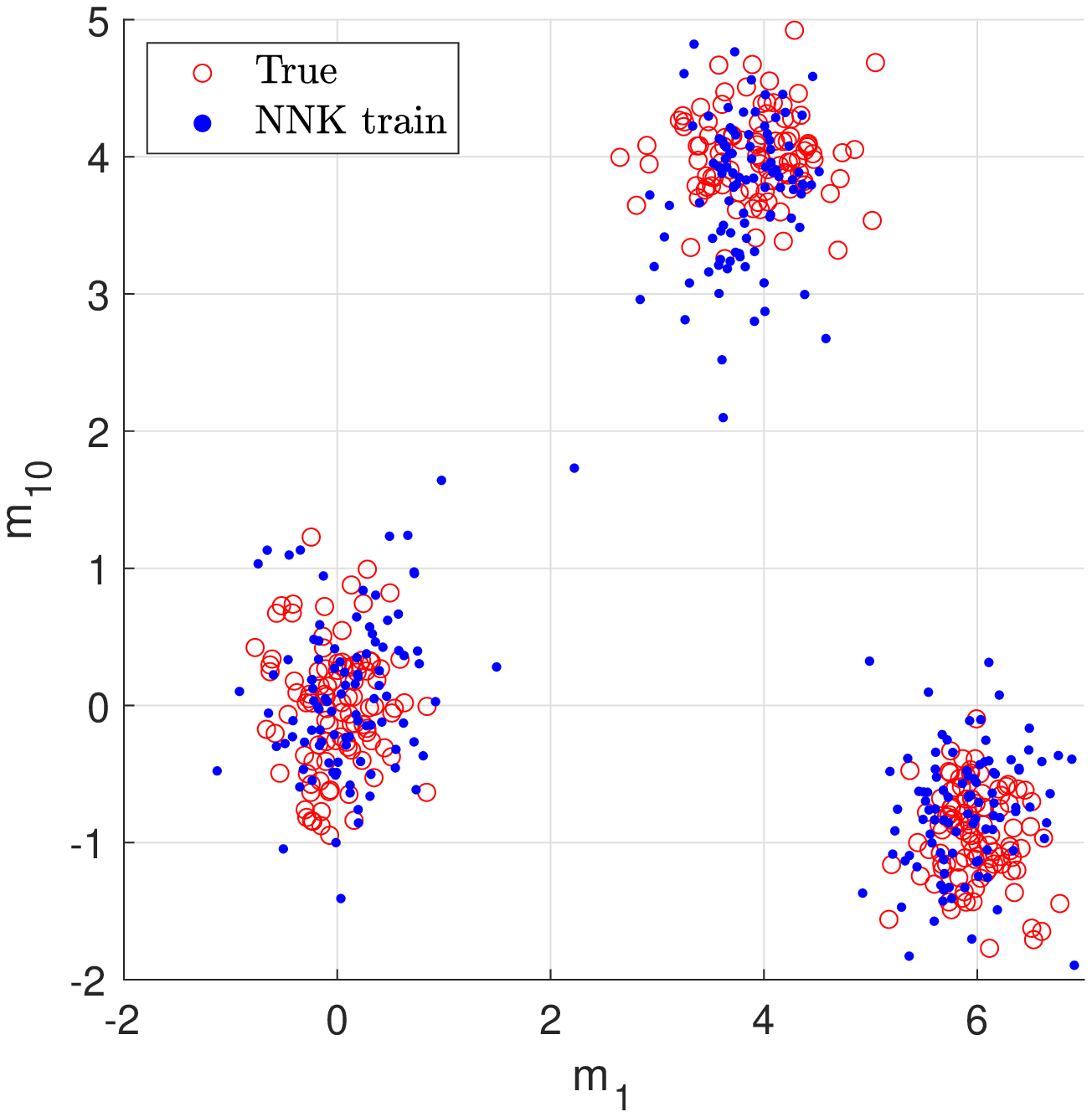}
\includegraphics[width=2.0in]{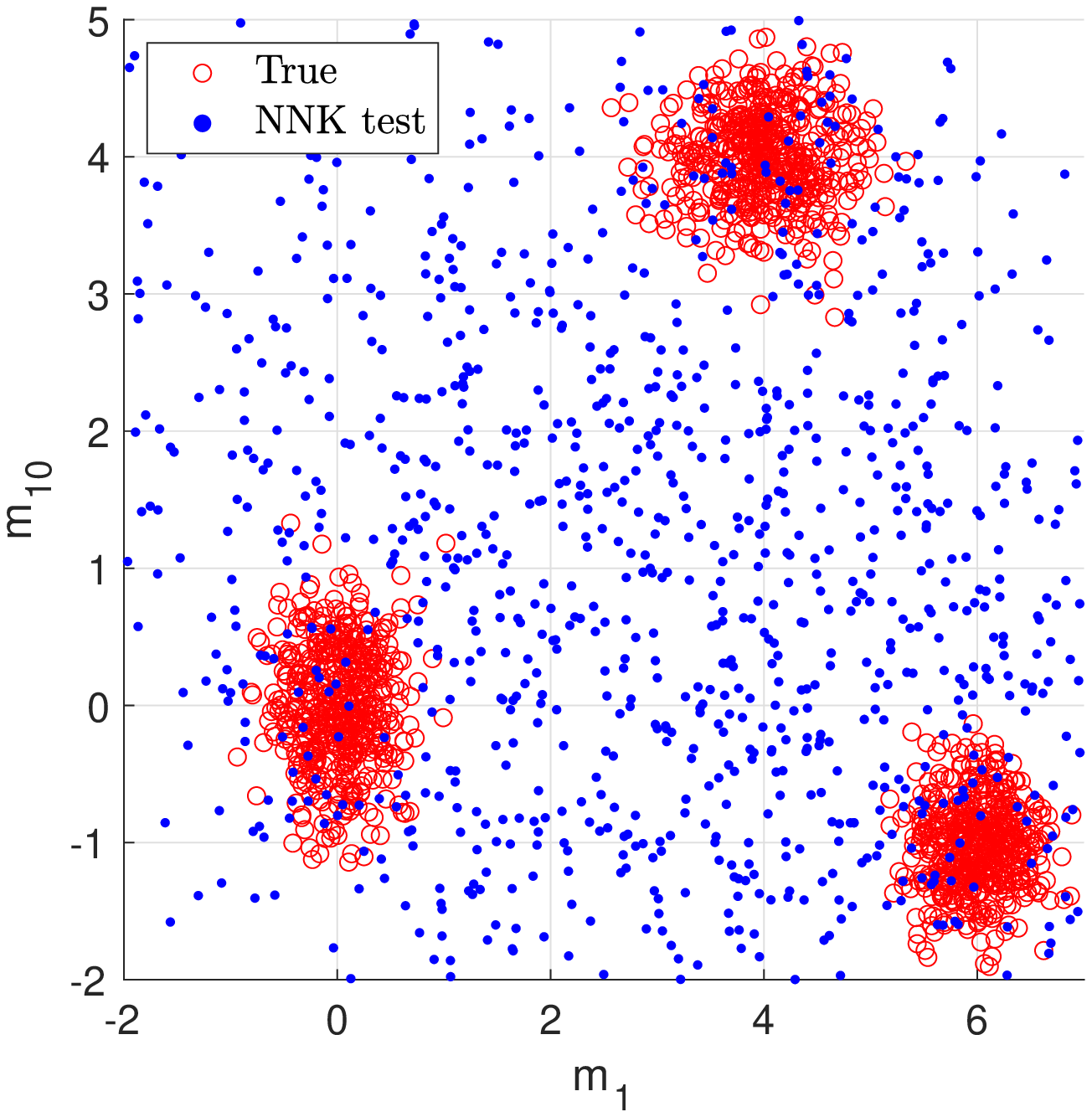}
\includegraphics[width=2.0in]{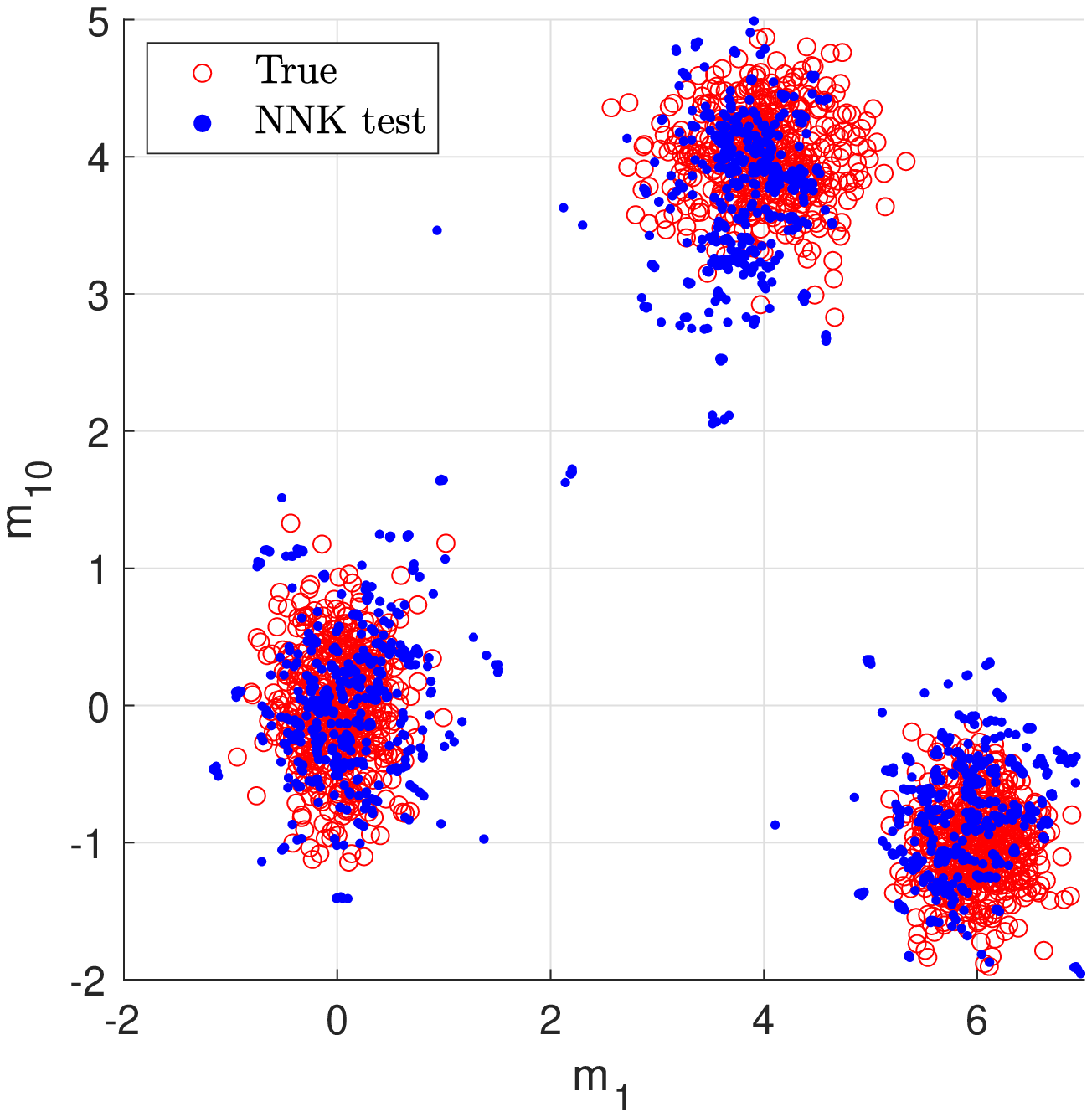}
\caption{\small{Inference in $10$ dimensions and trimodal uncertainty associated with the 3D topology optimization problem. Prior samples: training (top-left), randomly distributed samples for testing (top-middle), augmented training prior samples (top-right). Posterior samples, i.e. the prediction of the NNK map: training (bottom-left), prediction with randomly distributed prior samples (bottom-middle) and prediction with augmented training prior samples (bottom-right).}}\label{fig_2_2}
\end{figure}
We also investigate the predictions numerically on degrees of freedom that are almost in the middle line of the cylinder i.e. those nodes that admit $0 \leq x \leq 4,~-0.05 \leq y \leq 0.05, ~-0.05 \leq z \leq 0.05$. As observed before, the prediction with randomly distributed prior samples exhibits some difference with the exact solution obtained with Monte Carlo analysis but the one with augmented training samples is very similar to the true solution on this very challenging trimodal distribution. 
\begin{figure}[!h]
\centering
\includegraphics[width=2.0in]{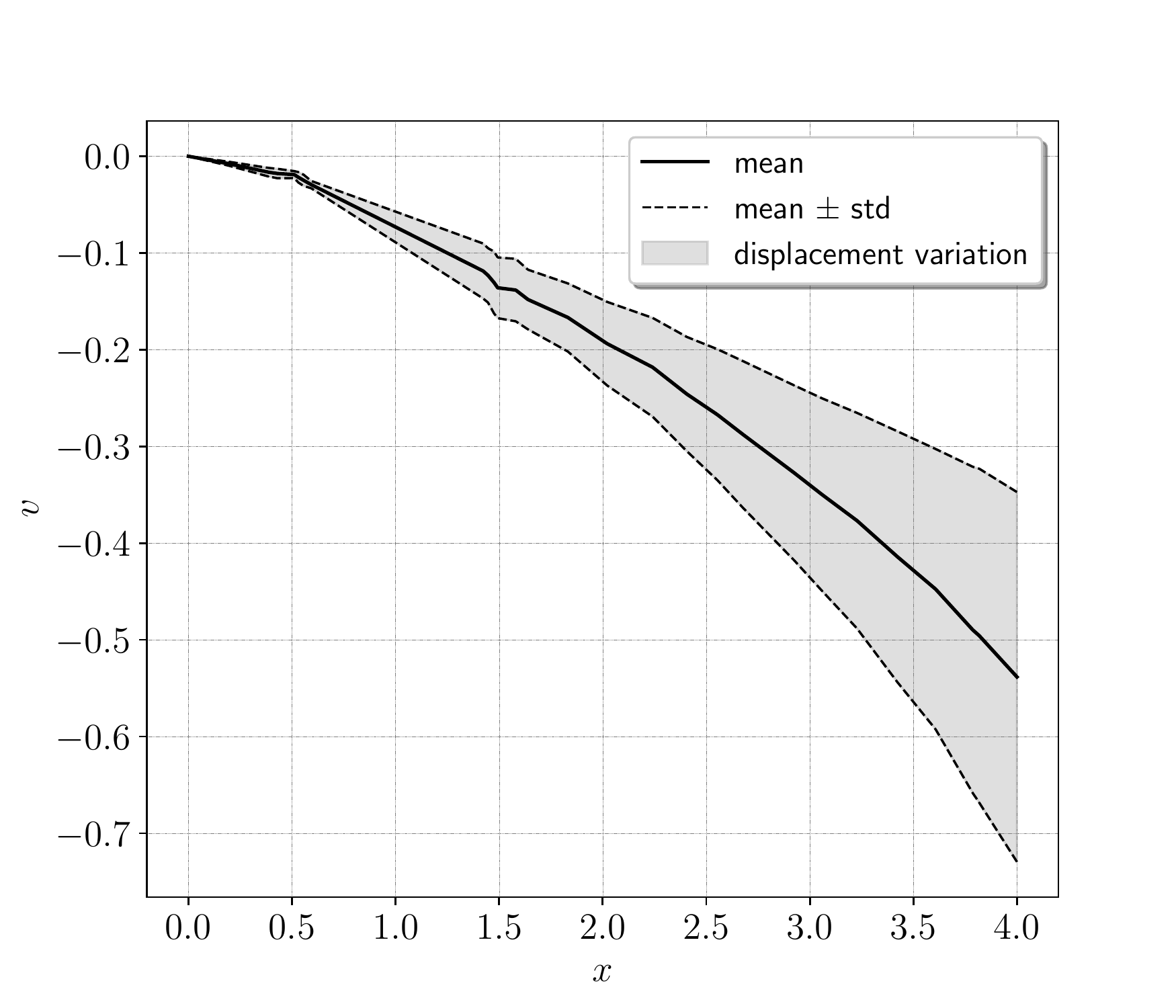}
\includegraphics[width=2.0in]{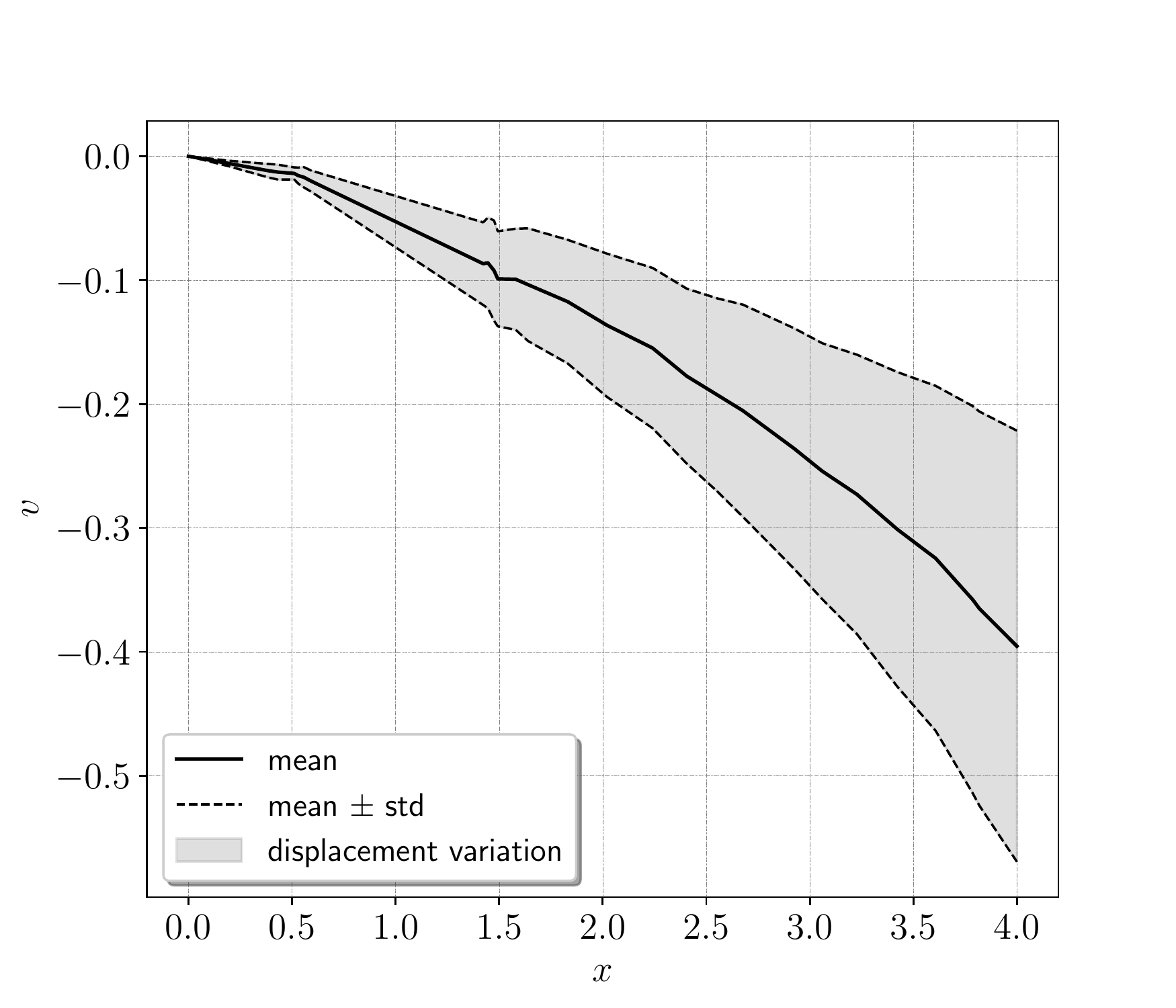}
\includegraphics[width=2.0in]{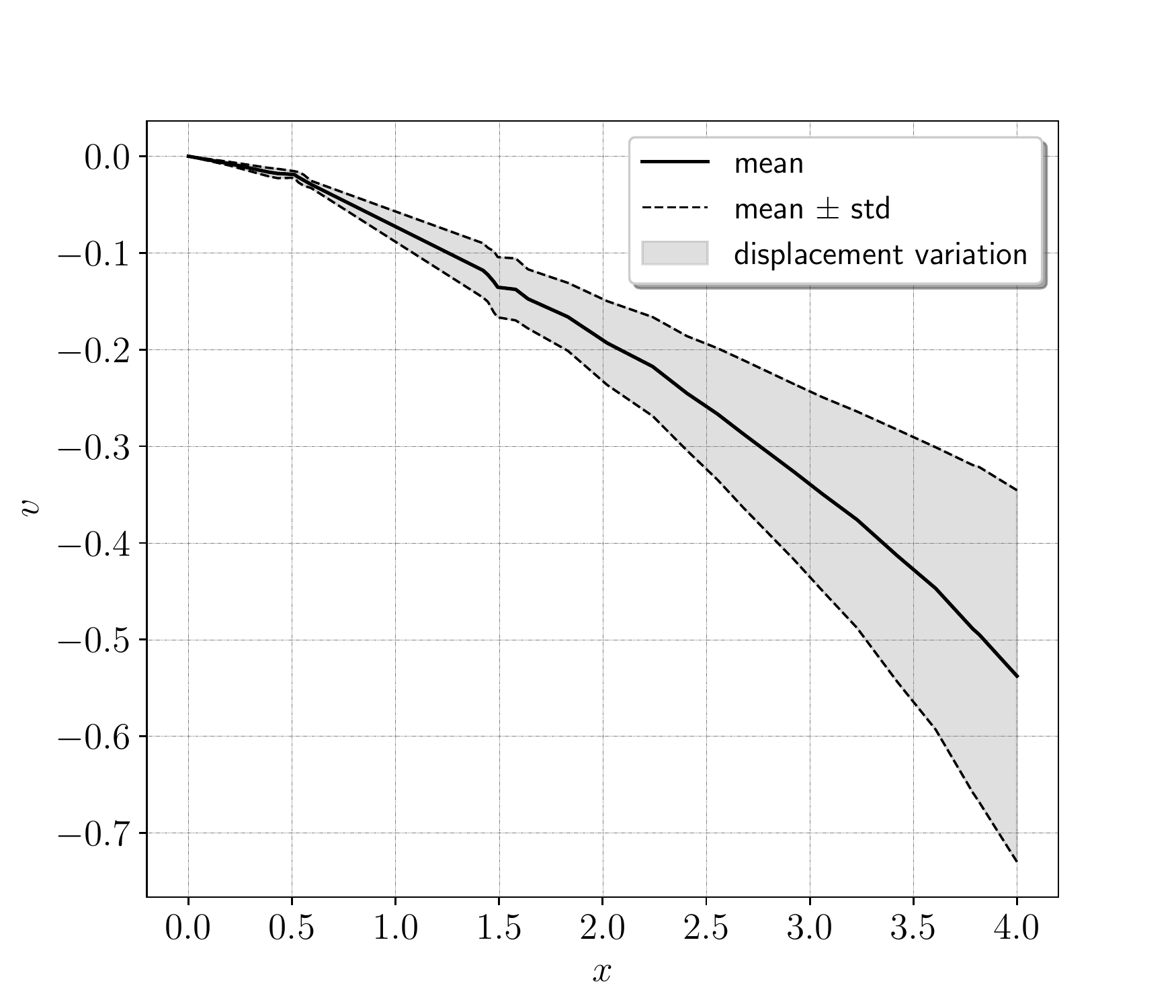}
\caption{\small{Statistical study on displacement in $y$ direction using Monte Carlo samples of the latent variables, i.e. true solution (left), posterior samples from the NNK map with randomly distributed prior samples (middle), posterior samples from the NNK map with augmented training prior samples (right).}}\label{fig_2_2}
\end{figure}

\section{Conclusions}\label{Sec5}
We present an inference procedure for identifying underlying uncertain parameters of general simulation models. At its heart, our approach involves inversion of the simulation model using a Newton-Raphson method. This inversion yields a limited number of true samples of the underlying parameter when the forward model is one-to-one and noise-free, or possibly close to true samples otherwise. The optimized samples are used for training of a specialized neural network, neural network kernel (NNK) which is subsequently used for generating large number of samples of the underlying parameter. The performance of the proposed procedure is compared with several standard statistical approaches via a simple numerical example. The following ideas can be pursued as directions for future research on the problem in this paper. 1) \emph{Development of a black-box probabilistic surrogate}:  The high level motivation behind this paper is to devise a procedure to replace Bayesian neural networks. The success of this approach relies on the knowledge of forward model and its sensitivity. In cases when only the information on $\bm x, \bm y$ is given, one can perform similar sample-wise training of a neural net which results in an ensemble of network hyperparameters. Proper regularization of these identified hyperparameters, e.g. minimizing their variance, may yield accurate probabilistic black-box surrogates outperforming standard BNNs. Such approaches require fast neural network computations/training which we discuss in the next remark.  
2) \emph{Large scale forward/inverse computations with fast linear algebra procedures}: The  approach in this paper depends on finite element computations as well as neural network training based on a Newton-Raphson method which both involve linear solves in the form of $\bm A\bm x = \bm b$. Developing fast linear solvers which exploit the sparsity or low-rankness of the stiffness or Jacobian matrices for similar problems is one of our main future research directions. In a similar vein, we plan to extend NNK in this paper to solution of partial differential equations, specifically linear elasticity problems, and subsequently use the solver for inverse problems and topology optimization. 3) \emph{Design optimization of probabilistic hyper-elastic structures}: Having real observations from a hyper-elastic material, different hyper-elastic models can be identified by incorporating nonlinear finite element analysis into the proposed framework~\cite{GOENEZEN20111406,FRANCK2016215,Gokhale_2008}. The identified models can then be further used for the task of design optimization with realistic probabilistic nonlinear models. 

\textbf{Acknowledgements}\\
This research was sponsored by ARL under Cooperative Agreement Number W911NF-12-2-0023. The views and conclusions contained in this document are those of the authors and should not be interpreted as representing the official policies, either expressed or implied, of ARL or the U.S. government. The U.S. government is authorized to reproduce and distribute reprints for government purposes notwithstanding any copyright notation herein. The first and third authors are partially supported by AFOSR FA9550-20-1-0338. The third author is partially supported by DARPA EQUiPS N660011524053.
\appendix
\section{Computing the Jacobian}\label{comp_J}
In cases where the number of output variables is small and the number of optimization variables are large, the adjoint sensitivity analysis is useful. We briefly explain this approach for computing the derivative of the scalar $\mathcal{P}(\bm m)$ cf. Equation~\eqref{opt_main} which is used in the first numerical example, and subsequently explain how to compute $\partial \mathcal{G}/\partial \bm m$ which involves a large number of outputs and a small number of inputs. The derivative of $\mathcal{P}(\bm m)$ is written as
\begin{align}\label{P_der}
\frac{\partial \mathcal{P}(\bm m)}{\partial \bm m} =\left (-\displaystyle \frac{\partial \mathcal{G}} {\partial \bm m}\right )^T \bm F_{\mathcal{G}} +\bm \Sigma^{-1}_0 (\bm m - \bm m_0).
\end{align}
where we define $\bm F_{\mathcal{G}}= \Gamma^{-1}_{nosie} (\bm y - \mathcal{G}(\bm m))$. Computing the second term is easy; the first term requires computation of the forward model $\mathcal{G}$ sensitivity with respect to the parameter $\bm m$. The solution of the forward problem $\mathcal{G}$ is typically obtained from a linear system of equations (as a result of e.g. finite element analysis) in the form of $\bm K \mathcal{G} = \bm F$ where $\bm K$ and $\bm F$ are stiffness matrix and force vectors. The first term in~\eqref{P_der} can be equivalently written as
\begin{equation}\label{eq_adj1}
\begin{array}{l}
\bm F_{\mathcal{G}}^T \left(\displaystyle\frac{\partial \mathcal{G}}{ \partial \bm m} \right) = \bm F_{\mathcal{G}}^T \left(\displaystyle\frac{\partial \mathcal{G}}{ \partial \bm m} \right) + \bm \Lambda^T \left(\displaystyle \frac{\partial \bm K}{ \partial \bm m }\mathcal{G} + \bm K \displaystyle \frac{\partial \mathcal{G}}{ \partial \bm m} - \displaystyle \frac{\partial \bm F}{ \partial \bm m }\right ) \\
\end{array}
\end{equation}
with the use of adjoint parameter $\bm \Lambda$ where the second term in the parenthesis is zero since $\bm K \mathcal{G} = \bm F$. The last term $\partial \bm F/\partial \bm m$ is also zero since we assume that the force vector $\bm F$ (in PDE/structural analysis) is independent of the parameter $\bm m$. The adjoint solution is obtained by grouping terms associated with the implicit derivative $\partial \mathcal{G}/\partial \bm m$ and solving for the zero coefficient, i.e. $\bm F_{\mathcal{G}} + \bm K \bm \Lambda = 0$ which yields $\bm \Lambda =- \bm K^{-1} \bm F_G$. As a result, Equation~\eqref{P_der} is written as
\begin{align}\label{P_der_adj}
\frac{\partial \mathcal{P}(\bm m)}{\partial \bm m} = (\bm K^{-1} \bm F_G )^T \frac{\partial \bm K}{ \partial \bm m} (\bm K^{-1} \bm F ) +\bm \Sigma^{-1}_0 (\bm m - \bm m_0).
\end{align}
In the above equation, note that if we remove $\bm F_G$ we can compute the derivative term $\partial \mathcal{G}/\partial \bm m$, i.e.
 \begin{align}\label{P_der_adj1}
\frac{\partial \mathcal{G}(\bm x, \bm m)}{\partial \bm m} = -\bm K^{-1}  \frac{\partial \bm K}{ \partial \bm m} (\bm K^{-1} \bm F ) 
\end{align}
The above equation can also be simply obtained by differentiating $\bm K \mathcal{G} = \bm F$ with respect to $\bm m$ which is a direct differentiation approach. The difference between Equations~\eqref{P_der_adj} and~\eqref{P_der_adj1} is that,  for every parameter $m_i \in \bm m$ the result of Equation~\eqref{P_der_adj} is a scalar and that of Equation~\eqref{P_der_adj1} is a vector. Concatanating these column vectors for different $m_i \in \bm m$ yields the Jacobian $\bm J$. 

Computation of $\partial \bm K/\partial \bm m$ for the first example on linear static springs is straightforward as each entry of the two by two matrix can be directly differentiated with respect to $\bm m$. This computation, however, in the examples involving KL expansion and large finite element matrices requires another matrix assembly.  The global stiffness matrix is obtained as
\begin{equation}
\bm K  =\sum_{i=1}^{n_{ele}} \bm K_e^{(i)} E(\bm x^{(i)}, \bm m)
\end{equation}
where $\bm K_e^{(i)}$ and $E(\bm x^{(i)}, \bm m)$ are the stiffness matrix of element $i$ and the elastic modulus at the centroid of element $i$. The derivative of the global stiffness matrix with respect to $\bm m$ is written as
\begin{equation}
\frac{\partial\bm K}{\partial \bm m}  =\sum_{i=1}^{n_{ele}} \bm K_e^{(i)} \frac{\partial E(\bm x^{(i)}, \bm m)}{\partial \bm m}
\end{equation}
where the derivative of the elastic modulus field with respect to each $m_i$ is obtained as ${\partial E(\bm x, \bm m)}/{\partial \bm m_i} =\sqrt{\lambda_i} E_i(\bm x) \exp(\tilde{E})$ cf. Equation~\eqref{KL_numerical}.

\section{Deep Kernels Implementation}~\label{many_layer}
\begin{lstlisting}[language=Matlab]
function [Kscell,backprop,KsWcell,Ksbcell,Wb]=many_layer(Ksi,Wnet,bnet,indzero)
nlayer=length(bnet);
Kscell=cell(1,nlayer);Ksinpcell=cell(1,nlayer);
KsWcell=cell(1,nlayer);Ksbcell=cell(1,nlayer);
for i=1:nlayer
    flag=(i==nlayer);
    [Kso,Kso_inp,Kso_W,Kso_b]=one_layer(Ksi,Wnet{i},bnet{i},real(flag),indzero);
    Kscell{i}=Kso;Ksinpcell{i}=Kso_inp;
    KsWcell{i}=Kso_W;Ksbcell{i}=Kso_b;
    Ksi=Kso;
end

backprop=cell(1,nlayer-1);
backprop{1}=Ksinpcell{nlayer};
for i=2:nlayer-1
    backprop{i}=multsumcell(backprop{i-1},Ksinpcell{nlayer-i+1});
end
    
Ws=cell(nlayer,1);bs=Ws;
for i=1:nlayer
    Ws{i,1}=Wnet{i}(:);
    bs{i,1}=bnet{i}(:);
end
Wb=cell2mat([Ws;bs]);
    

    



\end{lstlisting}

%\bibliography{references}

\end{document}